\documentclass[fleqn,10pt]{article}
\usepackage{latexsym, graphicx, epsfig, amsmath, amssymb,amsfonts}
\usepackage{natbib,amsthm,version}
\usepackage{amsbsy,bm,multirow,enumerate}
\usepackage[titletoc,page]{appendix}
\usepackage[mathscr]{eucal}
\usepackage{mathtools}
\usepackage{color}
\usepackage[utf8]{inputenc}
\usepackage[english]{babel}
\usepackage{amsthm}
\usepackage{enumerate}

\usepackage{amsfonts}
\usepackage{amsmath,version}
\usepackage{amssymb,graphicx,fancybox,mathrsfs,stmaryrd,relsize}
\usepackage[pdftex,hidelinks]{hyperref}
\usepackage{url}
\usepackage{subfigure}
\usepackage{color}
\usepackage{cases}

\catcode`\@=11 \theoremstyle{plain}
\@addtoreset{equation}{section}

\@addtoreset{figure}{section}
\renewcommand\thefigure{\thesection.\@arabic\c@figure}

\newcommand{\Order}{{\cal{O}}}             

\theoremstyle{proposition}

\newtheorem{theorem}{Theorem}[section]

\newtheorem{remark}{Remark}
\newtheorem{rem}{\bf Remark}[section]

\theoremstyle{definition}

\newcommand{\bs}[1]{\boldsymbol{#1}}

\graphicspath{{./figs/Chemo/} {./figs/GNKG/} {./figs/CHE/}  {./figs/VarCHE/}
{./figs/Chemo/Extra/} {./figs/GNKG/Extra/}
}

\title{
  A Roadmap for  Discretely Energy-Stable Schemes for Dissipative Systems
  Based on a Generalized Auxiliary Variable with Guaranteed Positivity
} 
\author{
  Zhiguo Yang, \
  Suchuan Dong\thanks{Author of correspondence.
    Email: sdong@purdue.edu} \\
  Center for Computational and Applied Mathematics \\
  Department of Mathematics \\
  Purdue University, USA 
 } 

\date{(March 29, 2019)} 
\begin{document}
\maketitle



\begin{abstract}

  We present a framework for devising discretely energy-stable
  schemes for general dissipative systems based on a generalized
  auxiliary variable. The auxiliary variable, a scalar
  number, can be defined in terms of the energy functional
  by a general class of functions, not
  limited to the square root function adopted in previous approaches.
  The current method has another remarkable property:
  the computed values for the generalized auxiliary variable
  are guaranteed to be positive on the discrete level,
  regardless of the time step sizes
  or the external forces. This
  property of guaranteed positivity
  is not available in previous approaches.
  A unified procedure for treating the dissipative governing equations
   and the generalized auxiliary variable
  on the discrete level has been presented.
  The discrete energy stability of the proposed numerical scheme
  and the positivity of the computed auxiliary variable have been proved
  for general dissipative systems.
  The current method, termed gPAV (generalized Positive Auxiliary Variable),
  requires only the solution of
  linear algebraic equations within a time step.
  With appropriate choice of the operator in the algorithm,
  the resultant linear algebraic systems upon discretization involve only
  constant and time-independent coefficient matrices,
  which only need to be computed once and can be pre-computed.
  Several specific dissipative systems are studied in relative detail using
  the gPAV framework. Ample numerical experiments are
  presented to demonstrate the performance of the method,
  and the robustness of the scheme at large time step sizes.

\end{abstract}


\vspace{0.05cm}

\section{Introduction}
\label{sec:intro}


Dissipative systems are of immense interest to
science and engineering. Physical systems encountered
in the real world are dissipative, thanks to the
second law of thermodynamics. In dissipative systems
there exists a storage function that is
bounded from below~\cite{Willems1972}. We will
refer to this function
as the energy in the current work. 
Dissipative systems are
distinguished from general dynamical systems by
the dissipation inequality, which basically states
that the increase in storage of the system over
a time interval cannot exceed the supply to
the system during that interval~\cite{Willems1972,Willems2007}.
The governing partial differential equations (PDE)
describing dissipative systems are typically
nonlinear, and they satisfy a balance equation
for the energy (or entropy) as an embodiment
of the dissipation
inequality~\cite{GrootM1984,Ottinger2005,AndersonMW1998,LowengrubT1998,AbelsGG2012,Dong2018}.


A highly desirable property for numerical algorithms
for dissipative systems is the preservation of the
energy dissipation (or conservation) on
the discrete level.
This not only preserves one important aspect of the
underlying structure of the continuous system~\cite{HairerLW2006},
but more practically also provides a control on
the numerical stability in
actual computer simulations.
The history for such strategies is long and they can be traced to at least
the work of \cite{CourantFL1928} on discrete energy conservation
for finite difference approximations in the 1920s.
While energy-stable schemes for specific domains of science
and engineering have been under intensive studies and
these efforts have borne invaluable fruits, 
the schemes and methods developed usually have only limited
applicability across domains. The energy-stable schemes
for one area are hardly transferable
to a different field, and they can hardly shed light on the development of
such types of schemes in new unexplored domains.
Unified techniques that can be broadly applied to treat different PDEs
from different domains for devising energy-stable schemes are
generally lacking.
The metaphor used in \cite{Iserles2008} (page 139) to compare
the motley collection of PDEs to a hugh unhappy family
(each unhappy in its own way; Tolstoy, ``Anna Karenina'') seems fitting in describing
this situation (see also~\cite{Celledonietal2012}).

Occasionally, certain methods appear  and seem
to be broadly applicable to a wide class of problems spanning different areas.
The average vector field (AVF) method~\cite{Celledonietal2012,QuispelM2008} and
the discrete variational derivative method (DVDM)~\cite{FurihataM2011},
both of which can be traced to the idea of discrete
gradients~\cite{Gonzalez1996,McLachlanQR1999},
are two such examples.
For gradient systems that can be expressed into the form
$
\frac{\partial\bs u}{\partial t} = \bs L\cdot\frac{\delta H}{\delta \bs u},
$
where $\bs L$ is an anti-symmetric or negative semi-definite matrix, 
$\bs u$ is the field variable, $H(\bs u)$ is the energy functional
and $\frac{\delta H}{\delta \bs u}$ denotes the variational derivative,
the AVF and DVDM methods can preserve
the energy conservation (resp.~energy dissipation) discretely.
We refer the reader to e.g.~\cite{Furihata1999,DahlbyO2011,MiyatakeM2014,CaiLW2018,EidnesOR2018} (among others)
for related and variants of these methods.
A potential drawback of these methods is their computational cost.
Because these are fully implicit schemes and
the governing PDEs  are in general nonlinear,
these methods will entail the solution of nonlinear algebraic
equations on the discrete level. Consequently, some
nonlinear algebraic solver (e.g.~Newton type methods) will be required
for computing the field functions, and the associated computational
cost can be substantial.

In the current work we present a framework for devising energy-stable
schemes for general dissipative systems that can potentially
be useful and
applicable to different domains. Our method does not require the governing PDEs
to be in any particular form, as long as they are dissipative (or conserving).
When devising the energy-stable numerical schemes, we are particularly mindful
of the computational cost involved therein.
The resultant energy-stable schemes from our method involve only the solution
of linear algebraic equations when computing the field functions
within a time step, and no nonlinear algebraic solver is needed.
Furthermore, with appropriate choice of the operator in the scheme,
the resultant linear algebraic systems upon discretization can involve only constant
and time-independent coefficient matrices, which only need to be
computed once and can be
pre-computed during pre-processing.
Thanks to these properties, the presented method and the resultant
energy-stable schemes are computationally very competitive and
attractive. In terms of the computational cost the presented method
enjoys a notable advantage when compared with the aforementioned methods.

The key to achieving the above useful properties for general dissipative systems
in the presented method
lies in the introduction of a generalized auxiliary variable.
%
The generalized auxiliary variable introduced here is inspired by
the scalar auxiliary variable (SAV) approach proposed by~\cite{ShenXY2018},
and to a lesser extent, by the invariant energy quadratization (IEQ)
method~\cite{Yang2016}, both of which are devised for gradient flows;
see also e.g.~\cite{ShenX2018,GongZYW2018,ChengS2018,Zhaoetal2018,KouSW2018,LiZW2019,YangLD2019,Yang2019} (among others) for 
extensions and applications of these techniques.
In SAV a scalar-valued auxiliary variable is defined, as
the square root of the shifted potential energy integral.
In IEQ an auxiliary field variable is defined, as the square
root of the shifted potential energy density function.
With these auxiliary variables, energy-stable schemes can be
devised for gradient flows and their discrete energy stability
can be proven in the SAV and IEQ methods.
In both SAV and IEQ, the use of the square root function is critical
to the proof of the discrete energy stability of the resultant
numerical schemes, due to the interesting property that
the square root is the only function form that satisfies the relation
\begin{equation*}
  2f(x)f'(x) = 1.
\end{equation*}
In the current work we will show that the square root function
is not essential
to devising energy-stable schemes. In the generalized auxiliary
variable method developed here, the auxiliary variable
(a scalar number) can be
defined by a rather general class of functions (conditions specifically given
in Section \ref{sec:formulation}) in terms of the energy functional,
which is why the method is termed ``generalized'', and
the resultant numerical schemes can be proven to be discretely
energy stable.

The method presented here is applicable to general dissipative
systems, which is another key difference
from previous auxiliary-variable approaches.
The ability to deal with general dissipative systems
hinges on how the governing PDEs are treated
based on the generalized auxiliary variable and how the generalized auxiliary
variable is numerically treated on the discrete level.
A unified procedure for treating discretely the dissipative governing equations
and the generalized auxiliary variable has been presented.
These numerical treatments have drawn inspirations from the recent developments 
in \cite{LinYD2019,YangD2018} for incompressible Navier-Stokes
equations and for the incompressible two-phase flows,
which are not gradient-type systems.

The generalized auxiliary variable method proposed herein has another
remarkable property: The computed values for the auxiliary
variable are guaranteed to be positive on the discrete level.
Such a property is not available in the SAV (or IEQ) method.
In both SAV and IEQ, as well as in the current method,
the auxiliary variable is computed discretely by solving an associated
dynamic equation, which is derived based on the definition
of the auxiliary variable in terms of the square root function
in SAV and IEQ or a general function in the current method.
The auxiliary variable physically should be positive according to
its definition. However, this positivity property is in general
not guaranteed in the computed values for
the auxiliary variable, because they are obtained by numerically solving
a differential equation. Indeed, in numerical experiments
we have observed negative values for the computed auxiliary variable
using the previous methods, especially at large time step sizes.
With the current method, on the other hand, we can prove that the computed values 
for the generalized auxiliary variable are guaranteed to be
positive, regardless of the time step sizes or the external forces.
The guaranteed positivity of the auxiliary variable
in the current method is intimately related to and
is critical to the proof of  discrete energy stability
of the proposed numerical schemes.

Because of these crucial properties, we will refer to
the framework  proposed herein as ``gPAV'',
which stands for the generalized Positive Auxiliary Variable
method.

In this paper we consider general dissipative systems
and outline the gPAV procedure for devising discretely energy-stable
schemes.
The discrete energy stability of
the proposed numerical scheme and the positivity property of the computed
auxiliary variable will be proven
for general dissipative systems.
As already mentioned, the gPAV method
requires only the solution of linear algebraic equations within
a time step, and with appropriate choice of
the operator in the algorithm, the resultant linear
algebraic systems involve only constant and time-independent
coefficient matrices that can be pre-computed.
We demonstrate the gPAV procedure by
looking into three specific dissipative systems:
a chemo-repulsion model~\cite{Gonzalez2019},
the Cahn-Hilliard equation~\cite{CahnH1958} with
constant and variable mobility, and the nonlinear
Klein-Gordon equation~\cite{Strauss1978}.
Ample numerical experiments are provided for each system to demonstrate
the performance of the algorithm and the effects of the parameters.


The current work contains several new aspects:
(i) the framework for developing discretely energy-stable schemes for general
dissipative systems;
(ii) the generalized auxiliary variable introduced herein; and
(iii) the guaranteed positivity of the computed auxiliary
variable on the discrete level.
Some other aspects, such as the generalization of the numerical
algorithm as discussed in Remarks~\ref{rem:rem_theta} and~\ref{rem:rem_partial},
are also  potentially useful to other researchers and the community.


The remainder of this paper is structured as follows.
In Section \ref{sec:adjstab} we introduce a generalized
auxiliary variable and present the gPAV framework
for devising discretely energy-stable 
schemes for general dissipative systems. The discrete energy
stability of the presented algorithm and
the positivity of the computed auxiliary variable 
will be proven. The solution algorithm for implementing
the proposed energy-stable scheme will be presented.
An alternative formulation for the energy-stable scheme will also
be discussed in this section.
Then in the three subsequent sections
(Sections \ref{sec:chemo}--\ref{sec:kg})
we apply the gPAV framework to  three
specific dissipative systems (a chemo-repulsion model,
Cahn-Hilliard equation with constant and variable mobility,
and Klein-Gordon equation). Ample numerical experiments
are provided to demonstrate the performance of the method
for each system,
and numerical results with large time step sizes
are presented to show 
the robustness of the proposed scheme.
Section \ref{sec:summary} concludes the discussions with
some closing remarks.
In Appendix A we provide a method for approximating the variables
for the first time step, which guarantees the positivity of
the computed auxiliary variable to start off.
This startup procedure is important for the proof of
discrete energy stability of the presented numerical scheme.


\section{The gPAV Framework for Energy-Stable Schemes for Dissipative Systems}
\label{sec:adjstab}

Consider a domain $\Omega$ in two or three dimensions and a dissipative system on this domain,
whose dynamics is described by,
\begin{equation}
  \label{eq:generalsys}
\frac{\partial \bs u}{\partial t}=\bs F(\bs u)+\bs f(\bs x,t)
\end{equation}
where $\bs x$ and $t$ denote the spatial coordinate and time, 
$\bs u(\bs x,t)$ denotes the state variables of the system and
can be a scalar- or vector-valued field function,
and $\bs f(\bs x,t)$ is an external source term (hereafter referred to
as the external force).
$\bs F(\bs u)$ is an operator that gives rise to
the dissipative dynamics of the system and can be nonlinear in general.
Equation \eqref{eq:generalsys} is supplemented by the
boundary condition
\begin{equation} \label{eq:bc}
\bs B(\bs u) = \bs f_b, \quad \text{on} \ \Gamma
\end{equation}
where $\Gamma$ denotes the domain boundary, $\bs f_b$ is an
external source term on the boundary, which will be referred to
as the external boundary force hereafter, and
$\bs B$ is assumed to be a linear operator for the sake of simplicity.
The initial condition is
\begin{equation}
  \bs u(\bs x, t=0) = \bs u_{in}(\bs x)
  \label{eq:ic}
\end{equation}
where $\bs u_{in}(\bs x)$ is the initial distribution of the state variable.

Because the system is dissipative, there exists a storage function
that is bounded from below~\cite{Willems1972},
which hereafter will be referred to as the energy,
\begin{equation}\label{eq:Energy1}
E_{tot}(t)=E_{tot}[\bs u]=\int_{\Omega}e(\bs u) d\Omega,
\end{equation}
where $e(\bs u)$ is the energy density function.
The evolution of the energy is described by
\begin{equation}\label{eq:energylaw1}
  \frac{dE_{tot}}{dt}
  =\int_{\Omega}\frac{\partial e}{\partial \bs u}\cdot \frac{\partial \bs u}{\partial t}d \Omega
  =\int_{\Omega}\frac{\partial e}{\partial  \bs u}\cdot \left[ \bs F(\bs u)+\bs f\right]    d \Omega, 
\end{equation}
where we have used equation \eqref{eq:generalsys}. 
With integration by part, the right-hand-side (RHS) of equation \eqref{eq:energylaw1} can be transformed into 
\begin{equation}\label{eq:VB}
  \int_{\Omega}\frac{\partial e}{\partial \bs u}\cdot [\bs F(\bs u)+\bs f] d \Omega
  =-\int_{\Omega}V(\bs u)d\Omega+\int_{\Omega}V_s(\bs f,\bs u)d\Omega+ \int_{\Gamma}B_s(\bs f_b,\bs u) d\Gamma,
\end{equation}
where
$V_s(\bs f, \bs u)=\frac{\partial e}{\partial \bs u}\cdot\mathbf{f}$
denotes the volume terms involving the external force $\bs f$,
which satisfies the property 
\begin{equation}
  V_s(\bs f, \bs u) = 0, \quad \text{if} \ \bs f=0.
  \label{eq:vs_cond}
\end{equation}
The rest of the volume terms are denoted by $-V(\bs u)$, not involving $\bs f$.
$B_s(\bs f_b, \bs u)$ denotes the boundary terms,
which may involve the boundary source term ($\bs f_b$) through the boundary
conditions. 

Substituting equation \eqref{eq:VB} into equation \eqref{eq:energylaw1},
we arrive at the following energy balance equation for the  system,
\begin{equation}
  \frac{dE_{tot}}{dt}=-\int_{\Omega}V(\bs u)d\Omega
  +\int_{\Omega}V_s(\bs f,\bs u) d\Omega+ \int_{\Gamma}B_s(\bs f_b,\bs u) d\Gamma.
\end{equation}
We assume that the boundary conditions \eqref{eq:bc}
satisfy the following property,
\begin{equation}
  B_s(\bs f_b, \bs u) = 0 \ \ \text{if} \ \bs f_b=0,
  \quad \text{on} \ \Gamma.
  \label{eq:bc_cond}
\end{equation}
The dissipative nature of the system ensures that
$\frac{dE_{tot}}{dt}\leqslant 0$ in the absence of the external forces (i.e.~$\bs f=0$ and $\bs f_b=0$).
Because the domain $\Omega$ can be arbitrary, it follows that
$V(\bs u)$ must be non-negative, i.e.
\begin{equation}\label{eq:Vcond}
  V(\bs u) \geqslant 0.
\end{equation}


\subsection{Reformulated Equivalent System}
\label{sec:formulation}

To facilitate energy-stable numerical approximations of the system \eqref{eq:generalsys},
we define a shifted energy of the following form 
\begin{equation}\label{eq:shitE}
E(t)=E[\bs u]=\int_{\Omega}e(\bs u)d\Omega+C_0,
\end{equation}
where $C_0$ is a chosen energy constant such that $E(t)>0$
for $0\leqslant t \leqslant T$, and
$T$ is the time interval  on which the computation is to be carried out.
Note that for a physical system the energy is bounded from below, and
thus $C_0$ can always be found.

Let $\mathscr{F}$ denote a one-to-one increasing differentiable function,
with its inverse $\mathscr{F}^{-1}=\mathscr{G}$, satisfying the property
\begin{equation}
  \left\{
  \begin{split}
    &
    \mathscr{F}(\chi)> 0, \quad \text{for} \ \chi > 0; \\
    &
    \mathscr{G}(\chi)> 0, \quad \text{for} \ \chi > 0.
  \end{split}
  \right.
  \label{eq:FG_property}
\end{equation}
We define a scalar variable $R(t)$ by
\begin{subequations}
  \begin{align}
    &
    \label{eq:FG}
    R(t)=\mathscr{G}(E), \\
    &
    E(t)=\mathscr{F}(R), \label{eq:FG_2}
  \end{align}
\end{subequations}
where $E(t)$ is the shifted energy given by \eqref{eq:shitE}.
$R(t)$ then satisfies the following evolution equation,
\begin{equation}\label{eq:dynR}
\mathscr{F}'(R)\frac{dR}{dt}=\int_{\Omega} e'(\bs u)\cdot \frac{\partial \bs u}{\partial t}d \Omega
\end{equation}
which is obtained by 
taking the time derivative of equation~\eqref{eq:FG_2} and
using equation \eqref{eq:shitE}.

\begin{remark} \label{remark_VF}
  
The choice for $\mathscr{F}$ and $\mathscr{G}$ is rather general. Some examples are,
\begin{equation}\label{eq:Fchoice}
\mathscr{F}(\chi)=\chi^{m},\quad \mathscr{G}(\chi)=\chi^{1/m},\quad m\in \mathcal{Z}^{+}=\{1,2,3,... \};
\end{equation}
or
\begin{equation}\label{eq:Fchoice2}
  \mathscr{F}(\chi)=\frac{e_0}{2}\ln \Big( \frac{\kappa_0+\chi}{\kappa_0-\chi}  \Big), \quad
\mathscr{G}(\chi)=\kappa_0 \tanh\left(\frac{\chi}{e_0}\right),
\end{equation}
where $\kappa_0$ and $e_0$ are positive constants. 
It is important to notice that a function like
$\mathscr{F}(\chi)=\chi^{2m+1}$ (with an integer $m\geqslant 0$)
or $\mathscr{F}(\chi)=\ln (1+\chi)$ does not automatically guarantee
that $\mathscr{F}(\chi)>0$ with arbitrary $\chi$. 
However, if one can ensure that the argument satisfies $\chi>0$,
the property $\mathscr{F}(\chi)>0$ can be guaranteed with
such choices of functions when defining $R(t)$.
This point is critical in the subsequent development of
the numerical algorithm.

\end{remark}

Noting that $\frac{\mathscr{F}(R)}{E}=1,$
we rewrite equation \eqref{eq:generalsys} into an equivalent form
\begin{equation}
  \frac{\partial \bs u}{\partial t}=\bs F_L(\bs u)
  +\frac{\mathscr{F}(R)}{E}\Big(\bs F(\bs u)-\bs F_L(\bs u) \Big)+\bs f, \label{eq:generaleq1}
\end{equation}
where $\bs F_L(\bs u)$ is a chosen linear operator about $\bs u$.
$\bs F_L(\bs u)$ should be of the same spatial order
as $\bs F(\bs u)$.
For improved accuracy $\bs F_L(\bs u)$ should be
an approximation of $\bs F(\bs u)$ in some way, such as the linear
component of $\bs F(\bs u)$ or a linearized approximation of
$\bs F(\bs u)$. For improved numerical efficiency $\bs F_L(\bs u)$
should be easy to compute and implement.
\begin{rem}

  $\bs F(\bs u)$ often consists of linear components and
  nonlinear components for many systems,
  and oftentimes one can choose the linear components
  as the $\bs F_L$ operator. One can also add/subtract certain
  linear operators, and treat one part freely and the other part
  together with $\frac{\mathscr{F}(R)}{E}$ as in equation \eqref{eq:generaleq1}.
  By choosing an $\bs F_L$ operator that involves only time-independent
  (or constant) coefficients, the resultant method will become computationally
  very efficient, because the coefficient matrices for the linear algebraic systems
  upon discretization will be time-independent and therefore
  can be pre-computed  when
  solving the field variables.
  This point will become clearer from later
  discussions.
  
\end{rem}

We reformulate equation \eqref{eq:dynR} as follows,
\begin{equation}
  \begin{split}
    \mathscr{F}'(R)\frac{dR}{dt} =&
    \int_{\Omega} e'(\bs u)\cdot \frac{\partial \bs u}{\partial t}d \Omega
    + \left[\frac{\mathscr{F}(R)}{E} - 1  \right]\int_{\Omega} e'(\bs u)\cdot
    \left[ \bs F_L(\bs u) + \bs f  \right]d\Omega \\
    & + \frac{\mathscr{F}(R)}{E} \left( \int_{\Omega}e'(\bs u)\cdot
      \left[\bs F(\bs u) - \bs F_L(\bs u) \right]d\Omega
      - \int_{\Omega}e'(\bs u)\cdot\left[\bs F(\bs u) - \bs F_L(\bs u) \right] d\Omega
      \right) \\
    & + \left[1 - \frac{\mathscr{F}(R)}{E} \right]\left| \int_{\Omega}V_s(\bs f,\bs u)d\Omega
      + \int_{\Gamma} B_s(\bs f_b,\bs u) d\Gamma  \right| \\
      =& \int_{\Omega} e'(\bs u)\cdot \frac{\partial \bs u}{\partial t}d \Omega
      - \int_{\Omega}e'(\bs u)  \cdot \left(
        \bs F_L(\bs u) + \frac{\mathscr{F}(R)}{E}\left[\bs F(\bs u) - \bs F_L(\bs u) \right] + \bs f
        \right)d\Omega \\
      &+ \frac{\mathscr{F}(R)}{E} \int_{\Omega} \frac{\partial e}{\partial \bs u}
        \cdot \left[ \bs F(\bs u) + \bs f  \right] d\Omega
        + \left[1 - \frac{\mathscr{F}(R)}{E} \right]\left| \int_{\Omega}V_s(\bs f,\bs u)d\Omega
      + \int_{\Gamma} B_s(\bs f_b,\bs u) d\Gamma  \right|
  \end{split}
  \label{eq:R_equ_trans}
\end{equation}
where it can be noted that a number of zero terms have been incorporated.
In the above equation $\big|(\cdot) \big|$ denotes the absolute value
of $(\cdot)$.
In light of \eqref{eq:VB}, we transform equation \eqref{eq:R_equ_trans}
into the final reformulated equivalent form
\begin{equation}
  \begin{split}
    \mathscr{F}'(R)\frac{dR}{dt} =&
     \int_{\Omega} e'(\bs u)\cdot \frac{\partial \bs u}{\partial t}d \Omega
      - \int_{\Omega}e'(\bs u)  \cdot \left(
        \bs F_L(\bs u) + \frac{\mathscr{F}(R)}{E}\left[\bs F(\bs u) - \bs F_L(\bs u) \right] + \bs f
        \right)d\Omega \\
        &+ \frac{\mathscr{F}(R)}{E} \left[ -\int_{\Omega} V(\bs u) d\Omega
          + \int_{\Omega} V_s(\bs f,\bs u) d\Omega
          + \int_{\Gamma} B_s(\bs f_b, \bs u) d\Gamma
            \right] 
        \\
        &+ \left[1 - \frac{\mathscr{F}(R)}{E} \right]\left| \int_{\Omega}V_s(\bs f,\bs u)d\Omega
      + \int_{\Gamma} B_s(\bs f_b,\bs u) d\Gamma  \right|.
  \end{split}
  \label{eq:generaleq2}
\end{equation}


The reformulated system consists of equations \eqref{eq:generaleq1}
and \eqref{eq:generaleq2}, the boundary conditions \eqref{eq:bc},
the initial condition \eqref{eq:ic} for $\bs u$, and
the following initial condition for $R(t)$,
\begin{equation}
  R(0) = \mathscr{G}(E(0)), \quad
  \text{where} \
  E(0) = \int_{\Omega} e(\bs u_{in}) d\Omega + C_0.
  \label{eq:ic_R}
\end{equation}
In the reformulated system, the dynamic variables are $\bs u$
and $R(t)$, which are coupled in the equations \eqref{eq:generaleq1}
and \eqref{eq:generaleq2}.
$E(t)$ is given by equation \eqref{eq:shitE}.
Note that in this system $R(t)$ is determined by
solving the coupled system of equations,
not by using the equation \eqref{eq:FG}.

\subsection{An Energy-Stable Scheme}
\label{sec:scheme}

We next present an energy-stable scheme for the reformulated
system consisting of \eqref{eq:generaleq1} and \eqref{eq:generaleq2}, together
with the boundary condition \eqref{eq:bc}
and the initial conditions \eqref{eq:ic} and \eqref{eq:ic_R}.

Let $n\geqslant 0$ denote the time step index, and $(\cdot)^n$
represent the variable $(\cdot)$ at time step $n$,
corresponding to the time $t=n\Delta t$, where $\Delta t$
is the time step size.
If a real-valued parameter $\theta$ is involved,
$(\cdot)^{n+\theta}$ represents the variable $(\cdot)$
at time step ($n+\theta$), corresponding to
the time $(n+\theta)\Delta t$.

Let $\chi$ denote a generic scalar or vector-valued
variable. We consider the following second-order approximations:
\begin{subequations}
  \begin{align}\label{eq:nplus12}
    &
\chi^{n+\frac{3}{2}}=\frac{3}{2}\chi^{n+1}-\frac{1}{2}\chi^n,\quad \chi^{n+\frac{1}{2}}=\frac{3}{2}\chi^{n}-\frac{1}{2}\chi^{n-1}, \\
\label{eq:tempdist}
&
\frac{\partial \chi}{\partial t}\Big|^{n+1}=\frac{\chi^{n+\frac{3}{2}} -\chi^{n+\frac{1}{2}}    }{{\Delta}t}=\frac{1}{\Delta{t}} \Big(\frac{3}{2}\chi^{n+1} -2 \chi^n +\frac{1}{2}\chi^{n-1}    \Big), \\
\label{eq:def_bar}
&
\bar \chi^{n+1}=2\chi^{n}- \chi^{n-1},
\end{align}
\end{subequations}
where \eqref{eq:tempdist} is the second-order backward differentiation
formula (BDF) and $\bar \chi^{n+1}$ is an explicit approximation
of $\chi^{n+1}$.
We also consider the following second-order approximation of
$\left.\frac{d\mathscr{F}(\chi)}{d\chi} \right|^{n+1}=\mathscr{F}'(\chi)\Big|^{n+1}$
based on the discrete directional derivative~\cite{Gonzalez1996},
\begin{equation}\label{eq:directdir}
  D_{\mathscr{F}}(\chi)\big|^{n+1}=\frac{
    \mathscr{F}(\chi^{n+\frac{3}{2}})-\mathscr{F}(\chi^{n+\frac{1}{2}})
    -\mathscr{F}'(\chi^{n+1})\cdot (\chi^{n+\frac{3}{2}} -\chi^{n+\frac{1}{2}})   }{\|  \chi^{n+\frac{3}{2}}-\chi^{n+\frac{1}{2}} \|^2} (\chi^{n+\frac{3}{2}}-\chi^{n+\frac{1}{2}} )
  +\mathscr{F}'(\chi^{n+1}),
\end{equation}
which satisfies the property
\begin{equation}\label{eq:dist_prop}
  D_{\mathscr{F}}(\chi)\Big|^{n+1}\cdot \left(\frac{3}{2}\chi^{n+1} -2 \chi^n +\frac{1}{2}\chi^{n-1}  \right)=
  D_{\mathscr{F}}(\chi)\Big|^{n+1}\cdot \left(\chi^{n+\frac{3}{2}}-\chi^{n+\frac{1}{2}} \right)
  = \mathscr{F}(\chi^{n+\frac{3}{2}})-\mathscr{F}(\chi^{n+\frac{1}{2}}).
\end{equation}
Note that in these equations $\chi^{n+3/2}$ and $\chi^{n+1/2}$
are given by \eqref{eq:nplus12}.
If $\chi$ represents a scalar-valued variable, one can also approximate
$\mathscr{F}'(\chi)\Big|^{n+1}$ by
\begin{equation}\label{eq:dirdiv_1}
  D_{\mathscr{F}}(\chi)\big|^{n+1} = \frac{\mathscr{F}(\chi^{n+\frac{3}{2}})-\mathscr{F}(\chi^{n+\frac{1}{2}})}{\chi^{n+\frac{3}{2}}-\chi^{n+\frac{1}{2}} }
  = \frac{\mathscr{F}(\chi^{n+\frac{3}{2}})-\mathscr{F}(\chi^{n+\frac{1}{2}})}{\frac{3}{2}\chi^{n+1} -2 \chi^n +\frac{1}{2}\chi^{n-1} },
\end{equation}
which satisfies the same property \eqref{eq:dist_prop}.

We propose the following scheme to approximate the reformulated system:
\begin{subequations}\label{eq:generalscheme}
\begin{align}
  &\frac{\partial \bs u}{\partial t}\Big|^{n+1}=
  \bs F_L(\bs u^{n+1})+\xi \Big[\bs F(\bar {\bs u}^{n+1})-\bs F_L(\bar {\bs u}^{n+1})\Big ]
  +\bs f^{n+1}, \label{eq:schemeeq1}\\
  & \xi=\frac{\mathscr{F}(R^{n+3/2})}{E[\tilde {\bs u}^{n+3/2}]},\label{eq:xiapprix}\\
  &
  E[\tilde {\bs u}^{n+3/2}]=\int_{\Omega}e(\tilde {\bs u}^{n+3/2})d\Omega+C_0,
  \label{eq:Eapprox} \\
  &
  \bs B(\bs u^{n+1}) = \bs f_b^{n+1}, \quad \text{on} \ \Gamma, \label{eq:bc_approx}
\end{align}
\begin{equation}
  \begin{split}
    D_{\mathscr{F}}(R)\big|^{n+1} & \left.\frac{dR}{dt}\right|^{n+1}
    =
    \int_{\Omega}
    e'({\bs u}^{n+1}) \cdot \left.\frac{\partial \bs u}{\partial t}\right|^{n+1}   d \Omega \\
    &-\int_{\Omega} e'(\bs u^{n+1})\cdot\left(
    \bs F_L(\bs u^{n+1}) + \xi\Big[\bs F(\bar{\bs u}^{n+1}) - \bs F_L(\bar {\bs u}^{n+1}) \Big]
    + \bs f^{n+1}
    \right) d\Omega \\
    &+ \xi\left[
      -\int_{\Omega} V(\tilde{\bs u}^{n+1})d\Omega
      + \int_{\Omega} V_s(\bs f^{n+1}, \tilde{\bs u}^{n+1}) d\Omega
      + \int_{\Gamma} B_s(\bs f_b^{n+1},\tilde{\bs u}^{n+1}) d\Gamma
      \right] \\
    &+(1-\xi)\left|
    \int_{\Omega} V_s(\bs f^{n+1}, \tilde{\bs u}^{n+1}) d\Omega
    + \int_{\Gamma} B_s(\bs f_b^{n+1},\tilde{\bs u}^{n+1}) d\Gamma
    \right|.
  \end{split}
  \label{eq:schemeeq2} 
\end{equation}
\end{subequations}
In the above equations, $\left.\frac{\partial {\bs u}}{\partial t}\right|^{n+1}$
and $\left.\frac{dR}{dt}\right |^{n+1}$ are defined by \eqref{eq:tempdist},
$\left.D_{\mathscr{F}}(R)\right|^{n+1}$ is defined by \eqref{eq:directdir}
(or \eqref{eq:dirdiv_1}),
$\bar{\bs u}^{n+1}$ is defined by \eqref{eq:def_bar}, and
$R^{n+3/2}$ is defined by \eqref{eq:nplus12}.
$\tilde{\bs u}^{n+1}$ and $\tilde{\bs u}^{n+3/2}$ are second-order
approximations of $\bs u^{n+1}$ and $\bs u^{n+3/2}$, respectively,
to be specifically defined later in \eqref{eq:defstar}.

\begin{rem}
\label{rem:rem_1}

It is critical to note that in the scheme \eqref{eq:schemeeq1}--\eqref{eq:schemeeq2},
$\frac{\mathscr{F}(R)}{E[\bs u]}$ is approximated at step ($n+\frac{3}{2}$)
while the other variables are approximated at step ($n+1$).
This feature, together with the approximation \eqref{eq:directdir},
allows $R^{n+1}$ to be computed from a linear algebraic
equation (no nonlinear algebraic
solver), and 
endows the scheme with the property that the computed $R^{n+1}$
and $\mathscr{F}(R^{n+1})$ (resp.~$R^{n+3/2}$ and $\mathscr{F}(R^{n+3/2})$,
for all $n\geqslant 0$) are
guaranteed to be positive. These points will become clear from
later discussions.
It should be noted that the approximation $\frac{\mathscr{F}(R^{n+3/2})}{E[\tilde{\bs u}^{n+3/2}]}$
at step ($n+3/2$)
is a second-order approximation of $\frac{\mathscr{F}(R)}{E}=1$.
In fact, the approximation involving any real parameter $\theta$,
\begin{equation}
\frac{\mathscr{F}(R^{n+\theta})}{E[\tilde{\bs u}^{n+\theta}]} = 1 + \Order(\Delta t)^2,
\end{equation}
is a second-order approximation of $\frac{\mathscr{F}(R)}{E}=1$,
as long as $R^{n+\theta}$ and $\tilde{\bs u}^{n+\theta}$ are
second-order approximations of $R(t)$ and $\bs u(t)$ at time
$(n+\theta)\Delta t$.
Therefore, the approximation in \eqref{eq:xiapprix} does not affect
the second-order accuracy of the scheme.

\end{rem}

The scheme given by \eqref{eq:schemeeq1}--\eqref{eq:schemeeq2}
has the following property.
\begin{theorem}\label{thm:thm_1}
  In the absence of the external force and external boundary
  force (i.e.~$\bs f=\bs 0$ and $\bs f_b=0$),
  the following relation holds with the scheme \eqref{eq:generalscheme}:
  \begin{equation}\label{eq:gen_eng_law}
    {\mathscr{F}(R^{n+\frac{3}{2}}) -\mathscr{F}(R^{n+\frac{1}{2}})}= -\Delta{t}  \frac{\mathscr{F}(R^{n+\frac{3}{2}})}{E[\tilde{\bs u}^{n+3/2}]} \int_{\Omega}V(\tilde{\bs u}^{n+1})
    \leqslant 0, \quad \text{for} \ n\geqslant 0,
  \end{equation}
  if the approximation of $R(t)$ at time step $\frac{1}{2}$ is positive, 
  i.e.~$Y_0=R^{n+1/2}\Big|_{n=0}>0$.
\end{theorem}
\begin{proof}
  
By equations \eqref{eq:tempdist} and \eqref{eq:directdir}, we have
\begin{equation}\label{eq:FRdt}
 D_{\mathscr{F}}(R) \Big|^{n+1} \left.\frac{dR}{dt}\right|^{n+1}=\frac{\mathscr{F}(R^{n+\frac{3}{2}}) -\mathscr{F}(R^{n+\frac{1}{2}})}{\Delta{t}}.
\end{equation}
Taking the $L^2$ inner product between equation \eqref{eq:schemeeq1}
and $e'({\bs u}^{n+1})$, and adding the resultant equation to equation \eqref{eq:schemeeq2} and noting equation \eqref{eq:FRdt}, we arrive at 
\begin{equation}\label{eq:enblaEq}
  \mathscr{F}(R^{n+\frac{3}{2}}) -\mathscr{F}(R^{n+\frac{1}{2}})
  = -\Delta t \frac{\mathscr{F}(R^{n+\frac{3}{2}})}{E[\tilde{\bs u}^{n+3/2}]}
  \int_{\Omega}V(\tilde{\bs u}^{n+1})
  +\left(1-\frac{\mathscr{F}(R^{n+\frac{3}{2}})}{E[\tilde{\bs u}^{n+3/2}]} \right) \left|S_0 \right|\Delta t
  +\frac{\mathscr{F}(R^{n+\frac{3}{2}})}{E[\tilde{\bs u}^{n+3/2}]} S_0\Delta t,
\end{equation}
where we have used equation \eqref{eq:xiapprix}, and
$S_0$ is defined by
\begin{equation}\label{eq:S0}
  S_0=\int_{\Omega}V_s(\bs f^{n+1},\tilde{\bs u}^{n+1}) d\Omega
  + \int_{\Gamma}B_s(\bs f_b^{n+1},\tilde{\bs u}^{n+1}) d\Gamma.
\end{equation}
Then it follows that, if $\bs f=0$ and $\bs f_b=0$,
\begin{equation}
  \mathscr{F}(R^{n+3/2}) = \frac{\mathscr{F}(R^{n+\frac{1}{2}}) }{
    1 + \frac{\Delta t}{E[\tilde{\bs u}^{n+3/2}]} \int_{\Omega}V(\tilde{\bs u}^{n+1})d\Omega
  }
  \label{eq:F_relation}
\end{equation}
where we have used the relations \eqref{eq:vs_cond} and \eqref{eq:bc_cond}.

Note that $E[\tilde{\bs u}^{n+3/2}]>0$ and $V(\tilde{\bs u}^{n+1}) \geqslant 0$,
in light of \eqref{eq:shitE} and \eqref{eq:Vcond}.
If $Y_0=R^{n+1/2}|_{n=0}>0$, then $\mathscr{F}(Y_0)>0$ based on
the property \eqref{eq:FG_property}.
By induction, we can conclude from equation \eqref{eq:F_relation}
that $\mathscr{F}(R^{n+3/2})>0$ for all $n\geqslant 0$.
The inequality in \eqref{eq:gen_eng_law} then holds.
%
%
%
We therefore conclude that, if $R^{n+1/2}|_{n=0}>0$,
\begin{equation}\label{eq:endisp}
  0 < \mathscr{F}(R^{n+\frac{3}{2}})\leqslant \mathscr{F}(R^{n+\frac{1}{2}}),
  \quad \text{for} \ n\geqslant 0.
\end{equation}

Thus, the scheme is unconditionally energy
stable with respect to the modified energy $\mathcal{F}(R)$,
if the approximation of $R(t)$ at time step $\frac 12$ is positive.
\end{proof}

There are many ways to approximate $R(t)$ to ensure that it is positive at
time step $\frac{1}{2}$ and that the overall scheme is second-order
accurate in time. One such method is given in the Appendix A.
Therefore we have the following result:
\begin{theorem}
  \label{thm:thm_2}

  With $\bs u^1$ and $R^1$ approximated using the method from Appendix A,
  in the absence of external forces ($\bs f=0$ and $\bs f_b=0$),
  the scheme represented by \eqref{eq:schemeeq1}--\eqref{eq:schemeeq2}
  is unconditionally energy-stable in the sense of
  the relation \eqref{eq:endisp}.
  
\end{theorem}

\begin{rem}

  If the functional form of $\mathscr{F}(\chi)$ is such that
  $\mathscr{F}(\chi) \geqslant 0$ for all $\chi\in (-\infty,\infty)$,
  e.g.~$\mathscr{F}(\chi)=\chi^{2m}$ (with an integer $m\geqslant 1$),
  then the scheme given by \eqref{eq:schemeeq1}--\eqref{eq:schemeeq2}
  is unconditionally energy stable regardless of the
  approximation of $R(t)$ at the time step $\frac{1}{2}$.

\end{rem}

\begin{rem} \label{rem:rem_theta}
  The scheme \eqref{eq:generalscheme} is devised by enforcing
  the system of equations consisting of \eqref{eq:generaleq1},
  \eqref{eq:generaleq2} and \eqref{eq:bc} at time step ($n+1$),
  approximating $\frac{\mathscr{F}(R)}{E}$ at time step ($n+\frac{3}{2}$),
  and employing the approximations \eqref{eq:nplus12}--\eqref{eq:directdir}.
  Inspired by the recent work~\cite{YangLD2019},
  we can generalize this scheme by enforcing the system of equations
  at time step ($n+\theta$), where $\theta$ is a real-valued
  parameter, to arrive at a family of energy-stable schemes.

  In brief, let us consider the following second-order approximations at
  time step ($n+\theta$) with $\theta\geqslant \frac12$:
  ($\chi$ denoting a generic variable, and $\beta\geqslant 0$ denoting
  a real parameter below)
  \begin{subequations}
\begin{equation}\label{eq:nplus12theta}
  \chi^{n+\theta+\frac{1}{2}}=\left(\theta+\frac{1}{2}\right)\chi^{n+1}
  -\left(\theta-\frac{1}{2}\right)\chi^n,
  \qquad
  \chi^{n+\theta-\frac{1}{2}}=\left(\theta+\frac{1}{2}\right)\chi^{n}
  -\left(\theta-\frac{1}{2}\right)\chi^{n-1};
\end{equation}
\begin{equation}\label{eq:ntheta}
  \begin{split}
  \chi^{n+\theta}=&\frac{1}{2}(\chi^{n+\theta+\frac{1}{2}}
  +  \chi^{n+\theta-\frac{1}{2}})+\beta(\chi^{n+1}-2\chi^n+\chi^{n-1}) \\
  =& \left(\beta + \frac{\theta}{2} + \frac{1}{4} \right)\chi^{n+1}
  + \left(\frac{1}{2} - 2\beta \right)\chi^n
  + \left(\beta - \frac{\theta}{2} + \frac{1}{4} \right)\chi^{n-1};
  \quad \text{(implicit approximation)}
  \end{split}
\end{equation}
\begin{equation}\label{eq:ntheta_1}
  \bar \chi^{n+\theta}=(1+\theta)\chi^{n}-\theta \chi^{n-1};
  \quad \text{(explicit approximation)}
\end{equation}
\begin{equation}\label{eq:tempdisttheta}
  \left.\frac{\partial \chi}{\partial t}\right|^{n+\theta}=\frac{\chi^{n+\theta+\frac{1}{2}} -\chi^{n+\theta-\frac{1}{2}}    }{{\Delta}t}
  = \frac{1}{\Delta{t}}\Big[ \Big(\theta+\frac{1}{2} \Big)\chi^{n+1}-2\theta \chi^n +\Big(\theta-\frac{1}{2} \Big)\chi^{n-1}    \Big];
\end{equation}
\end{subequations}
  and the following approximation of
  $\left.\frac{d\mathscr{F}(\chi)}{d\chi}\right|^{n+\theta}
  = \left.\mathscr{F}'(\chi)\right|^{n+\theta}$ based on discrete
  directional derivative,
\begin{equation}\label{eq:directdirtheta}
\begin{aligned}
  \left. D_{\mathscr{F}}(\chi)\right|^{n+\theta} =&
  \frac{ \mathscr{F}(\chi^{n+\theta+\frac{1}{2}})-\mathscr{F}(\chi^{n+\theta-\frac{1}{2}})
    -\mathscr{F}'(\chi^{n+\theta})\cdot(\chi^{n+\theta+\frac{1}{2}} -\chi^{n+\theta-\frac{1}{2}})   }{\|  \chi^{n+\theta+\frac{1}{2}}-\chi^{n+\theta-\frac{1}{2}} \|^2}
  (\chi^{n+\theta+\frac{1}{2}}-\chi^{n+\theta-\frac{1}{2}} )\\
&+\mathscr{F}'(\chi^{n+\theta}).
\end{aligned}
\end{equation}
These approximations satisfy the following properties:
\begin{subequations}
  \begin{equation}
    \begin{split}
       &\chi^{n+\theta}\left[\Big(\theta+\frac{1}{2} \Big)\chi^{n+1}-2\theta \chi^n +\Big(\theta-\frac{1}{2} \Big)\chi^{n-1} \right] 
    = \frac{1}{2}\left( \left|\chi^{n+\theta+\frac12} \right|^2
    - \left|\chi^{n+\theta-\frac12} \right|^2 \right) \\
    & \qquad\qquad\qquad
    + \frac{\beta}{2} \left(
    \left| \chi^{n+1} - \chi^n \right|^2
    - \left| \chi^{n} - \chi^{n-1} \right|^2
    \right) 
    + \theta\beta \left|\chi^{n+1}-2\chi^n + \chi^{n-1}  \right|^2;
    \end{split}
  \end{equation}
  \begin{equation}
    \left.D_{\mathscr{F}}(\chi)\right|^{n+\theta}
    \left[\Big(\theta+\frac{1}{2} \Big)\chi^{n+1}-2\theta \chi^n +\Big(\theta-\frac{1}{2} \Big)\chi^{n-1} \right] =
    \mathscr{F}(\chi^{n+\theta+\frac12}) - \mathscr{F}(\chi^{n+\theta-\frac12}).
  \end{equation}
\end{subequations}
Note that the parameter $\beta\geqslant 0$ in \eqref{eq:ntheta}
can often be used to control the numerical dissipation of the approximations,
which will be useful for approximating energy-conserving systems.
An example will be given with the Klein-Gordon equation in a later section.
The scheme given in \eqref{eq:generalscheme}
corresponds to $\theta=1$ and $\beta=\frac{1}{4}$.

By approximating the terms in
equations \eqref{eq:generaleq1}, \eqref{eq:generaleq2} and \eqref{eq:bc}
at time step ($n+\theta$), except for the term $\frac{\mathscr{F}(R)}{E}$,
which will be approximated at time step ($n+\theta+\frac12$),
and employing the approximations \eqref{eq:nplus12theta}--\eqref{eq:directdirtheta},
one can prove that the resultant family of schemes (with $\theta$
and $\beta$ as parameters) is unconditionally energy-stable.
The details will not be provided here.

\end{rem}

\subsection{Solution Algorithm}
\label{sec:alg}

Let us now consider how to implement the algorithm represented by equations \eqref{eq:schemeeq1}-\eqref{eq:schemeeq2}.
We first introduce some notations
($\chi$ again denoting a generic variable):
  \begin{equation}\label{equ:notation_1}
    \gamma_0 = \frac{3}{2}, 
    \quad  \hat{\chi} = 2\chi^n - \frac{1}{2} \chi^{n-1}. 
  \end{equation}
Then the approximation in \eqref{eq:tempdist} can be written as
\begin{equation}\label{eq:timedirnew}
\left.\frac{\partial\chi}{\partial t} \right|^{n+1}
  = \frac{\gamma_0\chi^{n+1}-\hat{\chi}}{\Delta t}.
\end{equation}

Inserting notation \eqref{eq:timedirnew} into equation \eqref{eq:schemeeq1}, we have
\begin{equation}\label{eq:Linearu}
  \frac{\gamma_0}{\Delta t}{\bs u}^{n+1}-\bs F_L(\bs u^{n+1})
  =\xi \Big[ \bs F(\bar {\bs u}^{n+1})-\bs F_L(\bar {\bs u}^{n+1}) \Big]+\bs f^{n+1}
  + \frac{\hat{\bs u}}{\Delta t}.
\end{equation}
Note that $\bar{\bs u}^{n+1}$ and $\hat{\bs u}$ are both explicitly known,
and $\xi$ is an unknown depending on $\bs u^{n+1}$. 
Taking advantage of the fact that $\xi$ is a scalar number
instead of a field function and the linearity of the
operator $\bs B$ in the boundary condition \eqref{eq:bc},
we introduce two field functions $(\bs u_1^{n+1},\bs u_2^{n+1})$
as solutions to the following two linear systems:
\begin{subequations}
\begin{align}
  & \frac{\gamma_0}{\Delta{t}}\bs u_1^{n+1}- \bs F_L(\bs u_1^{n+1})
  =\frac{\hat {\bs u}}{\Delta{t}}+\bs f^{n+1}, \label{eq:schemeu1}\\
  &
  \bs B(u_1^{n+1}) = \bs f_b^{n+1}, \quad \text{on} \ \Gamma.
  \label{eq:bc_u1}
\end{align}
\end{subequations}
\begin{subequations}
\begin{align}
  & \frac{\gamma_0}{\Delta{t}}\bs u_2^{n+1}-\bs F_L(\bs u_2^{n+1})
  = \bs F(\bar {\bs u}^{n+1})-\bs F_L(\bar {\bs u}^{n+1}).\label{eq:schemeu2} \\
  &
  \bs B(\bs u_2^{n+1}) = 0, \quad \text{on} \ \Gamma.
  \label{eq:bc_u2}
\end{align}
\end{subequations}
Since the operator $\bs F_L$ is chosen to be a linear operator
and relatively easy to compute, $\bs u_1^{n+1}$ and $\bs u_2^{n+1}$
can be solved efficiently from these equations. Then we have
the following result.
\begin{theorem}\label{thm:decouple}
  Given scalar value $\xi,$ the following function solves the system
  consisting of equations \eqref{eq:schemeeq1} and \eqref{eq:bc_approx}:
\begin{equation}\label{eq:soluu1u2}
\bs u^{n+1}= \bs u_1^{n+1}+\xi \bs u_2^{n+1},
\end{equation}
where $\bs u_1^{n+1}$ and $\bs u_2^{n+1}$ are given by
the equations \eqref{eq:schemeu1}-\eqref{eq:bc_u2}.
\end{theorem}

The scalar value $\xi$ still needs to be determined.
Define 
\begin{equation}\label{eq:defstar}
  \left\{
  \begin{split}
    &
    \tilde{\bs u}^{n+1}=\bs u_1^{n+1}+\bs u_2^{n+1}, \\
    &
    \tilde{\bs u}^{n+{3}/{2}}=\frac{3}{2} \tilde{\bs u}^{n+1}- \frac{1}{2}\bs u^n,
  \end{split}
  \right.
\end{equation}
which are second-order approximations of
$\bs u^{n+1}$ and $\bs u^{n+3/2}$.
These field variables can be explicitly computed
after ${\bs u}_1^{n+1}$ and $\bs u_2^{n+1}$ are obtained.
By equation \eqref{eq:xiapprix}, we have  
\begin{equation}\label{eq:Fr32}
\mathscr{F}(R^{n+\frac{3}{2}})=\xi E[\tilde{\bs u}^{n+\frac{3}{2}}].
\end{equation}
Note that equation \eqref{eq:schemeeq2} can be transformed into
equation \eqref{eq:enblaEq}. Inserting equation \eqref{eq:Fr32}
into equation \eqref{eq:enblaEq} leads to the solution for $\xi$, 
\begin{equation}\label{eq:soluxi}
\xi=\frac{\mathscr{F}(R^{n+1/2}) +{\Delta}t |S_0|}{ E[\tilde{\bs u}^{n+\frac{3}{2}}]+\Delta{t} \int_{\Omega}V(\tilde{\bs u}^{n+1}) +\Delta{t}(|S_0|-S_0) },
\end{equation}
where $\tilde{\bs u}^{n+1}$ and $\tilde{\bs u}^{n+3/2}$
are given by \eqref{eq:defstar},
$S_0$ is given by equation \eqref{eq:S0},
and $E[\tilde{\bs u}^{n+3/2}]$ is computed by
equation \eqref{eq:Eapprox}.

In light of equations \eqref{eq:Fr32} and \eqref{eq:nplus12},
we can then compute $R^{n+1}$ by
\begin{equation} \label{eq:RN1}
  \left\{
  \begin{split}
    &
    R^{n+3/2} = \mathscr{G}\left(\xi E[\tilde{\bs u}^{n+3/2}] \right), \quad n\geqslant 0;
    \\
    &
    R^{n+1} = \frac{2}{3}R^{n+3/2} + \frac{1}{3} R^n, \quad n\geqslant 0.
  \end{split}
  \right.
\end{equation}
The following result holds.

\begin{theorem}\label{thm:xi}
  The scalar value $\xi$ computed by equation \eqref{eq:soluxi}
  and the variable $R^{n+1}$ ($n\geqslant 0$) computed by
  equation \eqref{eq:RN1}
  are always positive, if the approximation of $R(t)$
  at time step $\frac{1}{2}$ is positive,
  i.e.~$Y_0=R^{n+1/2}|_{n=0}>0$.
\end{theorem}
\begin{proof}

  If $Y_0=R^{n+1/2}|_{n=0}>0$, then $\mathscr{F}(Y_0)>0$ based on
  \eqref{eq:FG_property}. Since $E(\bs u)$ is a positive function,
  $V(\bs u)\geqslant 0$ and $|S_0|-S_0 \geqslant 0$,
  we conclude by induction $\xi$ computed from \eqref{eq:soluxi}
  is always positive.

  Note that $R^0 = R(0)>0$ according to equation \eqref{eq:ic_R}.
  In light of the property \eqref{eq:FG_property}, we conclude that
  $R^{n+3/2}$ and $R^{n+1}$ computed from equation \eqref{eq:RN1}
  are both positive.
  \end{proof}

Using the method from the Appendix A can ensure the
positiveness of the approximation of $R(t)$ at
the time step $\frac{1}{2}$. We have the following result.

\begin{theorem}\label{thm:xi_1}

  With $\bs u^1$ and $ R^1$ computed based on the method
  from Appendix A, 
  the $\xi$ given by \eqref{eq:soluxi} and
  $R^{n+1}$ and $R^{n+3/2}$ given by \eqref{eq:RN1} satisfy 
  the property 
  \begin{equation}
    \xi >0, \quad R^{n+1}>0, \quad \text{and} \ \
    R^{n+3/2} > 0,
  \end{equation}
  for all $n\geqslant 0$, regardless of the external forces $\bs f$ and $\bs f_b$
  and the time step size $\Delta t$.

\end{theorem}

Combining the above discussions, we arrive at the solution procedure
for solving the system consisting of equations \eqref{eq:schemeeq1}-\eqref{eq:schemeeq2}.
Given $(\bs u^{n},R^{n})$, 
we compute
$(\bs u^{n+1}, R^{n+1})$ through the following steps:
\begin{enumerate}

\item
  Solve equations \eqref{eq:schemeu1}--\eqref{eq:bc_u1}  for $\bs u_1^{n+1}$; \\
  Solve equations \eqref{eq:schemeu2}--\eqref{eq:bc_u2}  for $\bs u_2^{n+1}$.

\item
  Compute $\tilde{\bs u}^{n+1}$ and $\tilde{\bs u}^{n+{3}/{2}}$
  based on equation \eqref{eq:defstar}; \\
  Compute $E[\tilde{\bs u}^{n+\frac{3}{2}}]$,
  $\int_{\Omega}V(\tilde{\bs u}^{n+1})$ and $S_0$ based on equations \eqref{eq:shitE}, \eqref{eq:VB} and \eqref{eq:S0}.

\item
      Compute $\xi$ based on equation \eqref{eq:soluxi}.

\item
  Compute $\bs u^{n+1}$ based on equation \eqref{eq:soluu1u2}.
  Compute $R^{n+1}$ based on equation \eqref{eq:RN1}.

\end{enumerate}


It can be noted that the numerical scheme and the solution algorithm
developed in this section has several attractive properties:
(i) Only linear systems need to be solved for the field variables $\bs u$
within a time step. Moreover, with appropriate choice for
the $\bs F_L$ operator, the system can involve only constant and
time-independent coefficient matrices, which can be pre-computed.
Therefore, the solution for $\bs u$ will be computationally very efficient.
(ii) The auxiliary variables $R$ and $\xi$ can be computed by
a well-defined explicit formula, and no nonlinear algebraic solver is involved.
Their computed values are guaranteed to be positive.
(iii) The auxiliary variable $R$ can be defined by a rather
general class of functions ($\mathscr{F}$ and $\mathscr{G}$)
using the method developed here. 
(iv) The scheme is unconditionally energy-stable for general
dissipative systems.


\subsection{An Alternative Formulation and Energy-Stable Scheme}
\label{sec:alter}

The numerical formulation presented in the previous
subsections is not the only way to devise energy-stable
schemes for dissipative systems.
In this subsection we outline an alternative formulation
and associated energy-stable scheme.
The process is analogous to the developments in
the sections \ref{sec:formulation}--\ref{sec:alg}.
So many details will be omitted in the following discussions.

The main idea with the alternative formulation is to
realize that $\frac{R(t)}{\mathscr{G}(E)}=1$ with the
auxiliary variable $R(t)$ defined in \eqref{eq:FG}.
Therefore, one can potentially employ $\frac{R}{\mathscr{G}(E)}$,
instead of $\frac{\mathscr{F}(R)}{E}$,
in the numerical formulations. With appropriate reformulation
and treatments of different terms, it turns out that
a discretely energy-stable
scheme can be obtained with similar attractive properties,
such as the guaranteed  positiveness
of the computed values for the variable $R(t)$.

Note that $R(t)$ is defined by \eqref{eq:FG}, where
$\mathscr{G}$ is a one-to-one increasing differentiable function
with $\mathscr{G}(\chi)>0$ and $\mathscr{G}'(\chi)>0$
for $\chi>0$.
$R(t)$ satisfies the following dynamic equation
\begin{equation}
  \frac{dR}{dt} = \mathscr{G}'(E)\int_{\Omega} e'(\bs u)\cdot\frac{\partial\bs u}{\partial t}
  d\Omega,
  \label{eq:Requ_alt}
\end{equation}
where $E(t)$ is defined by \eqref{eq:shitE}.

We reformulate equation \eqref{eq:generalsys} into
\begin{equation}
  \frac{\partial \bs u}{\partial t}=\bs F_L(\bs u)
  +\frac{R}{\mathscr{G}(E)}\Big(\bs F(\bs u)-\bs F_L(\bs u) \Big)+\bs f,
  \label{eq:gov_equ_alt}
\end{equation}
where the notations follow those defined in previous subsections.
Analogously, by incorporating  appropriate  zero terms
we can transform \eqref{eq:Requ_alt} into
\begin{equation}
  \begin{split}
    \frac{dR}{dt} =&
      \mathscr{G}'(E)\int_{\Omega} e'(\bs u)\cdot \frac{\partial \bs u}{\partial t}d \Omega
      - \mathscr{G}'(E)\int_{\Omega} \frac{\partial e}{\partial \bs u} \cdot \left(
        \bs F_L(\bs u) + \frac{R}{\mathscr{G}(E)}\left[\bs F(\bs u) - \bs F_L(\bs u) \right] + \bs f
        \right)d\Omega \\
        &+  \frac{R}{\mathscr{G}(E)} \mathscr{G}'(E)\left[ -\int_{\Omega} V(\bs u) d\Omega
          + \int_{\Omega} V_s(\bs f,\bs u) d\Omega
          + \int_{\Gamma} B_s(\bs f_b, \bs u) d\Gamma
            \right] 
        \\
        &+ \left[1 - \frac{R}{\mathscr{G}(E)} \right]\mathscr{G}'(E)\left| \int_{\Omega}V_s(\bs f,\bs u)d\Omega
      + \int_{\Gamma} B_s(\bs f_b,\bs u) d\Gamma  \right|.
  \end{split}
  \label{eq:Requ_alt_reform}
\end{equation}
The reformulated system now consists of equations \eqref{eq:gov_equ_alt}
and \eqref{eq:Requ_alt_reform}, the boundary condition \eqref{eq:bc},
and the initial conditions \eqref{eq:ic} and \eqref{eq:ic_R}.

We discretize the reformulated system as follows:
\begin{subequations}
\begin{align}
  &
  \frac{1}{\Delta t}\left(\frac{3}{2}\bs u^{n+1}-2\bs u^n+\frac{1}{2}\bs u^{n-1} \right)
  =
  \bs F_L(\bs u^{n+1})+\xi \Big[\bs F(\bar {\bs u}^{n+1})-\bs F_L(\bar {\bs u}^{n+1})\Big ]
  +\bs f^{n+1}, \label{eq:scheme_alt_equ_1}\\
  & \xi=\frac{R^{n+3/2}}{\mathscr{G}(E[\tilde {\bs u}^{n+3/2}])},\label{eq:scheme_alt_equ_2}\\
  &
  E[\tilde {\bs u}^{n+3/2}]=\int_{\Omega}e(\tilde {\bs u}^{n+3/2})d\Omega+C_0,
  \label{eq:scheme_alt_equ_3} \\
  &
  \bs B(\bs u^{n+1}) = \bs f_b^{n+1}, \quad \text{on} \ \Gamma, \label{eq:scheme_alt_equ_4}
\end{align}
\begin{equation}
  \begin{split}
    \frac{R^{n+3/2} - R^{n+1/2}}{\Delta t}
    =&
    \mathscr{G}'(E[\tilde{u}^{n+3/2}]) \left\{ \int_{\Omega} 
    e'({\bs u}^{n+1}) \cdot \frac{\frac{3}{2}\bs u^{n+1}-2\bs u^n + \frac{1}{2}\bs u^{n-1} }{\Delta t}   d \Omega \right. \\
    &-\int_{\Omega} e'(\bs u^{n+1})\cdot\left(
    \bs F_L(\bs u^{n+1}) + \xi\Big[\bs F(\bar{\bs u}^{n+1}) - \bs F_L(\bar {\bs u}^{n+1}) \Big]
    + \bs f^{n+1}
    \right) d\Omega \\
    &+ \xi\left[
      -\int_{\Omega} V(\tilde{\bs u}^{n+1})d\Omega
      + \int_{\Omega} V_s(\bs f^{n+1}, \tilde{\bs u}^{n+1}) d\Omega
      + \int_{\Gamma} B_s(\bs f_b^{n+1},\tilde{\bs u}^{n+1}) d\Gamma
      \right] \\
    &+(1-\xi) \left. \left|
    \int_{\Omega} V_s(\bs f^{n+1}, \tilde{\bs u}^{n+1}) d\Omega
    + \int_{\Gamma} B_s(\bs f_b^{n+1},\tilde{\bs u}^{n+1}) d\Gamma
    \right| \right\}.
  \end{split}
  \label{eq:scheme_alt_equ_5} 
\end{equation}
\end{subequations}
In these equations $\bar {\bs u}^{n+1}$ is defined by \eqref{eq:def_bar},
$R^{n+3/2}$ and $R^{n+1/2}$ are defined by \eqref{eq:nplus12}, and
$\tilde{\bs u}^{n+1}$ and $\tilde{\bs u}^{n+3/2}$
are second-order approximations of $\bs u^{n+1}$
and $\bs u^{n+3/2}$ respectively to be specified later.

Taking the $L^2$ inner product between $\mathscr{G}'(E[\tilde{u}^{n+3/2}])e'(\bs u^{n+1})$
and equation \eqref{eq:scheme_alt_equ_1}, and
summing up the resultant equation and equation \eqref{eq:scheme_alt_equ_5},
we get
\begin{equation}\label{eq:eng_alt_1}
  R^{n+3/2} - R^{n+1/2} =
  \Delta t \mathscr{G}'(E[\tilde{u}^{n+3/2}]) \left[ -\xi \int_{\Omega} V(\tilde{u}^{n+1}) d\Omega
  + (1-\xi)|S_0|
  + \xi S_0 \right]
\end{equation}
where $S_0$ is given by the equation \eqref{eq:S0}.
In the absence of external forces ($\bs f=0$ and $\bs f_b=0$),
$S_0=0$ and equation \eqref{eq:eng_alt_1} leads to
\begin{equation}\label{eq:R_alt_1}
  R^{n+3/2} = \frac{R^{n+1/2}}{1 + \Delta t\frac{\mathscr{G}'(E[\tilde{u}^{n+3/2}])}{\mathscr{G}(E[\tilde{u}^{n+3/2}]) }\int_{\Omega}V(\tilde{\bs u}^{n+1})d\Omega }
\end{equation}
where we have used \eqref{eq:scheme_alt_equ_2}.
Note that $E[\tilde{\bs u}^{n+3/2}] > 0$, $V(\tilde{\bs u}^{n+1})\geqslant 0$, and that
$\mathscr{G}(\chi)>0$ and $\mathscr{G}'(\chi)>0$ for $\chi>0$.
By induction we can conclude from \eqref{eq:R_alt_1} that
$R^{n+3/2} \geqslant 0$ (for all $n\geqslant 0$) if the approximation
of $R(t)$ at time step $\frac{1}{2}$ is non-negative. 
Equation \eqref{eq:eng_alt_1} then leads to the following
result.
\begin{theorem}\label{thm:thm_A}
  In the absence of external forces ($\bs f=0$ and $\bs f_b=0$),
  if the approximation of $R(t)$ at time step $\frac{1}{2}$
  is non-negative, the scheme given by
  \eqref{eq:scheme_alt_equ_1}--\eqref{eq:scheme_alt_equ_5} is
  unconditionally energy-stable in the sense that
  \begin{equation}\label{eq:eng_stab_alt}
    0\leqslant R^{n+3/2} \leqslant R^{n+1/2},
    \quad \text{for all} \ \ n \geqslant 0.
  \end{equation}

\end{theorem}

In the Appendix A, we have presented a method for computing
the first time step,
which can ensure that the approximation of
$R(t)$ at step $\frac{1}{2}$ is positive. This leads to
the following result.
\begin{theorem}\label{thm:thm_B}
  In the absence of external forces ($\bs f=0$ and $\bs f_b=0$),
  when the first time step is approximated using the method
  from Appendix A, the numerical scheme given by
  \eqref{eq:scheme_alt_equ_1}--\eqref{eq:scheme_alt_equ_5} is
  unconditionally energy-stable in the sense of equation
  \eqref{eq:eng_stab_alt}.
  
\end{theorem}


The scheme represented by \eqref{eq:scheme_alt_equ_1}--\eqref{eq:scheme_alt_equ_5}
can be implemented in a similar way to that of Section \ref{sec:alg}, with
the following steps:
\begin{itemize}

\item
Compute $\bs u_1^{n+1}$ and $\bs u_2^{n+1}$ by solving
equations \eqref{eq:schemeu1}--\eqref{eq:bc_u2}.

\item
  Define $\tilde{\bs u}^{n+1}$ and $\tilde{\bs u}^{n+3/2}$ again by
  equations \eqref{eq:defstar}. These variables can be computed.

\item
  Compute $\xi$ based on equation \eqref{eq:eng_alt_1}, specifically by
  \begin{equation}\label{eq:xi_alt}
    \xi = \frac{R^{n+1/2} + \Delta t |S_0|\mathscr{G}'(E[\tilde{\bs u}^{n+3/2}]) }
        {\mathscr{G}(E[\tilde{\bs u}^{n+3/2}]) +
          \Delta t\mathscr{G}'(E[\tilde{\bs u}^{n+3/2}])\left[
          \int_{\Omega}V(\tilde{\bs u}^{n+1})d\Omega + (|S_0| - S_0)  \right]
        }
  \end{equation}
  where $S_0$ is given by \eqref{eq:S0}.

\item
  Compute ${\bs u}^{n+1}$ by equation \eqref{eq:soluu1u2}. Compute $R^{n+1}$ by
  \begin{equation}\label{eq:RN1_alt}
    \left\{
    \begin{split}
      &
      R^{n+3/2} = \xi \mathscr{G}(E[\tilde{\bs u}^{n+3/2}]), \\
      &
      R^{n+1} = \frac{2}{3}R^{n+3/2} + \frac{1}{3}R^n,
    \end{split}
    \right.
  \end{equation}
  where we have used equations \eqref{eq:scheme_alt_equ_2} and \eqref{eq:nplus12}.
  
\end{itemize}

Noting the positiveness of energy $E(t)$ and the other functions
involved in equations \eqref{eq:xi_alt} and \eqref{eq:RN1_alt},
we have the following result.
\begin{theorem}\label{thm:thm_B1}
  If the first time step is approximated using the method from Appendix A,
  regardless of the external forces $\bs f$ and $\bs f_b$ and
  the time step size $\Delta t$,
  the computed values for $\xi$ and $R^{n+1}$ with the scheme
  \eqref{eq:scheme_alt_equ_1}--\eqref{eq:scheme_alt_equ_5}
  satisfy the property,
  \begin{equation}
    \xi>0, \ \ \text{and} \ \ R^{n+1}>0
  \end{equation}
  for all time steps.
  
\end{theorem}



\begin{rem}\label{rem:rem_partial}

  In the current paper we have used the total energy (shifted)
  $E_{tot}(t)$ (see equation \eqref{eq:shitE}) to define the auxiliary variable
  $R(t)$. One can also define an auxiliary variable based on
  a part of the total energy. Suppose the total energy of the system
  can be written as
  \begin{equation}
    E_{tot}(t) = E_1(t) + E_2(t), \quad \text{with} \ \
    E_1(t)=E_1[\bs u] = \int_{\Omega}e_1(\bs u)d\Omega, \ \
    E_2(t)=E_2[\bs u] = \int_{\Omega} e_2(\bs u)d\Omega
  \end{equation}
  where each of the energy components $E_1[\bs u]$ and $E_2[\bs u]$
  is bounded from below. One can define an auxiliary variable $R(t)$
  based on e.g.~$E_2(t)$ (shifted appropriately),
  \begin{equation}
    \left\{
    \begin{split}
      &
      \mathscr{F}(R) = E_s(t) = E_2(t) + C_0
      = \int_{\Omega} e_2(\bs u)d\Omega + C_0, \\
      &
      R(t) = \mathscr{G}(E_s),
    \end{split}
    \right.
  \end{equation}
  where the chosen energy constant $C_0$ is to ensure that $E_s(t)>0$.
  By appropriate reformulation of the system one can devise  energy-stable
  schemes in an analogous way. We refer the reader to
  \cite{YangD2018} for such an energy-stable scheme for incompressible
  two-phase flows with different densities and viscosities for the two fluids,
  which corresponds to a specific mapping function $\mathscr{F}(R)=R^2$.
  A drawback with this lies in that
  one needs to solve a nonlinear algebraic equation (or a quadratic equation),
  albeit about a scalar number,
  when computing the auxiliary variable, and that the property for guaranteed
  positiveness of the computed auxiliary-variable values will
  be lost.   

\end{rem}


\vspace{0.2in}
In the subsequent sections, we consider three  dissipative (or conserving)
systems (a chemotaxis model, Cahn-Hilliard equation, and Klein-Gordon equation)
as specific applications and demonstrations of the gPAV method
developed in this section.

\section{A Chemo-Repulsion Model}
\label{sec:chemo}

\subsection{Model and Numerical Scheme}

Consider the following repulsive-productive chemotaxis model
with a quadratic production term (see e.g.~\cite{Gonzalez2019}) in a
domain $\Omega$ (with boundary $\Gamma$):
\begin{subequations}\label{parabolicsystem}
\begin{align}
& \frac{\partial u}{\partial t}=\nabla^2 u+\nabla\cdot(u\nabla v)+f_1(\bs x,t),  \label{parabolicsystem1}\\
& \frac{\partial v}{\partial t}=\nabla^2 v-v+p(u)+f_2(\bs x,t), \label{parabolicsystem2} \\
&\bs n \cdot\nabla u=d_a(\bs x,t),~~\bs n\cdot\nabla v=d_b(\bs x,t),\;\;{\rm on}~\Gamma, \label{cheobd1} \\
&u(\bs x,0)= u_{in}(\bs x),~~v(\bs x,0)=v_{in}(\bs x),\label{cheoinitial}
\end{align}
\end{subequations}
where $p(u)=u^2$ is the quadratic production term, $u(\bs x,t)\geq 0$ is the cell density,
and $v(\bs x,t)\geq 0$ is the chemical concentration.
$f_1$, $f_2$, $d_a$ and $d_b$ denote the volume and boundary source terms,
respectively. $u_{in}$ and $v_{in}$ are the initial distributions of
the field variables.
This system is dissipative in the absence of the 
source terms, with
the total energy  given by (see \cite{Gonzalez2019})
 \begin{equation}\label{eq:chemoEn}
 E_{\rm tot}=\int_\Omega \Big( \frac{1}{2}|u|^2+\frac{1}{4}|\nabla v|^2\Big) d\Omega.
  \end{equation}
 By taking the $L^2$ inner products between \eqref{parabolicsystem1} and $u,$
 and between \eqref{parabolicsystem2} and $-\dfrac{1}{2} \nabla^2 v$, summing them up
 and
 performing integration by part and imposing boundary conditions in \eqref{cheobd1}, we can obtain the following energy balance equation:
 \begin{equation}\label{eq:chemoEnlaw}
 \begin{aligned}
 \int_\Omega\frac{\partial}{\partial t}\Big( \frac{1}{2}|u|^2+\frac{1}{4}|\nabla v|^2\Big) d\Omega=&-\int_\Omega \Big( |\nabla u|^2+\frac{1}{2} |\nabla^2 v|^2+\frac{1}{2} |\nabla v|^2  \Big) d\Omega+\int_\Omega \Big( f_1 u+\frac{1}{2}\nabla f_2\cdot\nabla v \Big) d\Omega \\
& +\int_{\Gamma}\Big( d_a u+\frac{1}{2}d_bu^2+\frac{1}{2}d_b\frac{\partial v}{\partial t}+\frac{1}{2}d_b v-\frac{1}{2}d_b f_2\Big) d\Gamma.
 \end{aligned}
 \end{equation}

 Following the gPAV procedure from section \ref{sec:adjstab},
 we define a shifted energy according to equation \eqref{eq:shitE}
\begin{equation}\label{eq:chemoshiftEn}
E(t) = E[u,v] =\int_\Omega \Big( \frac{1}{2}|u|^2+\frac{1}{4}|\nabla v|^2\Big) d\Omega+C_0,
\end{equation}
where $C_0$ is a chosen energy constant such that $E(t)>0$.
Define a scalar auxiliary variable $R(t)$
according to equation \eqref{eq:FG}.
Thus,  equation \eqref{eq:dynR} becomes
 \begin{equation}\label{eq:chemoauxeq}
   \mathscr{F}'(R)\frac{dR}{dt}=
   \int_\Omega \Big( u\frac{\partial u}{\partial t}
   +\frac{1}{2}\nabla v\cdot\nabla \frac{\partial v}{\partial t} \Big)d\Omega
   =\int_\Omega \Big( u\frac{\partial u}{\partial t}-\frac{1}{2}\nabla^2 v\frac{\partial v}{\partial t} \Big)d\Omega+\frac{1}{2} \int_{\Gamma}(\bs n\cdot\nabla v) \frac{\partial v}{\partial t} d\Gamma.
 \end{equation}

 Following equations \eqref{eq:generaleq1}-\eqref{eq:generaleq2},
 we reformulate equations \eqref{parabolicsystem1}-\eqref{parabolicsystem2}
 into the following equivalent form:
\begin{subequations}
\begin{align}
& \frac{\partial u}{\partial t}=\nabla^2 u+    \frac{\mathscr{F}(R)}{E}  \nabla\cdot(u\nabla v)+f_1,\label{eq:chemref1} \\
&\frac{\partial v}{\partial t}=\nabla^2 v-v+  \frac{\mathscr{F}(R)}{E} p(u)+f_2. \label{eq:chemoref2}
\end{align}
\end{subequations}
By incorporating the following zero terms into the right hand side
of equation \eqref{eq:chemoauxeq},
\begin{equation*}
  \begin{split}
    &
    \left(\frac{\mathscr{F}(R)}{E} -1 \right)\int_{\Omega}u(\nabla^2u+f_1)d\Omega
    + \frac{\mathscr{F}(R)}{E}\left[
      \int_{\Omega} u\nabla\cdot(u\nabla v)d\Omega - \int_{\Omega} u\nabla\cdot(u\nabla v)d\Omega
      \right] \\
    &
    -\left(\frac{\mathscr{F}(R)}{E} -1 \right)\int_{\Omega}\frac{1}{2}\nabla^2v
    (\nabla^2v -v + f_2)d\Omega
    + \frac{\mathscr{F}(R)}{E}\left[
      \int_{\Omega}\frac{1}{2}(\nabla^2v) p(u)d\Omega - \int_{\Omega}\frac{1}{2}(\nabla^2v) p(u)
      d\Omega\right] \\
    &
    + \left(\frac{\mathscr{F}(R)}{E} -1 \right)\int_{\Gamma}\frac{1}{2}(\bs n\cdot\nabla v)
    \frac{\partial v}{\partial t}d\Gamma \\
    &
    + \left(1-\frac{\mathscr{F}(R)}{E} \right)\left|
    \int_{\Omega} f_1u d\Omega + \frac{1}{2}\int_{\Omega}\nabla f_2\cdot\nabla v d\Omega
    + \int_{\Gamma}d_a u d\Gamma
    + \int_{\Gamma}\frac{d_b}{2}\left(\frac{\partial v}{\partial t} + v + u^2
    -f_2 \right)d\Gamma
    \right|,
  \end{split}
\end{equation*}
we can transform this equation into
\begin{equation}\label{eq:chemoFt}
\begin{split}
  &\mathscr{F}'(R)\frac{dR}{dt}= \int_\Omega \Big( u\frac{\partial u}{\partial t}-\frac{1}{2}\nabla^2 v\frac{\partial v}{\partial t} \Big)d\Omega \\
  &
  + \frac{\mathscr{F}(R)}{E} \left[ -\int_\Omega \Big( |\nabla u|^2+\frac{1}{2} |\nabla^2 v|^2+\frac{1}{2} |\nabla v|^2  \Big) d\Omega 
    +  \int_\Omega \Big( f_1 u+\frac{1}{2}\nabla f_2\cdot\nabla v \Big) d\Omega \right. \\
    & \left.\qquad\qquad
    +\int_{\Gamma}\Big( d_a u+\frac{1}{2}d_bu^2+\frac{1}{2}d_b\frac{\partial v}{\partial t}+\frac{1}{2}d_b v-\frac{1}{2}d_b f_2\Big)d\Gamma    \right] \\
&+\left(1-\frac{\mathscr{F}(R)}{E} \right) \left|  \int_\Omega \Big( f_1 u+\frac{1}{2}\nabla f_2\cdot\nabla v \Big) d\Omega  +\int_{\Gamma}\Big( d_a u+\frac{1}{2}d_bu^2+\frac{1}{2}d_b\frac{\partial v}{\partial t}+\frac{1}{2}d_b v-\frac{1}{2}d_b f_2\Big)d\Gamma   \right|\\
  &-\int_{\Omega} u\left( \nabla^2 u+    \frac{\mathscr{F}(R)}{E}  \nabla\cdot(u\nabla v)+f_1  \right) d\Omega
  +\int_{\Omega}\frac{1}{2}\nabla^2 v \left( \nabla^2 v-v+  \frac{\mathscr{F}(R)}{E} p(u)+f_2  \right)d\Omega,
\end{split}
\end{equation}
where we have used the fact $\frac{\mathscr{F}(R)}{E}=1$ and the boundary conditions
\eqref{cheobd1}.

The reformulated equivalent system consist of
equations \eqref{eq:chemref1}-\eqref{eq:chemoFt}
and \eqref{cheobd1}-\eqref{cheoinitial}.
The energy-stable scheme for this system is as follows:
\begin{subequations}
 \begin{align}
 &\frac{\partial u}{\partial t}\Big|^{n+1}=\nabla^2 u^{n+1}+\xi \nabla \cdot \big( \bar u^{n+1}\nabla \bar v^{n+1} \big)+f_1^{n+1}; \label{eq:chemsche1}\\
 & \frac{\partial v}{\partial t}\Big|^{n+1}=\nabla^2 v^{n+1}-v^{n+1}+  \xi p(\bar u^{n+1})+f_2^{n+1}; \label{eq:chemsche2}\\
   & \xi=\frac{\mathscr{F}(R^{n+\frac{3}{2}})}{E[\tilde{u}^{n+3/2},\tilde{v}^{n+3/2}]}; \label{eq:chemschexi}\\
   &
   E[\tilde{u}^{n+3/2},\tilde{v}^{n+3/2}]=\int_\Omega \Big( \frac{1}{2}|\tilde{u}^{n+3/2}|^2+\frac{1}{4}|\nabla \tilde{v}^{n+3/2}|^2\Big) d\Omega+C_0;\label{eq:chemsE}\\
 & \bs n \cdot\nabla u^{n+1}=d_a^{n+1},~~\bs n\cdot\nabla v^{n+1}=d_b^{n+1};\label{eq:chemsche4}
 \end{align}
 \end{subequations}
 and
  \begin{equation}\label{eq:chemsche3}
 \begin{split}
 &D_{\mathscr{F}}(R) \big|^{n+1}\left.\frac{dR}{dt}\right|^{n+1}=\int_\Omega \Big( u^{n+1}\frac{\partial u}{\partial t}\Big|^{n+1}-\frac{1}{2}\nabla^2 v^{n+1}\frac{\partial v}{\partial t}\Big|^{n+1} \Big)d\Omega\\
 &- \xi \int_\Omega \Big( |\nabla  \tilde{u}^{n+1}|^2+\frac{1}{2} |\nabla^2 \tilde{v}^{n+1}|^2+\frac{1}{2} |\nabla \tilde{v}^{n+1}|^2  \Big)d\Omega+\xi S_0+(1-\xi)|S_0|\\
 &-\int_{\Omega} u^{n+1}\Big( \nabla^2 u^{n+1}+ \xi \nabla\cdot(\bar u^{n+1}\nabla \bar v^{n+1})+f_1^{n+1}  \Big)d\Omega\\
 &+\int_{\Omega}\frac{1}{2}\nabla^2 v^{n+1} \Big( \nabla^2 v^{n+1}-v^{n+1}+ \xi p(\bar u^{n+1})+f_2^{n+1}  \Big)d\Omega.
 \end{split}
  \end{equation}
  In these equations, $\left.\frac{\partial u}{\partial t} \right|^{n+1}$,
  $\left.\frac{\partial v}{\partial t} \right|^{n+1}$ and
  $\left.\frac{d R}{d t} \right|^{n+1}$ are defined by equation \eqref{eq:tempdist}.
  $\bar u^{n+1}$ and $\bar v^{n+1}$ are defined by \eqref{eq:def_bar}.
  $\tilde{u}^{n+1}$ and $\tilde{v}^{n+1}$ are second-order approximations
  of $u^{n+1}$ and $v^{n+1}$ to be specified later in \eqref{eq:chemustar}.
  $\tilde{u}^{n+3/2}$ and $\tilde{v}^{n+3/2}$ are second-order approximations
  of $u^{n+3/2}$ and $v^{n+3/2}$ to be specified later in \eqref{eq:chemou32}.
$S_0$ in equation \eqref{eq:chemsche3} is given by
 \begin{equation}\label{eq:chemos0dis}
 \begin{split}
 S_0&= \int_\Omega \Big( f_1^{n+1} \tilde{u}^{n+1}+\frac{1}{2}\nabla f_2^{n+1}\cdot\nabla \tilde{v}^{n+1} \Big) d\Omega\\
 &+\int_{\Gamma}\Big( d_a^{n+1} \tilde{u}^{n+1}
 +\frac{1}{2}d_b^{n+1} (\tilde{u}^{n+1})^2+\frac{1}{2}d_b^{n+1}\frac{\partial v}{\partial t}\Big|^{*,n+1}+\frac{1}{2}d_b^{n+1}\tilde{v}^{n+1}-\frac{1}{2}d_b^{n+1} f_2^{n+1}\Big)d\Gamma,
 \end{split}
 \end{equation}
 where
\begin{equation}
  \left.\frac{\partial v}{\partial t}\right|^{*,n+1} =
  \frac{\frac{3}{2}\tilde{v}^{n+1} - 2v^n + \frac{1}{2}v^{n-1} }{\Delta t}.
  \label{eq:chemovtstar}
\end{equation}
  These equations are 
 supplemented by the following initial conditions
 \begin{equation}\label{eq:chemoint}
u^0=u_{\rm in}(\bs x),\quad v^0=v_{\rm in}(\bs x) ,\quad R^0=\mathscr{G}( E^0),\;\;\text{with}\;E^0=\int_\Omega \Big( \frac{1}{2}|u_{\rm in}|^2+\frac{1}{4}|\nabla v_{\rm in}|^2\Big) d\Omega+C_0.
 \end{equation}

 \begin{theorem}
   In the absence of the external force $f_1=f_2=0,$ and with homogeneous
   boundary conditions $d_a=d_b=0,$ the scheme consisting of \eqref{eq:chemsche1}-\eqref{eq:chemsche3} is unconditionally energy stable in the sense that:
 \begin{equation}\label{eq:chemodislaw}
   {\mathscr{F}(R^{n+\frac{3}{2}}) -\mathscr{F}(R^{n+\frac{1}{2}})}
   =-\xi \Delta{t}  \int_\Omega \Big( |\nabla  \tilde{u}^{n+1}|^2+\frac{1}{2} |\nabla^2 \tilde{v}^{n+1}|^2+\frac{1}{2} |\nabla \tilde{v}^{n+1}|^2  \Big)d\Omega\leqslant 0,
 \end{equation}
 if the approximation of $R(t)$ at the time step $\frac{1}{2}$
 is non-negative.
 \end{theorem}
This theorem can be proved in a way analogous to Theorem \ref{thm:thm_1}.
We can apply the method from Appendix A to this chemo-repulsion model
for the first time step, and this ensures that
$R^{n+1/2}|_{n=0}>0$.

 \subsection{Solution Algorithm and Implementation}


Using the notation \eqref{eq:timedirnew}, we rewrite equations \eqref{eq:chemsche1}-\eqref{eq:chemsche2} into
\begin{align}
&\frac{\gamma_0}{\Delta{t}}u^{n+1}-\nabla^2 u^{n+1}=\frac{\hat u}{\Delta{t}}+f_1^{n+1}+\xi \nabla \cdot \big( \bar u^{n+1}\nabla \bar v^{n+1} \big), \label{eq:chemous}\\
&\Big(\frac{\gamma_0}{\Delta{t}}+1  \Big)v^{n+1}-\nabla^2 v^{n+1}=\frac{\hat v}{\Delta{t}}+f_2^{n+1}+ \xi p(\bar u^{n+1}).\label{eq:chemovs}
\end{align}
Barring the unknown scalar $\xi,$ \eqref{eq:chemous} and \eqref{eq:chemovs} are two decoupled Helmholtz-type equations about $u^{n+1}$ and $v^{n+1},$ respectively.

Note that $\xi$ is a scalar number instead of a field function,
we define two sets of variables $(u_i^{n+1},v_i^{n+1})$ $(i=1,2)$ as the solutions to the following equations: 
\begin{align}
&\frac{\gamma_0}{\Delta{t}}u_1^{n+1}-\nabla^2 u_1^{n+1}=\frac{\hat u}{\Delta{t}}+f_1^{n+1},\quad \bs n \cdot\nabla u_1^{n+1}=d_a^{n+1}; \label{eq:chemou1s}\\
&\frac{\gamma_0}{\Delta{t}}u_2^{n+1}-\nabla^2 u_2^{n+1}=\nabla \cdot \big( \bar u^{n+1}\nabla \bar v^{n+1} \big),\quad \bs n \cdot\nabla u_2^{n+1}=0; \label{eq:chemou2s}\\
&\Big(\frac{\gamma_0}{\Delta{t}}+1  \Big)v_1^{n+1}-\nabla^2 v_1^{n+1}=\frac{\hat v}{\Delta{t}}+f_2^{n+1},\quad \bs n\cdot\nabla v_1^{n+1}=d_b^{n+1}; \label{eq:chemov1s}\\
&\Big(\frac{\gamma_0}{\Delta{t}}+1  \Big)v_2^{n+1}-\nabla^2 v_2^{n+1}=p(\bar u^{n+1}),\quad \bs n\cdot\nabla v_2^{n+1}=0.\label{eq:chemov2s}
\end{align}
Then we have the following result:
Given the scalar number $\xi,$ the following field functions solve the system consisting of equations \eqref{eq:chemous}-\eqref{eq:chemovs}:
\begin{equation}\label{eq:chemouvsolve}
u^{n+1}=u_1^{n+1}+\xi u_2^{n+1},\quad v^{n+1}=v_1^{n+1}+\xi v_2^{n+1},
\end{equation}
where $(u_i^{n+1},v_i^{n+1})$ $i=1,2$ is given by equations \eqref{eq:chemou1s}-\eqref{eq:chemov2s}, respectively. 

Once $(u_i^{n+1},v_i^{n+1})$ $i=1,2$ are known, we determine $\tilde{u}^{n+1}$, 
$\tilde{v}^{n+1}$, $\tilde{u}^{n+3/2}$ and $\tilde{v}^{n+3/2}$ according to
\eqref{eq:defstar}, specifically by
\begin{align}
&\tilde{u}^{n+1}=u_1^{n+1}+u_2^{n+1},\quad \tilde{v}^{n+1}=v_1^{n+1}+v_2^{n+1};\label{eq:chemustar}\\
  &\tilde{u}^{n+3/2}=\frac{3}{2}\tilde{u}^{n+1} - \frac{1}{2} u^n,\quad
  \tilde{v}^{n+3/2}=\frac{3}{2}\tilde{v}^{n+1} - \frac{1}{2} v^n.\label{eq:chemou32}
\end{align}
%
In light of equations \eqref{parabolicsystem2}, \eqref{eq:chemustar} and \eqref{eq:chemovtstar} , we compute $\nabla^2 \tilde{v}^{n+1}$ in equation \eqref{eq:chemsche3} by
\begin{equation}\label{eq:chemonabv}
  \nabla^2 \tilde{v}^{n+1}=\frac{\partial v}{\partial t}\Big|^{*,n+1}
  +\tilde{v}^{n+1}-p(\tilde{u}^{n+1})-f_2^{n+1},
\end{equation}
where $\left.\frac{\partial v}{\partial t} \right|^{*,n+1}$
is given by \eqref{eq:chemovtstar}.

Combining equations \eqref{eq:chemsche1}--\eqref{eq:chemsche2} and
\eqref{eq:chemsche3}, and using the property \eqref{eq:dist_prop},
we have
\begin{equation}\label{eq:chemosolver}
  \frac{\mathscr{F}(R^{n+\frac{3}{2}}) -\mathscr{F}(R^{n+\frac{1}{2}})}{\Delta{t}}
   =-\xi   \int_\Omega \Big( |\nabla  \tilde{u}^{n+1}|^2
   +\frac{1}{2} |\nabla^2 \tilde{v}^{n+1}|^2+\frac{1}{2} |\nabla \tilde{v}^{n+1}|^2
   \Big)d\Omega+\xi S_0+(1-\xi)|S_0|.
\end{equation}
This gives rise to
\begin{equation}\label{eq:chemoxif}
  \xi=\frac{\mathscr{F}(R^{n+1/2}) +{\Delta}t |S_0|}
           {
             E[\tilde{u}^{n+3/2},\tilde{v}^{n+3/2}]
             + \Delta{t} \left[
               \int_{\Omega}\Big( |\nabla  \tilde{u}^{n+1}|^2
   +\frac{1}{2} |\nabla^2 \tilde{v}^{n+1}|^2+\frac{1}{2} |\nabla \tilde{v}^{n+1}|^2
   \Big) d\Omega
               + (|S_0|-S_0)
               \right]
           },
\end{equation} 
in which $S_0$ is given by \eqref{eq:chemos0dis},
$\nabla^2\tilde{v}$ is to be computed by \eqref{eq:chemonabv},
and $E[\tilde{u}^{n+3/2},\tilde{v}^{n+3/2}]$ is given by \eqref{eq:chemsE}.
With $\xi$ known,
$R^{n+1}$ and $(u^{n+1},v^{n+1})$ can be evaluated directly
by \eqref{eq:RN1} and \eqref{eq:chemouvsolve}, respectively.

We employ $C^0$-continuous high-order spectral elements
for spatial discretizations in our implementation.
Note that equations \eqref{eq:chemou1s}--\eqref{eq:chemov2s}
involve Helmholtz type equations with Neumann type boundary conditions.
The weak formulations of these equations are: 
{Find $u_i^{n+1}$ and $v_i^{n+1}$ $\in H^1(\Omega)$ for $i=1,2,$ such that}
\begin{align*}
&\big(\nabla u_1^{n+1},\nabla\varphi \big)_{\Omega}+\frac{\gamma_0}{\Delta{t}}\big(u_1^{n+1},\varphi \big)_{\Omega}=\big(\frac{\hat u}{\Delta{t}}+f_1^{n+1},\varphi \big)_{\Omega}+\big \langle d_a^{n+1},\varphi \big\rangle_{\Gamma},\\
&\big(\nabla u_2^{n+1},\nabla \varphi \big)_{\Omega}+\frac{\gamma_0}{\Delta{t}}\big(u_2^{n+1},\varphi \big)_{\Omega}=-\big( \bar u^{n+1}\nabla \bar v^{n+1} ,\nabla \varphi \big)_{\Omega}+\big \langle \bs n\cdot \nabla \bar v^{n+1} \bar u^{n+1}, \varphi \big \rangle_{\Gamma},\\
&\big(\nabla v_1^{n+1},\nabla\varphi \big)_{\Omega} + \big(\frac{\gamma_0}{\Delta{t}}+1  \big)\big(v_1^{n+1},\varphi \big)_{\Omega}=\big(\frac{\hat v}{\Delta{t}}+f_2^{n+1},\varphi \big)_{\Omega}+\big \langle d_b^{n+1},\varphi \big \rangle_{\Gamma},\\
&\big(\nabla v_2^{n+1},\nabla\varphi \big)_{\Omega} + \big(\frac{\gamma_0}{\Delta{t}}+1  \big)\big(v_2^{n+1},\varphi \big)_{\Omega}=\big(p(\bar u^{n+1}),\varphi \big)_{\Omega},
\end{align*}
for $\forall \varphi \in H^1(\Omega)$, where
\begin{equation}
  (f,g)_{\Omega} = \int_{\Omega} f(\bs x)g(\bs x) d\Omega, \quad
  \langle f,g \rangle_{\Gamma} = \int_{\Gamma} f(\bs x)g(\bs x) d\Gamma.
\end{equation}
These weak forms can be discretized using $C^0$ spectral elements in the standard
way~\cite{KarniadakisS2005}.


\subsection{Numerical Results}

\subsubsection{Convergence Rate}

\begin{figure}[tbp]
 \subfigure[Errors vs Element order]{ \includegraphics[scale=.39]{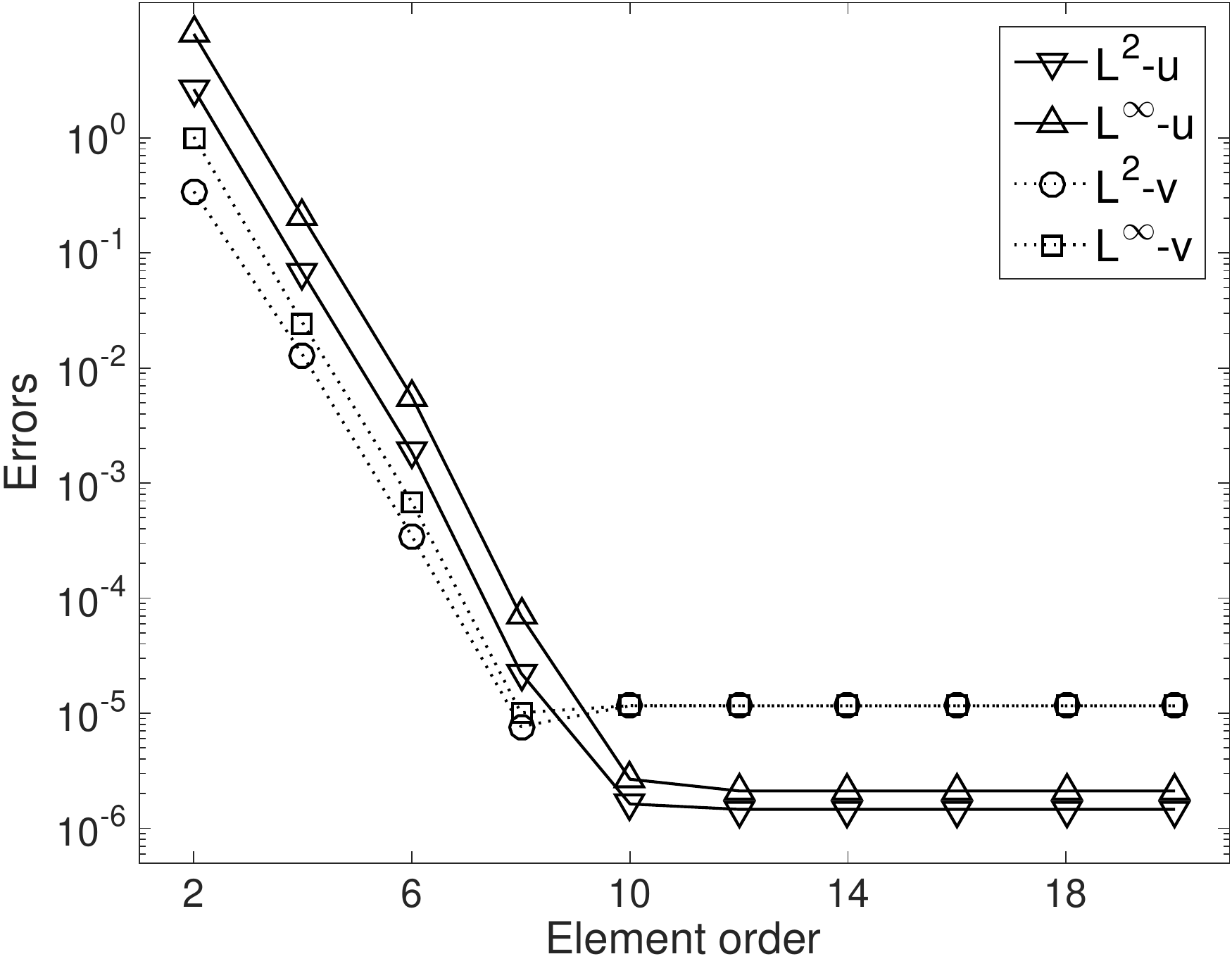}} \quad
 \subfigure[Errors vs ${\Delta}t$]{ \includegraphics[scale=.38]{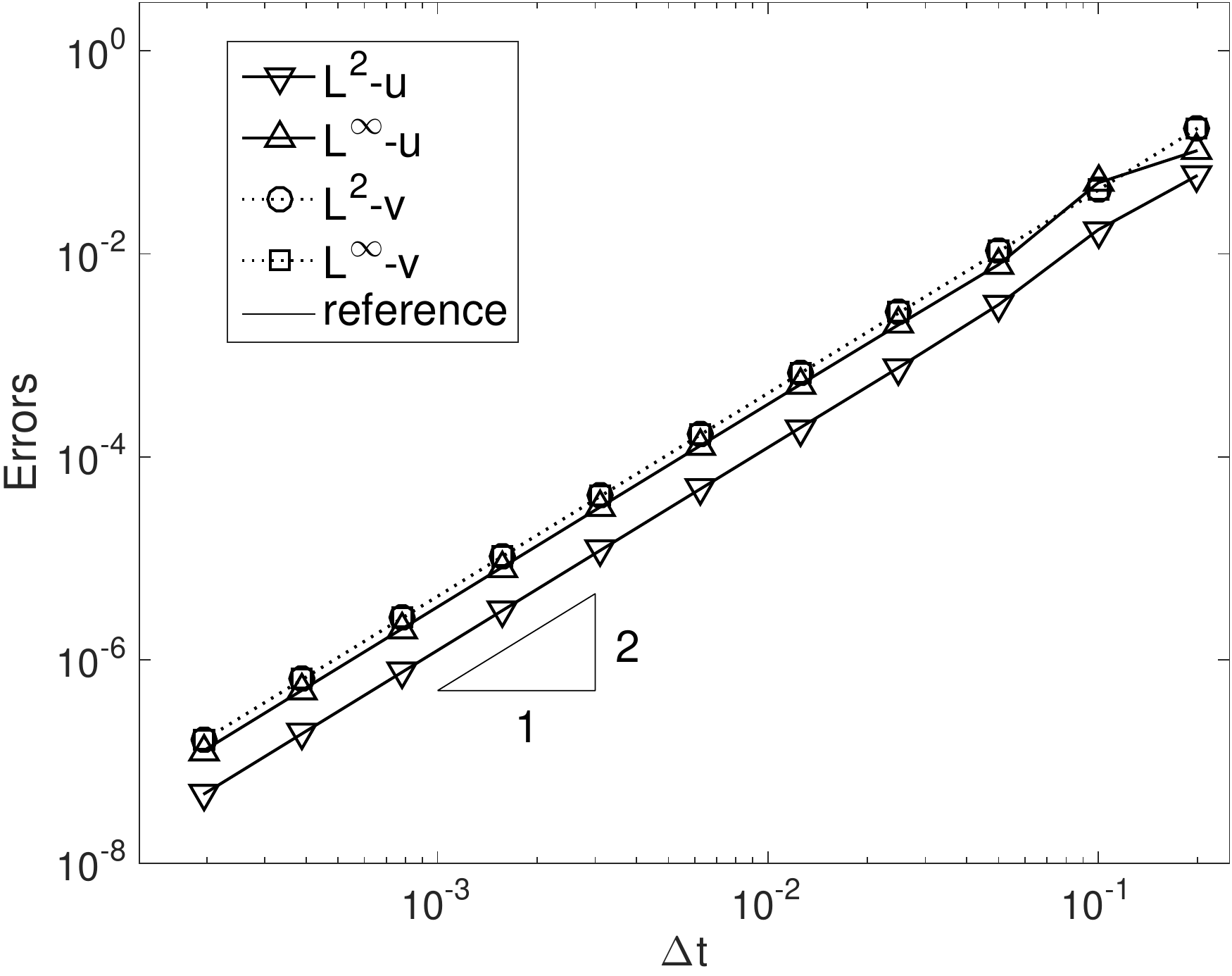}} \\
 \caption{Spatial/temporal convergence tests for chemo-repulsion model: $L^{2}$ and $L^{\infty}$ errors of $u$ and $v$ versus (a) element order (fixed $\Delta t=0.001$ and $t_f=0.1$),
   and (b) $\Delta t$ (fixed element order $18$ and $t_f=1$).}
\label{fig:chemostandtest}
\end{figure}

We first employ a manufactured analytical solution to the chemo-repulsion model
to demonstrate the spatial and temporal convergence rates of the proposed algorithm.

Consider the computational domain $\Omega=[0, 1]^2$ and the following contrived solution to the system \eqref{parabolicsystem} on this domain
\begin{equation}\label{eq:chemostd1}
u=\exp({-t})\big(\cos(2\pi x)\cos(2\pi y)+2\big),\quad v=\big(1+\sin(t)\big)\big(\cos(2\pi x)\cos(2\pi y)+2\big).
\end{equation}
The external forces $f_1(\bs x,t),$ $f_2(\bs x,t)$ and boundary forces $d_a(\bs x,t),$ $d_b(\bs x,t)$ therein are chosen such that the expressions in \eqref{eq:chemostd1} satisfy \eqref{parabolicsystem} .

The domain is discretized with four equal-sized quadrilateral elements. The initial cell density $u_{ in}$ and initial chemical concentration $v_{ in}$ are given according to the analytic expressions in \eqref{eq:chemostd1} by setting $t=0.$ We simulate this problem from $t=0$ to $t=t_f.$ Then we compare the numerical solutions of $u$ and $v$ at $t=t_f$  with the analytic solutions in \eqref{eq:chemostd1} and various norms of the errors are computed.
The element order and time step sizes are varied systematically in order to investigate their effects on the numerical errors. We employ the function $\mathscr{F}(R)=R$ for
defining the auxiliary variable $R(t)$
and the energy constant $C_0=1$ in the following convergence tests.

We first study the spatial convergence rate. A fixed $t_f=0.1$ and $\Delta{t}=0.001$ is employed and the element order is varied systematically between 2 and 20. We record the errors at $t=t_f$ between the numerical solution and the contrived solution \eqref{eq:chemostd1} in both $L^{\infty}$ and $L^2$ norms with respect to the element orders. Figure \ref{fig:chemostandtest}(a)
shows these numerical errors as a function of the element order. We observe an exponential decrease of the numerical errors with increasing element order, and a level-off of the error curves beyond element order 10 and 8, respectively for $u$ and $v$, due to the saturation of temporal errors.

The study of the temporal convergence rate is summarized by the results in
Figure \ref{fig:chemostandtest}(b). Here
we fix the integration time $t_f=1.0$ and the element order at a large value 18,
and vary $\Delta{t}$ systematically between $0.2$ and $1.953125\times 10^{-4}.$
This figure demonstrates the $L^{\infty}$ and $L^2$ errors of $u$ and $v$ as a function of $\Delta{t}$. It is evident that the proposed
scheme has a second-order convergence rate in time.

\subsubsection{Study of Unconditional Stability and Effect of Algorithmic Parameters}

\begin{figure}[tbp]
  \centering
 \subfigure[$t=0$]{ \includegraphics[scale=.36]{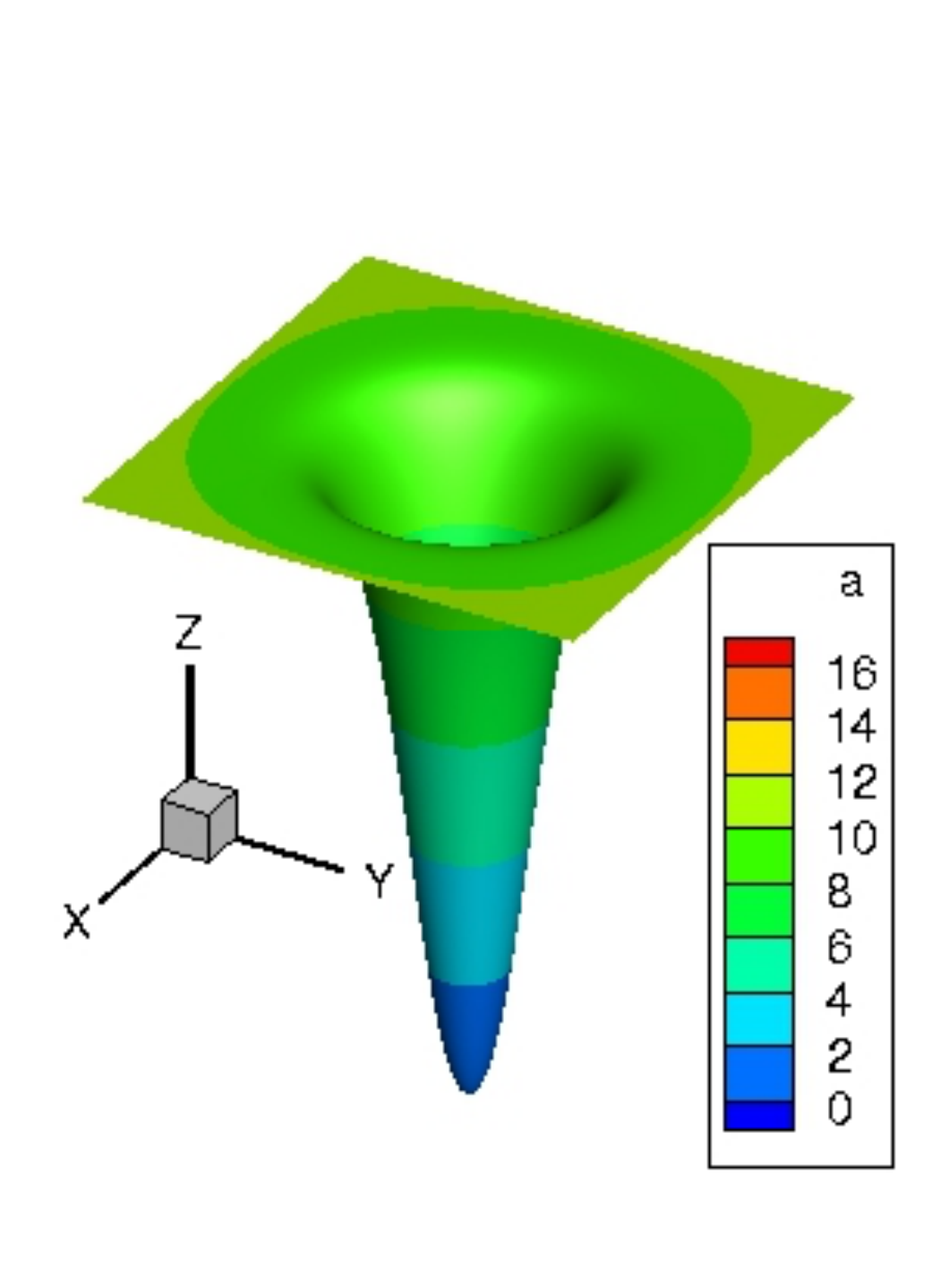}} \quad
 \subfigure[$t=10^{-2}$]{ \includegraphics[scale=.35]{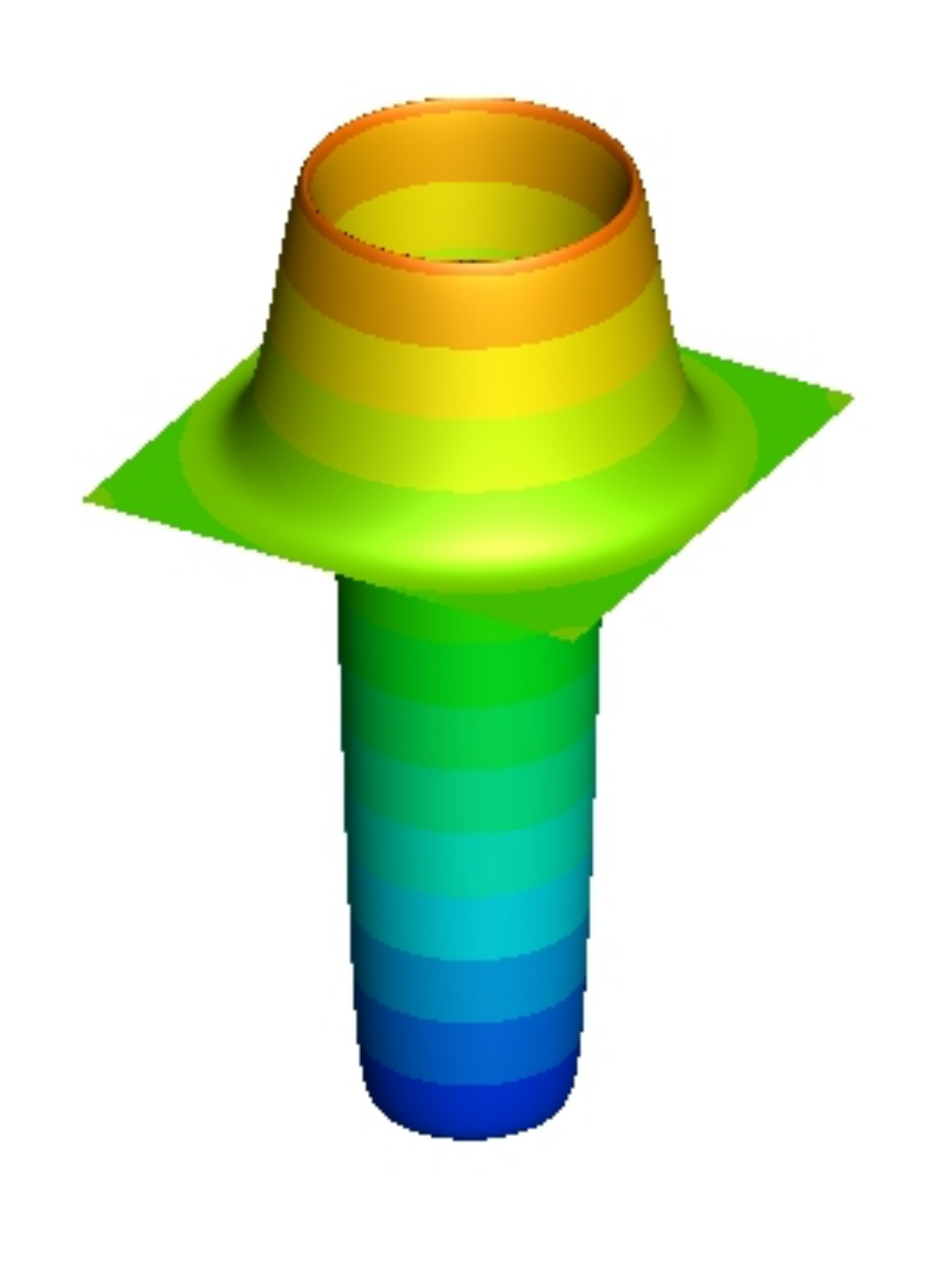}} \quad
  \subfigure[$t=2\times 10^{-2}$]{ \includegraphics[scale=.35]{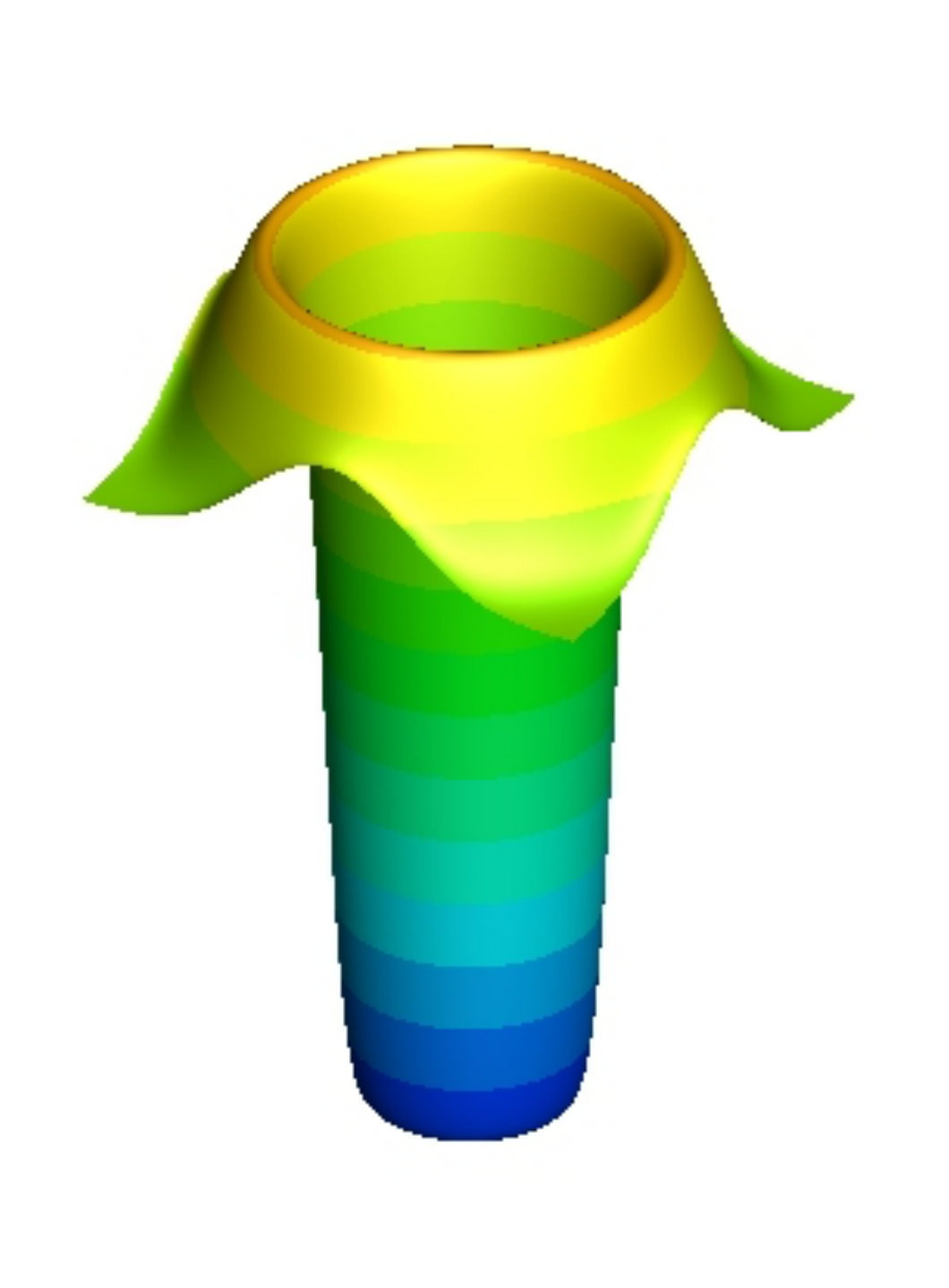}} \\
  \subfigure[$t=3\times 10^{-2}$]{ \includegraphics[scale=.35]{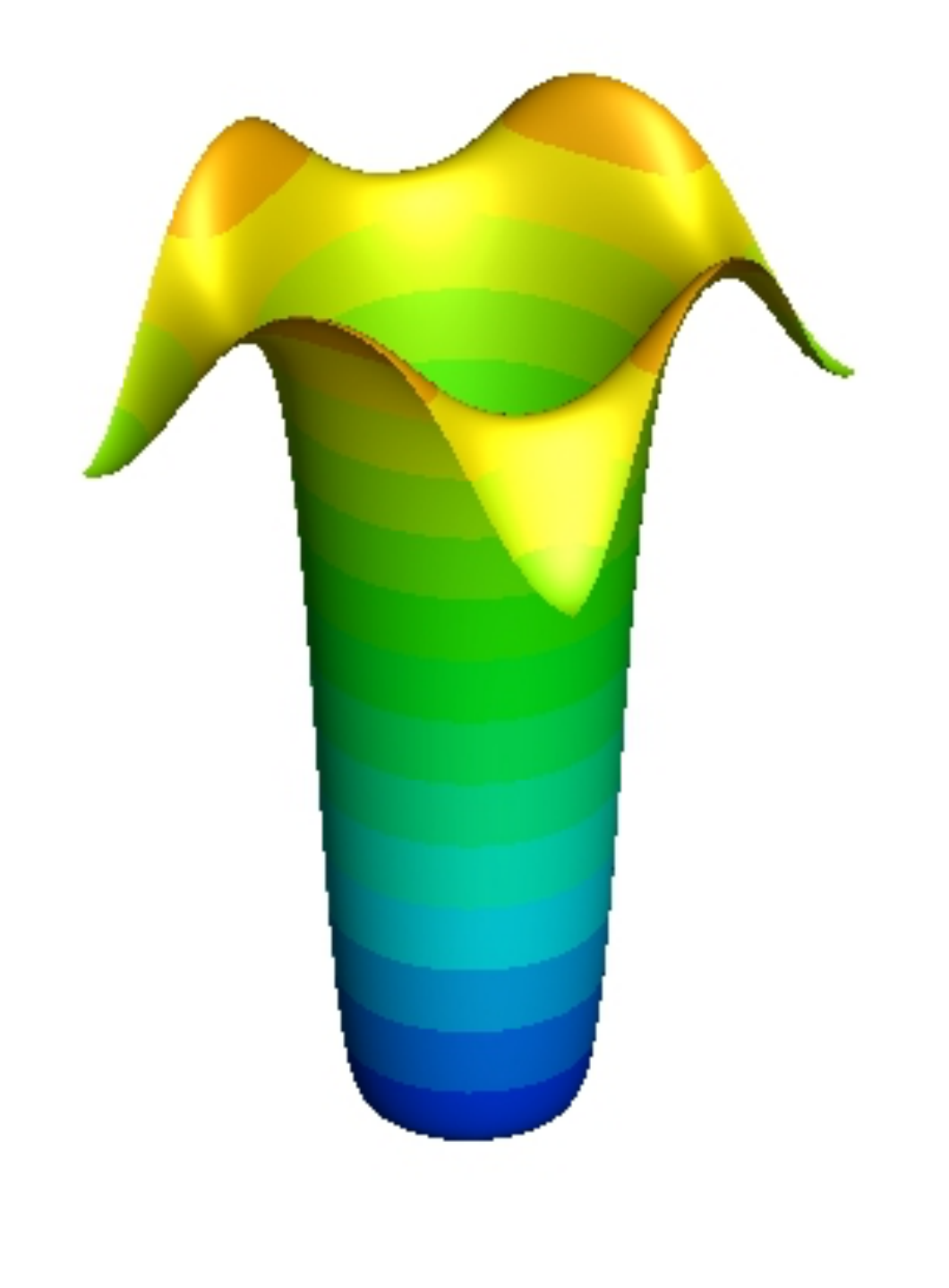}}\quad
  \subfigure[$t=5\times 10^{-2}$]{ \includegraphics[scale=.35]{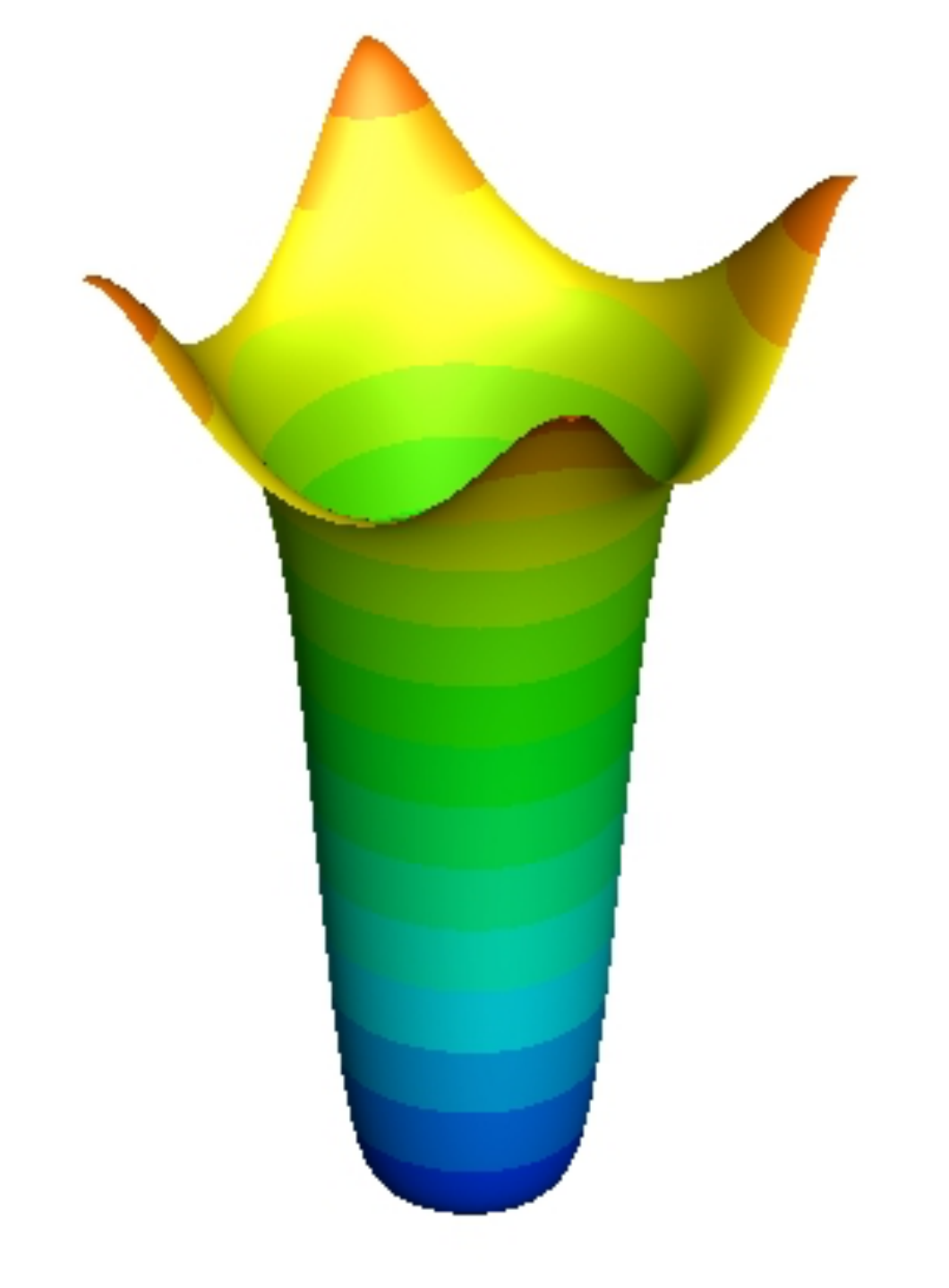}}\quad
  \subfigure[$t=7.5\times 10^{-2}$]{ \includegraphics[scale=.35]{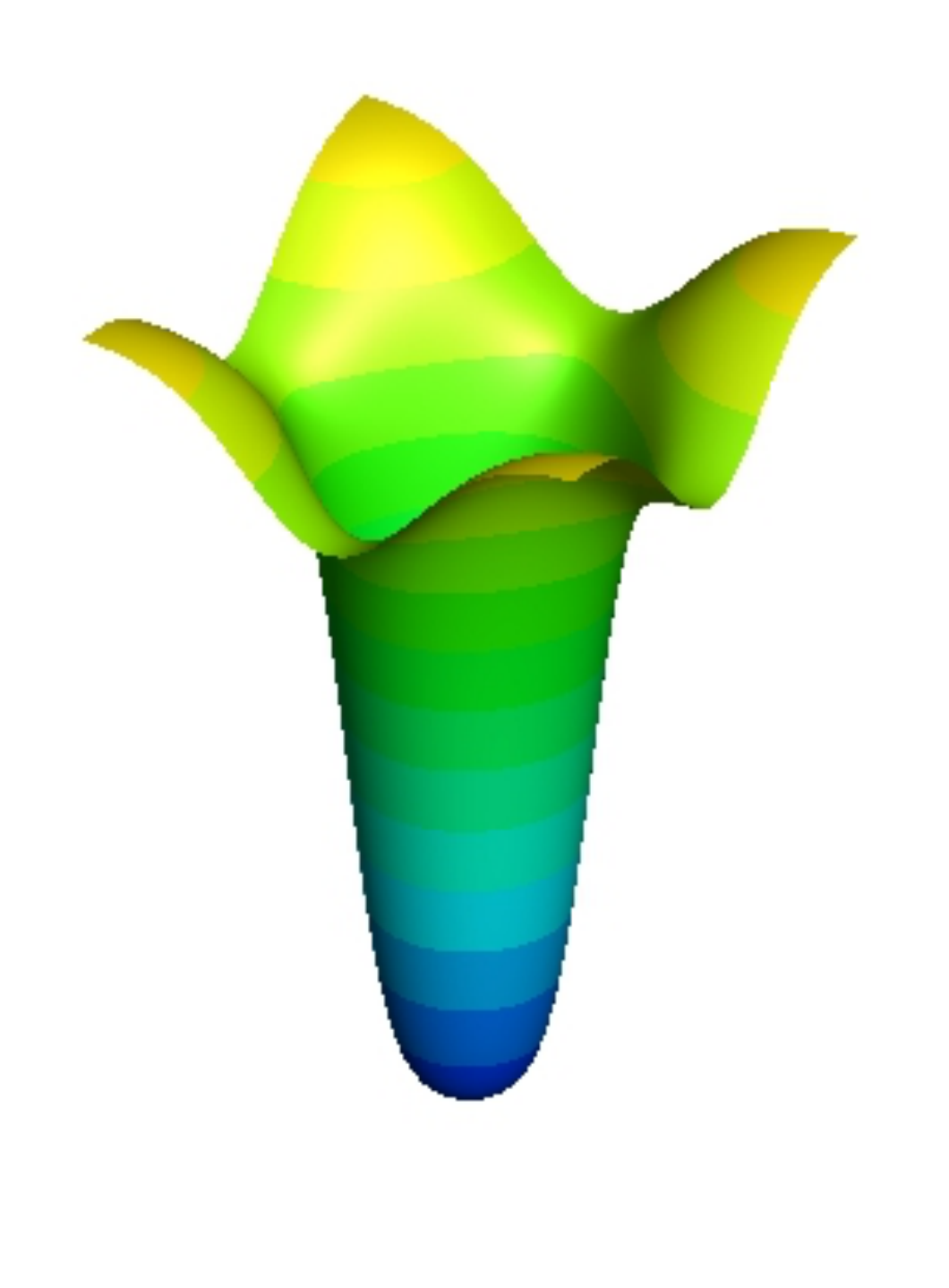}}\\
  \subfigure[$t=0.1$]{ \includegraphics[scale=.35]{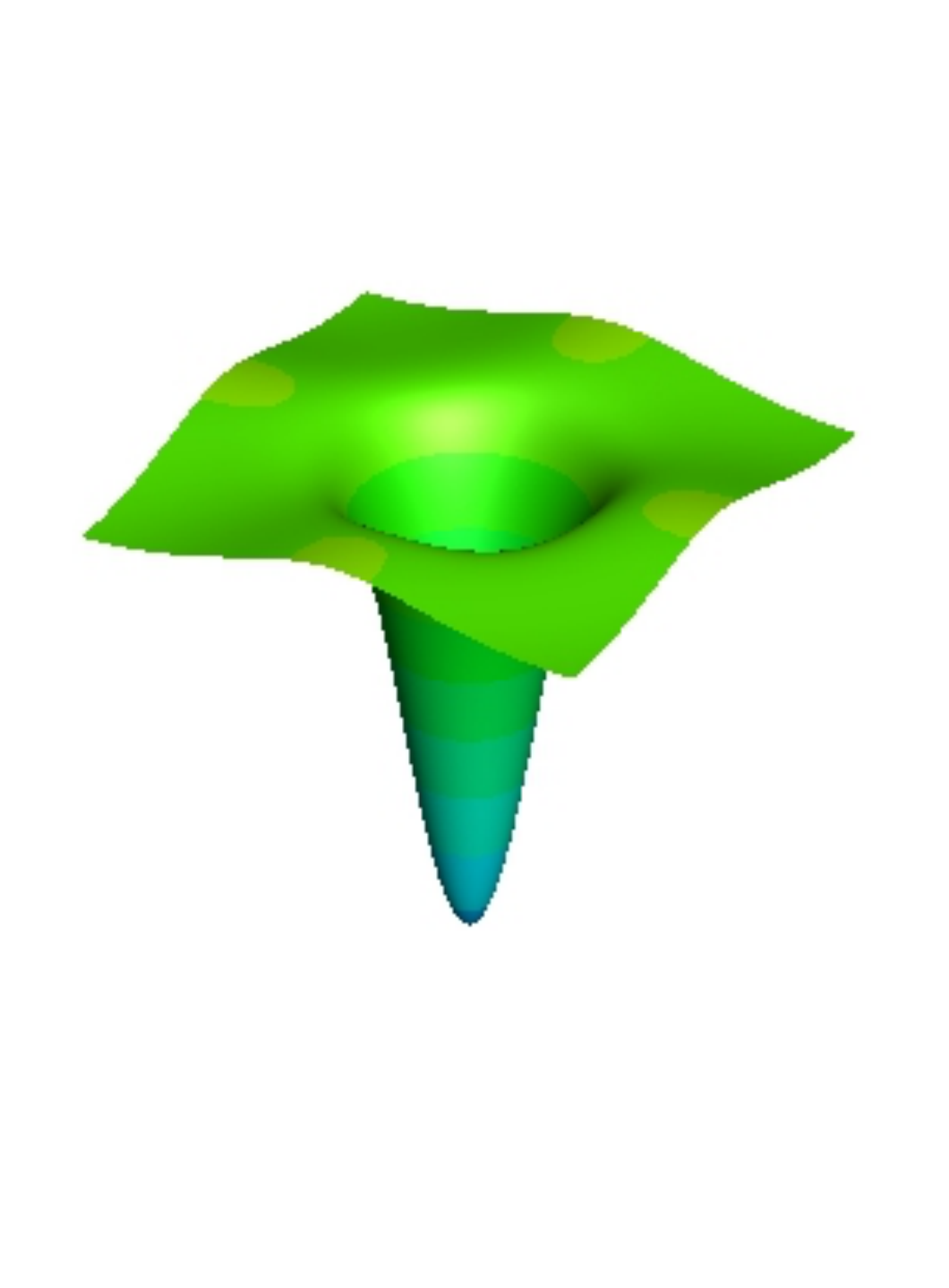}}\quad
  \subfigure[$t=0.2$]{ \includegraphics[scale=.35]{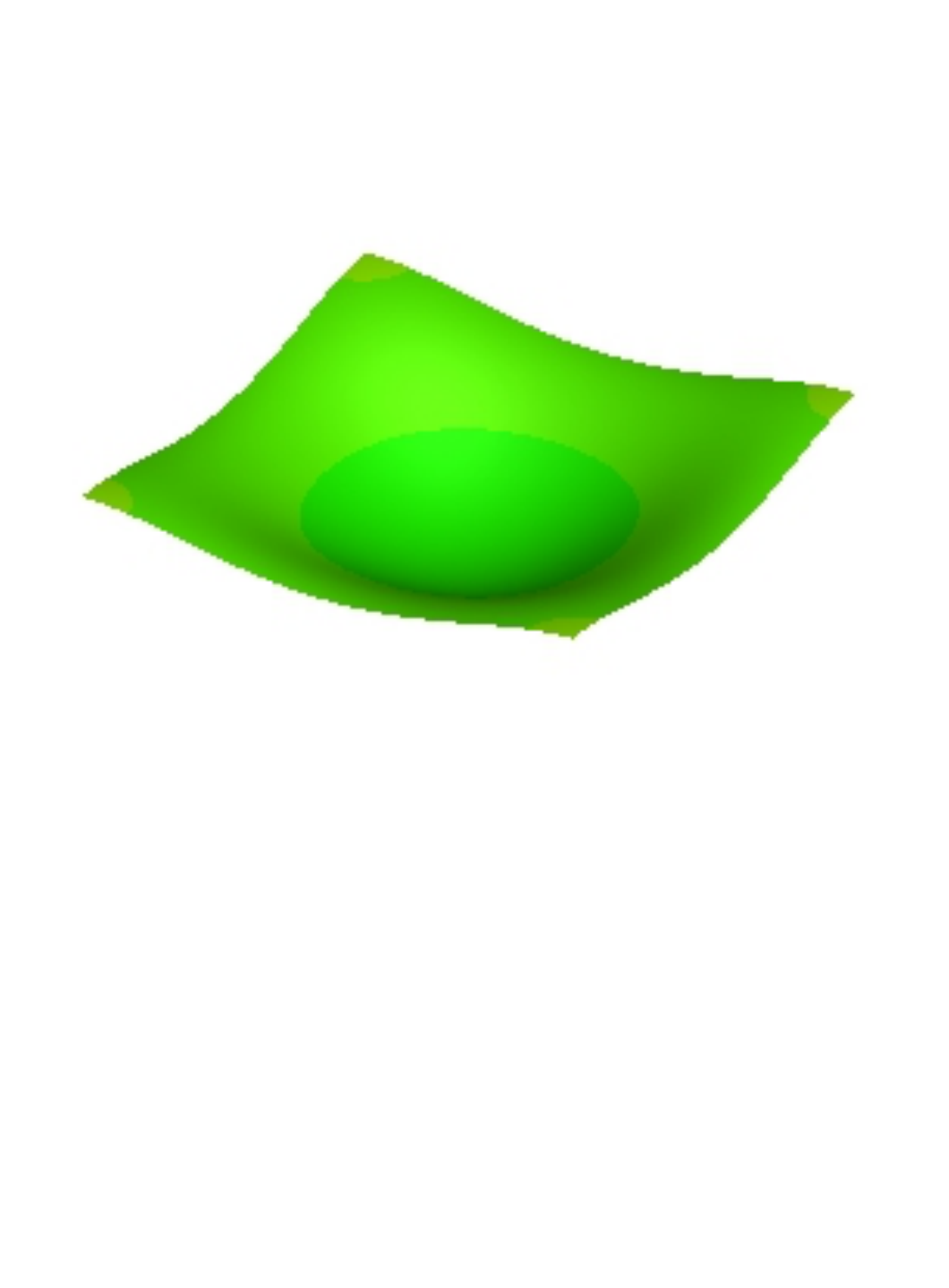}}\quad
  \subfigure[$t=0.5$]{ \includegraphics[scale=.35]{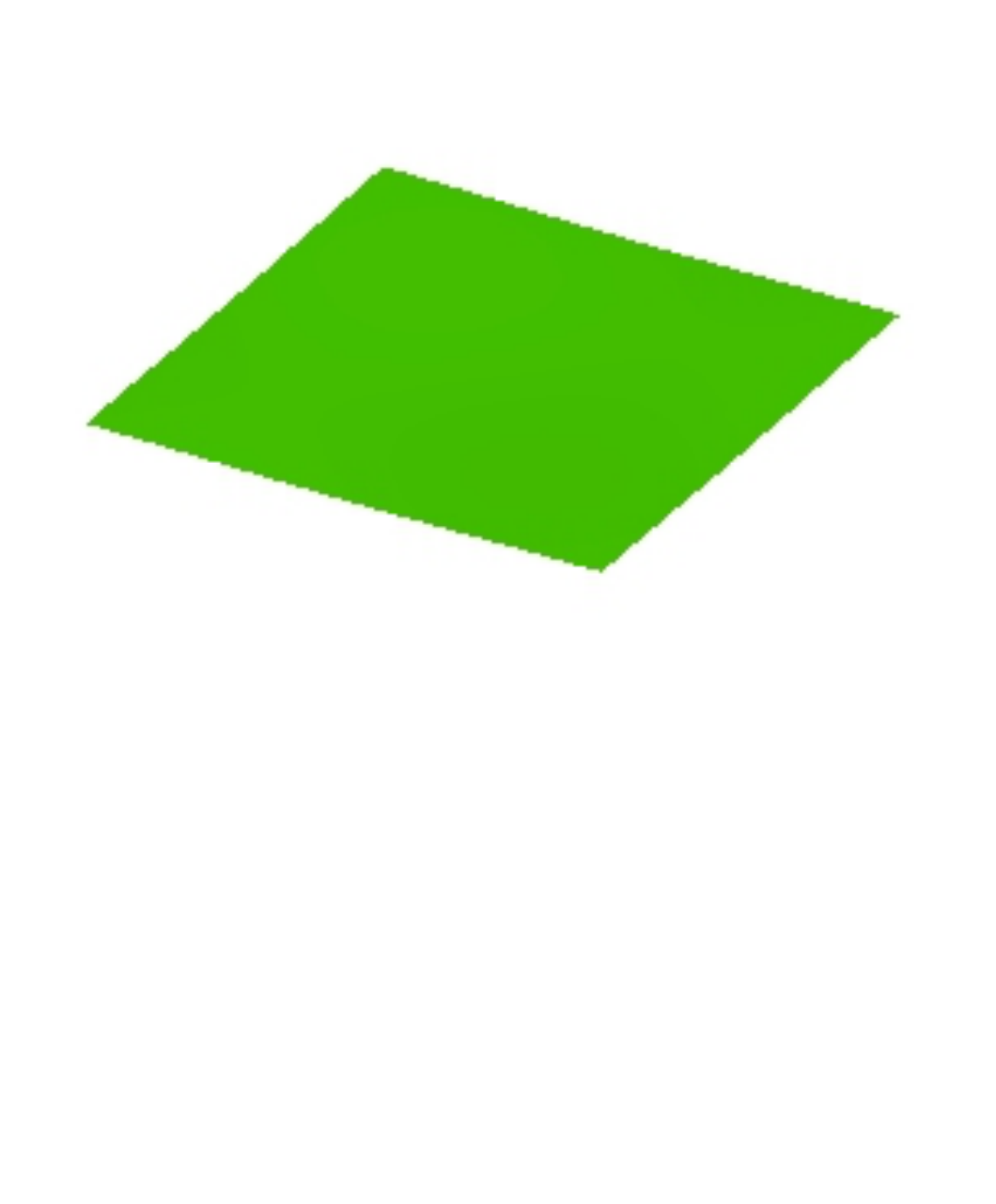}}\quad
  \caption{Chemo-repulsion model:
    Temporal sequence of snapshots of the cell density $u$ distribution
    visualized by its contours.
    The color map in (a) applies to all the plots.}
\label{fig:chemodynau}
\end{figure}

\begin{figure}[tbp]
  \centering
 \subfigure[$t=0$]{ \includegraphics[scale=.36]{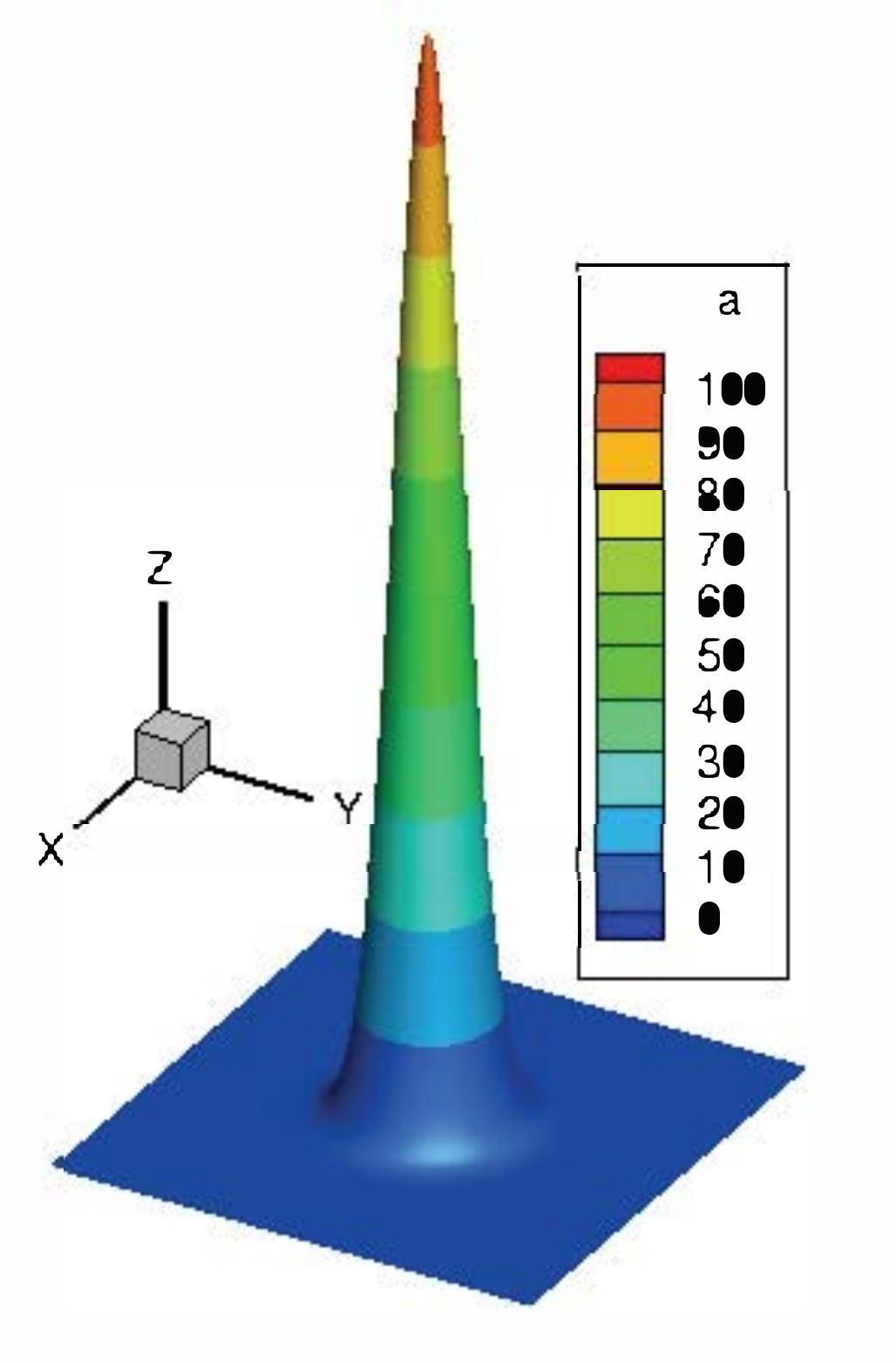}} \quad
 \subfigure[$t=10^{-2}$]{ \includegraphics[scale=.35]{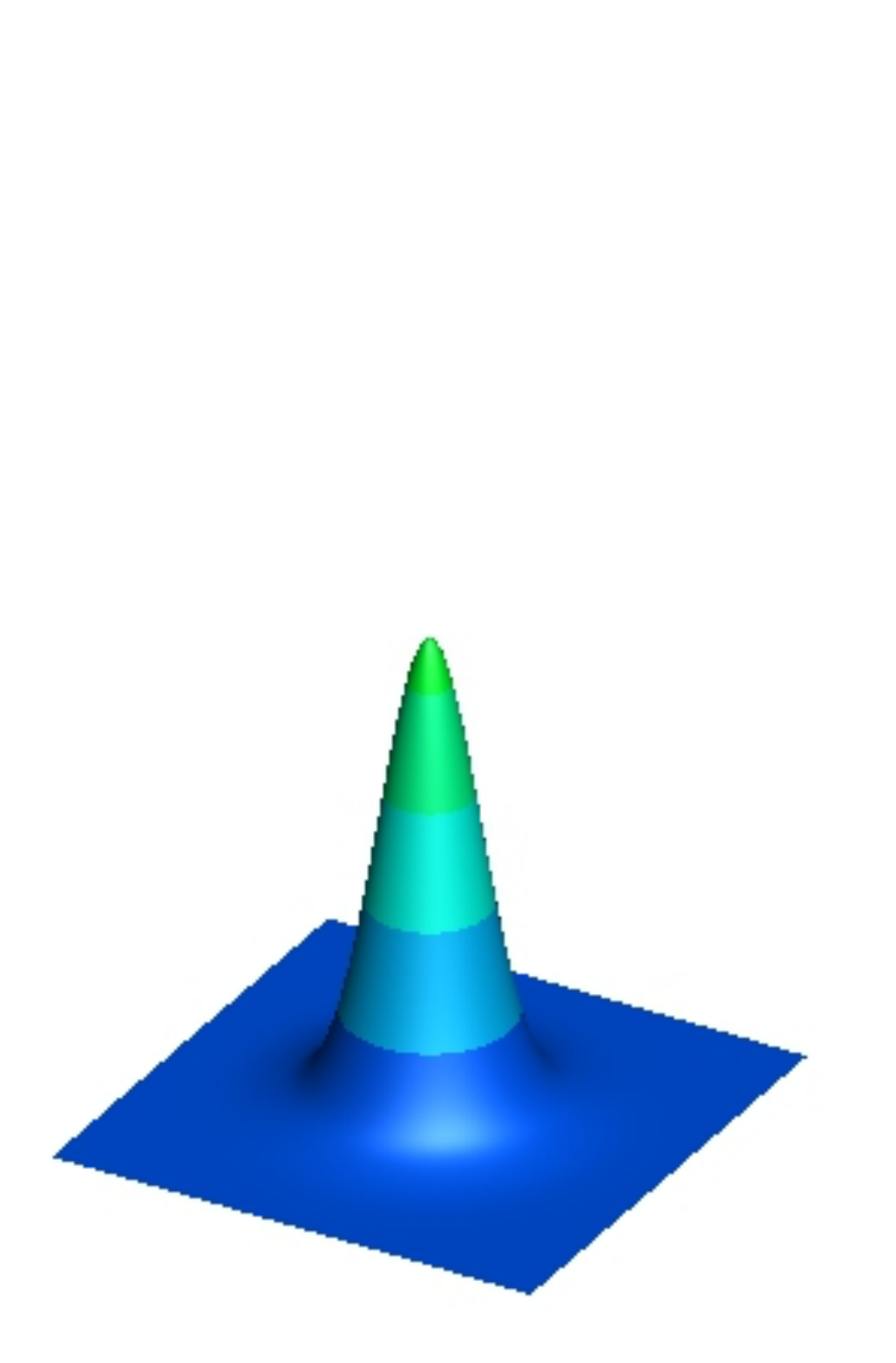}} \quad
  \subfigure[$t=2\times 10^{-2}$]{ \includegraphics[scale=.35]{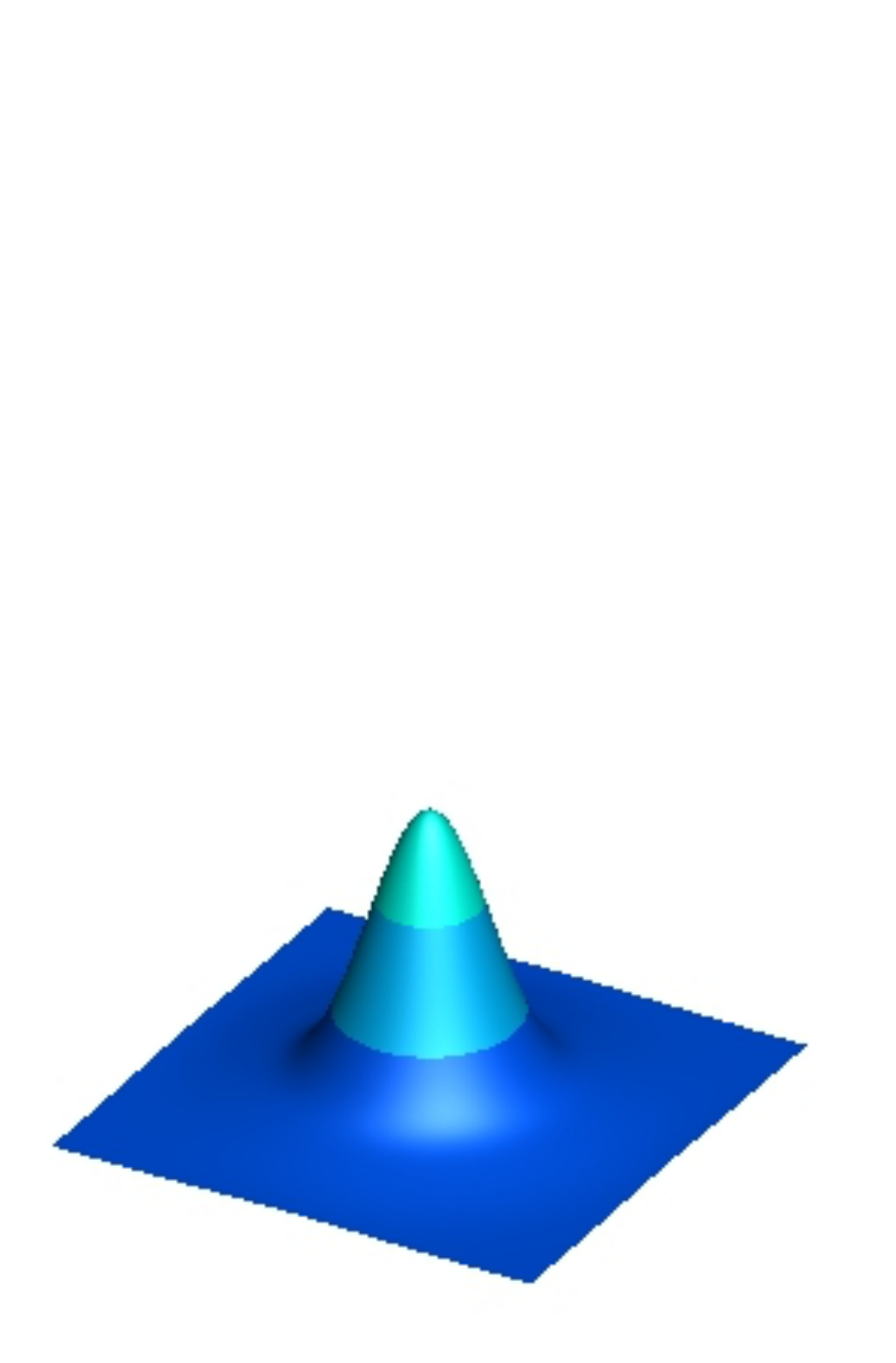}} \\
  \subfigure[$t=3\times 10^{-2}$]{ \includegraphics[scale=.35]{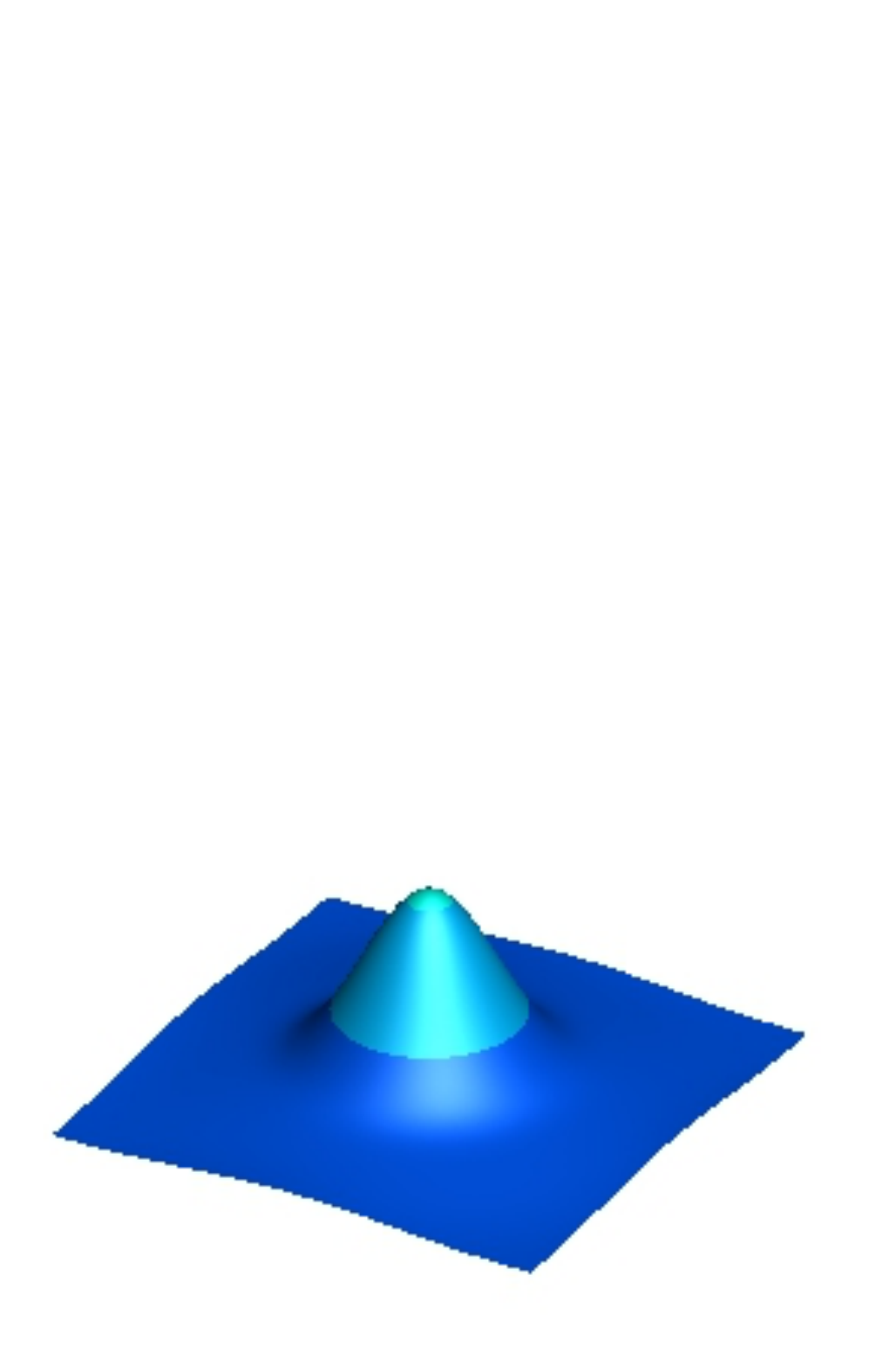}}\quad
  \subfigure[$t=5\times 10^{-2}$]{ \includegraphics[scale=.35]{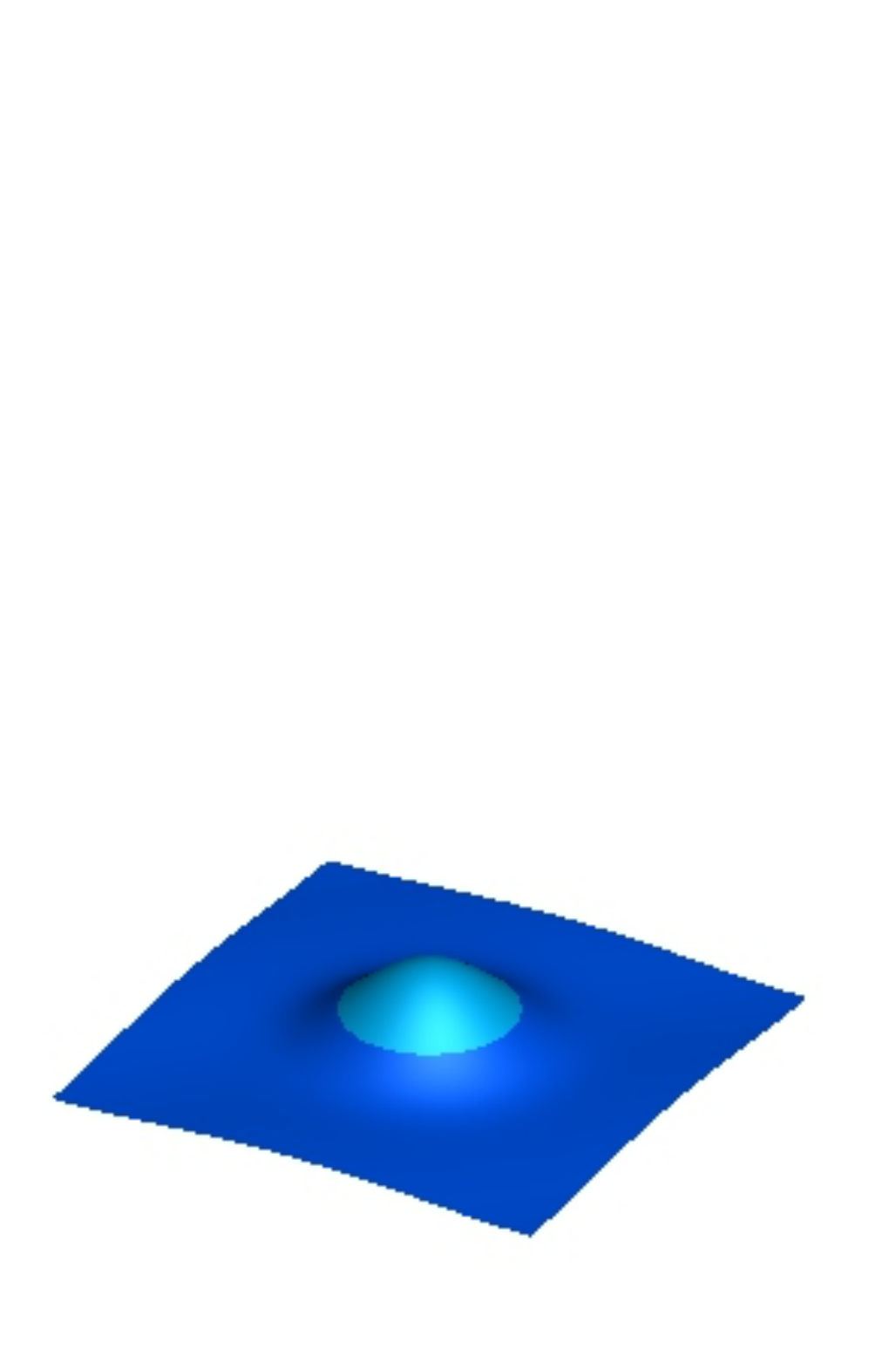}}\quad
  \subfigure[$t=7.5\times 10^{-2}$]{ \includegraphics[scale=.35]{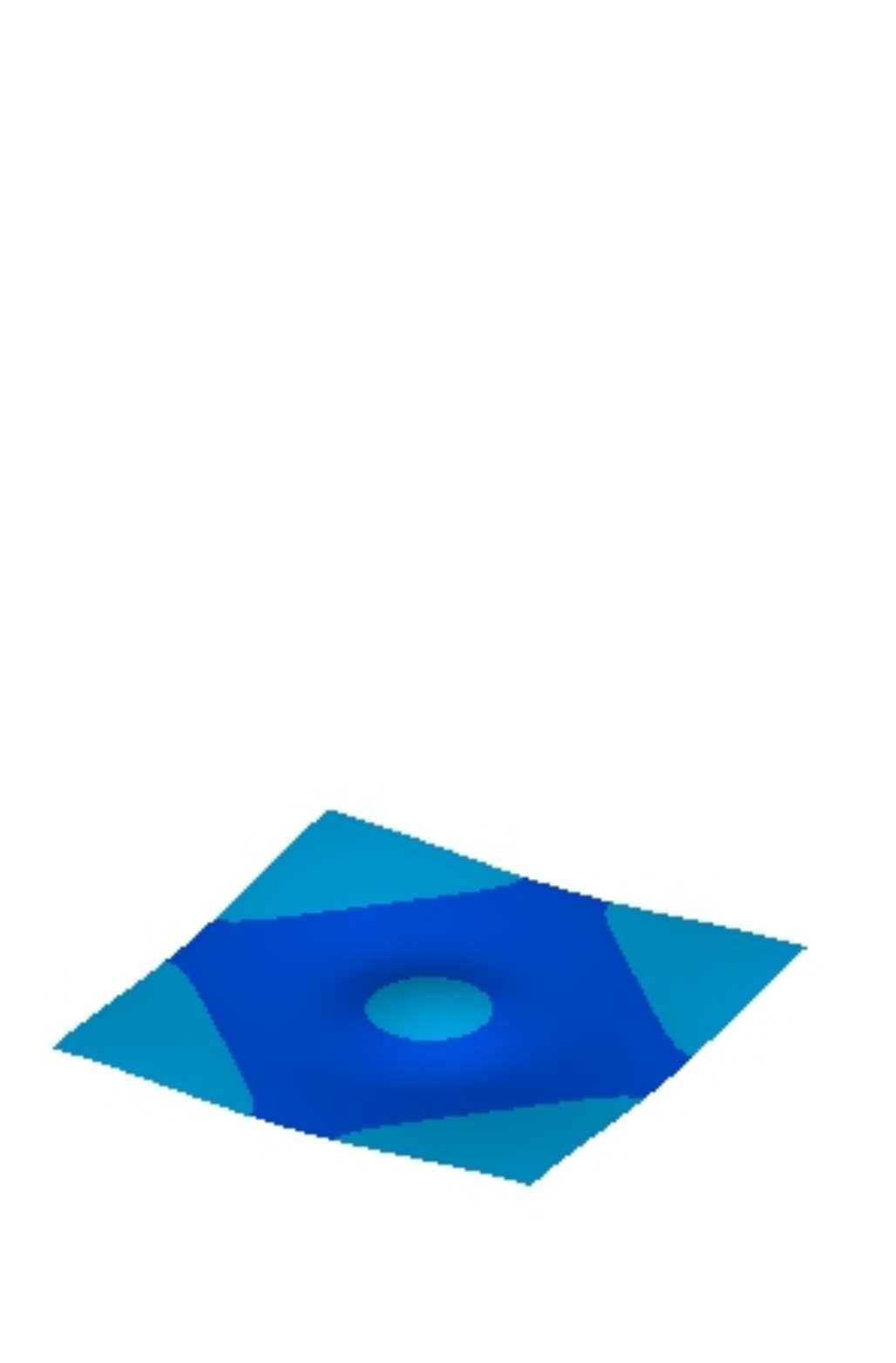}}\\
  \subfigure[$t=0.1$]{ \includegraphics[scale=.35]{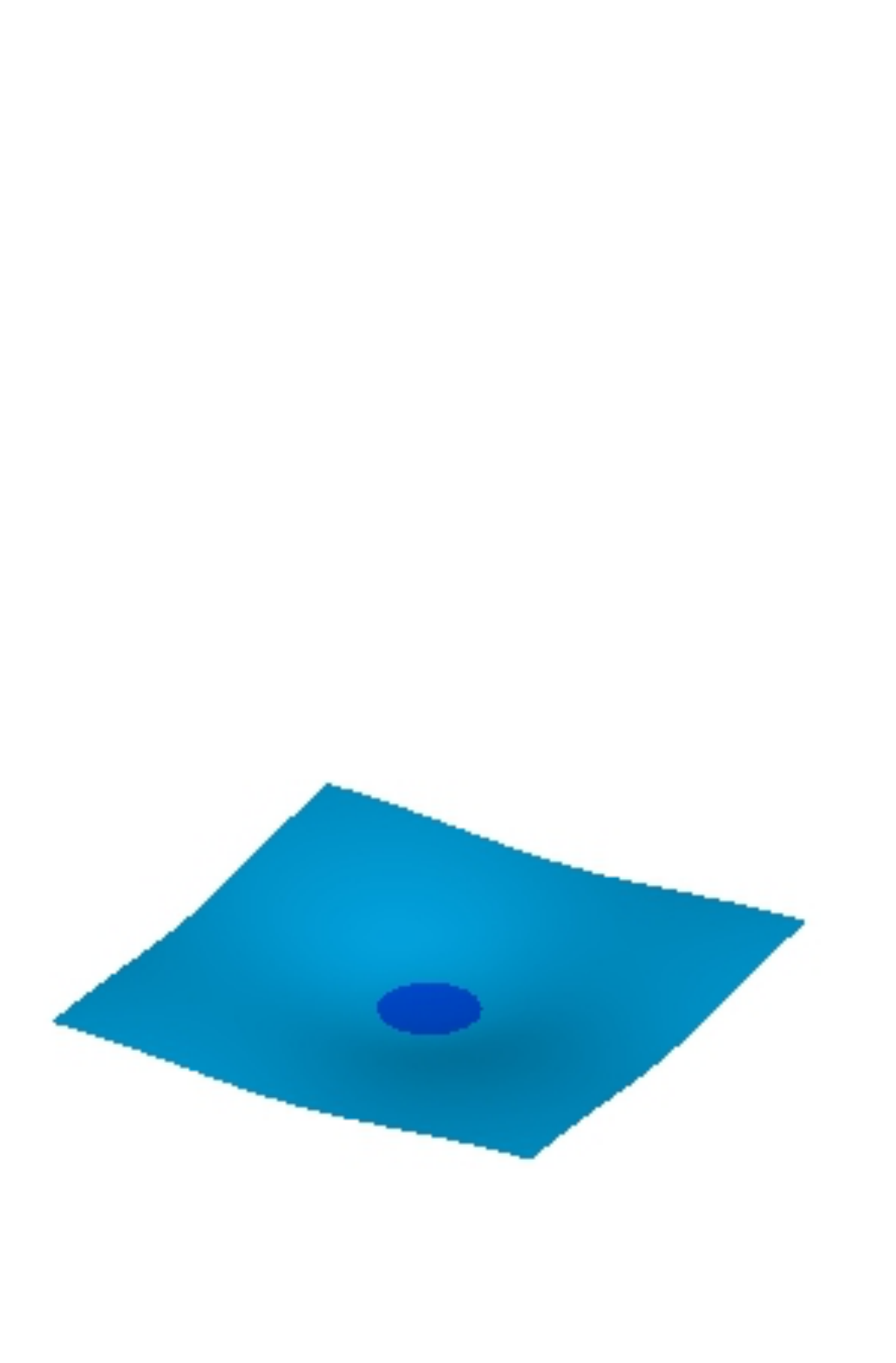}}\quad
  \subfigure[$t=0.2$]{ \includegraphics[scale=.35]{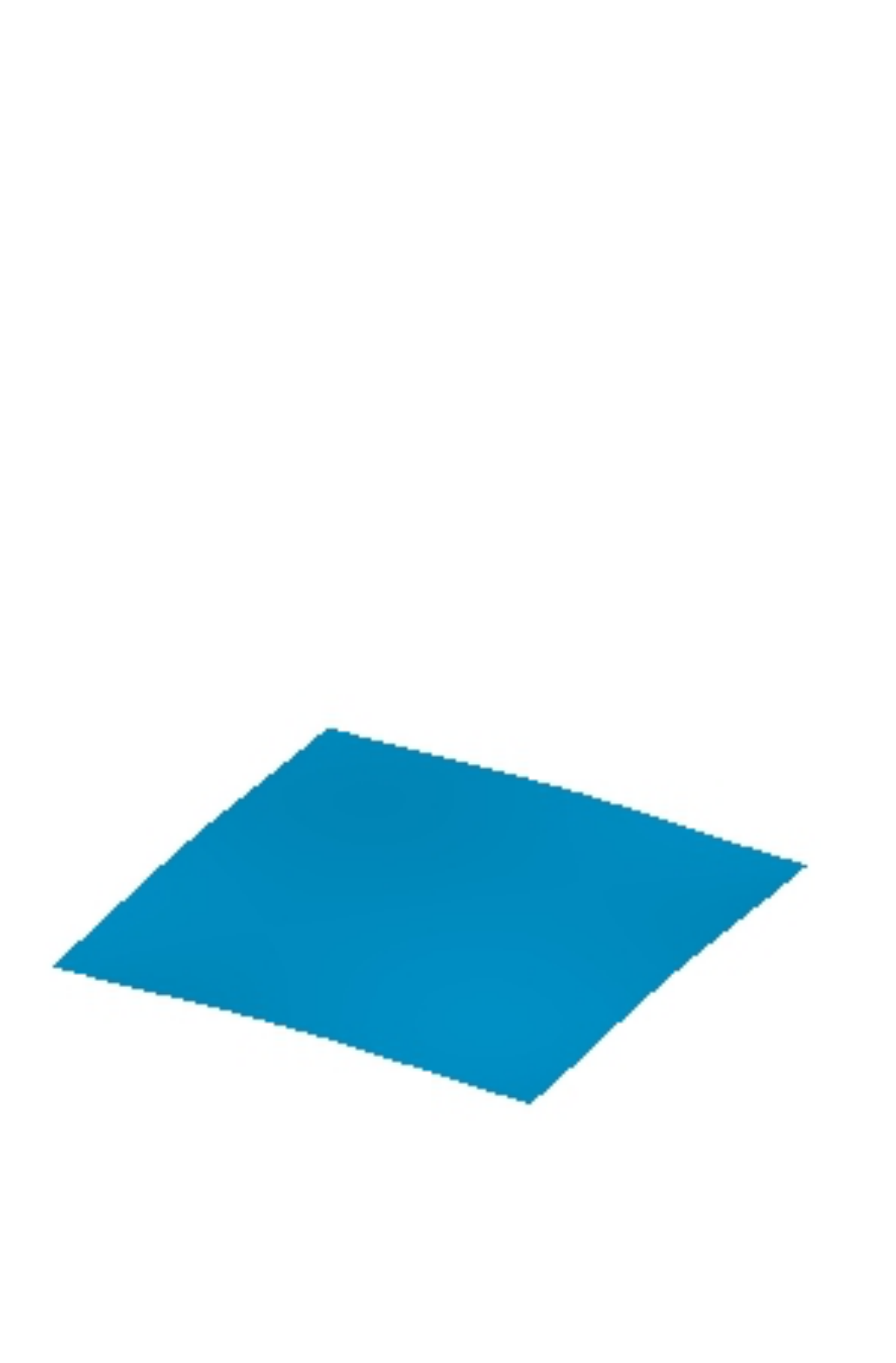}}\quad
  \subfigure[$t=0.5$]{ \includegraphics[scale=.35]{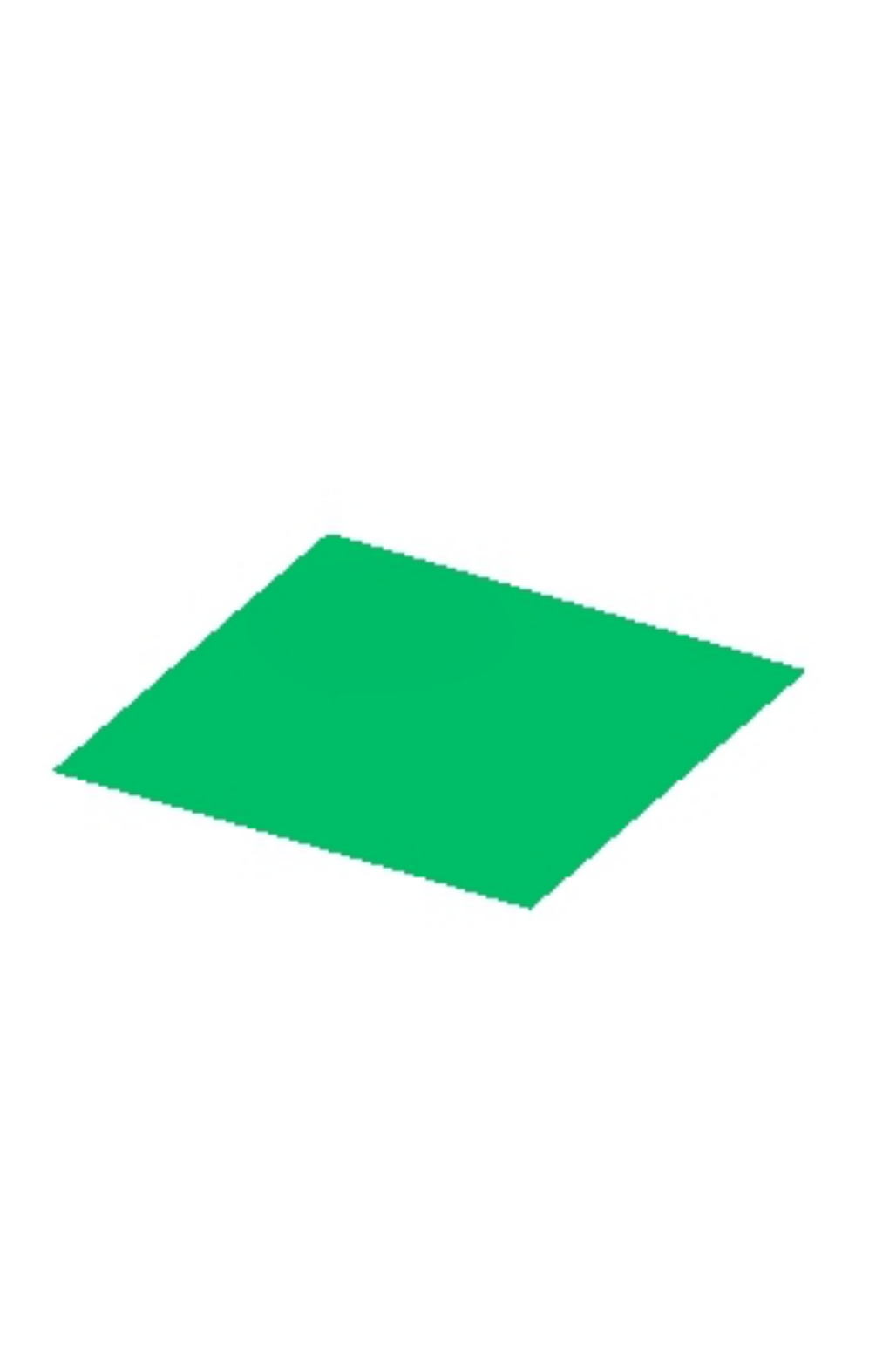}}\quad
  \caption{Chemo-repulsion model:
    Temporal sequence of snapshots of the chemical concentration $v$ visualized
    by its contours.
    The color map in (a) applies to all the plots herein.}
\label{fig:chemodynav}
\end{figure}

We next consider the test problem used in  \cite{Gonzalez2019}, and
show the efficiency and unconditional stability of the method proposed here.
Consider the domain $\Omega=[0,2]^2$ and the initial distributions for the
cell density $u$
and chemical concentration $v$ in this domain given by
\begin{subequations}\label{eq:chemoinitial}
\begin{align}
&u_{ in}(\bs x)=-10xy(2-x)(2-y)\exp(-10(y-1)^2-10(x-1)^2)+10.0001, \label{eq:chemouinitial}\\
&v_{ in}(\bs x)=100 xy(2-x)(2-y)\exp(-30(y-1)^2-30(x-1)^2)+0.0001.\label{eq:chemovinitial}
\end{align}
\end{subequations}
The external forces and boundary forces in \eqref{parabolicsystem} are set to $f_1=f_2=d_a=d_b=0.$ The computational domain is discretized with 400 equal-sized quadrilateral elements, and
the element order is fixed to be 10.

Figures \ref{fig:chemodynau} and \ref{fig:chemodynav} demonstrate
the dynamics of the system. These results are obtained with
$\Delta{t}=10^{-5}$, $\mathscr{F}(R)=R$ and $C_0=1$ in the numerical algorithm.
Figure \ref{fig:chemodynau} shows the evolution of the cell density $u(\bs x,t)$ 
with a temporal sequence of snapshots of the distribution
visualized by the contour plots. The $z$ coordinate corresponds to $u$ in these plots.
The system exhibits a very rapid dynamics.
The initial cell density has a Gaussian type distribution,
taking a minimal value 0.0001 at the domain center $\bs x_0=(1,1)$
and gradually approaching  the maximal value 10.0001 near the domain boundary.
In a very short time $t=10^{-2},$ the maximal density increases to around 16,
attained near the boundary of a circular region with radius 0.6
and center at $\bs x_0$; see Figure \ref{fig:chemodynau}(b). Then the maximal density gradually moves from the circular boundary to the domain boundary between $t=2\times 10^{-2}$ and
$t=7.5\times 10^{-2}$; see Figure \ref{fig:chemodynau}(c)-(f).
The high density near the domain boundary then appears to diffuse
to the low density region near the center $\bs x_0$, and the system
finally reaches an  equilibrium state between $t=0.1$ and $t=0.5$
with a constant density level;
see Figure \ref{fig:chemodynau}(g)-(i).
Figure \ref{fig:chemodynav} illustrates the evolution of the chemical
concentration $v(\bs x,t)$.
Figure \ref{fig:chemodynav}(a) shows the distribution
of the initial chemical concentration. It has also a Gaussian type distribution,
with a maximal value 100.0001 at the origin $\bs x_0$ and decreasing to 0.0001 gradually near the domain boundary. The concentration diffuses rapidly between $t=0$ to $t=5\times 10^{-2}$ (Figures \ref{fig:chemodynav}(a)-(e)), and the maximal concentration
decreases to around 10 at the origin. From $t=7.5\times 10^{-2}$ to $t=0.2,$ the contrast in the concentration levels in the domain becomes even smaller (Figure \ref{fig:chemodynav}(f)-(h)),
and the concentration reaches its equilibrium with a constant level around 36.6 (Figure \ref{fig:chemodynav}(i)).

\begin{figure}[tbp]
  \centering
 \subfigure[]{ \includegraphics[scale=.39]{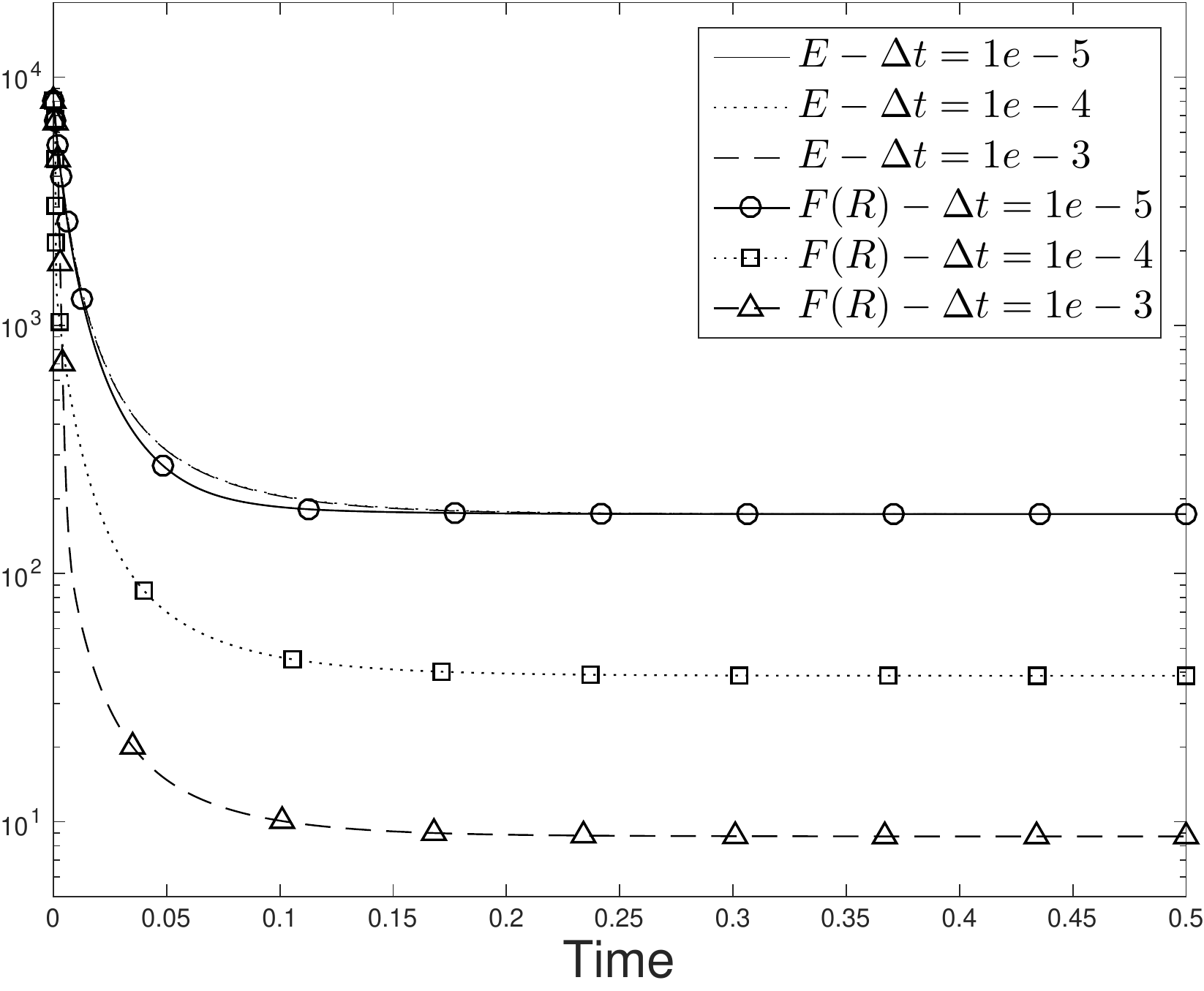}} \quad
 \subfigure[]{ \includegraphics[scale=.39]{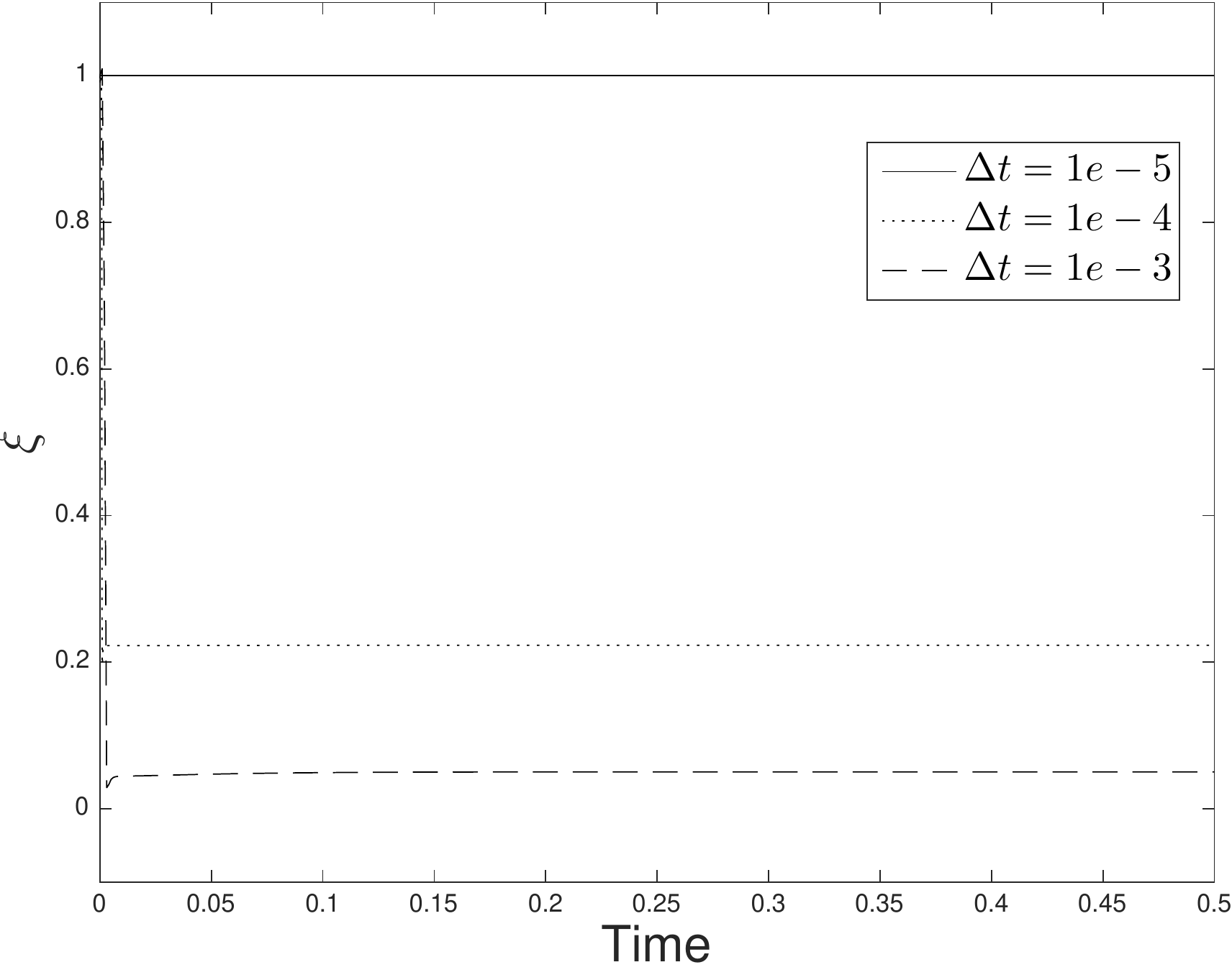}} \\
 \caption{Chemo-repulsion model: time histories of  (a) $E(t)$ and $\mathscr{F}(R)$,
   and (b) $\xi={\mathscr{F}(R)}/{E(t)}$,
   for several $\Delta{t}=10^{-3}, 10^{-4}, 10^{-5}.$
 }
\label{fig:chemovarydt}
\end{figure}

\begin{figure}[tbp]
  \centering
 \subfigure[]{ \includegraphics[scale=.39]{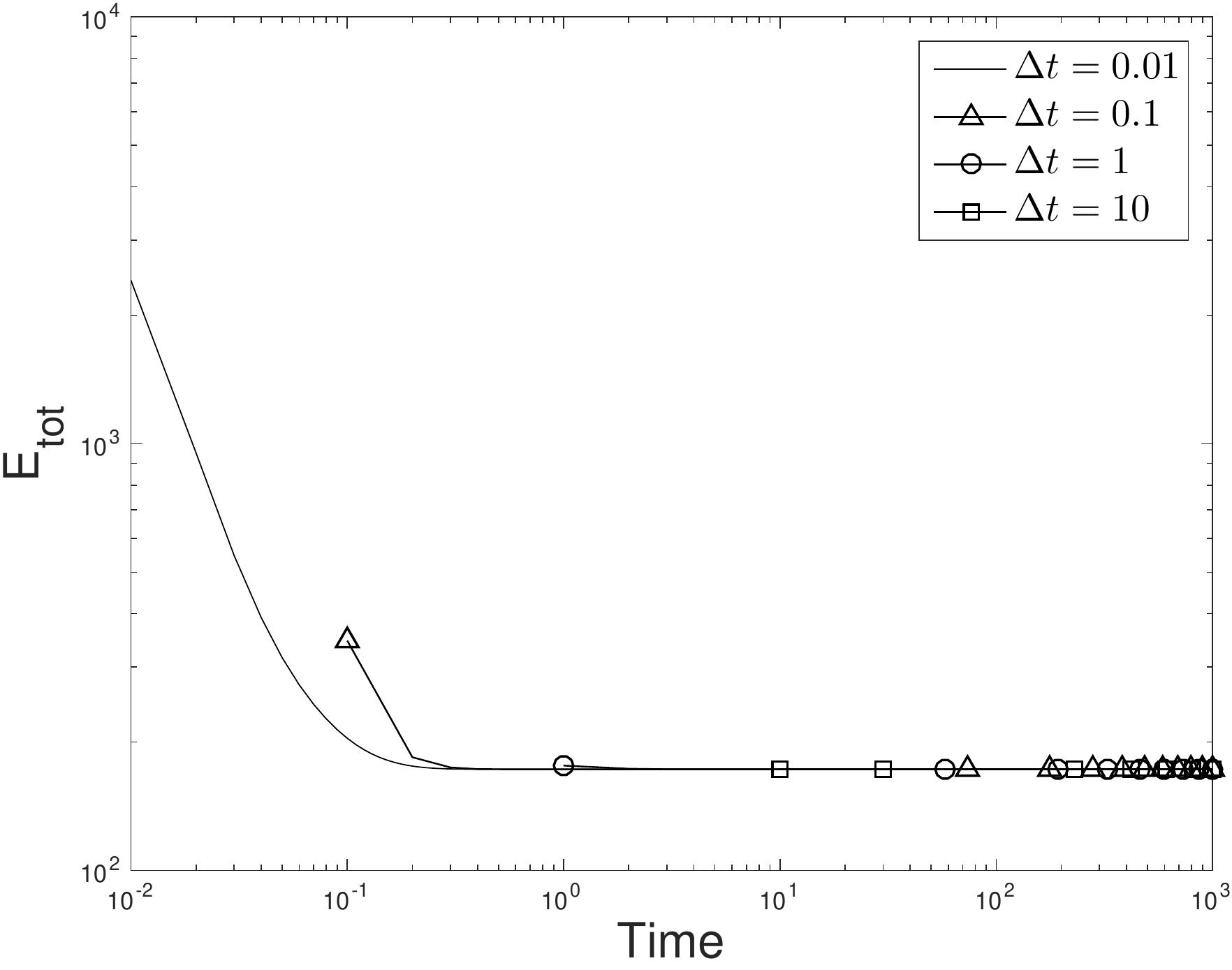}} \quad
 \subfigure[]{ \includegraphics[scale=.39]{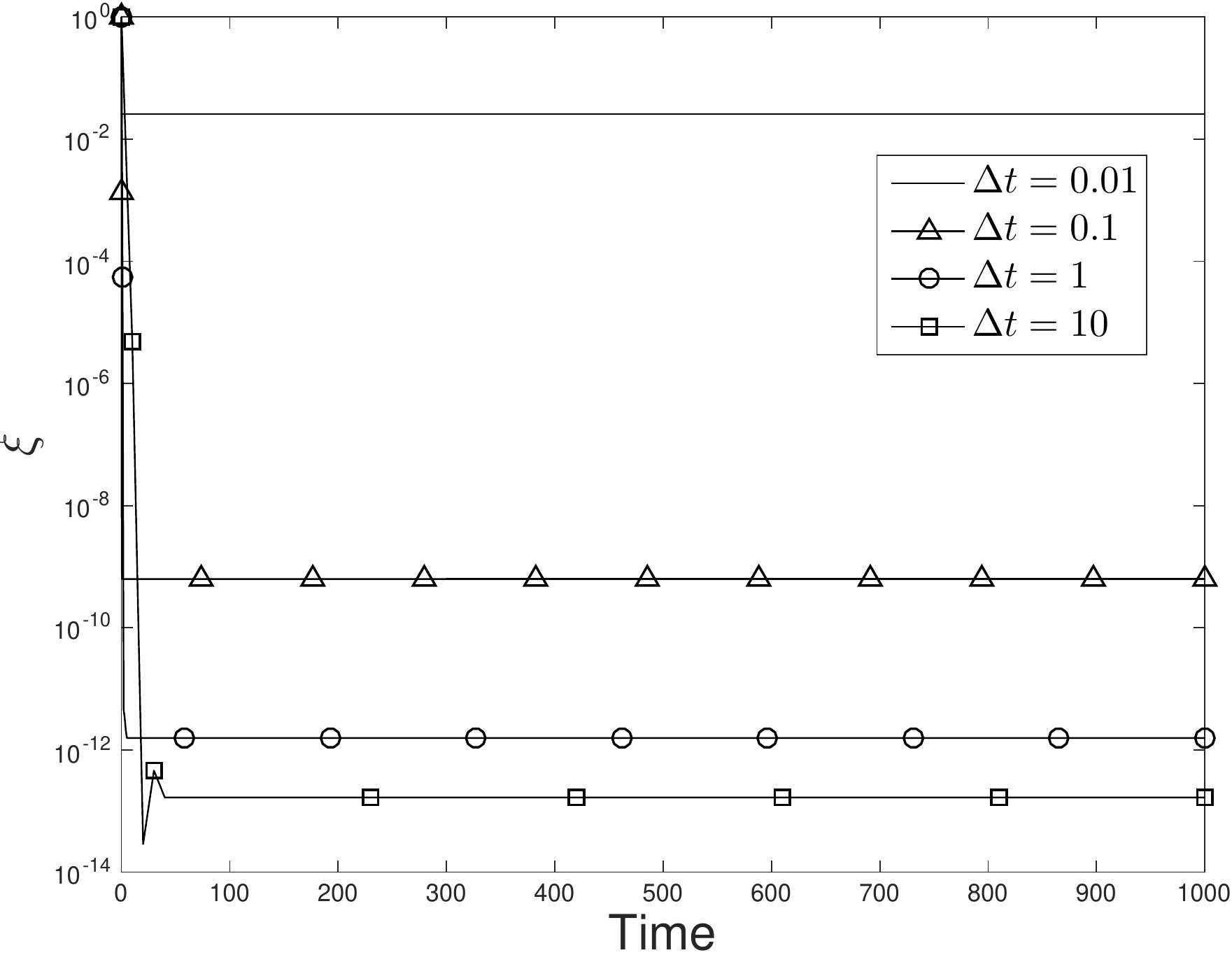}} \\
 \caption{Chemo-repulsion model: time histories of (a) $E_{tot}(t)$
   and (b) $\xi={\mathscr{F}(R)}/{E(t)}$ obtained with several
   large time step sizes $\Delta{t}=0.01,0.1,1,10$.
 }
\label{fig:chemovarylargedt}
\end{figure}

Figure \ref{fig:chemovarydt} shows time histories of three
quantities: $E(t)$, $\mathscr{F}(R)$, and $\xi=\frac{\mathscr{F}(R)}{E(t)}$,
corresponding to three time step sizes
$\Delta t=10^{-5}$, $10^{-4}$ and $10^{-3}$.
Note that $E(t)$ is computed based on equation \eqref{eq:chemoshiftEn},
$\mathscr{F}(R)$ is computed based on the $R(t)$ obtained from the algorithm,
and $\xi$ is computed based on equation \eqref{eq:chemoxif}.
These results are obtained with $\mathscr{F}(R)=R$ and $C_0=1$ in the algorithm.
It is observed from Figure \ref{fig:chemovarydt}(a) that
both $E(t)$ and $\mathscr{F}(R)$ decrease over time and gradually
level off at certain levels over time.
%
A comparison of the $E(t)$ histories obtained using different $\Delta{t}$ indicates that they
are quite close, with only some
slight difference on the interval between $t=0.002$ and $t=0.15$.
Note that $\mathscr{F}(R)$ is an approximation of $E(t)$ in the current method,
and the evolution equation for $R(t)$ stems from this relation;
see equations \eqref{eq:FG}--\eqref{eq:dynR}.
Therefore, the difference between $E(t)$ and $\mathscr{F}(R)$,
and also the quantity $\xi=\frac{\mathscr{F}(R)}{E(t)}$,
can serve as an indicator of the accuracy of the simulations.
If the difference between $E(t)$ and $\mathscr{F}(R)$ is small,
or the deviation of $\xi$ from the unit value is small, then
the simulation tends to be more accurate. On the other hand,
when the difference between $E(t)$ and $\mathscr{F}(R)$ is pronounced,
or the deviation between $\xi$ and the unit value is significant,
it implies that $\mathscr{F}(R)$ is no longer an accurate
approximation of $E(t)$ and the simulation will contain large numerical
errors.
Here it can be observed that $E(t)$ and $\mathscr{F}(R)$ computed with
$\Delta{t}=10^{-5}$ essentially overlap with each other,
indicating $\mathscr{F}(R)$ approximates well the quantity $E(t).$
However, the time histories for $E(t)$ and $\mathscr{F}(R)$ obtained with $\Delta{t}=10^{-4}$ and $10^{-3}$ exhibit noticeable discrepancies. This suggests that in these cases $\mathscr{F}(R)$ is no longer an accurate approximation of $E(t).$
We also observe from Figure \ref{fig:chemovarydt}(b) that
$\xi$ computed by $\Delta{t}=10^{-5}$ is essentially 1, while with larger values $\Delta{t}=10^{-4}$ and $\Delta{t}=10^{-3}$ the computed
$\xi$ attains values significantly smaller than 1. 
These results indicate that with the larger time step sizes
$\Delta t=10^{-4}$ and $10^{-3}$ the simulation results contain pronounced errors and
they are not accurate any more.
Because this problem exhibits very rapid dynamics (see Figures
\ref{fig:chemodynau} and \ref{fig:chemodynav}), to capture such dynamics accurately
the requirement
on $\Delta t$ is very stringent.

\begin{figure}[tbp]
  \centering
 \subfigure[]{ \includegraphics[scale=.38]{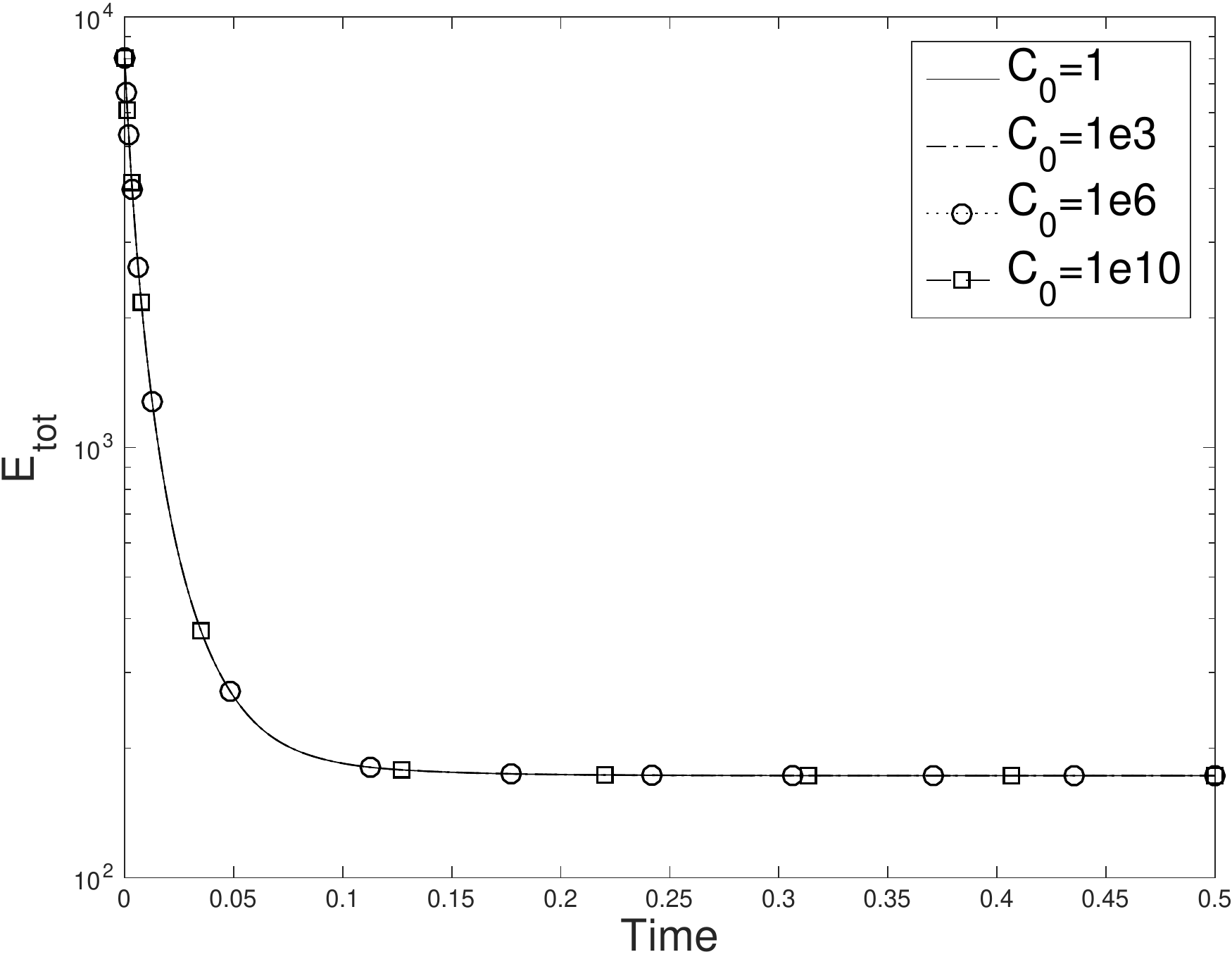}} \quad
 \subfigure[]{ \includegraphics[scale=.38]{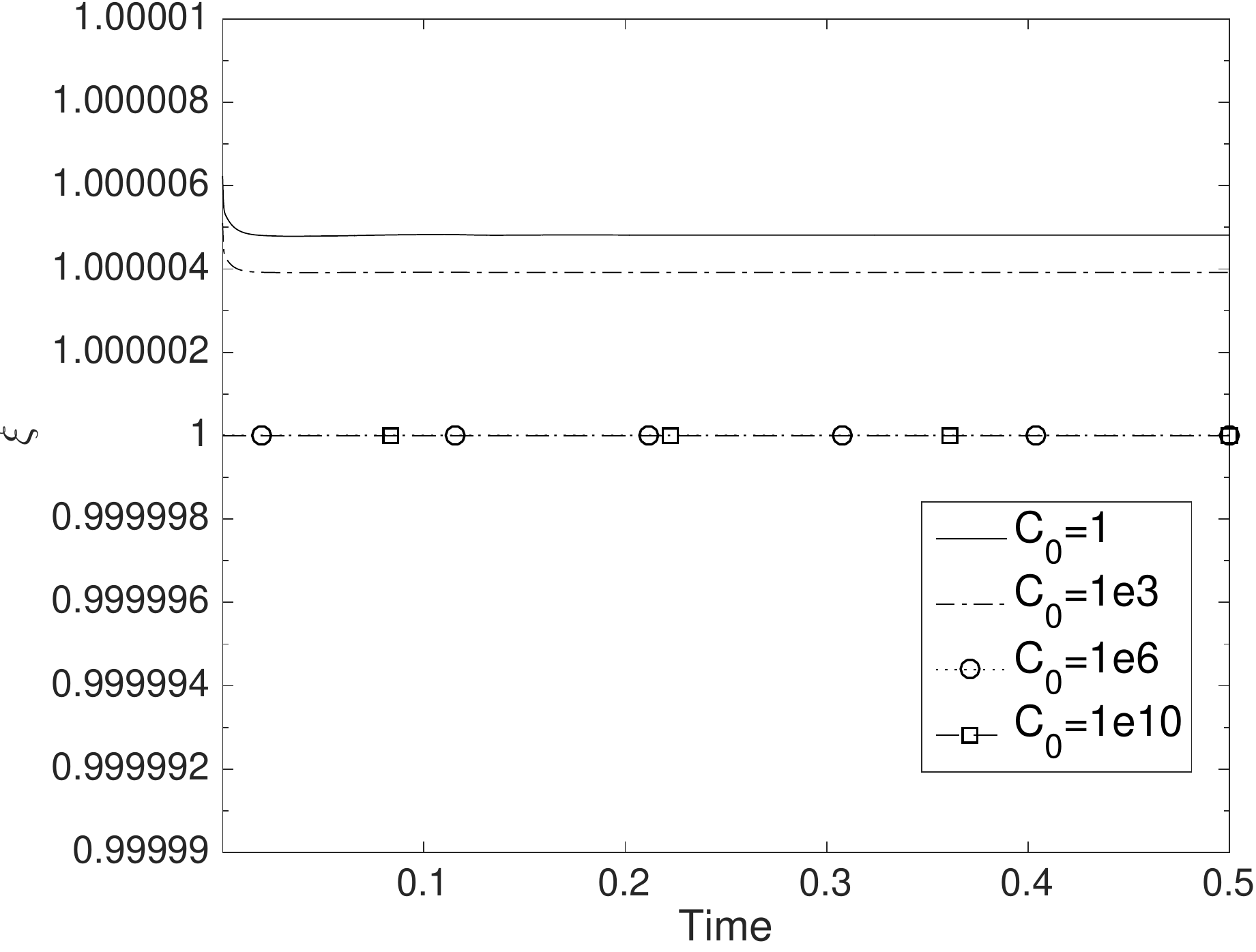}} \\
 \subfigure[]{ \includegraphics[scale=.39]{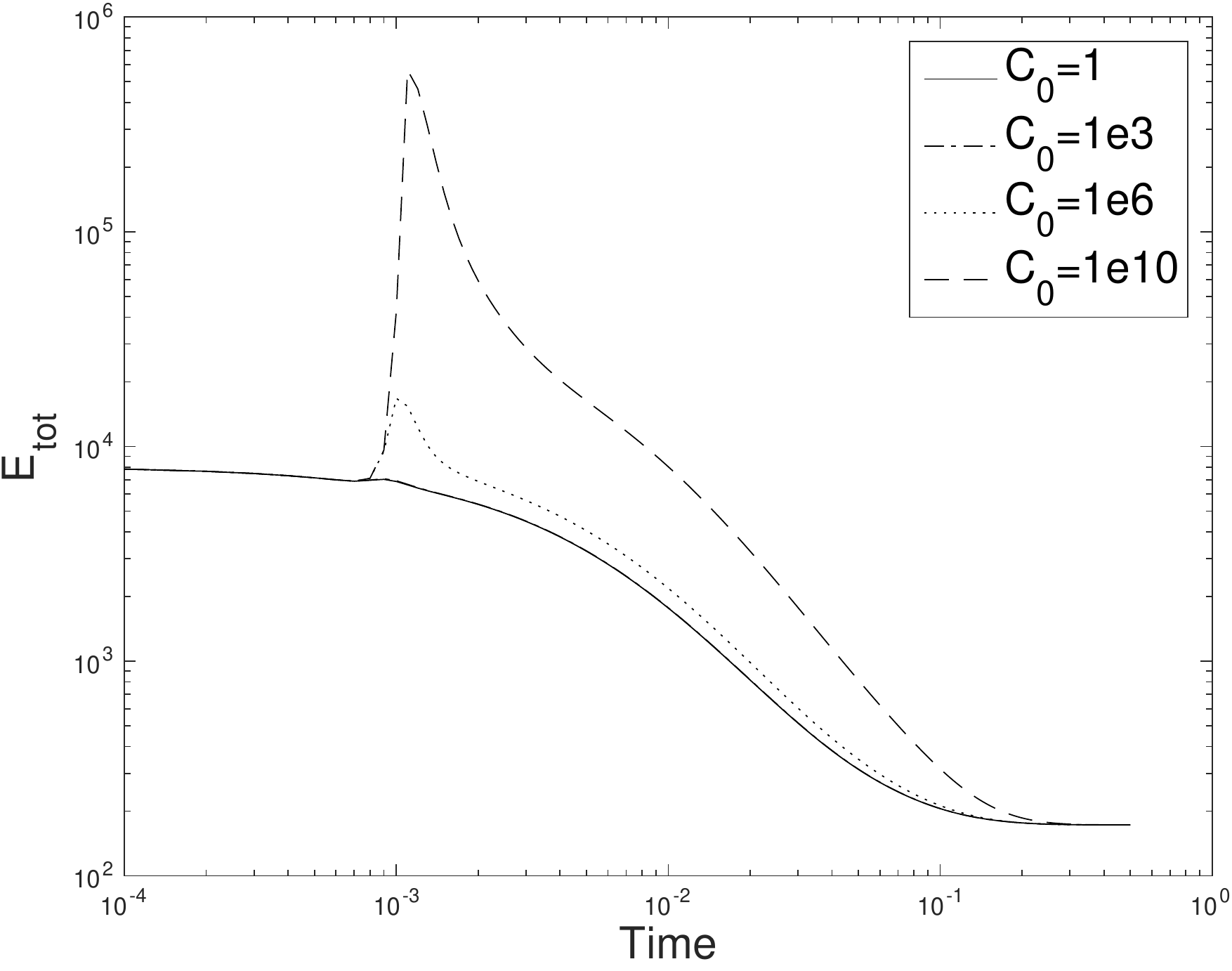}} \quad
 \subfigure[]{ \includegraphics[scale=.39]{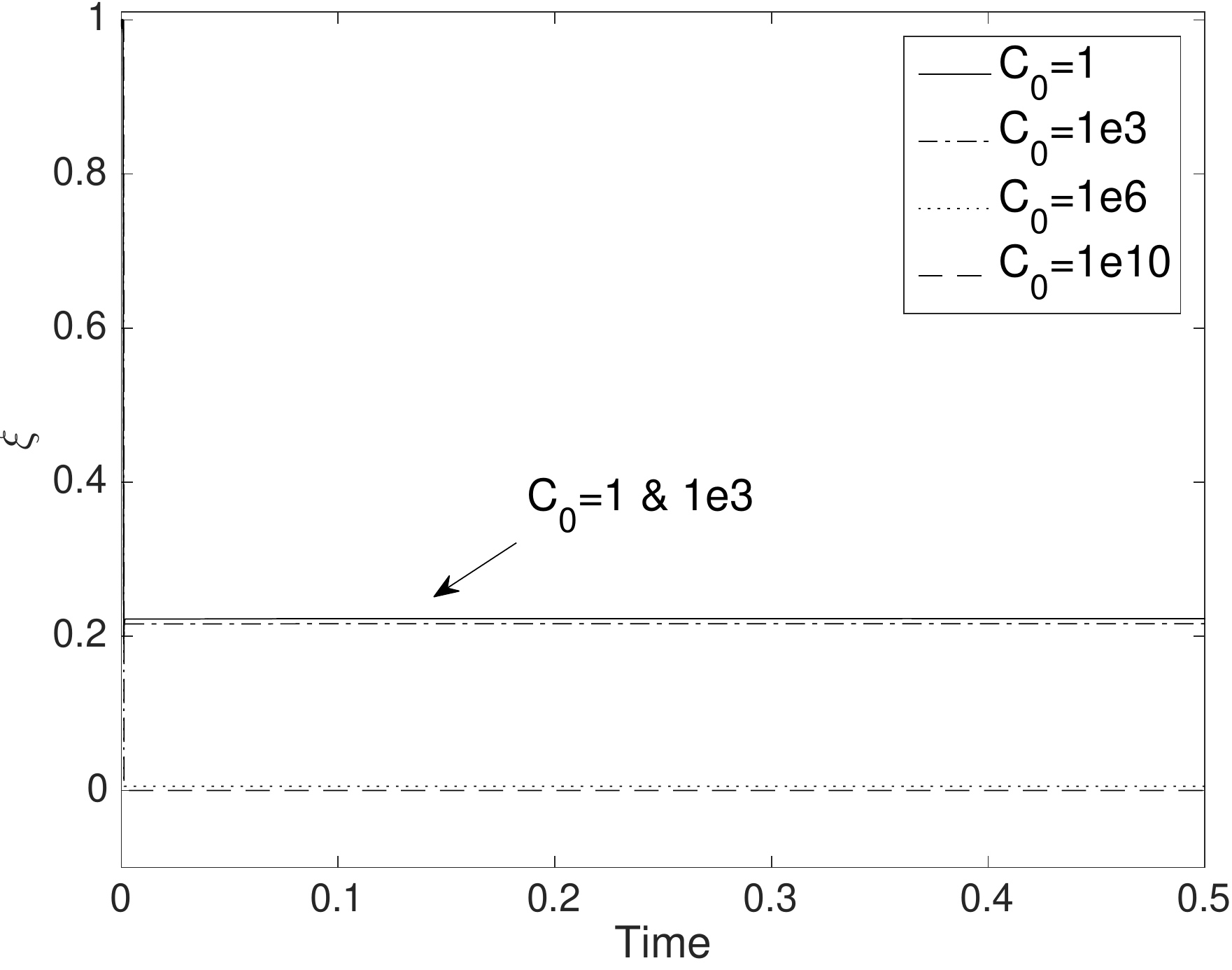}} \\
 \caption{Chemo-repulsion model: time histories of  $E_{tot}(t)$ (plots (a) and (c))
   and $\xi=\frac{\mathscr{F}(R)}{E(t)}$ (plots (b) and (d)) attained with
   various $C_0=1,1e3, 1e6, 1e10.$ The simulation results correspond to
   $\Delta t=10^{-5}$ in (a) and (b), and 
   $\Delta{t}=10^{-4}$ in (c) and (d).
 }
\label{fig:chemovaryC0}
\end{figure}

Thanks to its energy-stable nature, our algorithm can produce stable simulation results even with very large $\Delta{t}$ values.
This is demonstrated by Figure \ref{fig:chemovarylargedt} with several
large time step sizes, ranging from ${\Delta}t=0.01$ to $\Delta t=10$,
with $\mathscr{F}(R)=R$ and $C_0=1$ in the algorithm.
We show the time histories of the total energy $E_{tot}(t)$ (see equation \eqref{eq:chemoEn})
and the ratio $\xi=\frac{\mathscr{F}(R)}{E}$ for a much
longer simulation (up to $t = 1000$). The long time histories demonstrate that the computations with these large $\Delta{t}$ values are indeed stable using the current algorithm. On the other hand, because these $\Delta t$ values are very large,
we cannot expect that the results will be accurate. This is evident from the values of $\xi$ in Figure \ref{fig:chemovarylargedt}(b).
These time histories for $\xi$ tend to level off
at very small but positive values, with large deviations from the unit value.
It is noted that
the simulations are nonetheless stable, regardless of $\Delta t$.

\begin{figure}[tbp]
  \centering
 \subfigure[]{ \includegraphics[scale=.39]{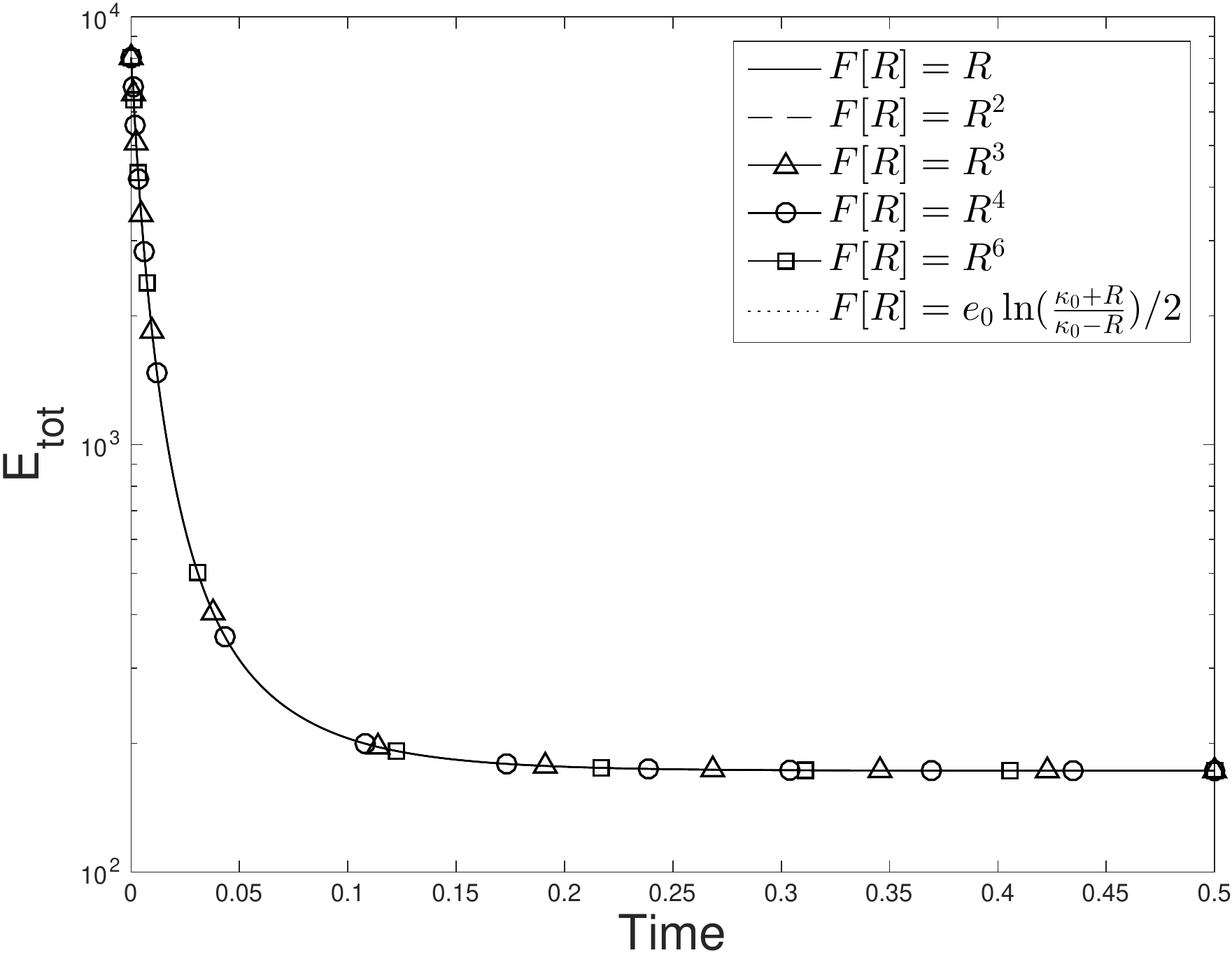}} \quad
 \subfigure[]{ \includegraphics[scale=.39]{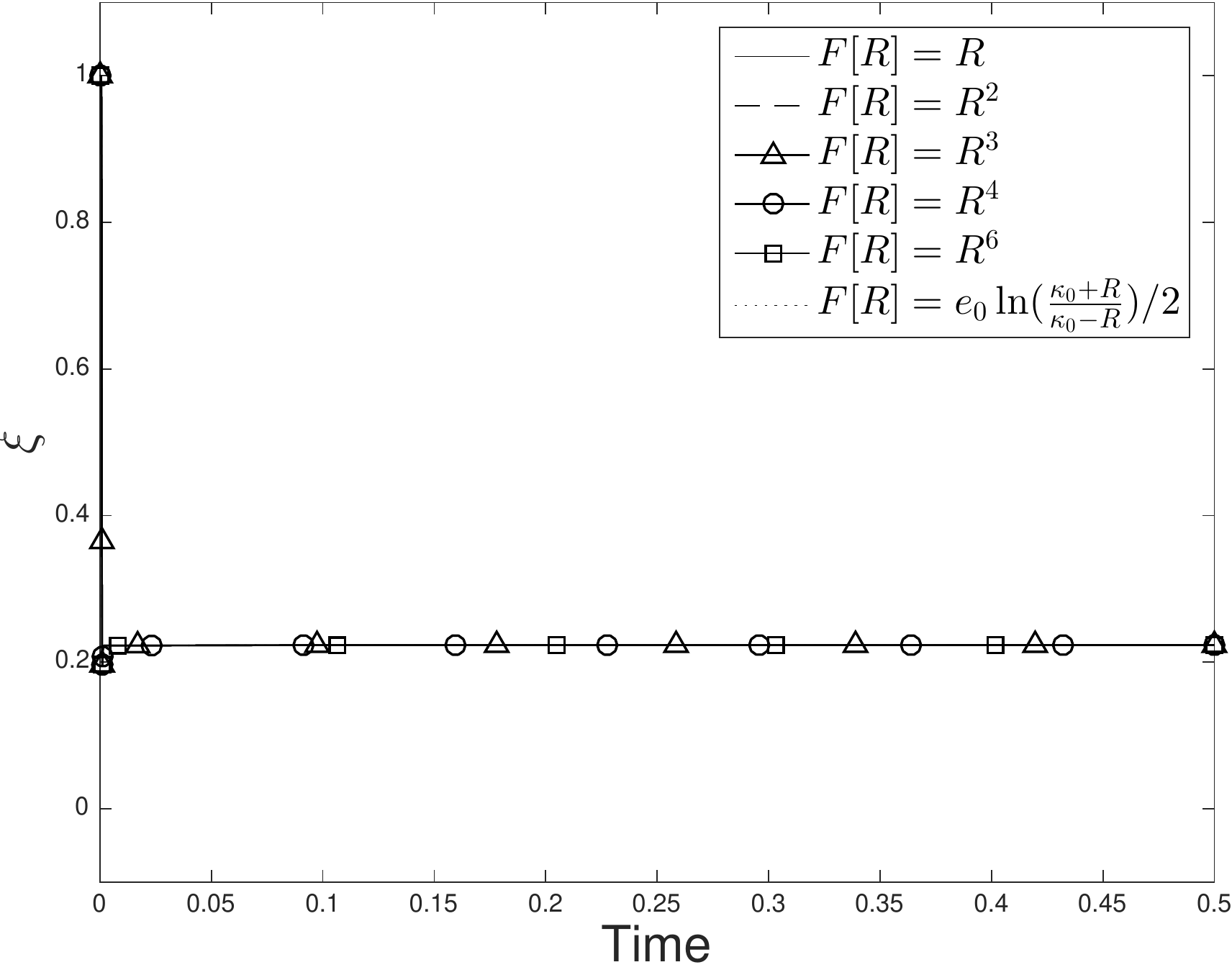}} \\
 \subfigure[]{ \includegraphics[scale=.38]{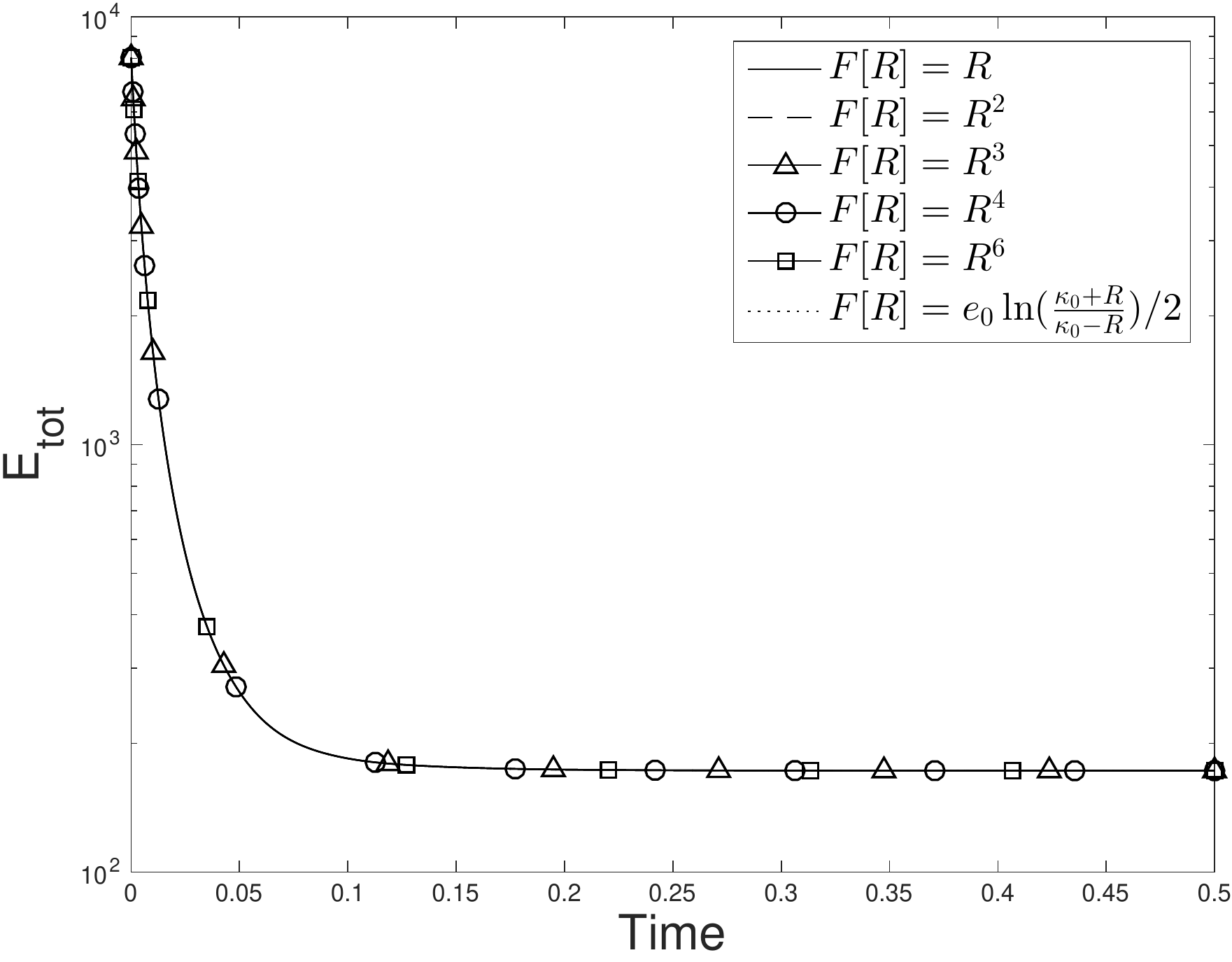}} \quad
 \subfigure[]{ \includegraphics[scale=.38]{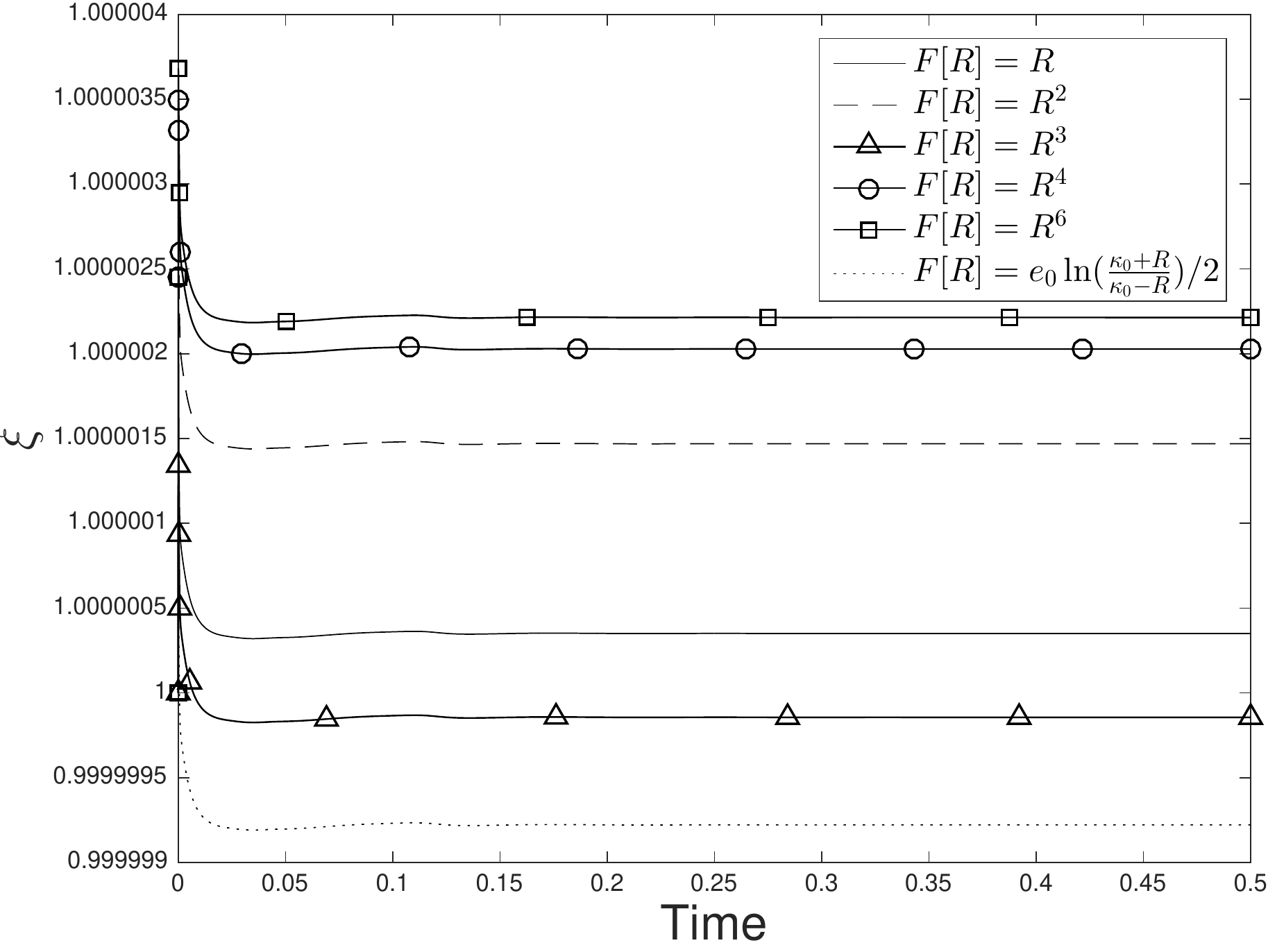}} \\
 \caption{Chemo-repulsion model: time histories of  $E_{tot}(t)$ (plots (a) and (c))
   and $\xi=\frac{\mathscr{F}(R)}{E}$ (plots (b) and (d)) obtained using
   several mapping functions
   $\mathscr{F}(R)$ as shown in the legend.
   Results in (a) and (b) correspond to $\Delta t=10^{-4}$ and
   those in (c) and (d) correspond to $\Delta t=10^{-5}$ in the simulations.
   Other parameters are fixed, with $C_0=1,$ $e_0=8040$ and $\kappa_0=1000.$
 }
\label{fig:chemovarymap}
\end{figure}

When defining the modified energy $E(t)$ (see equation \eqref{eq:chemoshiftEn})
we have incorporated an energy constant $C_0$.
The goal of $C_0$ is to ensure that $E(t)>0$ for all time,
even in certain extreme cases such as when $E_{tot}=0$, so that
$\frac{1}{E(t)}$ (as in $\frac{\mathscr{F}(R)}{E(t)}$) is always
well-defined. We observe that
the choice of the $C_0$ value seems to have some influence on the numerical results.
This effect is illustrated by Figure \ref{fig:chemovaryC0}. Here
we employ  $\mathscr{F}(R)=R$ and $\Delta{t}=10^{-5}$ and $10^{-4}$, and depict
the time histories of $E_{tot}(t)$ and $\xi$ obtained with several
$C_0$ values
($C_0=1$, $10^3$, $10^6$ and $10^{10}$).
With the smaller $\Delta t=10^{-5}$, the obtained $E_{tot}$ histories
corresponding to different $C_0$ values overlap with one another.
The computed $\xi$ values are essentially $1$, with a discrepancy
on the order of magnitude of $10^{-6}$.
This discrepancy between the computed $\xi$ and the unit value
is associated with the smaller $C_0=1$ and $10^3$.
With the larger $C_0=10^6$ and $10^{10}$, no difference can be
observed at this scale.
This suggests that with a small $\Delta t$ (so that the simulation result
is generally accurate) a larger $C_0$ value tends to give rise
to more accurate $\xi$ in terms of its discrepancy from
the unit value.
Figures \ref{fig:chemovaryC0}(c) and (d) are the corresponding
result obtained with a larger $\Delta t=10^{-4}$,
in which case the simulation result is no longer accurate.
In this case it is observed that with the larger $C_0=10^6$ and $10^{10},$
the energy $E_{tot}$ history curves exhibit a bump, apparently artificial;
see Figure \ref{fig:chemovaryC0}(c).
In contrast, with the smaller $C_0=1$ and $10^3$, such a bump is not quite obvious
from the energy history curves.
In addition, with the larger $C_0=10^6$ and $10^{10}$,
the computed $\xi$ attains a very small value (close to 0),
while $\xi$ attains a value around $0.2$ with the smaller
$C_0=1$ and $10^3$.
This indicates that, with larger $\Delta t$
(when simulation loses accuracy),
the simulation results obtained with a smaller $C_0$ 
may be better than those obtained with
a larger $C_0$, 
even though all the results become inaccurate.
The results of this group of tests suggest the following.
With small $\Delta t$ values, a larger $C_0$ tends to give rise to
more accurate results in the sense that the computed $\xi$ tends to be
closer to the unit value. However, a $C_0$ that is very large seems to
have an adverse effect when $\Delta t$ becomes large, because
it can lead to computed $\xi$ values that deviate
from the unit value more severely.
The majority of
simulations in this section are performed using $C_0=1$.

\begin{figure}[tbp]
  \centering
 \subfigure[]{ \includegraphics[scale=.39]{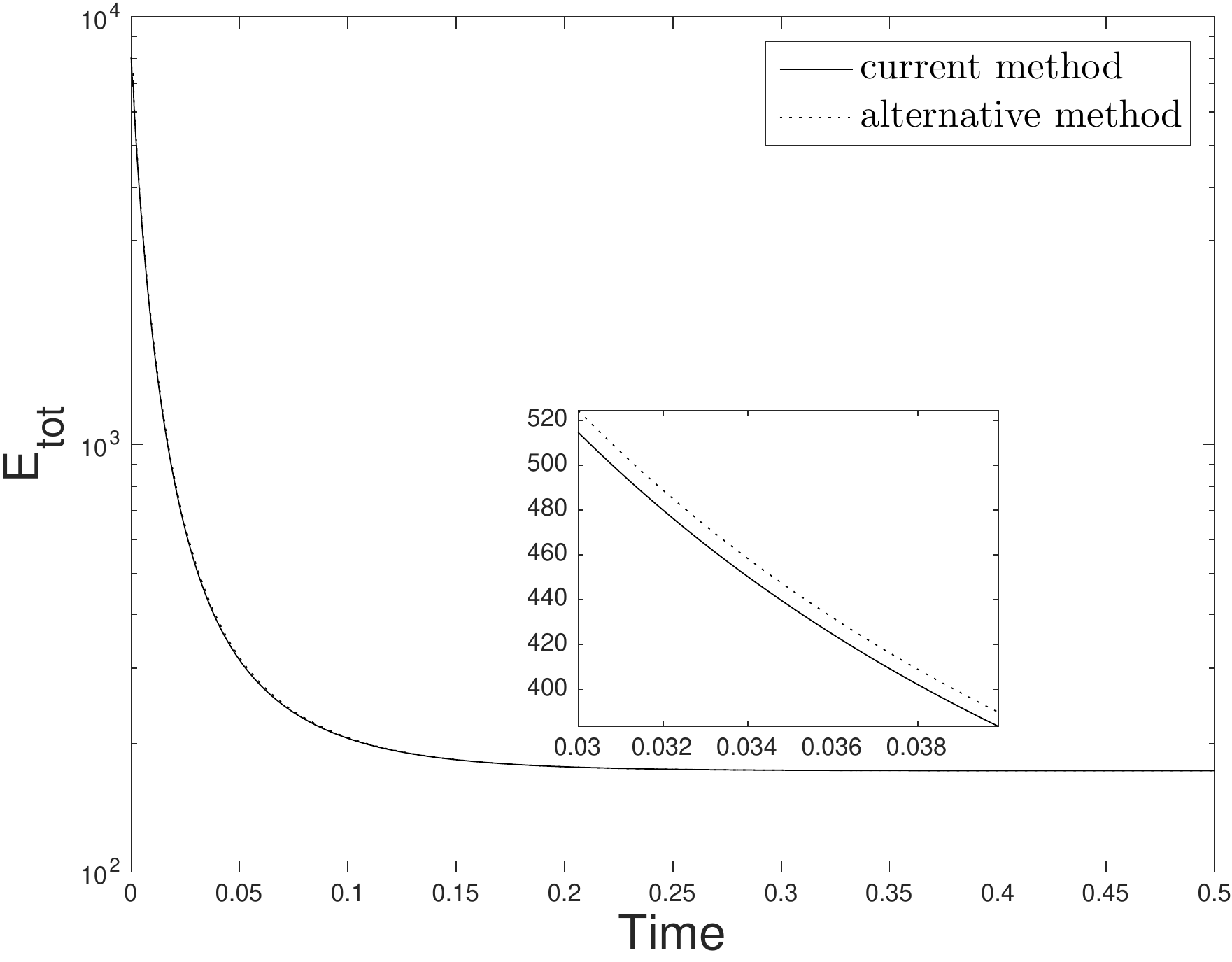}} \quad
 \subfigure[]{ \includegraphics[scale=.39]{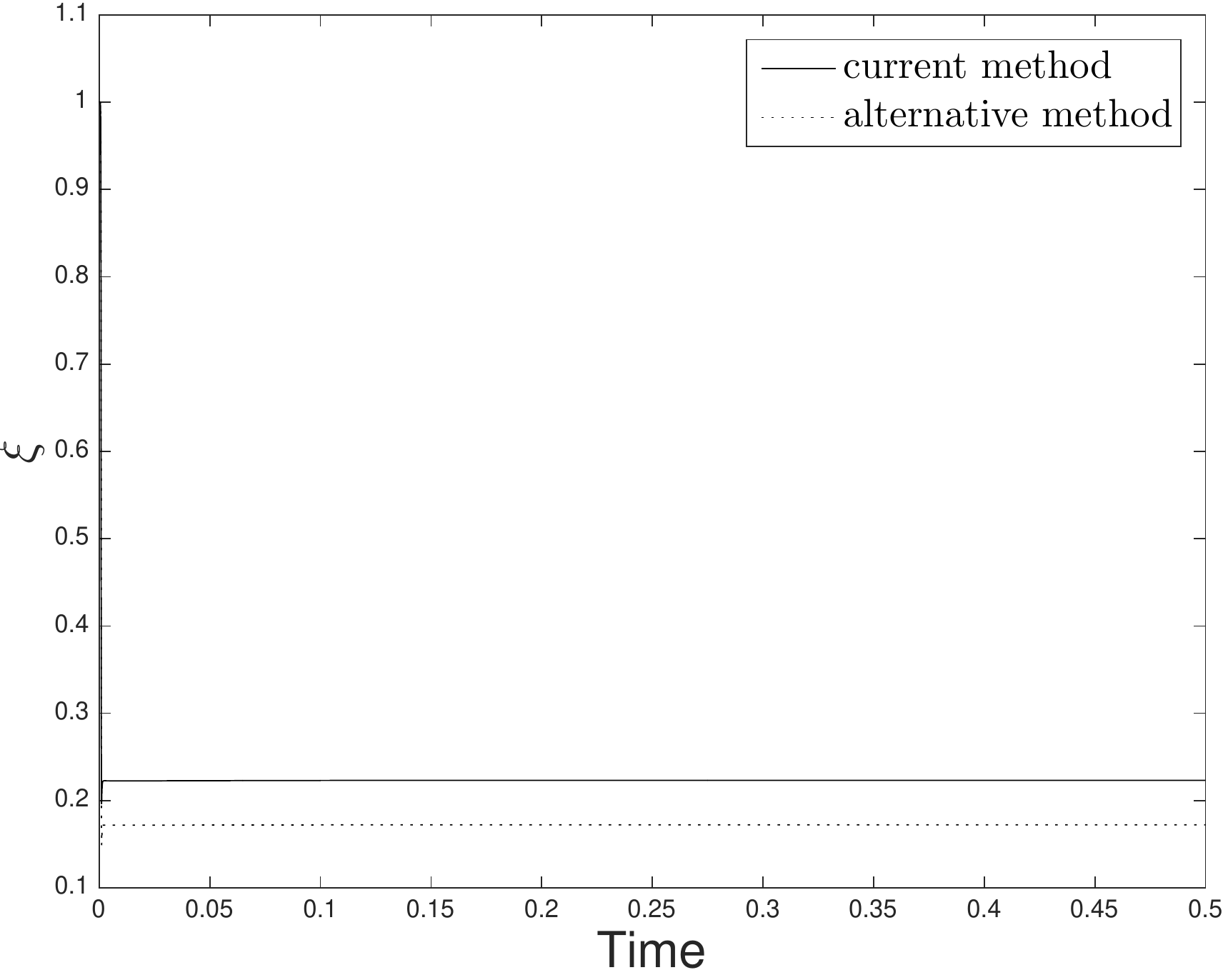}} \\
 \subfigure[]{ \includegraphics[scale=.39]{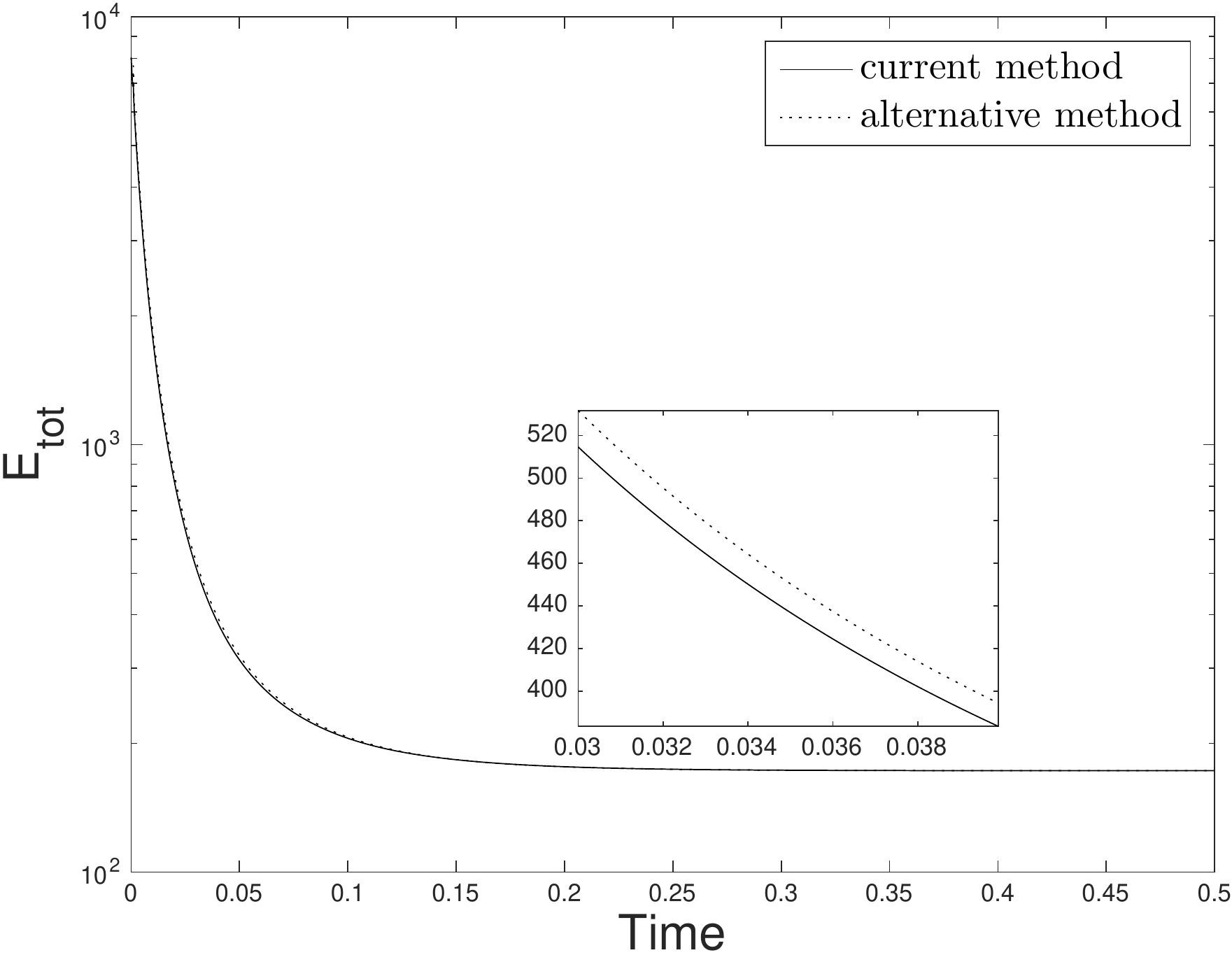}} \quad
 \subfigure[]{ \includegraphics[scale=.39]{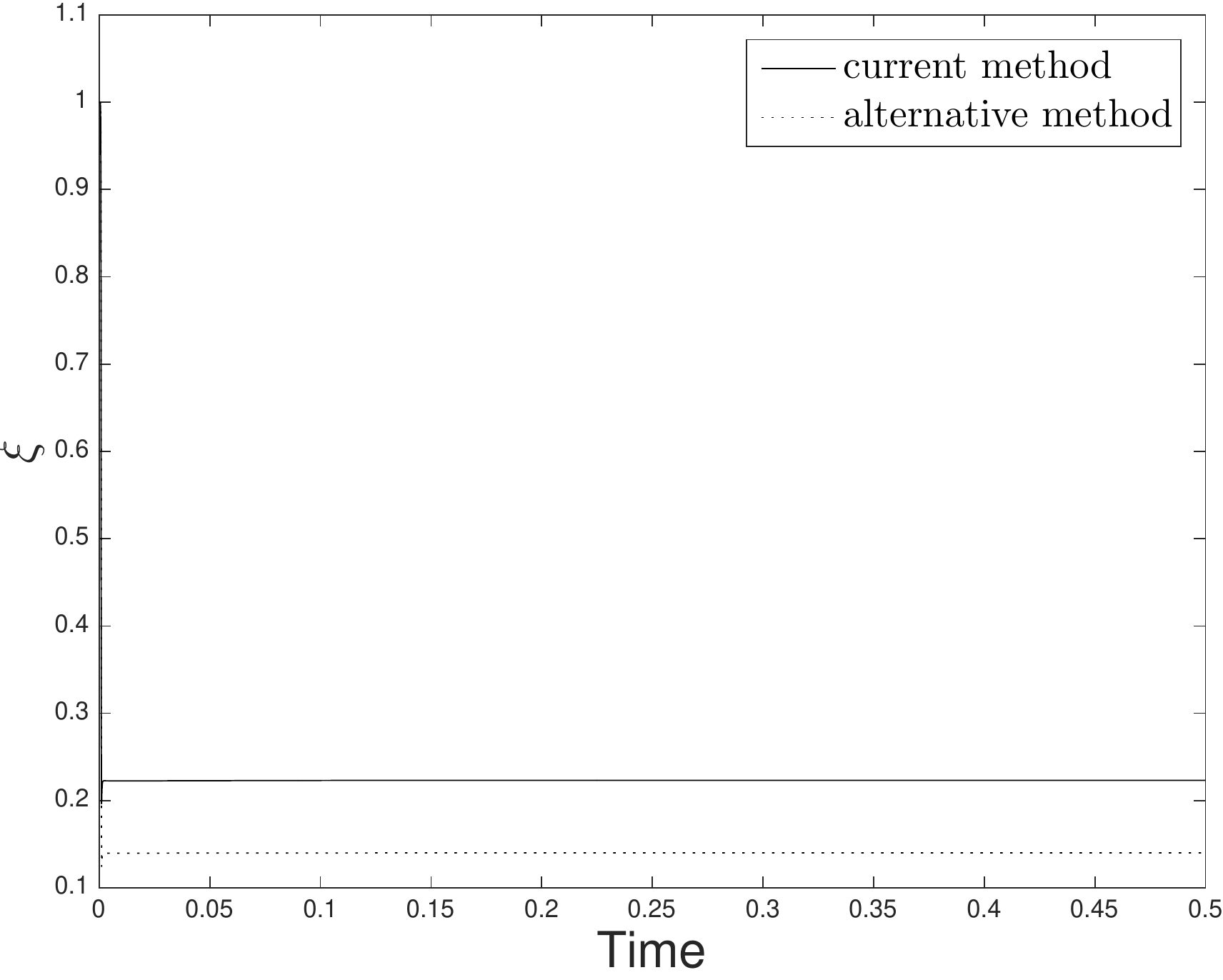}} \\
 \caption{Chemo-repulsion model: comparison of the time histories of $E_{tot}(t)$
   (plots (a) and (c))
   and $\xi$ (plots (b) and (d)) computed using the current method and the alternative
   method from Section \ref{sec:alter}. 
   In the current method $\xi={\mathscr{F}(R)}/{E}$,
   and in the alternative method $\xi={R}/{\mathscr{G}(E)}.$
   Plots (a) and (b) are obtained with the mapping
   $\mathscr{F}(R)=R^2$ (i.e.~$\mathscr{G}(E)=\sqrt{E}$),
   and plots (c) and (d) are obtained with
   $\mathscr{F}(R)=R^3$ (i.e.~$\mathscr{G}(E)=\sqrt[3]{E}$).
   Other parameters are fixed with $\Delta{t}=10^{-4}$ and  $C_0=1.$
 }
\label{fig:chemovarymethod}
\end{figure}

The method developed in the current work can employ
a general function $\mathscr{F}(R)$ (with inverse $\mathscr{G}$)
to define the auxiliary variable $R(t)$, as long as $\mathscr{F}$ is
a one-to-one increasing differentiable function satisfying \eqref{eq:FG_property}.
We observe that the choice for the specific mapping $\mathscr{F}$ seems to
have very little or no influence on the simulation results using
the current method. This point is demonstrated by Figure \ref{fig:chemovarymap}.
Here we have considered several functions,
$\mathscr{F}(R)=R^m$ ($m=1,2,3,4,6$) and
$\mathscr{F}(R)=\frac{e_0}{2}\ln (\frac{\kappa_0+R}{\kappa_0-R})$
with $e_0=8040$ and $\kappa_0=10^3$.
Figure \ref{fig:chemovarymap} shows the time histories of $E_{tot}(t)$
and $\xi$ obtained using these mappings, together 
with a fixed $C_0=1$ and two time step sizes $\Delta{t}=10^{-4}$ and $10^{-5}$.
It can be observed that the time history curves for both $E_{tot}(t)$ and $\xi$
corresponding to different $\mathscr{F}$ functions
overlap with one another, suggesting no or very little
difference in the simulation results.
In particular, Figure \ref{fig:chemovarymap}(d) shows
the $\xi$ history curves corresponding to different
$\mathscr{F}$ obtained with the smaller $\Delta t$,
with the vertical axis $\xi$ magnified around the unit value.
It can be observed that the difference between various curves
is on the order of magnitude $10^{-6}$.
Since little difference in the numerical results
is observed with different mapping functions $\mathscr{F}(R)$ using
the current method, the majority of numerical tests
reported in this and subsequent sections will be carried out
using the simplest mapping $\mathscr{F}(R)=R$.

In Section \ref{sec:alter} we have discussed another unconditionally
energy-stable scheme
(referred to as ``alternative method''), which is based on an alternative
formulation with $\xi=\frac{R}{\mathscr{G}(E)}$. The dynamic equation
for the auxiliary variable $R(t)$ is accordingly
replaced by equation \eqref{eq:Requ_alt}.
Figure \ref{fig:chemovarymethod} is a comparison of the time histories
for $E_{tot}(t)$ and $\xi$ obtained using these two methods.
The results in
Figure \ref{fig:chemovarymethod}(a) and (b) are obtained with
a mapping function $\mathscr{F}(R)=R^2$ (or equivalently $\mathscr{G}(E)=\sqrt{E}$),
and those in (c) and (d) correspond to $\mathscr{F}(R)=R^3$
(or $\mathscr{G}(E)=\sqrt[3]{E}$).
We observe that there seems to be little difference in the computed
total energy $E_{tot}(t)$. But some difference can be noted with
the $\xi$ histories.
The computed $\xi$ values using the current method (with
$\frac{\mathscr{F}(R)}{E}$) seem to be consistently larger than
those using the alternative method (with $\frac{R}{\mathscr{G}(E)}$).
While all these values deviate from the unit value substantially
because of the time step size $\Delta t=10^{-4}$,
the deviation with the current method appears
noticeably smaller than that with the alternative method.
This seems to suggest that,
while the simulation results using these methods are
not very much different,
the formulation using
$\frac{\mathscr{F}(R)}{E}$ may be somewhat better than
the alternative formulation using $\frac{R}{\mathscr{G}(E)}$.


\section{Cahn-Hilliard Equation with Constant and Variable Mobility}
\label{sec:che}


We apply the gPAV method to simulate the Cahn-Hilliard equation~\cite{CahnH1958}
in this section. This equation has widespread applications in
the phase-field modeling of materials science, two-phase
and multiphase flows (see e.g.~\cite{LowengrubT1998,Chen2002,LiuS2003,YueFLS2004,KimL2005,DingSS2007,DongS2012,Dong2012,Dong2014,LiuSY2015,WuX2017,XuLWB2019}, among others).
Consider the Cahn-Hilliard equation
on a domain $\Omega$ (with boundary $\Gamma$):
\begin{subequations}\label{eq:chsys}
\begin{align}
&\frac{\partial \phi}{\partial t}=\nabla \cdot\big(m(\phi)\nabla \mu \big)+f(\bs x,t),\;\;  \label{eq:cheq1}\\
  & \mu = \frac{\delta E_{tot}}{\delta \phi}
  =-\lambda \nabla^2 \phi +h(\phi),\;\;\label{eq:cheq2}\\
  &m(\phi) \bs n \cdot \nabla \mu=d_a(\bs x,t),\;\;{\rm on}~\Gamma,\label{eq:cheq3a} \\
  &\bs n \cdot \nabla \phi =d_b(\bs x, t)\;\; {\rm on}~\Gamma,\label{eq:cheq3}
\end{align}
\end{subequations}
supplemented by the initial condition 
\begin{equation}\label{eq:chinitial}
\phi(\bs x,0)=\phi_{\rm in}(\bs x).
\end{equation}
In these equations, $\phi(\bs x,t)\in [-1,1]$ is the phase field function,
$f(\bs x,t)$, $d_a(\bs x,t)$ and $d_b(\bs x,t)$ are prescribed source terms for the purpose of convergence testing only, and will be set to $f(\bs x,t)=d_a(\bs x,t)=d_b(\bs x,t)=0$ in actual simulations.
$E_{tot}$ is the free energy functional,
\begin{equation}\label{eq:che_free_eng}
  E_{tot}(t) = E_{tot}[\phi,\nabla\phi] = \int_{\Omega}\left[
    \frac{\lambda}{2}\nabla\phi\cdot\nabla\phi + H(\phi)
    \right]d\Omega,
  \quad
  \text{with} \ \ H(\phi) = \frac{\lambda}{4\eta^2}(\phi^2-1)^2
\end{equation}
in which $\eta$ is the characteristic interfacial thickness scale, and
$\lambda$ is referred to as the mixing energy density
coefficient and is related to other physical parameters. For example,
for two-phase flow problems $\lambda$ is given by $\lambda=\frac{3}{2\sqrt{2}}\sigma \eta,$ where $\sigma$ is the surface tension.
$\mu$ is referred to as the chemical potential, and
the nonlinear term $h(\phi)$ is given by $h(\phi)=H'(\phi)$.
$H(\phi)$ is referred to as the potential free energy density function,
which can take many different forms. In this paper we only consider
the double-well form as given in \eqref{eq:che_free_eng}.
$m\geqslant 0$ is the mobility, and in this work we consider two cases:
(i) $m=m_0$, and (ii) $m=m(\phi)=\max(m_0(1-\phi^2), 0 )$,
with $m_0$ being a given positive constant.

We take the $L^2$ inner product between \eqref{eq:cheq1} and $\mu$, perform integration by part and impose the boundary condition \eqref{eq:cheq3}. This leads to 
the energy balance equation,
\begin{multline}\label{eq:chenglaw}
  \frac{\partial}{\partial t}\int_\Omega \Big(\frac{\lambda}{2}|\nabla \phi|^2+H(\phi)\Big)
  d\Omega=
  -\int_\Omega m(\phi)|\nabla \mu|^2d\Omega 
  +\int_\Omega f\mu d\Omega \\
  +\int_{\Gamma} m(\phi)(\bs n\cdot\nabla \mu) \mu d\Gamma
  +\lambda \int_{\Gamma} (\bs n \cdot\nabla\phi) \frac{\partial \phi}{\partial t}
  d\Gamma.
\end{multline}
Based on equations \eqref{eq:shitE} and \eqref{eq:chenglaw}, we define the shifted total energy by 
\begin{equation}\label{eq:chshiftEn}
E(t)= E[\phi]=\int_\Omega \Big( \frac{\lambda}{2}|\nabla \phi|^2+H(\phi)\Big) d\Omega+C_0,
\end{equation}
where $C_0$ is chosen to ensure $E(t)>0$. Let us define $\mathscr{F}$ and $\mathscr{G}$ and $R(t)$ based on equations \eqref{eq:FG}--\eqref{eq:FG_2}.
 Following equation \eqref{eq:dynR} and using \eqref{eq:chshiftEn}, we have
 \begin{equation}\label{eq:che_auxeq}
\mathscr{F}'(R)\frac{dR}{dt}=\int_\Omega \Big[-\lambda\nabla^2\phi +  h(\phi) \Big]\frac{\partial \phi}{\partial t} d\Omega+\lambda \int_{\Gamma} d_b \frac{\partial \phi}{\partial t} d\Gamma,
 \end{equation}
 where the boundary condition \eqref{eq:cheq3} has been used.

\subsection{Constant Mobility}\label{subsec:constmob}

Assume that $m(\phi)=m_0>0$ is a constant.
We reformulate equations \eqref{eq:cheq1}--\eqref{eq:cheq3a} as
follows,
\begin{subequations}
  \begin{align}
    &
    \frac{\partial\phi}{\partial t} = m_0\nabla^2\left[
      -\lambda\nabla^2\phi + S(\phi-\phi)
      + \frac{\mathscr{F}(R)}{E}h(\phi)
      \right] + f,
    \label{eq:che_trans_1} \\
    &
    m_0\bs n\cdot\nabla\left[
      -\lambda\nabla^2\phi + S(\phi-\phi)
      + \frac{\mathscr{F}(R)}{E}h(\phi)
      \right] = d_a,
    \quad \text{on} \ \ \Gamma,
    \label{eq:che_trans_2}
  \end{align}
\end{subequations}
where $S$ is chosen constant satisfying a condition to be
specified later. Note that a zero term $S(\phi-\phi)$ is
added in these equations.
We reformulate equation \eqref{eq:che_auxeq} as follows,
\begin{equation}\label{eq:che_auxeq_reform}
  \begin{split}
  \mathscr{F}'(R)\frac{dR}{dt}=&
  \int_\Omega \mu\frac{\partial \phi}{\partial t} d\Omega
  -\int_{\Omega} \mu\left[
  m_0\nabla^2\left(
  -\lambda\nabla^2\phi  + S(\phi-\phi)
      + \frac{\mathscr{F}(R)}{E}h(\phi)
  \right) + f
  \right]d\Omega \\
  &
  +\frac{\mathscr{F}(R)}{E}\left[
    -\int_{\Omega}m_0|\nabla\mu|^2d\Omega
    + \int_{\Omega} f\mu d\Omega
    + \int_{\Gamma}d_a\mu d\Gamma
    + \int_{\Gamma}\lambda d_b\frac{\partial\phi}{\partial t} d\Gamma
    \right] \\
  &
  +\left(1- \frac{\mathscr{F}(R)}{E} \right)\left|
  \int_{\Omega} f\mu d\Omega
    + \int_{\Gamma}d_a\mu d\Gamma
    + \int_{\Gamma}\lambda d_b\frac{\partial\phi}{\partial t} d\Gamma
  \right|
  \end{split},
\end{equation}
where $\mu$ is given by \eqref{eq:cheq2}, and
the following zero terms have been incorporated into the RHS,
\begin{equation}
  \begin{split}
    &
  \left(\frac{\mathscr{F}(R)}{E}-1 \right)
  \int_{\Omega}\mu\left[
    m_0\nabla^2(-\lambda\nabla^2\phi) + f
    \right]d\Omega
  + \frac{\mathscr{F}(R)}{E}\left[
    \int_{\Omega}\mu[m_0\nabla^2h(\phi)]d\Omega
    - \int_{\Omega}\mu[m_0\nabla^2h(\phi)]d\Omega
    \right] \\
  &
  +\left(\frac{\mathscr{F}(R)}{E}-1 \right)
  \int_{\Gamma}\lambda d_b\frac{\partial\phi}{\partial t}d\Gamma
  + \left(1- \frac{\mathscr{F}(R)}{E} \right)\left|
  \int_{\Omega} f\mu d\Omega
    + \int_{\Gamma}d_a\mu d\Gamma
    + \int_{\Gamma}\lambda d_b\frac{\partial\phi}{\partial t} d\Gamma
  \right|.
  \end{split}
\end{equation}

The energy-stable scheme for the equations \eqref{eq:che_trans_1}--\eqref{eq:che_trans_2},
\eqref{eq:cheq3} and \eqref{eq:che_auxeq_reform} is
as follows:
\begin{subequations}\label{eq:chscheme}
\begin{align}
& \frac{\partial \phi}{\partial t}\Big|^{n+1}=m_0\nabla^2\Big[-\lambda \nabla^2 \phi^{n+1}+   S(\phi^{n+1}-\bar \phi^{n+1})+\xi h(\bar \phi^{n+1})   \Big]+f^{n+1},\label{eq:chsch1}\\
  & m_0 \bs n \cdot\nabla \big[-\lambda \nabla^2 \phi^{n+1}+S(\phi^{n+1}-\bar \phi^{n+1})+\xi h(\bar \phi^{n+1}) \big]=d_a^{n+1},
  \quad \text{on} \ \Gamma,
  \label{eq:chsch4}\\
  & \bs n\cdot\nabla \phi^{n+1}=d_b^{n+1}, \quad \text{on} \ \Gamma,
  \label{eq:chbd2}\\
  & \xi=\frac{\mathscr{F}(R^{n+\frac{3}{2}})}{E[\tilde{\phi}^{n+3/2}]},\label{eq:chsch3} \\
  & E[\tilde{\phi}^{n+3/2}]=
  \int_\Omega \left[ \frac{\lambda}{2}|\nabla \tilde{\phi}^{n+3/2}|^2
  +H(\tilde{\phi}^{n+3/2}) \right] d\Omega+C_0, \label{eq:chsch3a}
 \end{align}
\end{subequations}
and
\begin{equation}\label{eq:chsch2}
\begin{split}
  & D_{\mathscr{F}}(R) \Big|^{n+1} \frac{dR}{dt}\Big|^{n+1}
  =\int_\Omega \Big[-\lambda\nabla^2\phi^{n+1} +  h(\phi^{n+1}) \Big]\left.\frac{\partial \phi}{\partial t}\right|^{n+1} d\Omega \\
  &-\int_{\Omega}  \Big[-\lambda\nabla^2\phi^{n+1} +  h(\phi^{n+1}) \Big]\Big\{  m_0\nabla^2\Big[-\lambda \nabla^2 \phi^{n+1}+S(\phi^{n+1}-\bar \phi^{n+1})+\xi h(\bar \phi^{n+1})   \Big]+f^{n+1} \Big\}d\Omega \\
  & 
  +\xi \Big\{ -\int_\Omega m_0|\nabla \tilde{\mu}^{n+1}|^2d\Omega +
  \int_\Omega f^{n+1}\tilde{\mu}^{n+1} d\Omega
  +\int_{\Gamma} \left(d_a^{n+1} \tilde{\mu}^{n+1}
  +\lambda d_b^{n+1} \left.\frac{\partial \phi}{\partial t}\right|^{*,n+1} \right)d\Gamma  \Big\}\\
  &+(1-\xi)\left|\int_\Omega f^{n+1}\tilde{\mu}^{n+1}d\Omega
  +\int_{\Gamma} \left(d_a^{n+1} \tilde{\mu}^{n+1}+\lambda d_b^{n+1} \left.\frac{\partial \phi}{\partial t}\right|^{*,n+1} \right)d\Gamma   \right|.
 \end{split}
\end{equation}
These are supplemented by the initial conditions
\begin{equation}\label{eq:chsch5}
\phi^0(\bs x)=\phi_{\rm in}(\bs x),\quad R^0=\mathscr{G}( E^0),\;\;\text{with}\;\;E^0=\int_\Omega \Big( \frac{1}{2}|\nabla \phi_{\rm in}|^2+H(\phi_{\rm in})\Big) d\Omega+C_0.
\end{equation}
In the above equations, $\left.\frac{\partial\phi}{\partial t} \right|^{n+1}$
and $\left.\frac{dR}{dt} \right|^{n+1}$ are defined by \eqref{eq:tempdist}, and
${\bar\phi}^{n+1}$ is defined by \eqref{eq:def_bar}.
$\tilde{\phi}^{n+1}$, $\tilde{\phi}^{n+3/2}$ and $\tilde{\mu}^{n+1}$
are second-order approximations
of $\phi^{n+1}$, $\phi^{n+3/2}$ and $\mu^{n+1}$,
respectively, to be specified later in \eqref{eq:chphistar}--\eqref{eq:phistar32}.
$\left.\frac{\partial \phi}{\partial t} \right|^{*,n+1}$
is an approximation of $\left.\frac{\partial \phi}{\partial t} \right|^{n+1}$
to be specified later in \eqref{eq:chmustar}.

 \begin{theorem}\label{thm:chdislaw}
 In the absence of the external force $f=0,$ and with zero boundary conditions $d_a=d_b=0,$ the scheme consisting of \eqref{eq:chscheme}-\eqref{eq:chsch2} is unconditionally energy stable in the sense that
 \begin{equation}\label{eq:chdislaw}
{\mathscr{F}(R^{n+\frac{3}{2}}) -\mathscr{F}(R^{n+\frac{1}{2}})}=-\xi \Delta{t} \int_\Omega m_0|\nabla \tilde{\mu}^{n+1}|^2 \leq 0,
 \end{equation}
 if the approximation of $R(t)$ at time step $\frac{1}{2}$ is positive. 
 \end{theorem}

\begin{proof}
  Multiplying $-\lambda\nabla^2\phi^{n+1} +  h(\phi^{n+1}) $ to
  equation \eqref{eq:chsch1}, integrating over the domain,
  and adding the resultant equation to equation \eqref{eq:chsch2},
  we obtain the energy balance relation as follows:
 \begin{equation}\label{eq:chsolver}
 \begin{split}
   & \frac{\mathscr{F}(R^{n+\frac{3}{2}}) -\mathscr{F}(R^{n+\frac{1}{2}})}{\Delta{t}}
   =\xi \Big\{ \int_\Omega f^{n+1}\tilde{\mu}^{n+1}d\Omega
   +\int_{\Gamma} \left(d_a^{n+1} \tilde{\mu}^{n+1}+\lambda d_b^{n+1} \left.\frac{\partial \phi}{\partial t}\right|^{*,n+1} \right)d\Gamma  \Big\}\\
   &-\xi \int_\Omega m_0|\nabla \tilde\mu^{n+1}|^2d\Omega
   +(1-\xi)\bigg|\int_\Omega f^{n+1}\tilde\mu^{n+1}d\Omega
   +\int_{\Gamma} \left(d_a^{n+1} \tilde\mu^{n+1}+\lambda d_b^{n+1} \left.\frac{\partial \phi}{\partial t}\right|^{*,n+1} \right)d\Gamma \bigg|    ,
 \end{split}
\end{equation}
 where we have used the relation \eqref{eq:FRdt}.
 If $f=0$ and $d_a=d_b=0$, then
 \begin{equation}
   \xi = \frac{\mathscr{F}(R^{n+1/2}) }
           {E[\tilde\phi^{n+3/2}] + \Delta t \int_\Omega m_0|\nabla \tilde\mu^{n+1}|^2  }.
 \end{equation}
 If $R^{n+1/2}|_{n=0}>0$, one can conclude by induction that $\xi>0$ for any $n\geqslant 0$.
 This leads to \eqref{eq:chdislaw}.
\end{proof}
The method from the Appendix A can be employed to
compute the first time step, which can
ensure that the approximation of $R(t)$
at the step $\frac{1}{2}$ is positive.

To implement the scheme we note that
equation \eqref{eq:chsch1} can be transformed into
\begin{equation}\label{eq:ch4order}
\nabla^2(\nabla^2 \phi^{n+1})-\frac{S}{\lambda}\nabla^2 \phi^{n+1}+\frac{\gamma_0}{m_0 \lambda \Delta{t}}\phi^{n+1}=\frac{1}{m_0\lambda}\Big[ \frac{\hat \phi}{\Delta{t}}+f^{n+1} \Big]-\frac{S}{\lambda}\nabla^2 \bar \phi^{n+1}+\xi \frac{1}{\lambda}\nabla^2 h(\bar \phi^{n+1}),
\end{equation}
where we have used the notation in equation \eqref{eq:timedirnew}.
This equation can be reformulated into the following two Helmholtz type
equations that are de-coupled from each other (barring 
the unknown scalar number $\xi$), (see e.g.~\cite{DongS2012,YangLD2019}
for details)
\begin{subequations}\label{eq:chpsiphigov}
\begin{align}
&\nabla^2 \psi^{n+1}-\Big( \alpha+\frac{S}{\lambda} \Big)\psi^{n+1}=\frac{1}{m_0\lambda}\Big[ \frac{\hat \phi}{\Delta{t}}+f^{n+1} \Big]-\frac{S}{\lambda}\nabla^2 \bar \phi^{n+1}+\xi \frac{1}{\lambda}\nabla^2 h(\bar \phi^{n+1}), \label{eq:chpsi1}\\
&\nabla^2 \phi^{n+1}+\alpha \phi^{n+1}=\psi^{n+1}, \label{eq:chphi}
\end{align}
\end{subequations}
where $\psi^{n+1}$ is an auxiliary field variable defined by \eqref{eq:chphi}, and
the constant $\alpha$ is given by and the chosen constant $S$ must satisfy
\begin{equation}\label{eq;chalpha}
  \alpha=-\frac{S}{2\lambda}\Big( 1-\sqrt{1-\frac{4\gamma_0 \lambda}{m_0 \Delta{t} S^2}}  \Big);\quad
  S\geqslant \sqrt{\frac{4\lambda \gamma_0}{m_0\Delta{t}}}.
\end{equation}
In light of \eqref{eq:chphi} and \eqref{eq:chbd2},
the boundary condition \eqref{eq:chsch4} can be transformed into
\begin{equation}\label{eq:chbdnew1}
\bs n\cdot \nabla \psi^{n+1}=\Big[\Big( \alpha+\frac{S}{\lambda}  \Big)d_b^{n+1}-\frac{1}{m_0\lambda}d_a^{n+1} \Big]-\frac{S}{\lambda}\bs n \cdot \nabla \bar \phi^{n+1}+\xi \frac{1}{\lambda}\bs n\cdot \nabla h(\bar \phi^{n+1}).
\end{equation}

To solve equations \eqref{eq:chpsi1}-\eqref{eq:chphi} together with the boundary conditions \eqref{eq:chbdnew1} and \eqref{eq:chbd2}, we take advantage of
the fact that $\xi$ is a scalar number and
introduce two sets of field functions $(\psi_i^{n+1},\phi_i^{n+1})$ $(i=1,2)$ as solutions of the following equations:\\
\noindent\underline{For $\psi_{1}^{n+1}$}: 
\begin{subequations}\label{eq:chpsi1_ind}
  \begin{align}
    \label{eq:chpsi1eq1}
    &
    \nabla^2 \psi_1^{n+1}-\Big( \alpha+\frac{S}{\lambda} \Big)\psi_1^{n+1}=\frac{1}{m_0\lambda}\Big[ \frac{\hat \phi}{\Delta{t}}+f^{n+1} \Big]-\frac{S}{\lambda}\nabla^2 \bar \phi^{n+1},
    \\
%
    \label{eq:chpsi1eq2}
    &
\bs n\cdot \nabla \psi_1^{n+1}=\Big[\Big( \alpha+\frac{S}{\lambda}  \Big)d_b^{n+1}-\frac{1}{m_0\lambda}d_a^{n+1} \Big]-\frac{S}{\lambda}\bs n \cdot \nabla \bar \phi^{n+1}.
\end{align}
\end{subequations}
\noindent\underline{For $\psi_{2}^{n+1}$}: 
\begin{equation}\label{eq:chpsi2}
\nabla^2 \psi_2^{n+1}-\Big( \alpha+\frac{S}{\lambda} \Big)\psi_2^{n+1}=\frac{1}{\lambda}\nabla^2 h(\bar \phi^{n+1}),\quad \bs n\cdot \nabla \psi_2^{n+1}=\frac{1}{\lambda}\bs n\cdot \nabla h(\bar \phi^{n+1}).
\end{equation}
\noindent\underline{For $\phi_{1}^{n+1}$}: 
\begin{equation}\label{eq:chphi1}
\nabla^2 \phi_1^{n+1}+\alpha \phi_1^{n+1}=\psi_1^{n+1},\quad  \bs n\cdot\nabla \phi_1^{n+1}=d_b^{n+1}.
\end{equation}
\noindent\underline{For $\phi_{2}^{n+1}$}: 
\begin{equation}\label{eq:chphi2}
\nabla^2 \phi_2^{n+1}+\alpha \phi_2^{n+1}=\psi_2^{n+1},\quad  \bs n\cdot\nabla \phi_2^{n+1}=0.
\end{equation}
Then for
given scalar number $\xi,$ the following field functions solve the system consisting of equations \eqref{eq:chpsiphigov}, \eqref{eq:chbdnew1} and \eqref{eq:chbd2}:
\begin{equation}\label{eq:chpsiphi}
\psi^{n+1}=\psi_1^{n+1}+\xi \psi_2^{n+1},\quad \phi^{n+1}=\phi_1^{n+1}+\xi \phi_2^{n+1},
\end{equation}
where $(\psi_i^{n+1}, \phi_i^{n+1})$ $(i=1,2)$ are given by equations \eqref{eq:chpsi1eq1}-\eqref{eq:chphi2}.

Now we are ready to determine the unknown scalar $\xi.$
Following equations \eqref{eq:defstar},
we define
\begin{equation}\label{eq:chphistar}
\tilde\phi^{n+1}=\phi_1^{n+1}+\phi_2^{n+1},\quad  \tilde\psi^{n+1}=\psi_1^{n+1}+\psi_2^{n+1},\quad  \nabla^2 \tilde\phi^{n+1}=\tilde\psi^{n+1}-\alpha \tilde\phi^{n+1}
\end{equation}
where equation \eqref{eq:chphi} has been used.
Accordingly, in light of equations \eqref{eq:cheq2} and \eqref{eq:timedirnew},
we define
\begin{equation}\label{eq:chmustar}
  \left\{
  \begin{split}
    &
  \tilde\mu^{n+1}=-\lambda \nabla^2 \tilde\phi^{n+1}+h(\tilde\phi^{n+1})
  = -\lambda(\tilde\psi^{n+1}-\alpha\tilde\phi^{n+1}) + h(\tilde\phi^{n+1}), \\
  &
  \left.\frac{\partial \phi}{\partial t}\right|^{*,n+1}=\frac{\gamma_0 \tilde\phi^{n+1}-\hat \phi}{\Delta{t}}.
  \end{split}
  \right.
\end{equation}
We further define
\begin{equation}\label{eq:phistar32}
\tilde\phi^{n+\frac{3}{2}}=\frac32 \tilde\phi^{n+1}-\frac12 \phi^n.
\end{equation}
Combining equations \eqref{eq:chsch3} and \eqref{eq:chsolver},
we obtain the formula for $\xi$,
\begin{equation}\label{eq:chxide}
\xi=\frac{\mathscr{F}(R^{n+1/2}) +{\Delta}t |S_0|}{ E[\tilde\phi^{n+\frac{3}{2}}]+ \Delta{t}m_0\int_\Omega |\nabla \tilde\mu^{n+1}|^2d\Omega +\Delta{t}(|S_0|-S_0) },
\end{equation} 
where $S_0$ is given by 
\begin{equation}\label{eq:chS0}
  S_0= \int_\Omega f^{n+1}\tilde\mu^{n+1}d\Omega
  +\int_{\Gamma} \left(d_a^{n+1} \tilde\mu^{n+1}+\lambda d_b^{n+1} \left.\frac{\partial \phi}{\partial t}\right|^{*,n+1} \right)d\Gamma.
\end{equation}
Once $\xi$ is known, $\phi^{n+1}$ and $\psi^{n+1}$ can be obtained directly by equation \eqref{eq:chpsiphi} and $R^{n+1}$ can be computed based on equation \eqref{eq:RN1}.

Equations \eqref{eq:chpsi1}-\eqref{eq:chphi2} are Helmholtz type equations
with Neumann type boundary conditions. They can be implemented with
$C^0$ spectral elements in a straightforward fashion.

\begin{rem}
  In equation \eqref{eq:chsch1}, we have treated the nonlinear term
  explicitly by $ h(\bar \phi^{n+1})$. When $\Delta{t}$ becomes large,
  $\bar \phi^{n+1}$ can no longer approximate $\phi^{n+1}$ well. Thus, although the scheme \eqref{eq:chscheme}-\eqref{eq:chsch2} is unconditionally stable,
  the simulation will lose accuracy for large time steps.
  One possible approach to improve the accuracy
  is to replace $\xi h(\bar \phi^{n+1})$ in equation \eqref{eq:chsch1} by 
\begin{equation*}
\frac{\lambda}{\eta^2}\Big( \phi_0^2-1 \Big)\phi^{n+1}+\xi\Big[ h(\bar \phi^{n+1})- \frac{\lambda}{\eta^2}\Big( \phi_0^2-1 \Big)\bar \phi^{n+1}  \Big],
\end{equation*}
where $\phi_0$ is a chosen field function close to $\phi^{n+1}$,
e.g.~a snapshot of the $\phi$ field in the recent past.
The first term in the above equation serves as a linearized
approximation of $h(\phi^{n+1})$ and the second term serves as
a correction to this approximation. 
By doing so, equation \eqref{eq:chsch1} with the mentioned modification
is still linear, but can no longer be decoupled straightforwardly.
One needs to solve either a fourth-order linear equation
or a coupled linear system. However, this treatment can result
in improved accuracy besides unconditional stability.  We will demonstrate
this in the forthcoming case for the
Cahn-Hilliard equation with variable mobility.
\end{rem}

\subsection{Variable Mobility}\label{subsec:varmob}

Next, we consider the case with a variable mobility,
$m(\phi)=\max(m_0(1-\phi^2), 0)$.
We reformulate the equations \eqref{eq:cheq1}--\eqref{eq:cheq3a} into
\begin{subequations}
  \begin{align}
    &
    \frac{\partial\phi}{\partial t} = \nabla\cdot\left[
      m_c(\phi_0)\nabla C
      \right]
    + \frac{\mathscr{F}(R)}{E}\nabla\cdot\left[
      m(\phi)\nabla\mu - m_c(\phi_0)\nabla C
      \right] + f, \label{eq:varch_1_trans}
    \\
    &
    m_c(\phi_0)\bs n\cdot\nabla C +
    \frac{\mathscr{F}(R)}{E}\bs n\cdot\left[
      m(\phi)\nabla\mu - m_c(\phi_0)\nabla C
      \right] = d_a. \label{eq:varch3_trans}
  \end{align}
\end{subequations}
In these equations, $\mu$ is given by \eqref{eq:cheq2},
$\phi_0$ is a chosen field distribution corresponding to $\phi(\bs x,t)$
at a certain time instant or at some time instants, and
\begin{equation}\label{eq:def_C}
  \left\{
  \begin{split}
    &
    C = -\lambda\nabla^2\phi + S(\phi-\phi) + \kappa(\phi_0)\phi; \\
    &
    m_c(\phi_0) = m(\phi_0), \quad \text{or} \ \ m_c(\phi_0)=m_0; \\
    &
    \kappa(\phi_0) = \frac{\lambda}{\eta^2}(\phi_0^2 -1),
    \quad \text{or} \ \ \kappa(\phi_0) = 0;
  \end{split}
  \right.
\end{equation}
where $S\geqslant 0$ is a chosen constant.
By incorporating the following zero terms into the RHS of
\eqref{eq:che_auxeq},
\begin{equation}
  \begin{split}
    &
  \left( \frac{\mathscr{F}(R)}{E} -1 \right)
  \int_{\Omega}\mu\left[
    \nabla\cdot(m_c(\phi_0)\nabla C) + f
    \right]d\Omega \\
  &
  + \frac{\mathscr{F}(R)}{E} \left[
    \int_{\Omega}\mu\nabla\cdot\left[ m(\phi)\nabla\mu - m_c(\phi_0)\nabla C \right]d\Omega
    - \int_{\Omega}\mu\nabla\cdot\left[ m(\phi)\nabla\mu - m_c(\phi_0)\nabla C \right]d\Omega
    \right] \\
  &
  + \left( \frac{\mathscr{F}(R)}{E} -1 \right)
  \int_{\Gamma} \lambda d_b\frac{\partial\phi}{\partial t} d\Gamma
  + \left(1- \frac{\mathscr{F}(R)}{E}\right) \left|
  \int_{\Omega}f\mu d\Omega + \int_{\Gamma}d_a\mu d\Gamma
  + \int_{\Gamma}\lambda d_b \frac{\partial\phi}{\partial t}d\Gamma
  \right|,
  \end{split}
\end{equation}
we can transform this equations into,
\begin{equation}\label{eq:che_eng_trans}
  \begin{split}
    \mathscr{F}'(R)\frac{dR}{dt} =&
    \int_{\Omega} \mu\frac{\partial\phi}{\partial t}d\Omega
    -\int_{\Omega}\mu\left[
      \nabla\cdot(m_c(\phi_0)\nabla C)
      + \frac{\mathscr{F}(R)}{E}\nabla\cdot\left[
      m(\phi)\nabla\mu - m_c(\phi_0)\nabla C
      \right] + f
      \right] d\Omega \\
    &
    + \frac{\mathscr{F}(R)}{E}\left[
      -\int_{\Omega}m(\phi)\nabla\mu\cdot\nabla\mu
      + \int_{\Omega} f\mu d\Omega + \int_{\Gamma}d_a\mu d\Gamma
      + \int_{\Gamma}\lambda d_b\frac{\partial \phi}{\partial t} d\Gamma
      \right] \\
    &
    +\left(1- \frac{\mathscr{F}(R)}{E}\right)\left|
    \int_{\Omega}f\mu d\Omega + \int_{\Gamma}d_a\mu d\Gamma
    + \int_{\Gamma}\lambda d_b \frac{\partial\phi}{\partial t}d\Gamma
    \right|.
  \end{split}
\end{equation}

Following equations \eqref{eq:schemeeq1}-\eqref{eq:schemeeq2},
we propose the following scheme:
\begin{subequations}\label{eq:varCHgov}
\begin{align}
  & \frac{\partial \phi}{\partial t}\Big|^{n+1}
  =\nabla\cdot \big(m_c(\phi_0)\nabla C^{n+1} \big)+\xi \nabla \cdot \Big[m(\bar {\phi}^{n+1})\nabla \bar \mu^{n+1}-m_c(\phi_0)\nabla \bar C^{n+1}   \Big]+f^{n+1},\label{eq:varch1}\\
& C^{n+1}=-\lambda \nabla^2 \phi^{n+1} +S(\phi^{n+1}-\bar \phi^{n+1})  +       \kappa(\phi_0)\phi^{n+1},\label{eq:chC}\\
  &
  \xi = \frac{\mathscr{F}(R^{n+3/2})}{E[\tilde\phi^{n+3/2}]}, \\
  &
  E[\tilde\phi^{n+3/2}] = \int_{\Omega}\left[
    \frac{\lambda}{2}\left|\nabla\tilde\phi^{n+3/2} \right|^2
    + H(\tilde\phi^{n+3/2})
    \right]d\Omega + C_0,
  \\
  &m_c(\phi_0){\bs n} \cdot \nabla C^{n+1}+\xi  \bs n \cdot \Big[  m(\bar {\phi}^{n+1})\nabla \bar \mu^{n+1}-m_c(\phi_0)\nabla \bar C^{n+1}   \Big] =d_a^{n+1}\;\text{on}\;\;\partial \Omega,\label{eq:varchsch3} 
\end{align}
\end{subequations}
and
 \begin{equation}\label{eq:varch2}
\begin{split}
  & D_{\mathscr{F}}(R) \Big|^{n+1} \frac{dR}{dt}\Big|^{n+1}
  =\int_\Omega \big[-\lambda\nabla^2\phi^{n+1} +  h(\phi^{n+1}) \big]\frac{\partial \phi}{\partial t}\Big|^{n+1} d\Omega
  -\xi \int_\Omega m(\tilde\phi^{n+1})\left|\nabla \tilde\mu^{n+1}\right|^2d\Omega
  \\
  &+\int_{\Omega}  \big[\lambda\nabla^2\phi^{n+1} - h(\phi^{n+1}) \big]
  \Big\{ \nabla\cdot \big(m_c(\phi_0)\nabla C^{n+1} \big)
  + \xi \nabla \cdot \Big[m(\bar {\phi}^{n+1})\nabla \bar \mu^{n+1}-m_c(\phi_0)\nabla \bar C^{n+1}   \Big]+f^{n+1}\Big \}d\Omega \\
  &+\xi \left[ \int_\Omega f^{n+1}\tilde\mu^{n+1}d\Omega
    +\int_{\Gamma} \Big(d_a^{n+1} \tilde\mu^{n+1}
    +\lambda d_b^{n+1}\left. \frac{\partial \phi}{\partial t}\right|^{*,n+1} \Big)d\Gamma  \right]\\
  &+(1-\xi)\left|\int_\Omega f^{n+1}\tilde\mu^{n+1}d\Omega
  +\int_{\Gamma} \left(d_a^{n+1} \tilde\mu^{n+1}
  +\lambda d_b^{n+1} \left.\frac{\partial \phi}{\partial t}\right|^{*,n+1} \right)d\Gamma
  \right|,
 \end{split}
\end{equation}
 together with 
 the boundary condition \eqref{eq:chbd2} and
 the initial condition \eqref{eq:chsch5}.
 In these equations,
 $\left.\frac{\partial\phi}{\partial t}\right|^{n+1}$
 and $\left.\frac{dR}{d t}\right|^{n+1}$ are defined in \eqref{eq:tempdist},
 $\bar\phi^{n+1}$ is given by \eqref{eq:def_bar}, and
 $\bar C^{n+1}$ and $\bar \mu^{n+1}$ are computed by
\begin{equation}\label{eq:varCHCbar}
\bar C^{n+1}=-\lambda \nabla^2 \bar \phi^{n+1}+\kappa(\phi_0)\bar \phi^{n+1},\quad \bar \mu^{n+1}=-\lambda \nabla^2 \bar \phi^{n+1}+h(\bar \phi^{n+1}).
\end{equation}
$\tilde\phi^{n+1}$, $\tilde\phi^{n+3/2}$, $\tilde\mu^{n+1}$,
and $\left.\frac{\partial\phi}{\partial t} \right|^{*,n+1}$
are approximations to be specified later.

 \begin{theorem}
   In the absence of the external source term ($f=0$),
   and with zero boundary conditions ($d_a=d_b=0$),
   the scheme consisting of \eqref{eq:varCHgov}-\eqref{eq:varch2}
   is unconditionally energy stable in the sense that
 \begin{equation}\label{eq:varchdislaw}
   {\mathscr{F}(R^{n+\frac{3}{2}}) -\mathscr{F}(R^{n+\frac{1}{2}})}
   =-\xi \Delta{t} \int_\Omega m(\tilde\phi^{n+1})|\nabla \tilde\mu^{n+1}|^2 \leq 0,
 \end{equation}
 if the approximation of $R(t)$ at time step $\frac{1}{2}$ is positive.
 \end{theorem}

\begin{proof}
  We take the  $L^2$ inner product between $\big(-\lambda\nabla^2\phi^{n+1} +  h(\phi^{n+1}) \big)$ and equation \eqref{eq:varch1}, and add the resultant equation to equation \eqref{eq:varch2}. This leads to
 \begin{equation}\label{eq:varchsolver}
 \begin{split}
   & \frac{\mathscr{F}(R^{n+\frac{3}{2}}) -\mathscr{F}(R^{n+\frac{1}{2}})}{\Delta{t}}
   =\xi \Big\{ \int_\Omega f^{n+1}\tilde{\mu}^{n+1}d\Omega
   +\int_{\Gamma} \big(d_a^{n+1} \tilde{\mu}^{n+1}+\lambda d_b^{n+1} \left.\frac{\partial \phi}{\partial t}\right|^{*,n+1} \big)d\Gamma  \Big\}\\
   &-\xi \int_\Omega m(\tilde\phi^{n+1})|\nabla \tilde\mu^{n+1}|^2d\Omega
   +(1-\xi)\bigg|\int_\Omega f^{n+1}\tilde\mu^{n+1}+\int_{\Gamma} \big(d_a^{n+1} \tilde\mu^{n+1}+\lambda d_b^{n+1} \left.\frac{\partial \phi}{\partial t}\right|^{*,n+1} \big) d\Gamma\bigg|.
 \end{split}
\end{equation}  
 By the same arguments as in the proof of Theorem \ref{thm:chdislaw}, we
 arrive at the relation \eqref{eq:varchdislaw} based on the above equation.
\end{proof}

For implementation of the scheme, one notes that
equation \eqref{eq:varch1} can be transformed into
\begin{equation}\label{eq:varchtran1}
\frac{\gamma_0}{\Delta{t}} \phi^{n+1}-\nabla \cdot \big[ m_c(\phi_0)\nabla C^{n+1} \big]=\Big( \frac{\hat \phi}{\Delta{t}}+f^{n+1}\Big)+\xi \nabla \cdot \Big[m(\bar {\phi}^{n+1})\nabla \bar \mu^{n+1}-m_c(\phi_0)\nabla \bar C^{n+1}   \Big].
\end{equation}
Barring the unknown scalar $\xi,$
equations \eqref{eq:varchtran1}, \eqref{eq:chC}, \eqref{eq:varchsch3} and \eqref{eq:chbd2} can be solved as follows. 
Introduce two pairs of field functions $(\phi_i^{n+1},C_i^{n+1})$ $(i=1,2),$ as the solution of the following equations:\\
\noindent\underline{For $(\phi_1^{n+1},C_1^{n+1})$}: 
\begin{subequations}\label{eq:varchphiC}
\begin{align}
&\frac{\gamma_0}{\Delta{t}} \phi_1^{n+1}-\nabla \cdot \big[ m_c(\phi_0)\nabla C_1^{n+1} \big]=\frac{\hat \phi}{\Delta{t}}+f^{n+1}, \label{eq:varC1eq1}\\
&\big(\kappa(\phi_0)+S\big)\phi_1^{n+1}  -\lambda \nabla^2 \phi_1^{n+1}-C_1^{n+1}=S\bar \phi^{n+1} ,\label{eq:varC1eq2}\\
& m_c(\phi_0){\bs n} \cdot \nabla C_1^{n+1}=d_a^{n+1},\;\;\text{on}\; \Gamma  \label{eq:varC1eq3}\\
& \bs n \cdot \nabla \phi_1^{n+1}=d_b^{n+1},\;\;\text{on}\; \Gamma. \label{eq:varC1eq4}
\end{align}
\end{subequations}
\noindent\underline{For $(\phi_2^{n+1},C_2^{n+1})$}: 
\begin{subequations}\label{eq:varchphic2}
\begin{align}
&\frac{\gamma_0}{\Delta{t}} \phi_2^{n+1}-\nabla \cdot \big[ m_c(\phi_0)\nabla C_2^{n+1} \big]= \nabla \cdot \Big[m(\bar {\phi}^{n+1})\nabla \bar \mu^{n+1}-m_c(\phi_0)\nabla \bar C^{n+1}   \Big], \label{eq:varC2eq1}\\
&\big(\kappa(\phi_0)+S\big)\phi_2^{n+1}  -\lambda \nabla^2 \phi_2^{n+1}-C_2^{n+1}=0 ,  \label{eq:varC2eq2}\\
& m_c(\phi_0){\bs n} \cdot \nabla C_2^{n+1}= -\bs n \cdot \Big[  m(\bar {\phi}^{n+1})\nabla \bar \mu^{n+1}-m_c(\phi_0)\nabla \bar C^{n+1}   \Big] ,\;\;\text{on}\; \Gamma,  \label{eq:varC2eq3} \\
& \bs n \cdot \nabla \phi_2^{n+1}=0,\;\;\text{on}\; \Gamma.  \label{eq:varC2eq4}
\end{align}
\end{subequations}
Then for
 given scalar value $\xi,$ the following field functions solve the system consisting of equations \eqref{eq:varch1}-\eqref{eq:varchsch3} and \eqref{eq:chbd2}:
\begin{equation}\label{eq:varCphi}
C^{n+1}=C_1^{n+1}+\xi C_2^{n+1},\quad \phi^{n+1}=\phi_1^{n+1}+\xi \phi_2^{n+1}
\end{equation}
where $(C_i^{n+1}, \phi_i^{n+1})$ $(i=1,2)$ are given by equations \eqref{eq:varchphiC}-\eqref{eq:varchphic2}.

The unknown scalar value $\xi$ remains to be determined.
Following equation \eqref{eq:defstar},
$\tilde\phi^{n+1},$ $\tilde\mu^{n+1}$ and $\left.\frac{\partial \phi}{\partial t}\right|^{*,n+1}$ are again given by equations \eqref{eq:chphistar} and \eqref{eq:chmustar},
where based on  equation \eqref{eq:chC} we compute $\nabla^2 \tilde\phi^{n+1}$ by
\begin{equation}
  \nabla^2 \tilde\phi^{n+1}=\frac{1}{\lambda}\Big[ \kappa(\phi_0)\tilde\phi^{n+1}
    +S(\tilde\phi^{n+1}-\bar \phi^{n+1})-(C_1^{n+1}+C_2^{n+1})  \Big].
\end{equation}
The approximation $\tilde\phi^{n+\frac{3}{2}}$ is given by \eqref{eq:phistar32}.
As a result, $\xi$ can be computed by, 
\begin{equation}
  \xi=\frac{\mathscr{F}(R^{n+1/2}) +{\Delta}t |S_0|}
           { E[\tilde\phi^{n+\frac{3}{2}}]
             + \Delta{t}\int_\Omega m(\tilde\phi^{n+1}) |\nabla \tilde\mu^{n+1}|^2d\Omega +\Delta{t}(|S_0|-S_0) },
\end{equation}
where $S_0$ is given by \eqref{eq:chS0}, 
and $\phi^{n+1}$ and $R^{n+1}$ can be evaluated by equations \eqref{eq:varCphi} and \eqref{eq:RN1}, respectively.

Equations 
\eqref{eq:varchphiC}-\eqref{eq:varchphic2} can be discretized in space by
$C^0$ spectral elements, and their weak forms are: \\
\noindent\underline{For $(\phi_{1}^{n+1}, C_{1}^{n+1})$}: Find $\phi_{1}^{n+1}\;, C_{1}^{n+1} \in H^1(\Omega)$ such that
\begin{align}
  &
\frac{\gamma_0}{\Delta{t}}\big( \phi_{1}^{n+1}, \varphi  \big)_{\Omega}+\Big( m_c(\phi_{0})\nabla C_{1}^{n+1}, \nabla \varphi   \Big)_{\Omega}=\Big( \frac{\hat \phi}{\Delta{t}}+f^{n+1},\varphi  \Big)_{\Omega}+\langle d_{a}^{n+1}\varphi \rangle_{\Gamma}, \\
%
&
\Big(\big[\kappa(\phi_{0})+S\big]\phi_{1}^{n+1},\varphi   \Big)_{\Omega}+\lambda \big(  \nabla \phi_{1}^{n+1},\nabla \varphi \big)_{\Omega}-\big( C_{1}^{n+1},\varphi \big)_{\Omega}=S\big( \bar \phi^{n+1},\varphi \big)_{\Omega}+\lambda \langle d_{b}^{n+1}\varphi \rangle_{\Gamma},
\end{align}
for all $ \varphi \in H^{1}(\Omega).$
\\
\noindent\underline{For $(\phi_{2}^{n+1}, C_{2}^{n+1})$}: Find $\phi_{2}^{n+1}\;, C_{2}^{n+1} \in H^1(\Omega)$ such that
\begin{align}
  &
\frac{\gamma_0}{\Delta{t}}\big( \phi_{2}^{n+1}, \varphi  \big)_{\Omega}+\Big( m_c(\phi_{0})\nabla C_{2}^{n+1}, \nabla \varphi   \Big)_{\Omega}=-\Big( m(\bar {\phi}^{n+1})\nabla \bar \mu^{n+1}-m_c(\phi_{0})\nabla \bar C^{n+1} , \nabla \varphi  \Big)_{\Omega}, \\
%
&
\Big(\big[\kappa(\phi_{0})+S\big]\phi_{2}^{n+1},\varphi   \Big)_{\Omega}+\lambda \big(  \nabla \phi_{2}^{n+1},\nabla \varphi \big)_{\Omega}-\big( C_{2}^{n+1},\varphi \big)_{\Omega}=0,
\end{align}
for all $  \varphi \in H^1(\Omega).$

\begin{rem}

  If one chooses $\kappa(\phi_0)=0$ and $m_c(\phi_0)=m_0>0$,
  then the scheme \eqref{eq:varch1}--\eqref{eq:varch2} can also be implemented
  by solving four de-coupled Helmholtz type equations in
  a way similar to the constant mobility case in Section \ref{subsec:constmob}.

\end{rem}


\subsection{Numerical Results}


We next provide numerical examples to demonstrate
the accuracy and unconditional stability of the proposed schemes \eqref{eq:chscheme}-\eqref{eq:chsch2} and \eqref{eq:varCHgov}-\eqref{eq:varch2} for Cahn-Hilliard equation
with constant and variable mobilities.
For cases with variable mobility we employ
$m_c(\phi_0) = m(\phi_0) = \max(m_0(1-\phi_0^2),0)$ in the algorithm
with these tests,
where $m_0$ and $\phi_0$ will be specified below.

\subsubsection{Convergence Rates}

\begin{figure}[tbp]
  \centering
 \subfigure[Errors vs Element order (constant mobility)]{ \includegraphics[scale=.39]{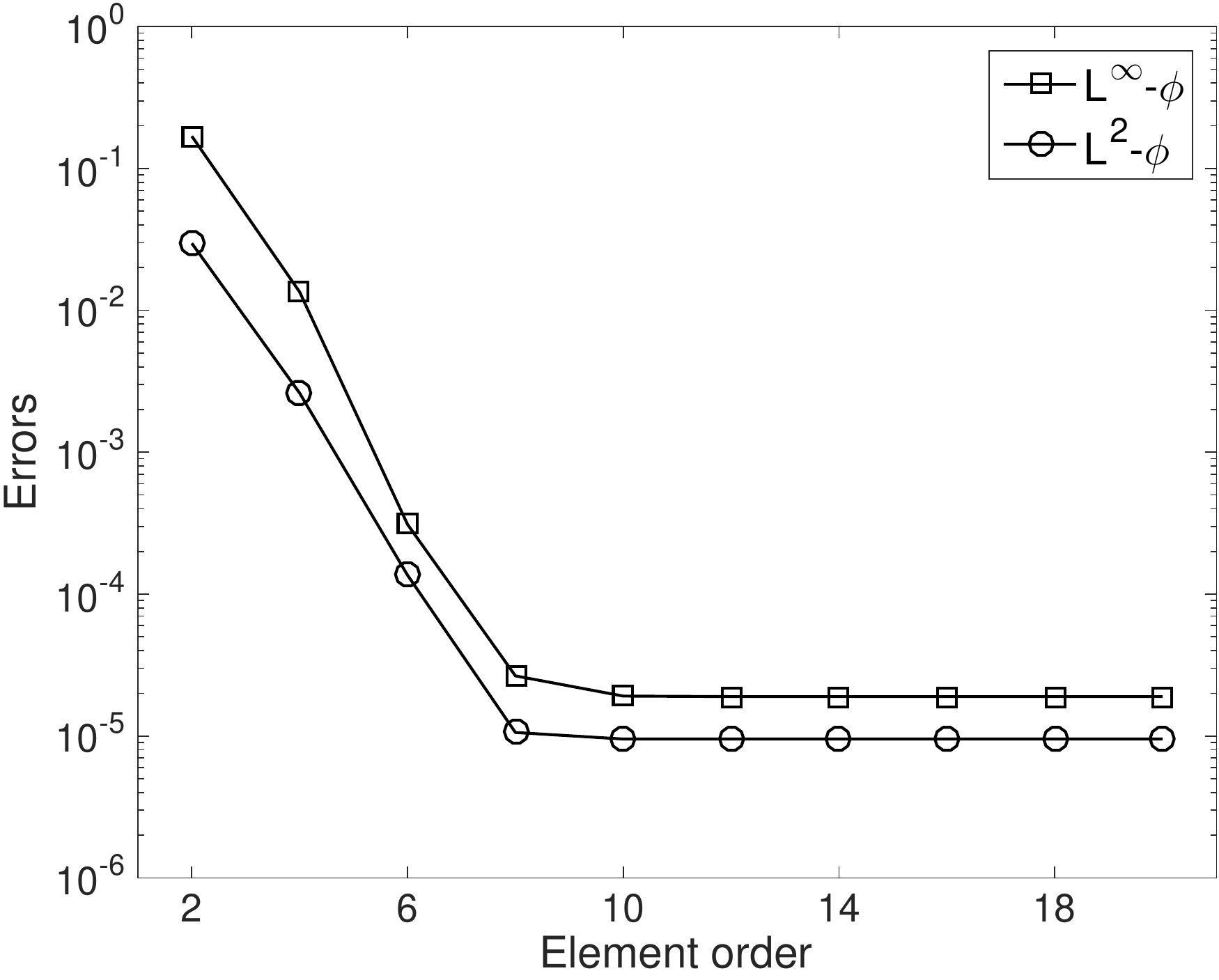}} \quad
  \subfigure[Errors vs Element order (variable mobility)]{ \includegraphics[scale=.39]{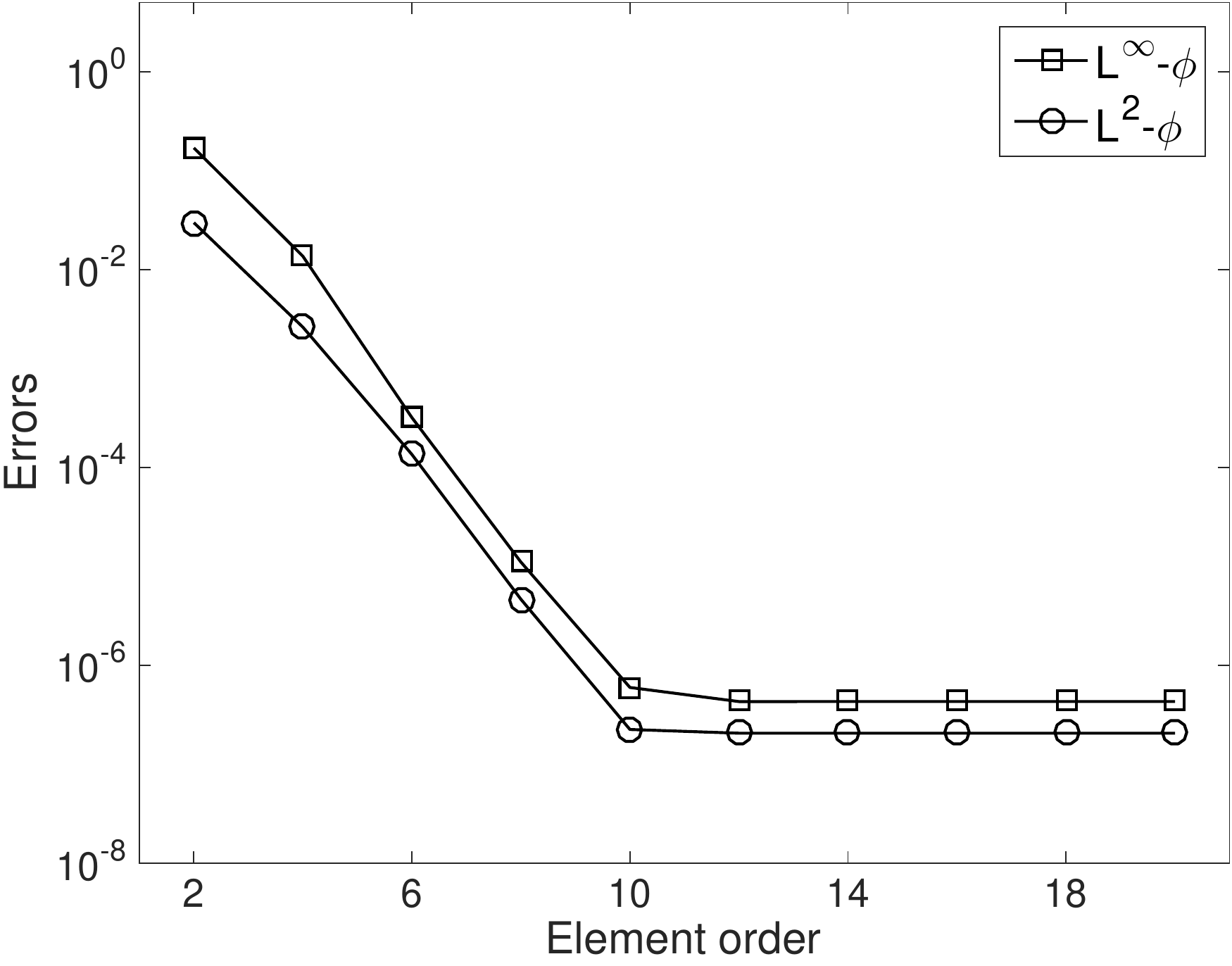}} \\
 \subfigure[Errors vs ${\Delta}t$ (constant mobility)]{ \includegraphics[scale=.38]{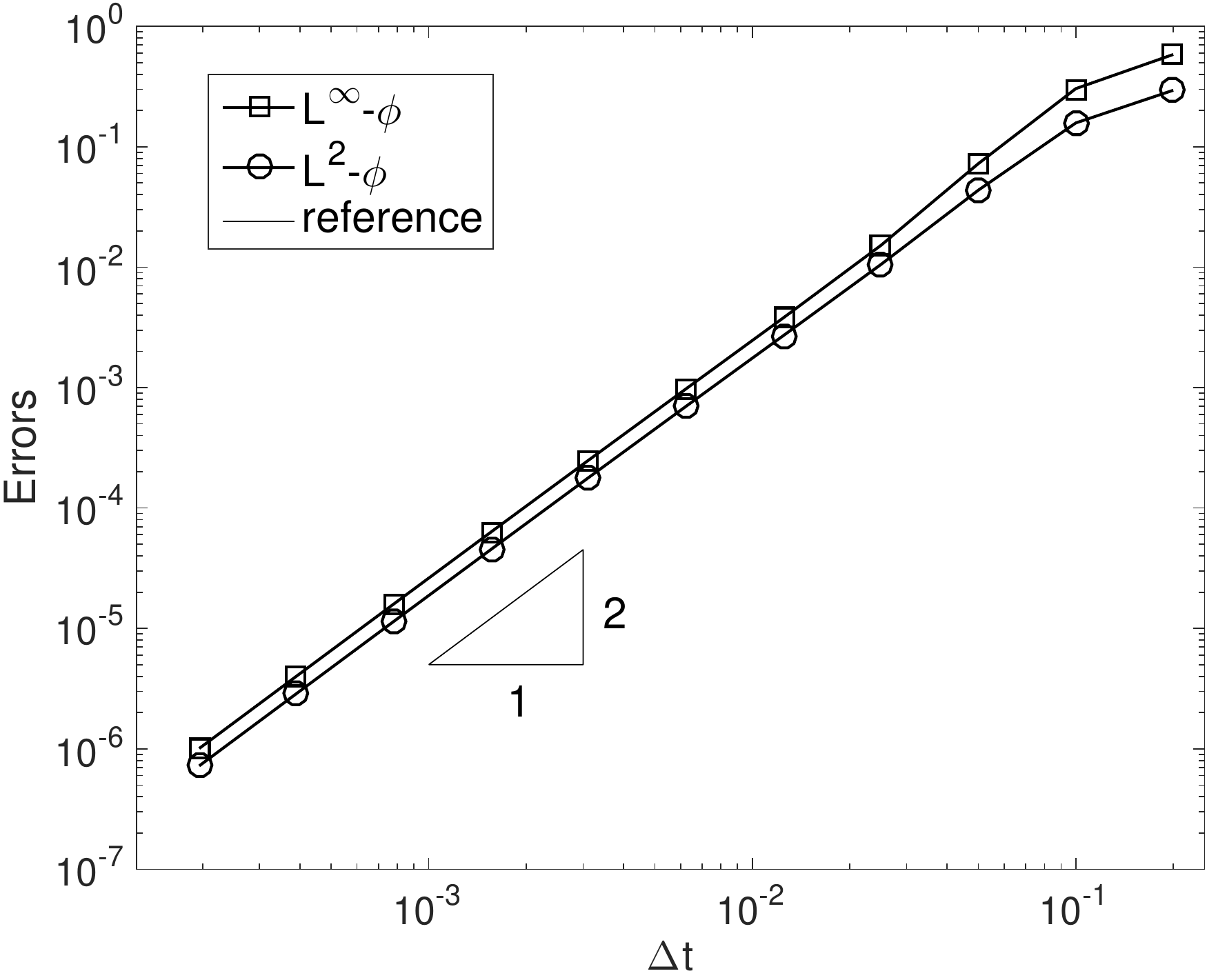}} \quad
 \subfigure[Errors vs ${\Delta}t$ (variable mobility)]{ \includegraphics[scale=.38]{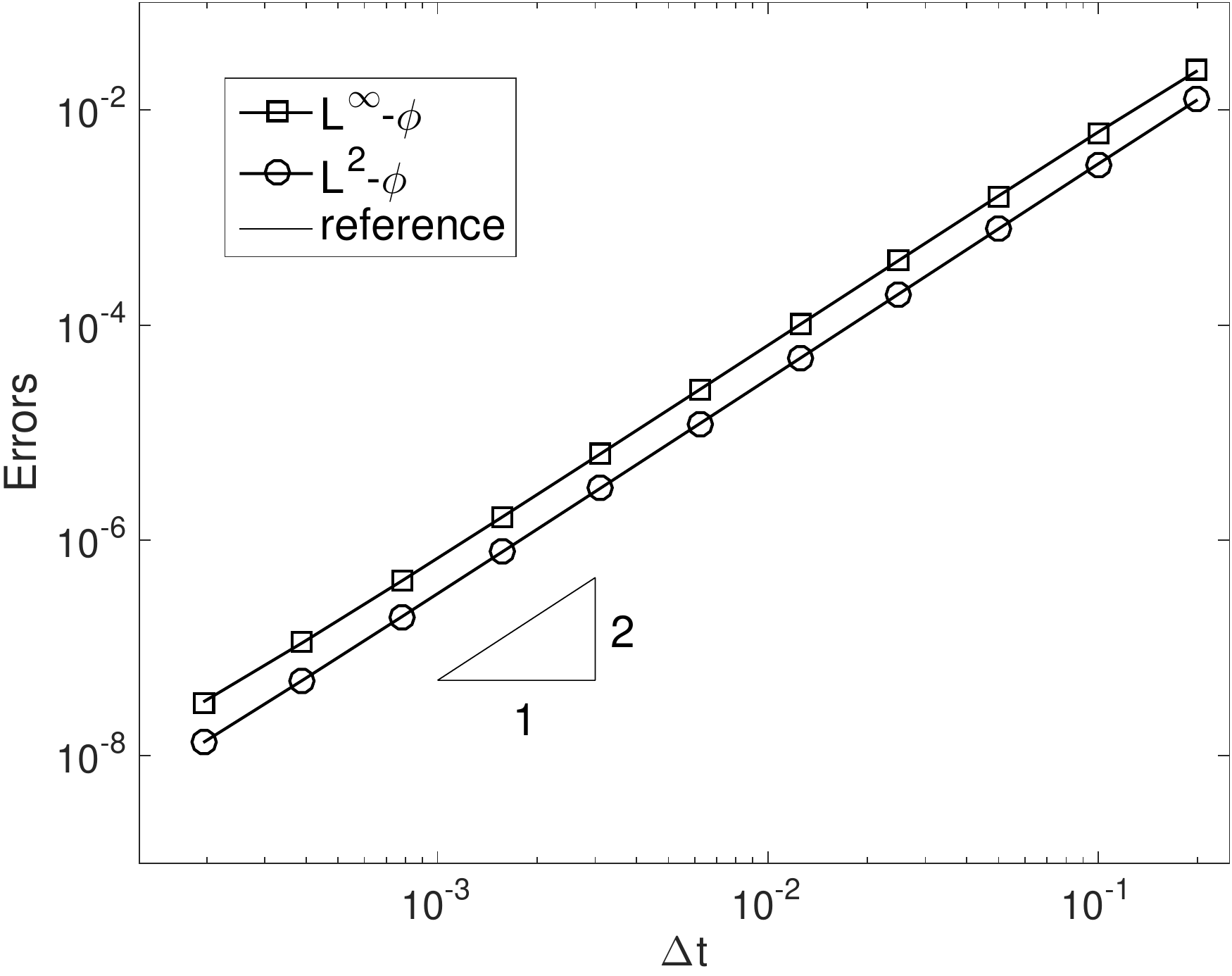}} \\
 \caption{Spatial/temporal convergence tests for Cahn-Hilliard equation. $L^{2}$ and $L^{\infty}$ errors of $\phi$ versus element order for (a) constant mobility, (b) variable mobility (fixed $\Delta t=0.001,$ $t_0=0.1,$ $t_f=0.2$). $L^{2}$ and $L^{\infty}$ errors of $\phi$ versus $\Delta t$ for (c) constant mobility, (d) variable mobility (fixed element order $18$ and $t_0=0.1,$ $t_f=1.1$). Numerical results correspond to
   $\phi_0=\phi_{in}(\bs x)$ in \eqref{eq:varCHgov} for cases with variable mobility.
}
\label{fig:CHEstandtest}
\end{figure}

Consider domain $\Omega=[0,2]\times[-1,1]$ and a contrived solution in this domain:
\begin{equation}\label{eq:chstd}
\phi(\bs x,t)=\cos(\pi x)\cos(\pi y)\sin(t).
\end{equation}
The external force and boundary source terms
$f(\bs x, t)$, $d_a(\bs x,t)$ and $d_b(\bs x, t)$ in \eqref{eq:cheq1},
\eqref{eq:cheq3a} and \eqref{eq:cheq3} are chosen such that the analytic expression \eqref{eq:chstd} satisfies \eqref{eq:chsys}. 

The computational domain $\Omega$ is discretized with
two equal-sized quadrilateral elements. The algorithms \eqref{eq:chscheme}-\eqref{eq:chsch2} for the constant-mobility case and \eqref{eq:varCHgov}-\eqref{eq:varch2} for the
variable-mobility case are employed to numerically integrate the Cahn-Hilliard equation from $t=t_0$ to $t=t_f$. The initial field function $\phi_{in}$ is obtained by setting $t=t_0$ in the contrived solution \eqref{eq:chstd}. The numerical errors are computed by
comparing the numerical solution against the analytic solution \eqref{eq:chstd} at $t=t_f.$
In the following convergence tests, we fix $\mathscr{F}(R)=R,$ $C_0=1,$
and $\phi_0=\phi_{in}(\bs x)$ in \eqref{eq:varCHgov}.
The values for the simulation parameters are summarized in Table \ref{tab:CHsimu_param}.
\begin{table}
  \centering
  \begin{tabular}{ll| ll}
    \hline
    parameter & value & parameter & value \\
    $C_0$ & $1$ & $\lambda$ & $0.01$  \\
    $m_0$ & $0.01$ & $\eta$ & $0.1$  \\
    $t_0$ & $0.1$ & $t_f$ & $0.2$ (spatial tests) or $1.1$ (temporal tests) \\
    Element order & (varied) & Elements & $2$ \\
    $\Delta t$ & (varied) &$\Delta t_{\min}$ & $1e-4$  \\
    S & $1$ (variable mobility solver) & $S$ & $\sqrt{\frac{4\gamma_0\lambda}{m_0\Delta t}}$
    or $\sqrt{\frac{4\gamma_0\lambda}{m_0\Delta t_{\min}}}$ (constant mobility solver)  \\
    $\mathscr{F}(R)$ & $R$ &  $\phi_0$ & $\phi_{in}$     \\
    $m_c(\phi_0)$ & $m(\phi_0)$ \\
    \hline
  \end{tabular}
  \caption{Simulation parameter values for convergence tests of Cahn-Hilliard equation.}
  \label{tab:CHsimu_param}
\end{table}

In the spatial convergence test, we fix $\Delta{t}=0.001,$ $t_0=0.1$ and $t_f=0.2$,
and vary the element order systematically from 2 to 20. The numerical errors
in $L^{\infty}$ and $L^2$ norms at $t=t_f$ are then recorded.
For the algorithm with constant mobility, $S$ in equation \eqref{eq:chscheme} is chosen as $S=\sqrt{\frac{4\gamma_0\lambda}{m_0\Delta t}},$ while for the algorithm
with variable mobility
we use $S=1.$ Figures \ref{fig:CHEstandtest}(a) and (b) show the numerical
errors as a function of the element order from these tests.
It can be observed that the errors decrease exponentially with increasing element order
and that the error curves level off at around $10^{-5}$ and $10^{-6}$ beyond element order 8 and 10, respectively for these two solvers, due to the saturation of temporal errors.

In the temporal convergence test, we fix the element order at a large value 18, $t_0=0.1,$ and $t_f=1.1$,
and vary  $\Delta{t}$ systematically from $0.2$ to $1.953125\times 10^{-4}$
to study the behavior of numerical errors.
For the constant-mobility case, $S=\sqrt{\frac{4\gamma_0\lambda}{m_0\Delta t_{\min}}}$
(where $\Delta t_{\min}=10^{-4}$),
while for the variable-mobility case $S=1.$
Figures \ref{fig:CHEstandtest}(c) and (d) show the numerical errors as
a function of $\Delta t$ for these cases.
We observe a second-order convergence rate in time for both cases.

\subsubsection{Constant Mobility: Coalescence of Two Drops}

\begin{figure}[tbp]
  \centering
 \subfigure[$t=0$]{ \includegraphics[scale=.2]{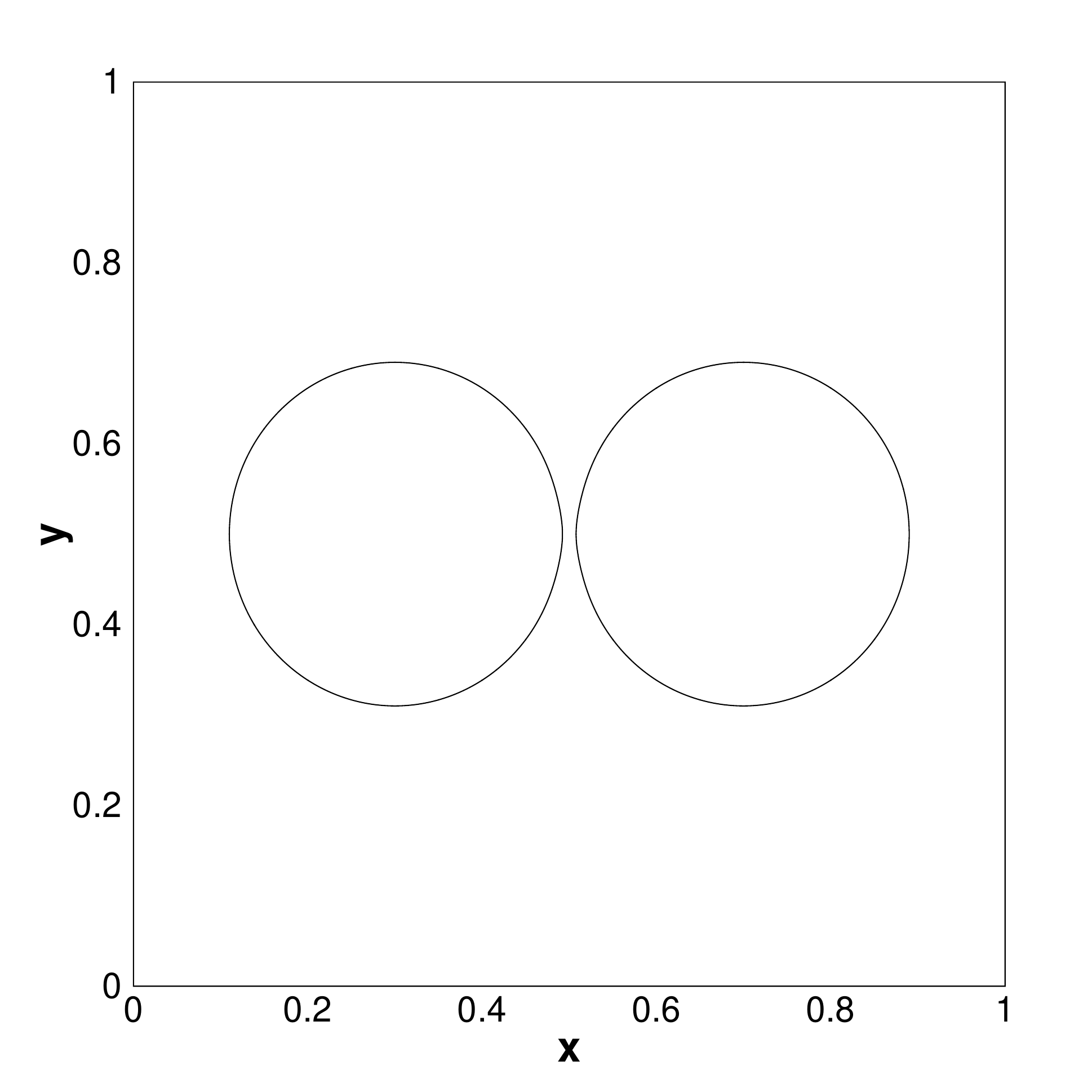}} \quad
 \subfigure[$t=1$]{ \includegraphics[scale=.2]{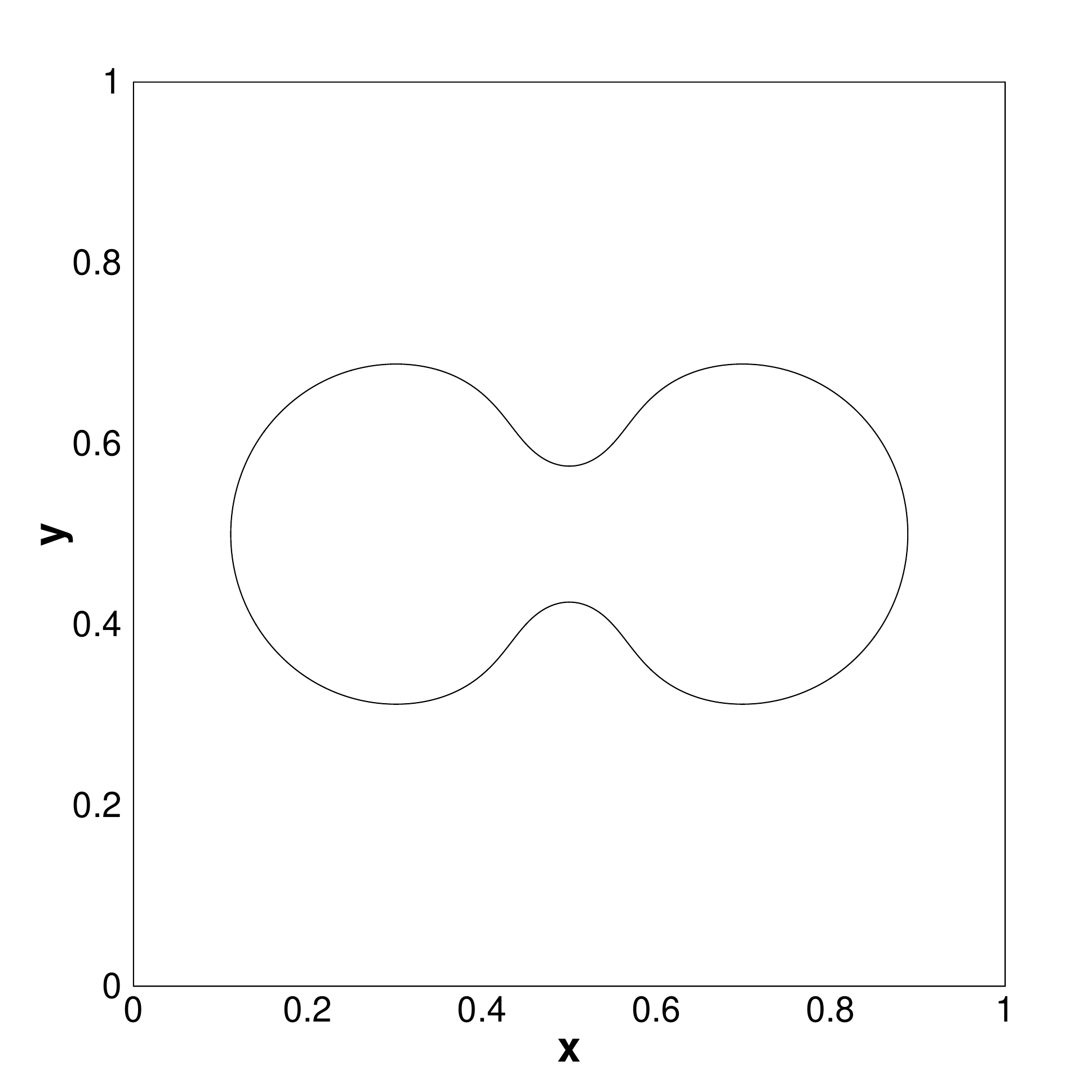}} \quad
 \subfigure[$t=5$]{ \includegraphics[scale=.2]{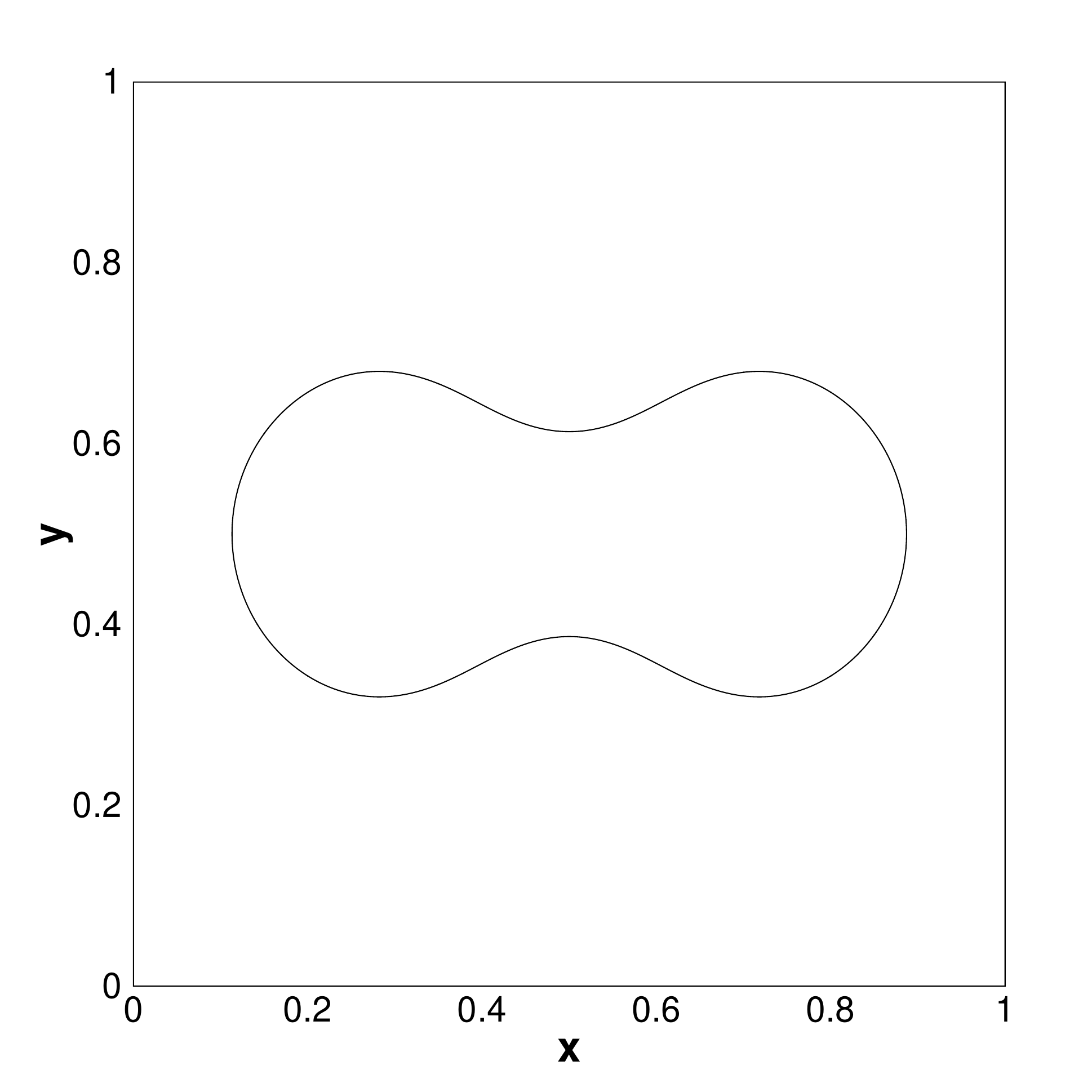}} \\
 \subfigure[$t=10$]{ \includegraphics[scale=.2]{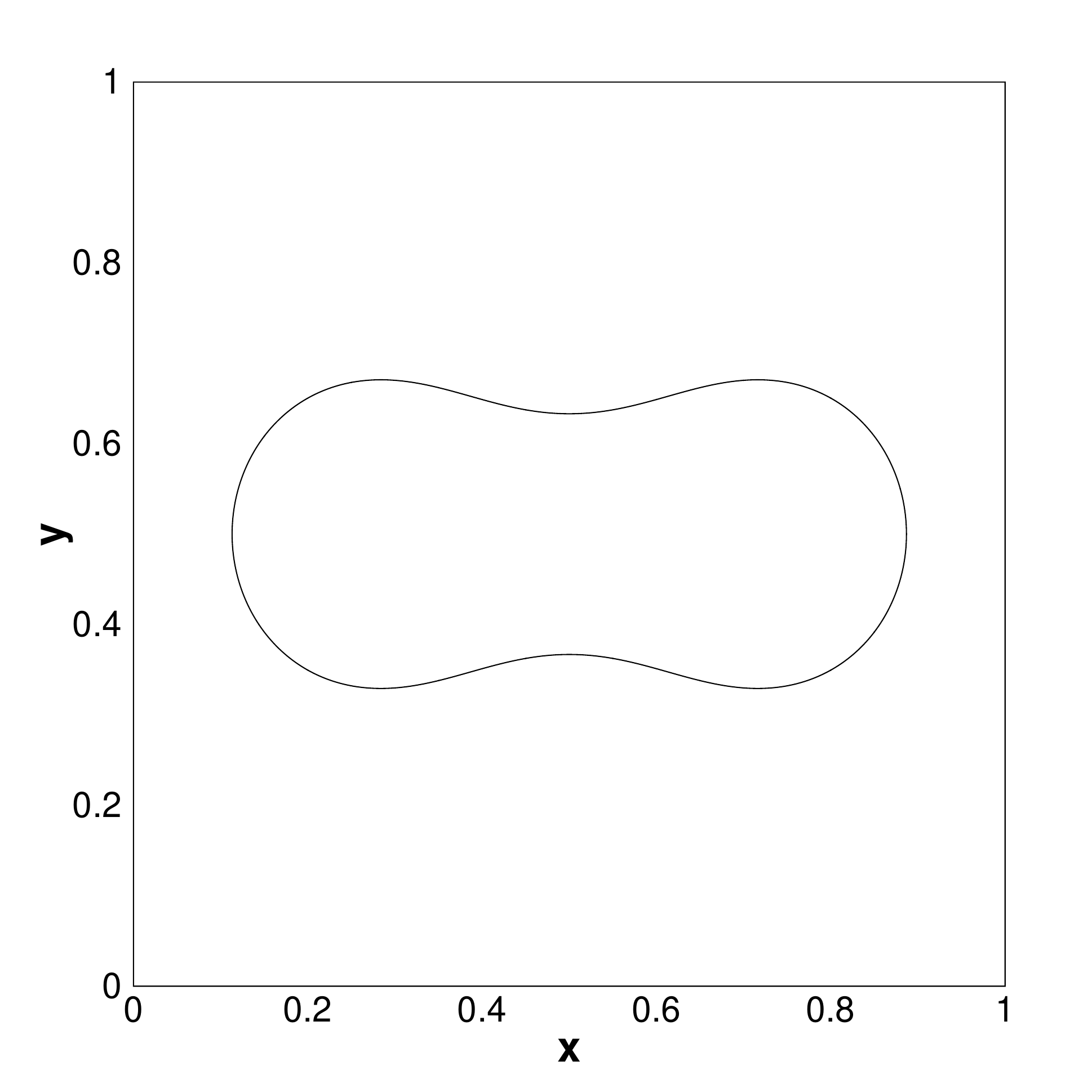}} \quad
 \subfigure[$t=25$]{ \includegraphics[scale=.2]{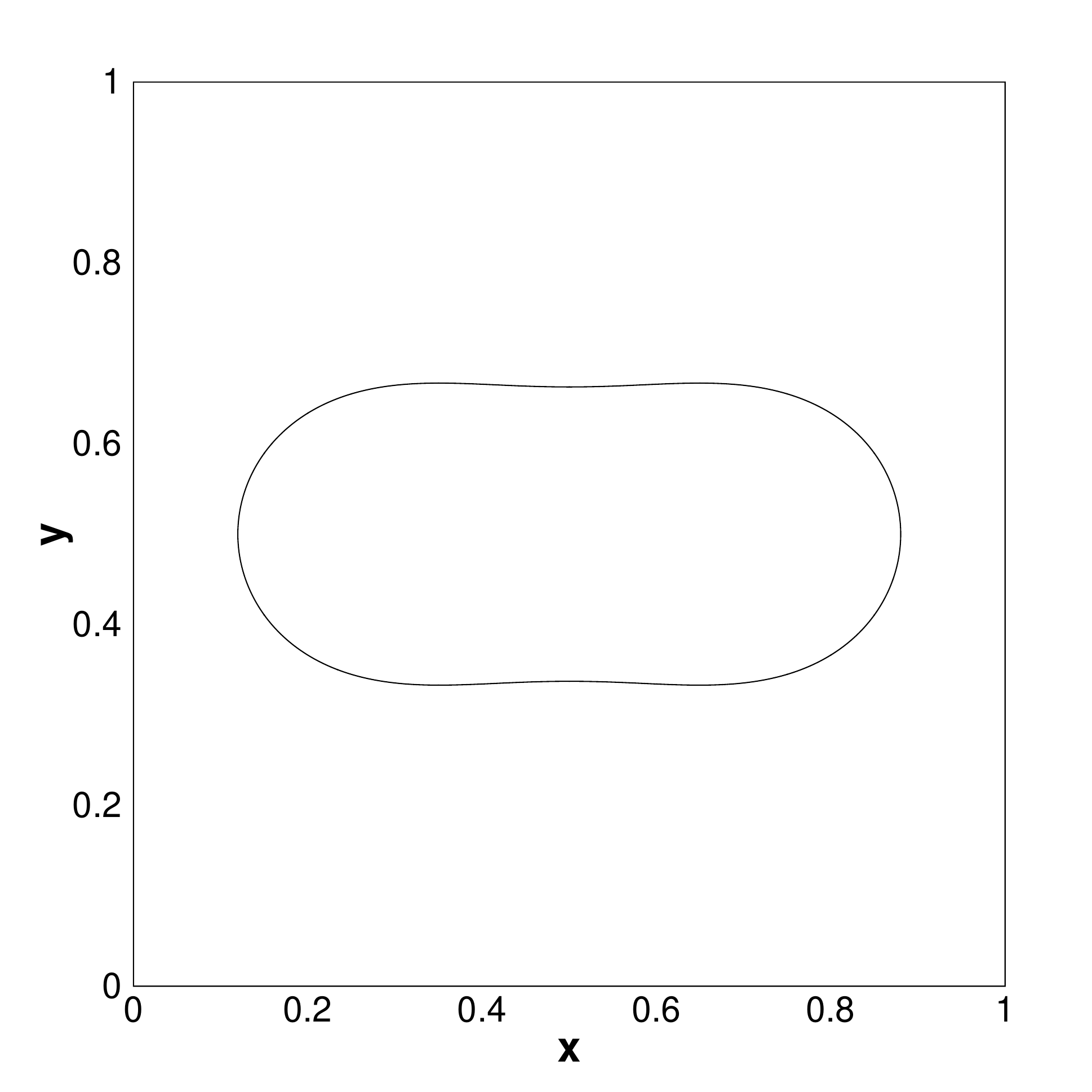}} \quad
 \subfigure[$t=50$]{ \includegraphics[scale=.2]{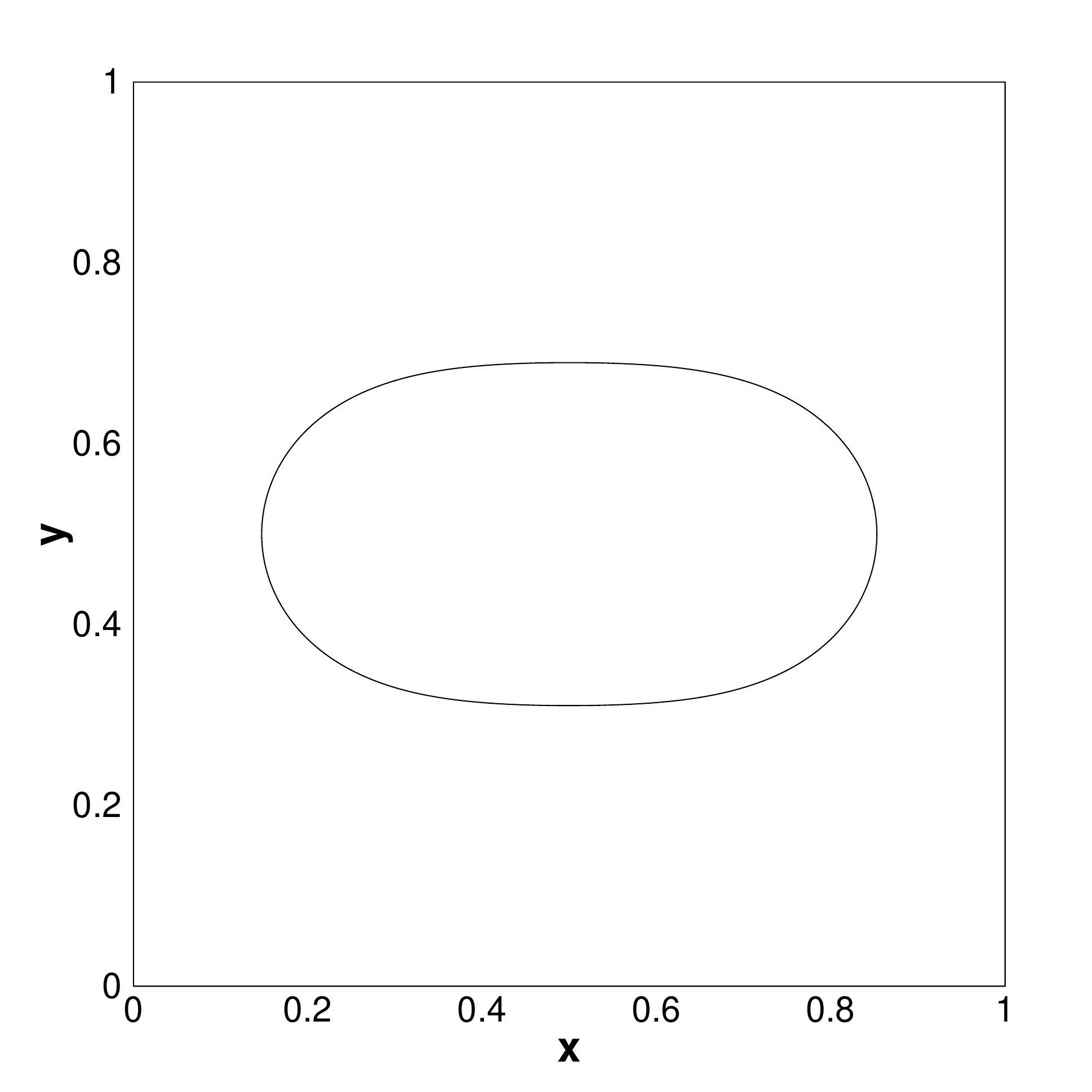}} 
 \caption{Temporal sequence of snapshots  showing the coalescence of two circular drops visualized by the contour level $\phi=0$ governed by Cahn-Hilliard
   equation with constant mobility.
 }
\label{fig:CHEmergeevol}
\end{figure}

\begin{figure}[tbp]
  \centering
 \subfigure[$\Delta{t}=10^{-1}$]{ \includegraphics[scale=.18]{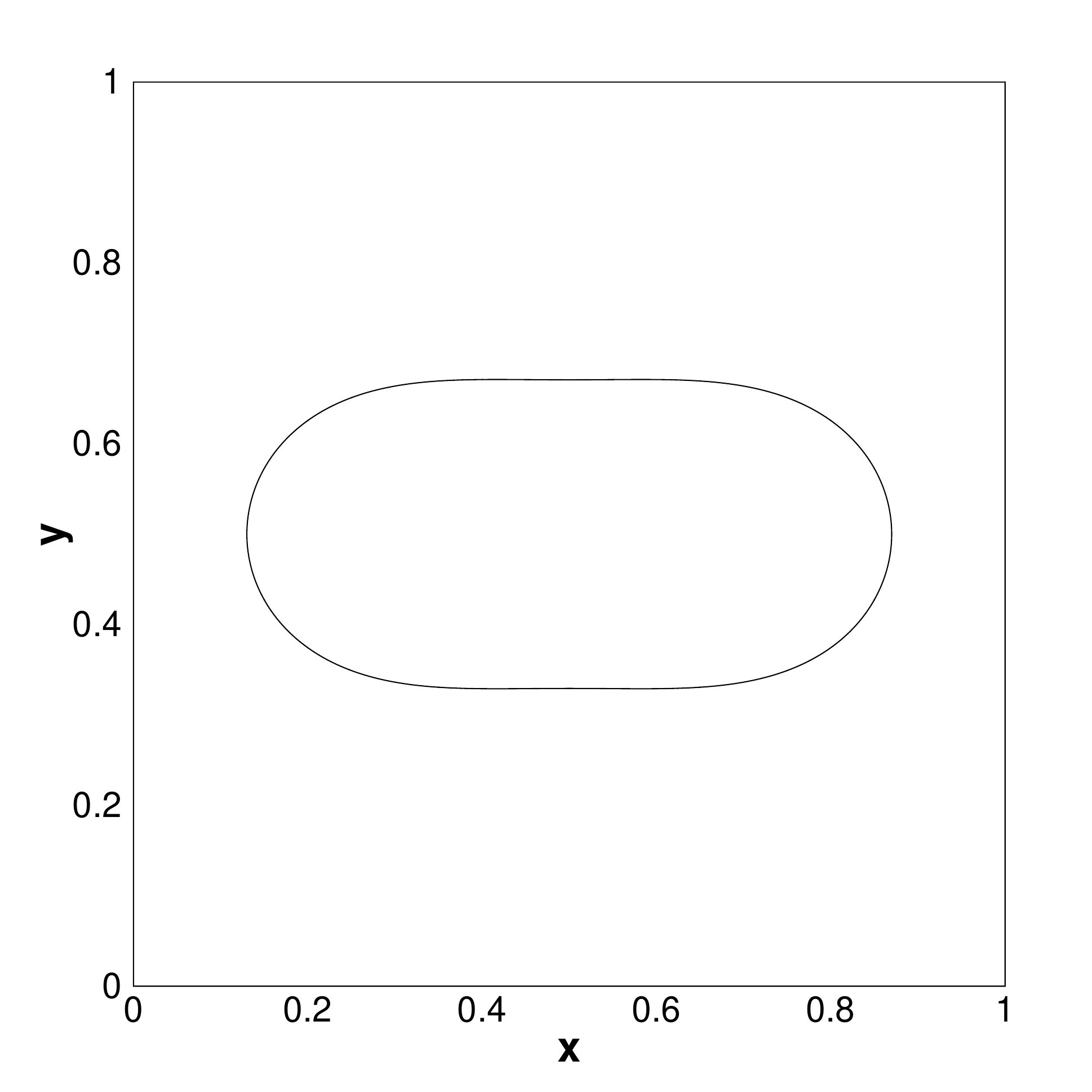}} 
 \subfigure[$\Delta{t}=10^{-2}$]{ \includegraphics[scale=.18]{CHEsnap6.pdf}}
 \subfigure[$\Delta{t}=10^{-3}$]{ \includegraphics[scale=.18]{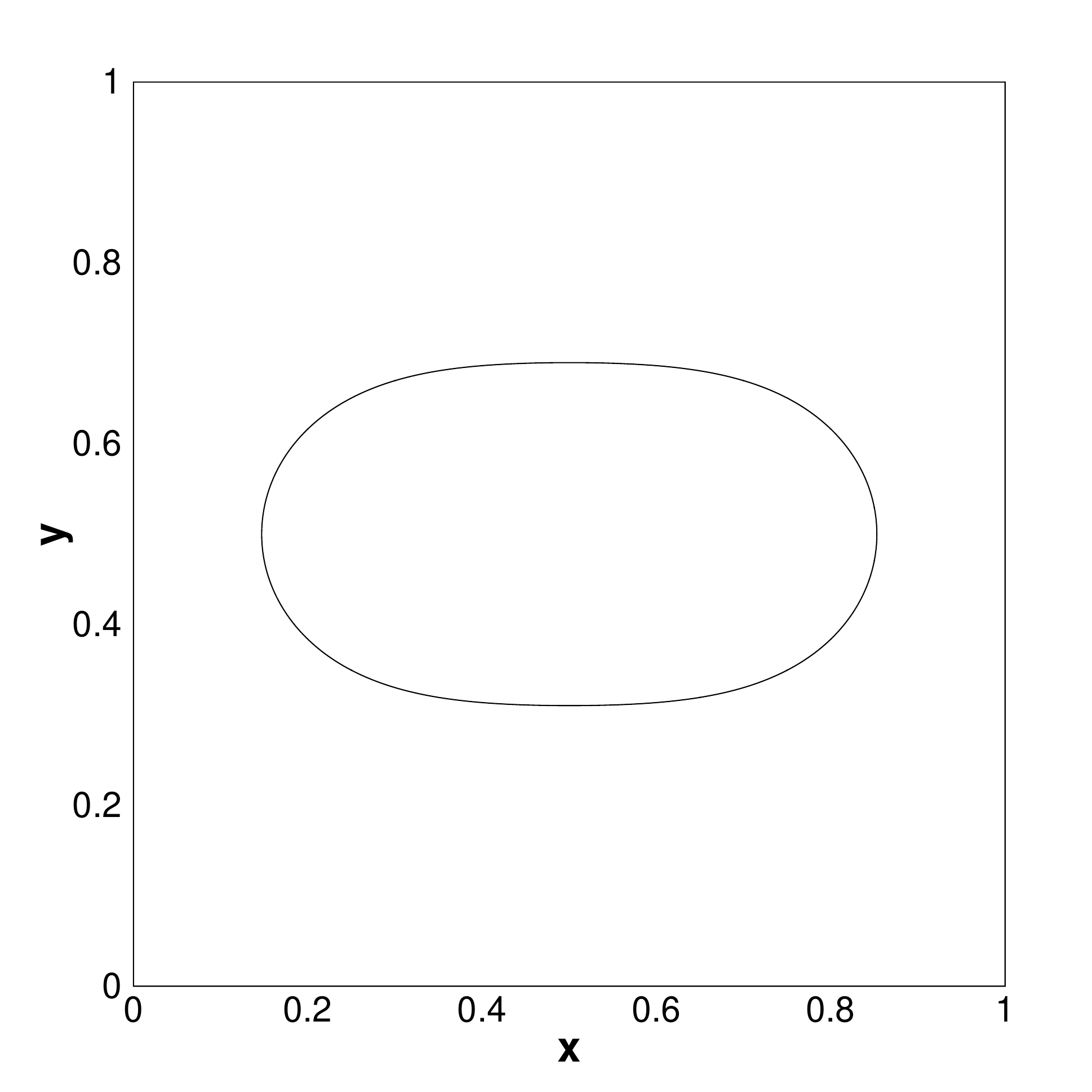}} 
 \subfigure[$\Delta{t}=10^{-4}$]{ \includegraphics[scale=.18]{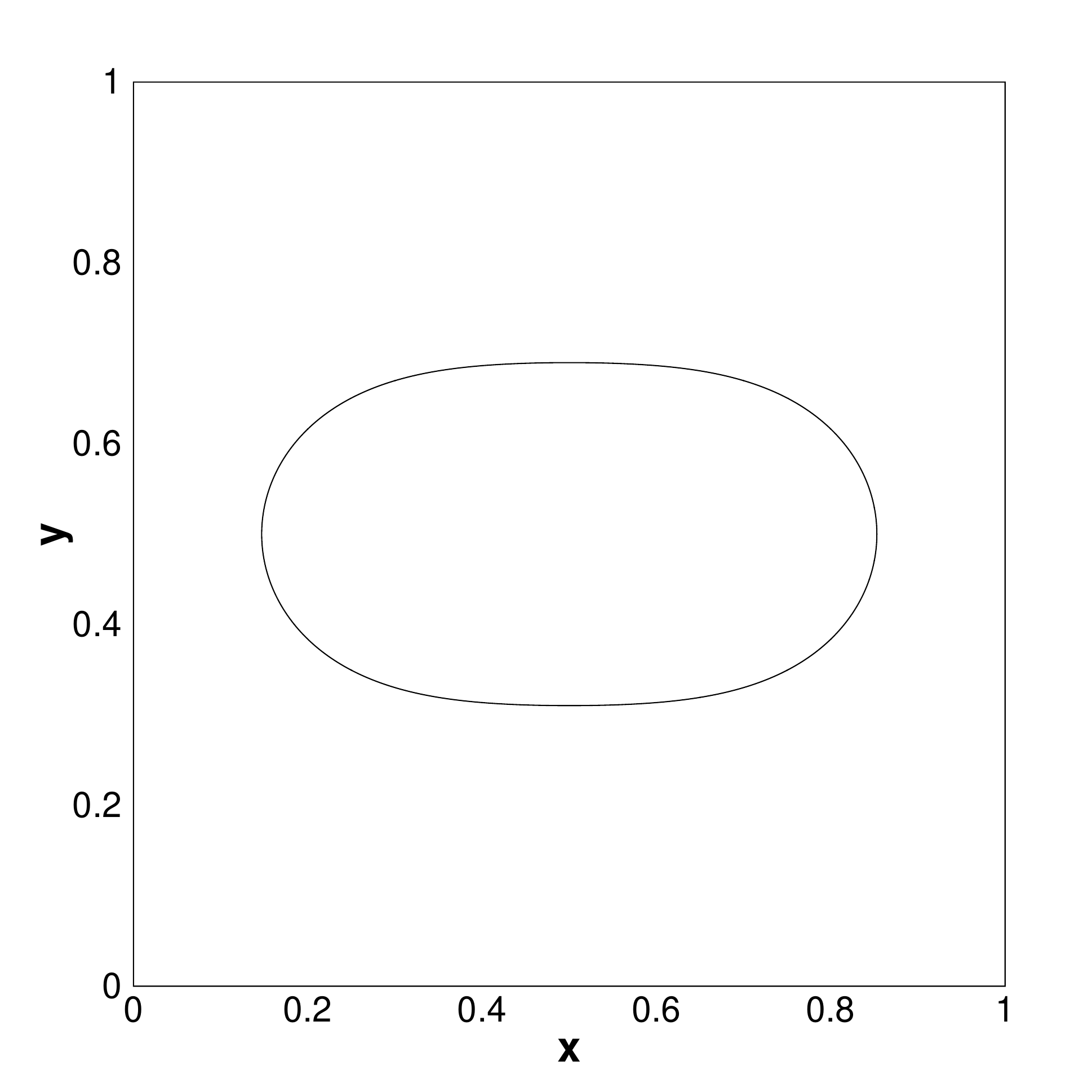}} 
 \caption{Coalescence of two drops:
   snapshots of material interfaces at $t=50$ computed using
   (a) $\Delta{t}={10^{-1}}$, (b) $\Delta{t}={10^{-2}}$, (c) $\Delta{t}={10^{-3}}$, (d) $\Delta{t}={10^{-4}}.$
 }
\label{fig:CHEmergecompare}
\end{figure}

We next consider the  coalescence of two drops  to demonstrate the numerical properties of the proposed scheme  \eqref{eq:chscheme}-\eqref{eq:chsch2} for problems with constant mobility.
Consider a square domain $\Omega=[0,1]^2$ and two materials contained in this domain. It is assumed that the dynamics of the material regions is governed by the Cahn-Hilliard equation
with a constant mobility, $m(\phi)=m_0>0$,
and that $\phi=1$ and $\phi=-1$ correspond to the bulk of the first and second materials, respectively.  We assume that at $t=0$ the first material occupies two circular regions that are right next to each other and the the rest of the domain is filled by the second material.

To be more specific, the initial distribution of the material takes the form
\begin{equation}\label{eq:chphiin}
\phi_{in}(\bs x)=1-\tanh\frac{|\bs x-\bs x_0|-R_0}{\sqrt{2}\eta}-\tanh\frac{|\bs x-\bs x_1|-R_0}{\sqrt{2}\eta},
\end{equation} 
where $\bs x_0=(x_0, y_0)=(0.3, 0.5)$ and $\bs x_1=(0.7, 0.5)$ are the centers of the circular regions for the first material, and $R_0=0.19$ is the radius of these circles.  The external force and the boundary source terms in \eqref{eq:chsys} are set to $f(\bs x, t)=d_a(\bs x,t)=d_b(\bs x,t)=0.$ We discretize the domain using 400 equal-sized quadrilateral elements with element order 10. We employ a mapping
function $\mathscr{F}(R)=R^2$ for this problem.
The simulation parameters are listed as follows:
\begin{equation}\label{eq:chepapramcoal}
\eta=0.01,\quad \sigma=151.15, \quad \lambda=\frac{3}{2\sqrt{2}}\sigma \eta, \quad m_0=\frac{10^{-6}}{\lambda},\quad S=\sqrt{\frac{4\gamma_0\lambda}{m_0\Delta t}},\quad C_0=10^6.
\end{equation}

Figure \ref{fig:CHEmergeevol} shows the evolution of the
two material regions  with a temporal sequence of snapshots of the interfaces
between these two materials visualized by the contour level $\phi=0.$
It can be observed that the two separate regions of the first material gradually coalescence with each other to form a single drop under the Cahn-Hilliard dynamics.

To investigate the effect of time step size on
the accuracy of the simulation results, in Figure \ref{fig:CHEmergecompare} we compare the distributions of the material interfaces at $t=50$ obtained with several time step sizes, ranging from $\Delta{t}=10^{-1}$ to $\Delta{t}=10^{-4}.$ The distribution computed with $\Delta{t}=10^{-2},$ $10^{-3}$ and $10^{-4}$ are essentially the same. With the larger time step size $\Delta{t}=10^{-1},$ some difference can be noticed in the material
distribution compared with
those obtained using smaller $\Delta{t}$ values.
This suggests the simulation is starting to lose accuracy
 with time step sizes $\Delta t=10^{-1}$ and larger.

\begin{figure}[tbp]
  \centering
 \subfigure[]{ \includegraphics[scale=.39]{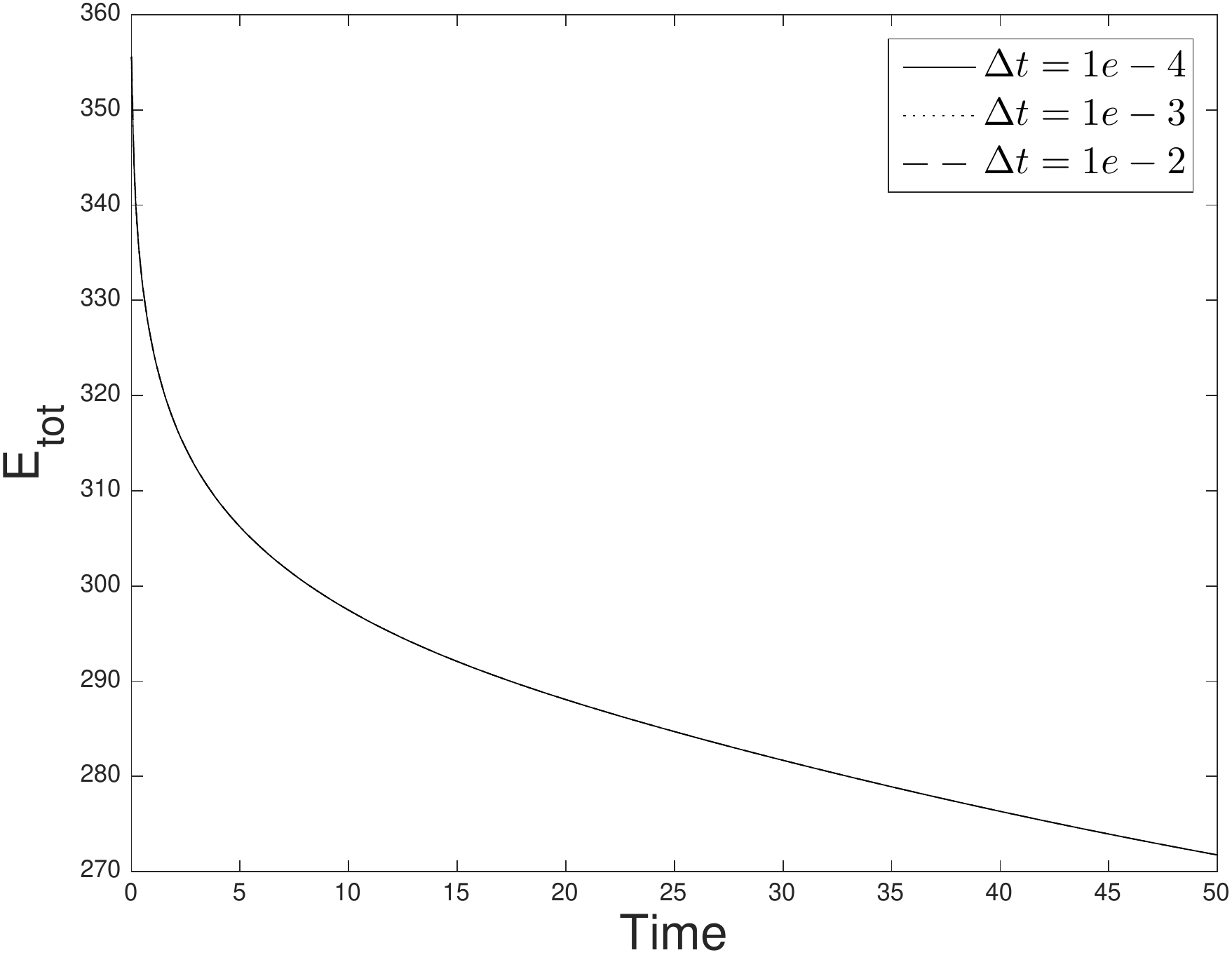}} 
 \subfigure[]{ \includegraphics[scale=.39]{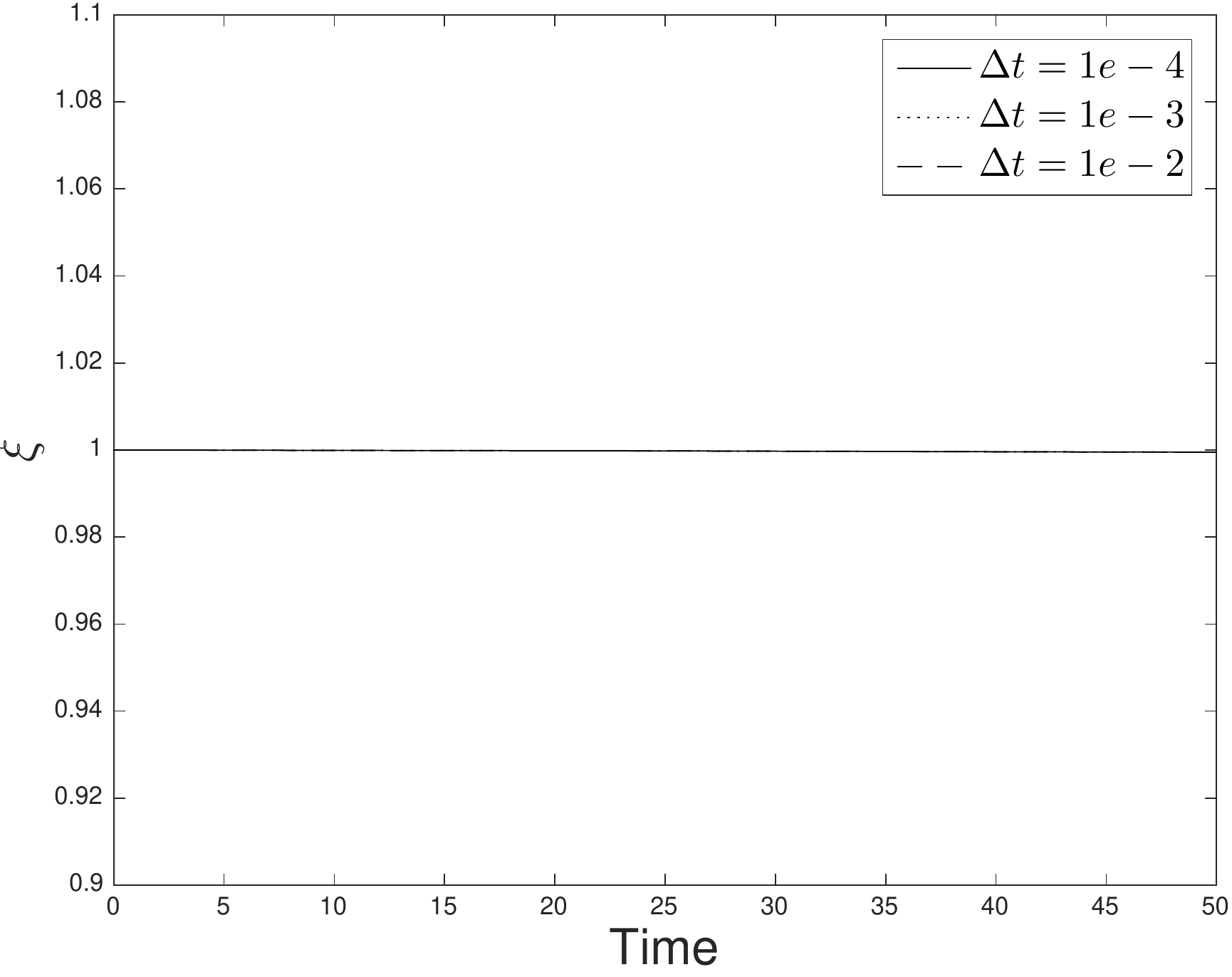}}
 \caption{Coalescence of two drops: time histories of (a) $E_{tot}(t)$ and
   (b) $\xi$ corresponding to
   a range of smaller time step sizes $\Delta{t}=10^{-2}, 10^{-3}, 10^{-4}$.
 }
\label{fig:CHEmergeEn}
\end{figure}

\begin{figure}[tbp]
  \centering
 \subfigure[]{ \includegraphics[scale=.39]{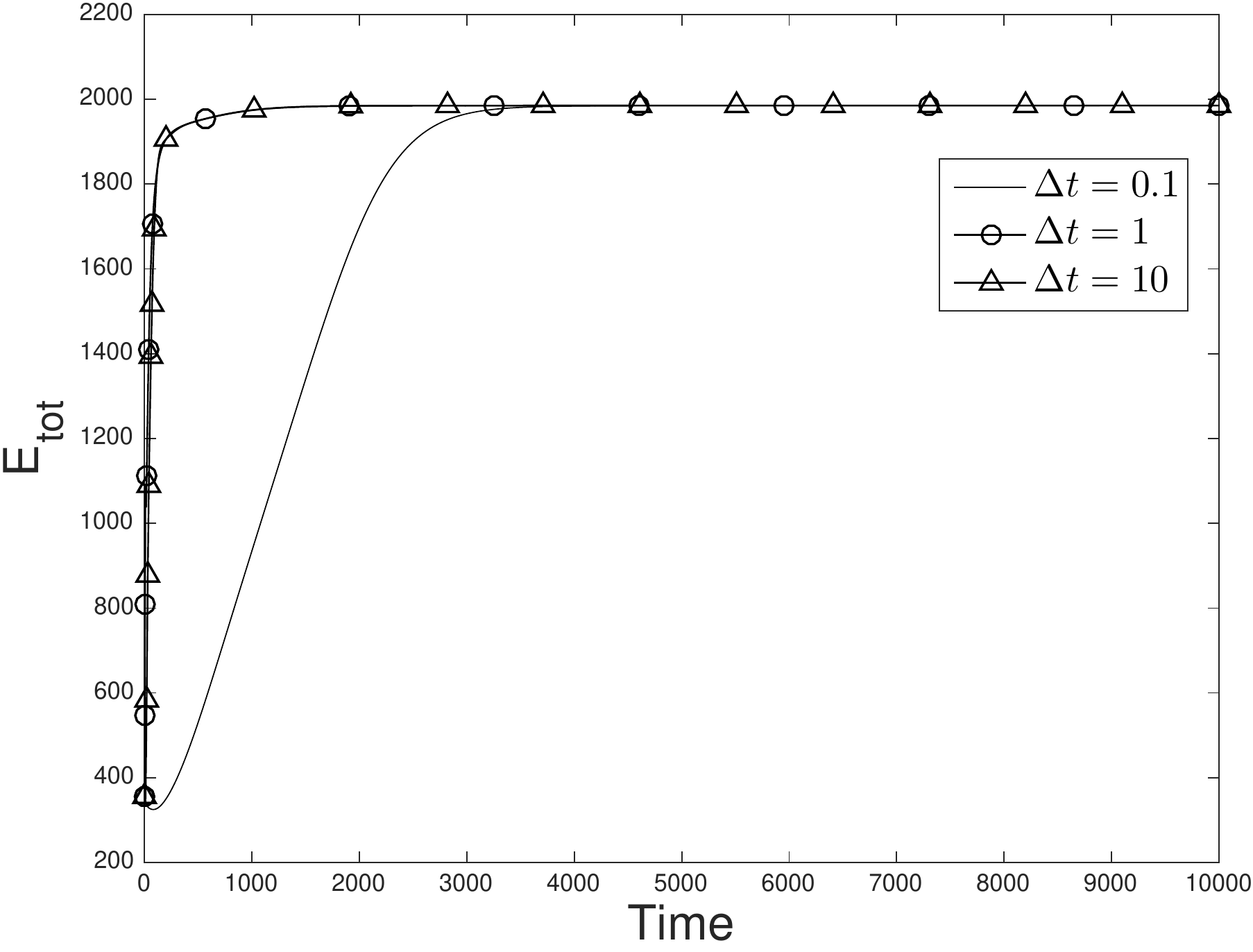}} \quad
 \subfigure[]{ \includegraphics[scale=.39]{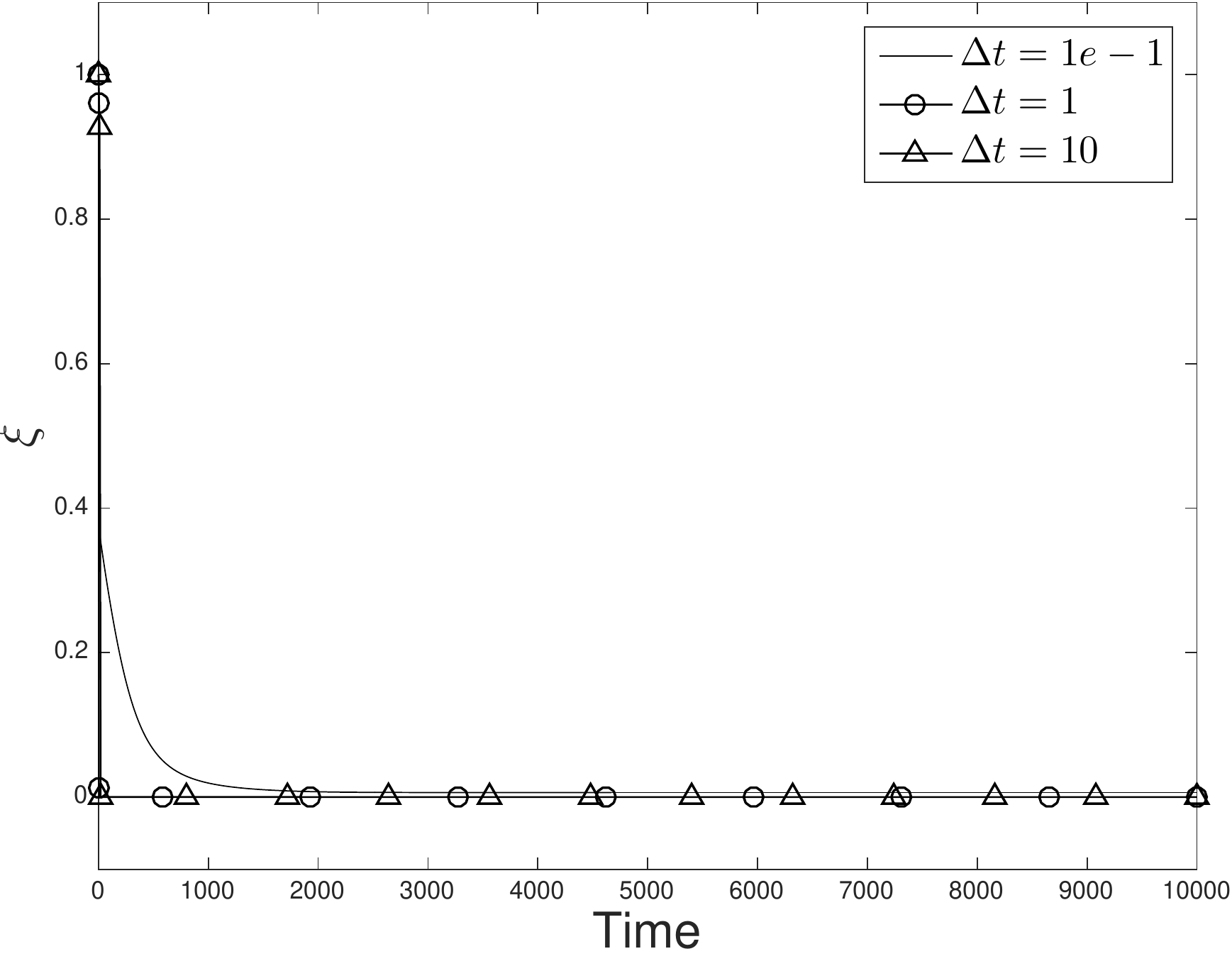}}
 \caption{Coalescence of two drops: time history of (a) $E_{tot}(t)$ and
   (b) $\xi$ for several large time step sizes $\Delta{t}=0.1, 1, 10$.
 }
\label{fig:CHEmergeEnlarge}
\end{figure}

Figure \ref{fig:CHEmergeEn} shows the time histories of the total energy
$E_{tot}(t)$
(see equation \eqref{eq:che_free_eng})
and the ratio $\xi=\frac{\mathscr{F}(R)}{E}$ obtained using
time step sizes $\Delta{t}=10^{-2}$ to $\Delta{t}=10^{-4}.$ It can be observed that
the history curves essentially overlap with one another for different time step sizes.
The computed values for $\xi=\frac{\mathscr{F}(R)}{E}$ are
very close to 1 for each $\Delta{t}$, suggesting that $\mathscr{F}(R)$ is a good approximation for $E(t)$ and the numerical approximation is accurate with these time steps.

\begin{figure}[tbp]\centering
 \subfigure[Current method]{ \includegraphics[scale=.39]{CHERn.pdf}} 
 \subfigure[SAV method]{ \includegraphics[scale=.39]{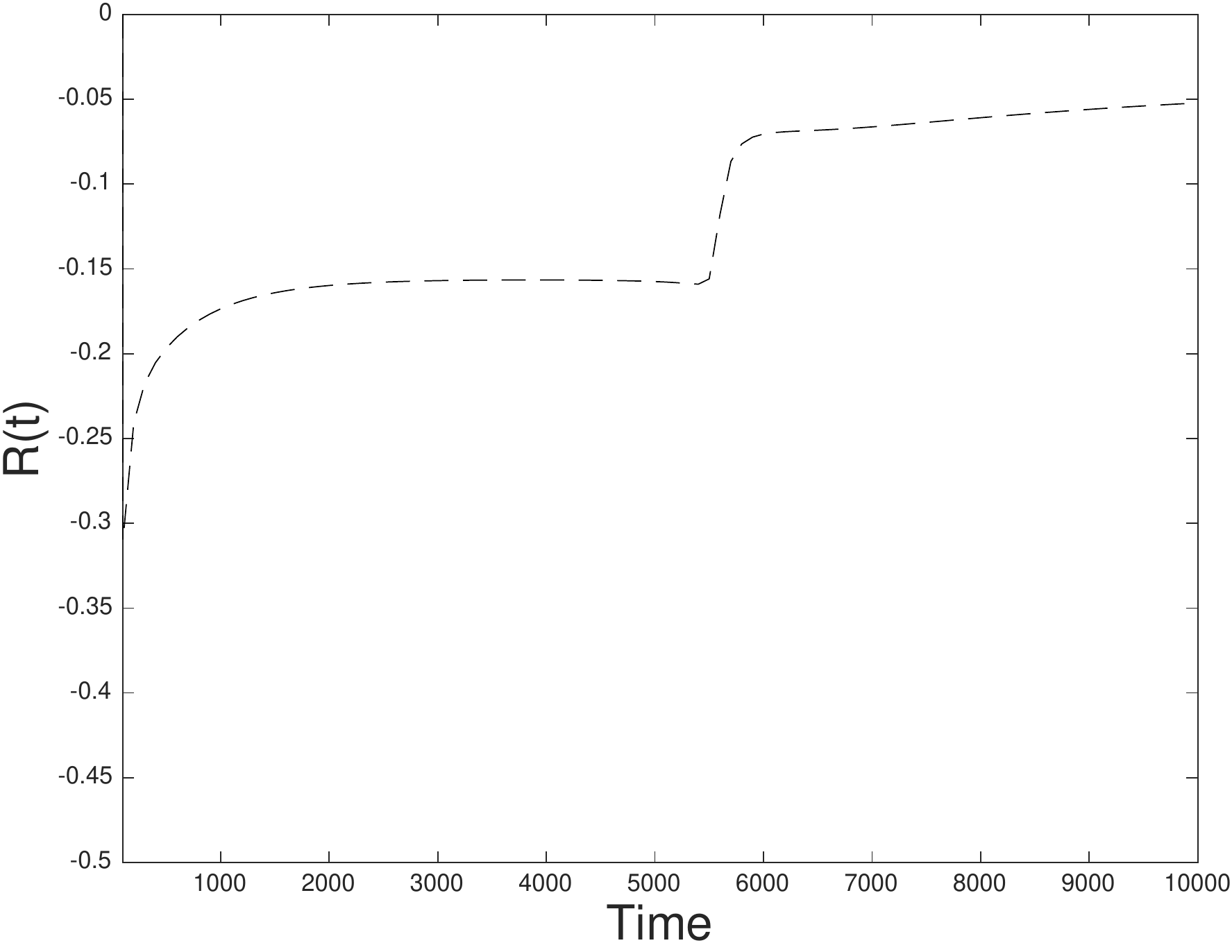}}
 \caption{Coalescence of two drops (Cahn-Hilliard equation with constant mobility):
   time history of $R(t)$ obtained by (a) the current method,
   and (b) the SAV method of \cite{ShenXY2018}.
   $\Delta t=1$ in the simulations.
 }
\label{fig:CHESAVcompare}
\end{figure}

Thanks to the energy stability property of the current method,
we can use fairly large time step sizes for the simulations. In Figure \ref{fig:CHEmergeEnlarge}, we depict some longer time histories (up to $t=10000$) of the total energy $E_{tot}(t)$
and the ratio $\xi=\frac{\mathscr{F}(R)}{E}$ obtained using several
large time step sizes $\Delta{t}=0.1, 1, 10.$ At these large $\Delta{t}$ values we can no longer expect the results to be accurate. Indeed, in Figure \ref{fig:CHEmergeEnlarge}(a), $E_{tot}$ increases initially, and levels off over time at around $E_{tot}\approx 2000.$ Meanwhile, $\xi$ decreases rapidly to a smaller number close to 0, suggesting that there is a large discrepancy between $\mathscr{F}(R)$ and $E(t).$
While these computation results are not accurate,
they nonetheless demonstrate the proposed method is stable and robust
with large time steps.

As discussed in previous sections,
the current scheme 
guarantees the positivity of the computed $\xi$ and $R(t)$ values,
regardless of
the time step size or the external forces.
In Figure \ref{fig:CHESAVcompare}, we compare the time histories of
the computed auxiliary variable $R(t)$ obtained using the current method
and the scalar auxiliary variable (SAV) method from \cite{ShenXY2018}.
In the SAV method, the auxiliary variable $R(t)$ is computed by a dynamic equation
stemming from the relation 
$R(t)=\sqrt{E_1(t)}$, where $E_1(t)=\int_{\Omega}H(\phi)d_{\Omega}+C_0>0$.
Therefore, $R(t)$ is expected to be positive on the continuous level.
In reality, however, the discrete solutions for $R(t)$ computed by
the SAV method can become negative.
This is evident from Figure \ref{fig:CHESAVcompare}(b), where the result obtained using
the SAV method
with a large $\Delta{t}=1$ is shown.
On the other hand, the discrete solutions for $R(t)$ from the current method
are guaranteed to be positive, which is evident from
Figure \ref{fig:CHESAVcompare}(a).

\subsubsection{Variable Mobility: Evolution of a Drop}

\begin{figure}[tbp] \centering
 \subfigure[$t=0$]{ \includegraphics[scale=.18]{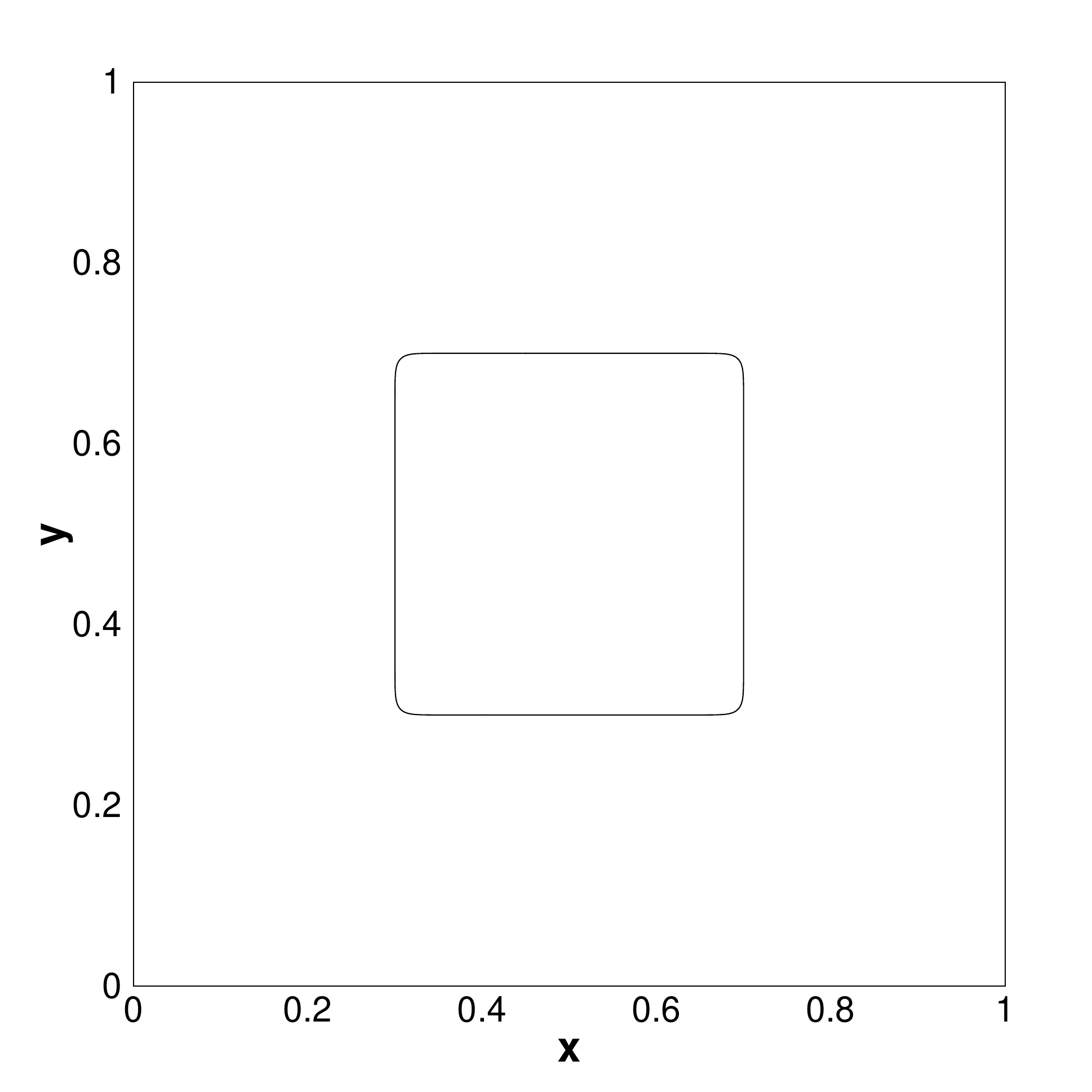}} 
 \subfigure[$t=1$]{ \includegraphics[scale=.18]{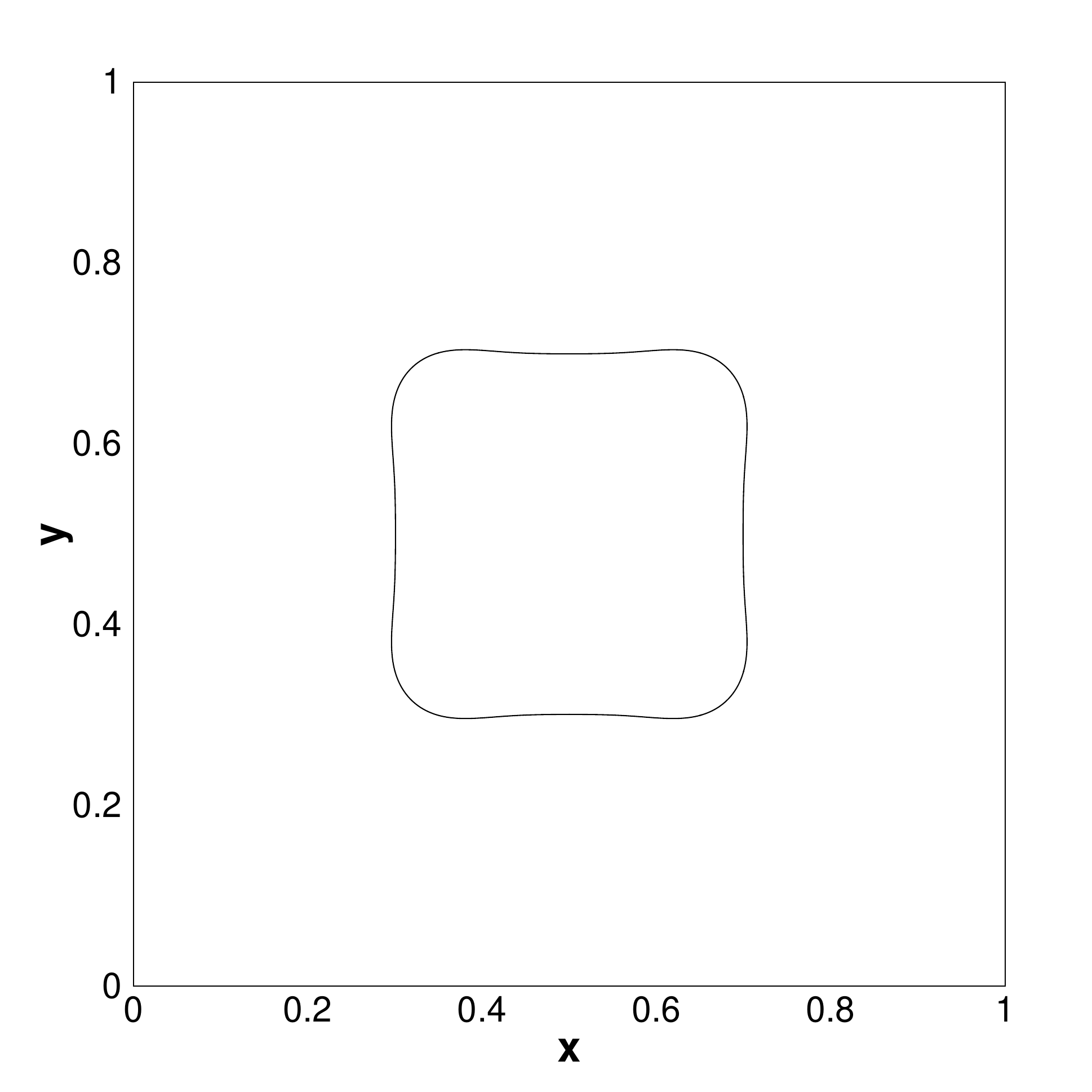}}
 \subfigure[$t=10$]{ \includegraphics[scale=.18]{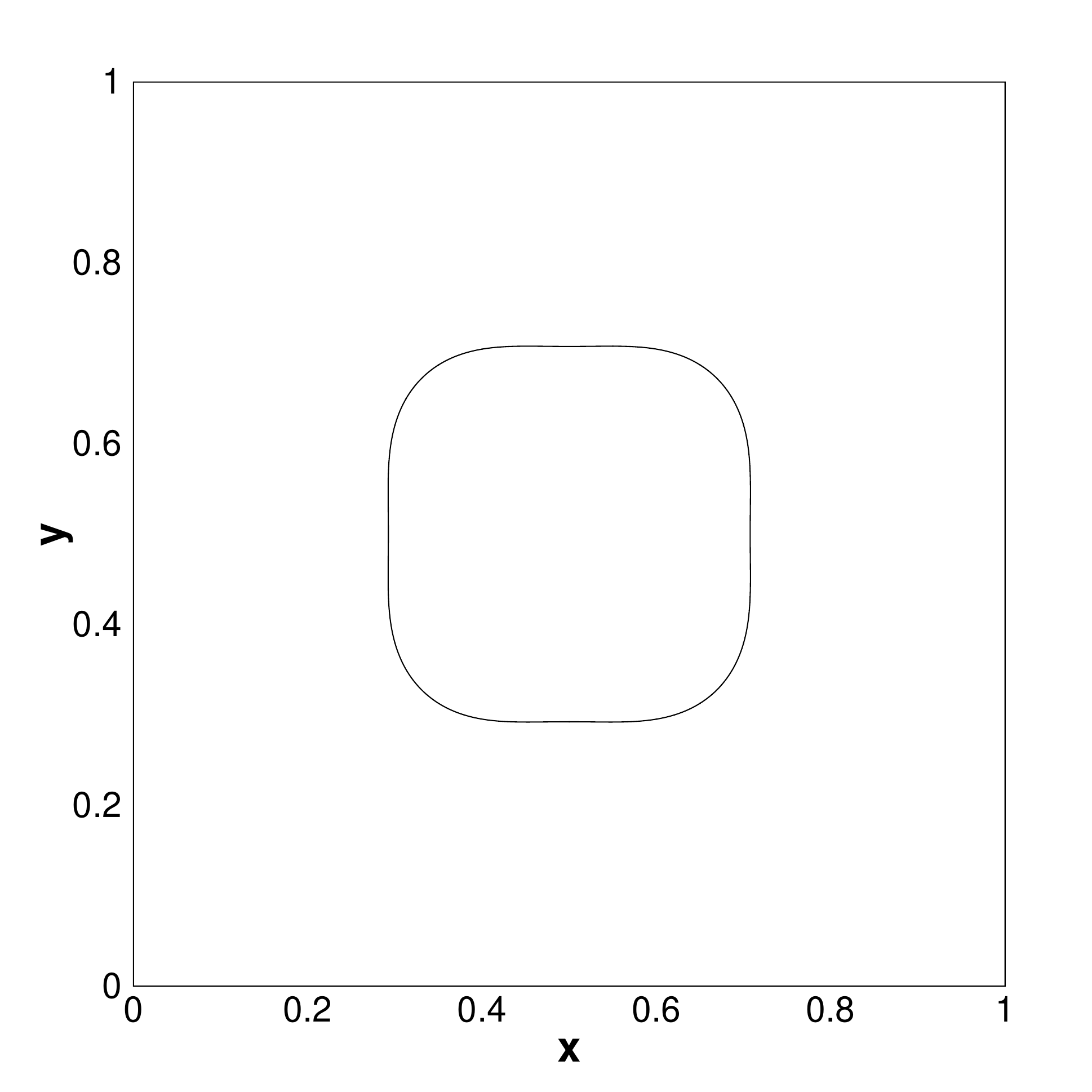}} 
 \subfigure[$t=30$]{ \includegraphics[scale=.18]{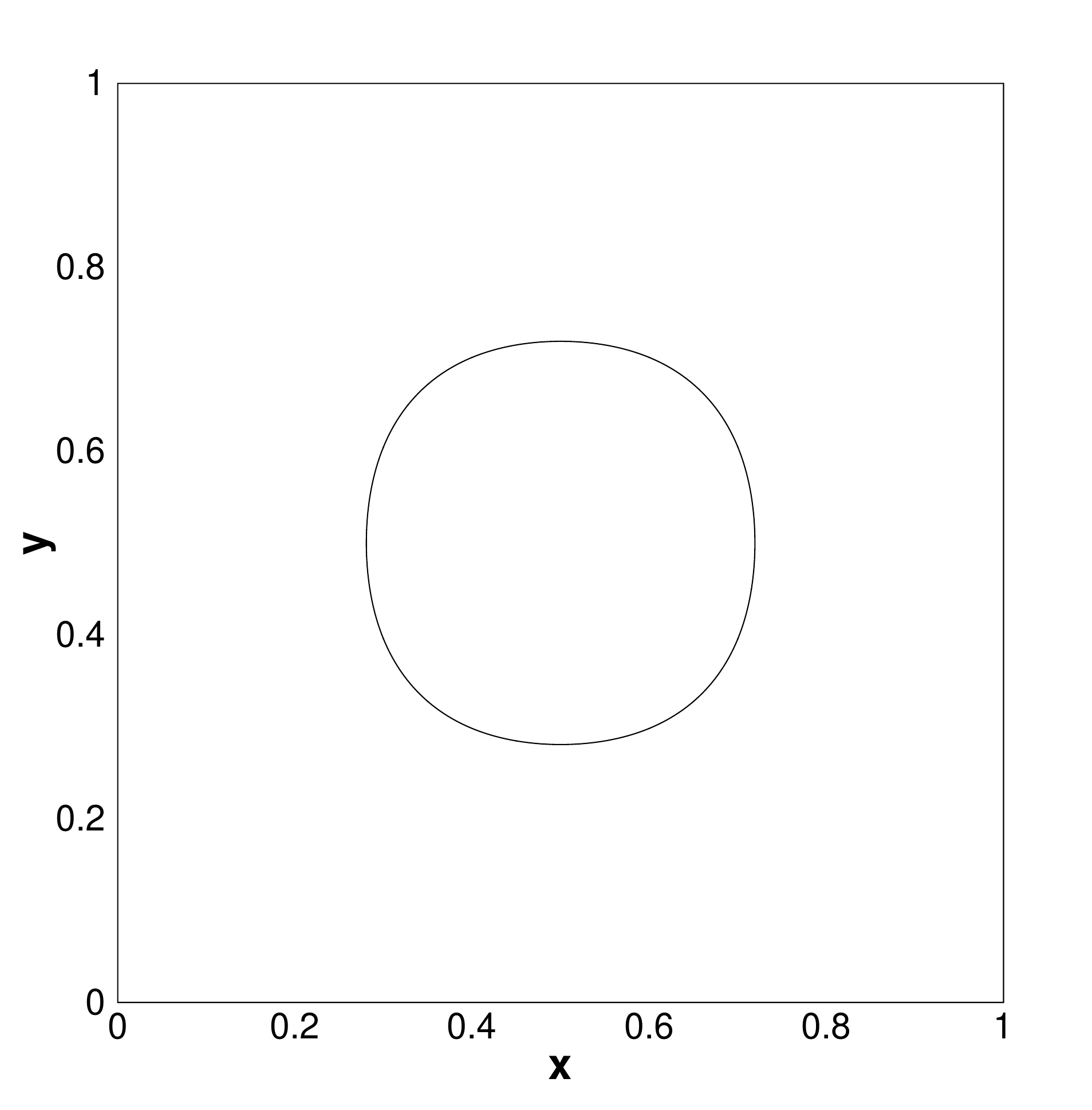}} 
 \caption{Evolution of a square drop (Cahn-Hilliard equation with variable mobility):
   Temporal snapshots of the material interface visualized by $\phi=0.$
   $\Delta t=10^{-2}$ in the simulations.
 }
\label{fig:CHEvaritest1}
\end{figure}

We next consider the evolution of a square drop
governed by the Cahn-Hilliard equation with a variable mobility.
The computational domain and the settings follow those for the coalescence of two drops
discussed above.  The difference lies in the initial distribution of the materials.
To be precise, the initial distribution of field function is set  as follows:
\begin{equation}\label{eq:varcheintial}
\phi_{in}(\bs x)=\frac{1}{2}\Big[ \tanh\frac{x-x_0+h_0}{\sqrt{2}\eta}-\tanh \frac{x-x_0-h_0}{\sqrt{2}\eta}  \Big]\cdot \Big[ \tanh\frac{y-y_0+h_0}{\sqrt{2}\eta}-\tanh \frac{y-y_0-h_0}{\sqrt{2}\eta}  \Big]-1,
\end{equation}
where $(x_0,y_0)=(0.5,0.5)$ is the center of the domain and $h_0=0.2.$ 

Figure \ref{fig:CHEvaritest1} shows the evolution of the system
with a temporal sequence of snapshots of the interfaces between the two materials.
These results are computed with a time step size $\Delta{t}=0.01$,
$S=1$, $C_0=10^6$, and the mapping function $\mathscr{F}(R)=R^2$.
The $\phi_0$ in the algorithm is taken
as the field $\phi(\bs x,t)$ at every fifth time step, i.e.~$\phi_0(\bs x)=\phi^{5k}(\bs x)$
($k=0,1,2\dots$). In other words, the $\phi_0$ field and
also the coefficient matrices of the system are updated every $5$ time steps
in this set of tests.
These results illustrate the process for the evolution of
the initial square region into a circular region
under the Cahn-Hilliard dynamics.

\begin{figure}[tbp] \centering
 \subfigure[]{ \includegraphics[scale=.39]{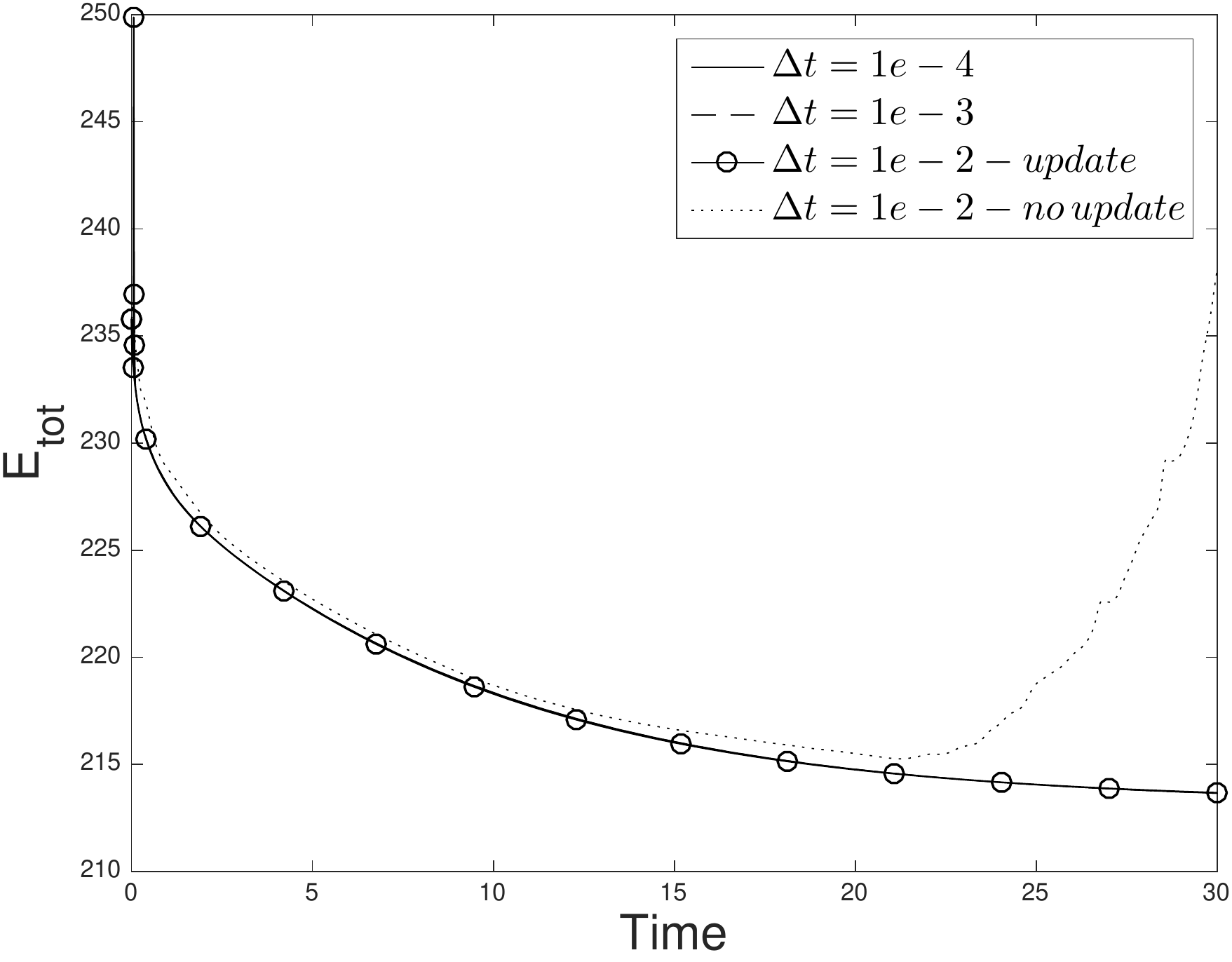}} 
 \subfigure[]{ \includegraphics[scale=.39]{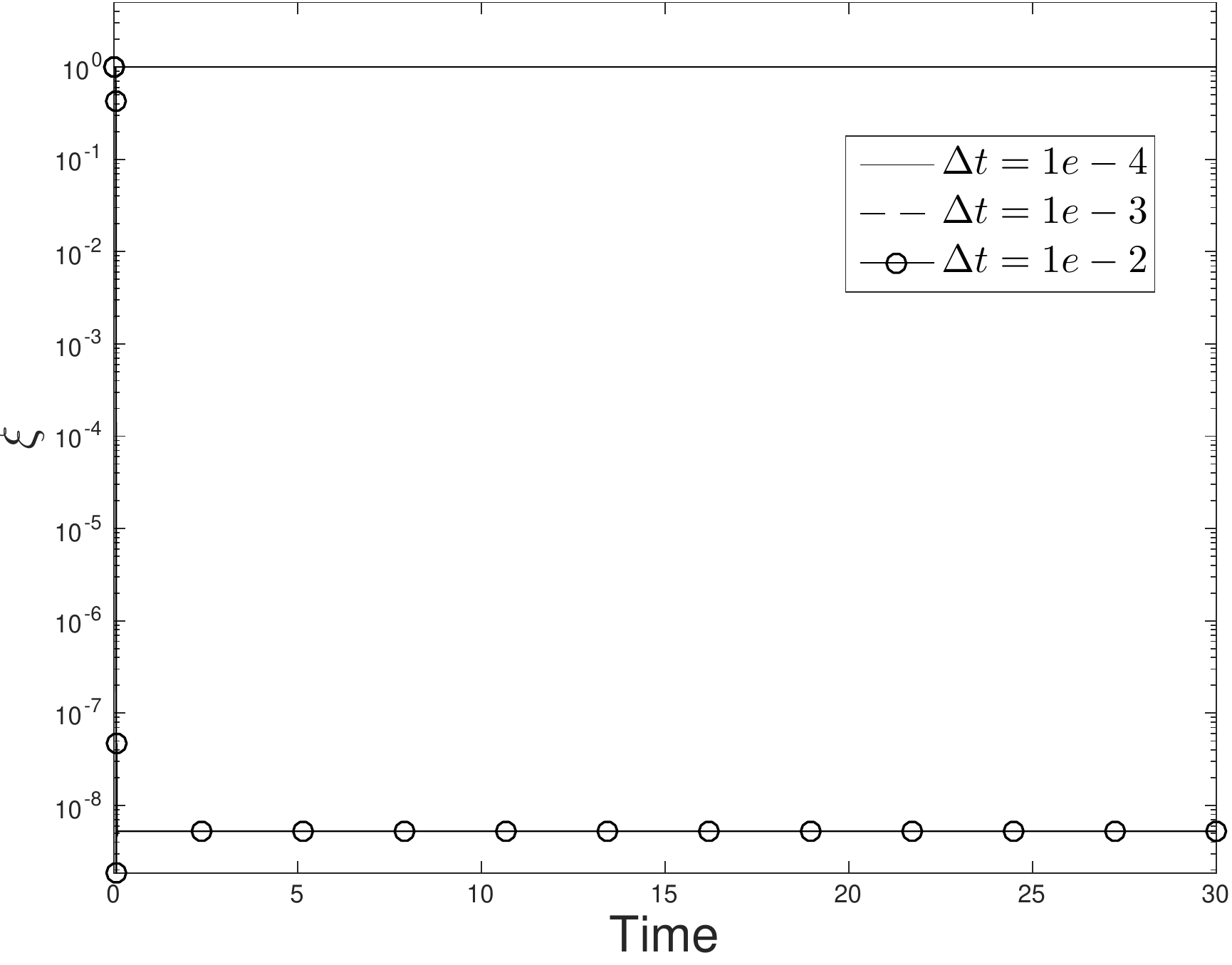}}
 \caption{Evolution of a square drop (Cahn-Hilliard equation
   with variable mobility): time histories of (a) $E_{tot}(t)$
   and (b) $\xi$ for various time step sizes $\Delta{t}=10^{-2}, 10^{-3}, 10^{-4}$.
   In these tests, 
   $\phi_0=0$ for $\Delta{t}=10^{-3}$ and $10^{-4},$ while for $\Delta{t}=10^{-2},$ we set $\phi_0=0,$ referred to as ``no update'' in the legend of (a),
   and also update $\phi_0$ to $\phi^n$ every 5 time steps, referred to as ``update''.
   The case $\Delta t=10^{-2}$ in (b) corresponds to the ``update'' case.
}
\label{fig:CHEvaritest2}
\end{figure}

In Figure \ref{fig:CHEvaritest2}, we show the time histories of the total
energy $E_{tot}(t)$ and $\xi=\frac{\mathscr{F}(R)}{E(t)}$
obtained with  several time step sizes
ranging from $\Delta{t}=10^{-2}$ to $\Delta{t}=10^{-4}.$
Note that the variable mobility is $m(\phi)=\max(m_0(1-\phi^2), 0)$.
Here we have considered two ways to simulate the problem:
\begin{itemize}
\item
  by setting $\phi_0=0$ in the algorithm. This leads to
  $m_c(\phi_0)=m(\phi_0)=m_0$ and
  $\kappa(\phi_0) = -\frac{\lambda}{\eta^2}$,
  and a time-independent coefficient matrix for the system, which
  can be pre-computed. We refer to this setting as
  the standard way.

\item
  by setting $\phi_0=\phi^{5k}$ ($k=0,1,2,\dots$) in the algorithm.
  The $\phi_0$ field and the coefficient matrix are quasi time-independent, and they 
  are updated
  every $5$ time steps.
\end{itemize}
With the smaller time step sizes $\Delta{t}=10^{-3}$ and $10^{-4}$,
we set $\phi_0=0$ in the algorithm (the standard way) when performing simulations.
With the larger $\Delta{t}=10^{-2}$, we have conducted simulations
in both ways with the algorithm. 
In Figure \ref{fig:CHEvaritest2}(a) the results from these two settings
are marked by ``no update'' (standard way) and ``update'' (second way)
in the legend corresponding to $\Delta t=10^{-2}$.
It is observed that the energy histories corresponding to $\Delta{t}=10^{-4}$ and $10^{-3}$,
and $\Delta t=10^{-2}$ with $\phi_0$ updated periodically,
essentially overlap with each other.
However, the energy history corresponding to $\Delta{t}=10^{-2}$ with $\phi_0=0$
exhibits a pronounced discrepancy compared with the other cases.
These results indicate that with the standard way (by setting $\phi_0=0$) in
the algorithm the simulation result would cease to be accurate
when the time step size increases to $\Delta t=10^{-2}$.
However, if one uses the second way (by updating $\phi_0$ periodically),
accurate simulation result can be obtained even with $\Delta t=10^{-2}$.
In other words, by updating $\phi_0$ in the algorithm from time to time,
one can improve the accuracy of the simulations even at larger time
step sizes.
We depict in Figure \ref{fig:CHEvaritest2}(b) the time histories of
$\xi=\frac{\mathscr{F}(R)}{E}$ corresponding to these time step sizes.
Shown for $\Delta{t}=10^{-2}$ in this plot is the result with $\phi_0$ updated
periodically. It is observed that the computed $\xi$ is essentially 1
with $\Delta{t}=10^{-3}$ and $10^{-4}.$ 
With $\Delta{t}=10^{-2}$ (and $\phi_0$ updated periodically),
the computed $\xi$ is substantially smaller than 1.
But interestingly, the simulation results for the field function $\phi$
are still quite accurate with this larger $\Delta t$.
This group of tests suggests that one possible way to improve the accuracy of
the proposed energy-stable scheme is to update the $\phi_0$ in
the algorithm periodically, e.g.~every $N$ time steps. By choosing an
appropriate $N$ for a given problem, one can enhance
the simulation accuracy even at large or fairly large time step
sizes. Because $\phi_0$ and the coefficient matrix for the system
only needs to be updated infrequently, the cost associated with
updating the coefficient matrix can be manageable.
There is a drawback with this, however. The computations using
the second way (updating
$\phi_0$ periodically) seems not as robust as
the standard way (by setting $\phi_0=0$) for large $\Delta t$.
Because of the non-zero $\phi_0$ field in the algorithm,
the conditioning of the system coefficient matrix using the second way
seems to become worse
for large $\Delta t$.
We observe that for larger $\Delta{t}\geqslant 0.1$ the system coefficient matrix
using the second way can become singular and the computation may break down.


\section{Nonlinear Klein-Gordon Equation}
\label{sec:kg}

We consider an energy-conserving system, the nonlinear
Klein-Gordon equation, in this section and
apply the gPAV method to this system.
Consider the  nonlinear Klein-Gordon equation~\cite{Strauss1978}
on a domain $\Omega$ (with boundary $\Gamma$)
\begin{align}
&  \frac{\partial u}{\partial t}=v,  \label{eq:kgsys1}\\
& \varepsilon^2\frac{\partial v}{\partial t}-\alpha^2 \nabla^2 u + \varepsilon_1^2 u+g(u)=f(\bs x,t),  \label{eq:kgsys2} \\
& u=d_a(\bs x,t), \;\; {\rm on}~\Gamma, \label{eq:kgbd1}
\end{align}
where $\varepsilon$, $\alpha$ and $\varepsilon_1$ are positive constants.
These equations are supplemented by
the initial conditions
\begin{equation}\label{eq:kginitial}
u(\bs x,0)=u_{in}(\bs x),\quad v(\bs x,0)=v_{in}(\bs x)\;\; {\rm in}~\Omega.
\end{equation}
In these equations $g(u)=G'(u)$ and $G(u)$ is a potential energy function
with $G(u)\geqslant 0.$
The above system satisfies the following energy balance law:
\begin{equation}\label{eq:kgengbalaw}
  \frac{\partial}{\partial t}\int_\Omega\Big(  \frac{\alpha^2}{2} |\nabla u|^2+\frac{\varepsilon_1^2}{2}|u|^2+\frac{\varepsilon^2}{2}|v|^2+ G(u) \Big)d\Omega
  =\int_\Omega fv d\Omega+\alpha^2 \int_{\partial\Omega} (\bs n\cdot\nabla u) v d\Gamma.
\end{equation}

We define a shifted total energy according to equation \eqref{eq:shitE}, 
\begin{equation}\label{eq:kgEn}
E(t) = E[u,v] =\int_\Omega \Big( \frac{\alpha^2}{2} |\nabla u|^2+\frac{\varepsilon_1^2}{2}|u|^2+\frac{\varepsilon^2}{2}|v|^2+ G(u) \Big) d\Omega+C_0,
\end{equation}
where $C_0$ is chosen such that $E(t)>0$.
Choose $\mathscr{F}$ and $\mathscr{G}$, and define the auxiliary variable
$R(t)$ based on equation \eqref{eq:FG}.
Following equation \eqref{eq:dynR}, we have
\begin{equation}\label{eq:kgFeq}
  \mathscr{F}'(R)\frac{d R}{d t}
  =\int_\Omega \Big(-\alpha^2 \nabla^2 u +\varepsilon_1^2 u +g(u)\Big)\frac{\partial u}{\partial t} d\Omega+\int_{\Omega} \varepsilon^2 v\frac{\partial v}{\partial t}d\Omega+\alpha^2 \int_{\Gamma} (\bs n\cdot\nabla u) \frac{\partial u}{\partial t} d\Gamma,
\end{equation}
where integration by part has been used.

Following equations \eqref{eq:generaleq1}-\eqref{eq:generaleq2},
we reformulate equations \eqref{eq:kgsys2} and \eqref{eq:kgFeq} into
\begin{subequations}
\begin{equation}\label{eq:kgvnew}
\frac{\partial v}{\partial t}=\Big( \frac{\alpha}{\varepsilon} \Big)^2 \nabla^2 u-\Big(  \frac{\varepsilon_1}{\varepsilon} \Big)^2 u- \frac{\mathscr{F}(R)}{E}   \frac{1}{\varepsilon^2} g( u) +\frac{1}{\varepsilon^2}f,
\end{equation}
\begin{equation}\label{eq:kgfnew}
\begin{split}
  & \mathscr{F}'(R)  \frac{dR}{dt}
  =\int_{\Omega}\Big(-\alpha^2 \nabla^2 u +\varepsilon_1^2 u +g(u)\Big)\frac{\partial u}{\partial t}
  +\int_{\Omega} \varepsilon^2 v\frac{\partial v}{\partial t}d\Omega\\
  &+\frac{\mathscr{F}(R)}{E} \Big(  \int_\Omega fvd\Omega
  +\alpha^2 \int_{\Gamma} (\bs n\cdot\nabla u) v d\Gamma\Big)
  +\left[1-\frac{\mathscr{F}(R)}{E} \right]\left|  \int_\Omega fvd\Omega
  +\alpha^2 \int_{\Gamma} d_avd\Gamma\right|\\
  &-\int_{\Omega} \Big(-\alpha^2 \nabla^2 u +\varepsilon_1^2 u+g(u)\Big)vd\Omega
  -\int_{\Omega} \varepsilon^2 v\left[ \Big( \frac{\alpha}{\varepsilon} \Big)^2 \nabla^2 u-\Big(  \frac{\varepsilon_1}{\varepsilon} \Big)^2 u-\frac{\mathscr{F}(R)}{E}  \frac{1}{\varepsilon^2} g( u) +\frac{1}{\varepsilon^2}f  \right]d\Omega.
\end{split}
\end{equation}
\end{subequations}
Note that when deriving \eqref{eq:kgfnew} we have incorporated
the following zero terms to the RHS,
\begin{equation*}
  \begin{split}
    &
  \left(\frac{\mathscr{F}(R)}{E}-1 \right)
  \int_{\Omega}  \Big(-\alpha^2 \nabla^2 u +\varepsilon_1^2 u+g(u)\Big)v d\Omega
  + \left(\frac{\mathscr{F}(R)}{E}-1 \right)
  \int_{\Omega}v\left(
    \alpha^2\nabla^2u - \varepsilon^2u + f
    \right) d\Omega \\
    &
    +\frac{\mathscr{F}(R)}{E} \left(\int_{\Omega}vg(u)d\Omega -  \int_{\Omega}vg(u)d\Omega  \right)
    + \left(\frac{\mathscr{F}(R)}{E}-1 \right) \int_{\Gamma} (\bs n\cdot\nabla u) v d\Gamma \\
    &
    + \left(1-\frac{\mathscr{F}(R)}{E} \right)\left|  \int_\Omega fvd\Omega
  +\alpha^2 \int_{\Gamma} d_avd\Gamma\right|.
    \end{split}
\end{equation*}
The reformulated system consists of equations \eqref{eq:kgsys1}, \eqref{eq:kgvnew}-\eqref {eq:kgfnew} and \eqref{eq:kgbd1}-\eqref{eq:kginitial},
which is equivalent to the original system \eqref{eq:kgsys1}-\eqref{eq:kginitial}.

Since the Klein-Gordon equation is conservative (in the absence
of external source term and with appropriate boundary condition),
we will employ the Crank-Nicolson method for time
discretization of the field variables, by enforcing
the discretized equations at step $(n+1/2)$.
This corresponds
to the approximations \eqref{eq:nplus12theta}--\eqref{eq:directdirtheta}
with $\theta=\frac12$ and $\beta=0$.
So the method here is slightly different than the one
presented in Section \ref{sec:scheme}, which corresponds to
$\theta=1$ and $\beta=\frac14$ in the approximations
\eqref{eq:nplus12theta}--\eqref{eq:directdirtheta}.
The energy-stable scheme for the nonlinear Klein-Gordon equation
is then as follows:
\begin{subequations}\label{eq;kgscheme}
\begin{align}
  & \frac{u^{n+1}-u^n}{\Delta t}
  =v^{n+\frac{1}{2}}, \label{eq:kgscheme1}\\
  & \frac{v^{n+1}-v^n}{\Delta t}
  =\Big( \frac{\alpha}{\varepsilon} \Big)^2 \nabla^2 u^{n+\frac{1}{2}}-\Big(  \frac{\varepsilon_1}{\varepsilon} \Big)^2 u^{n+\frac{1}{2}}-\xi \frac{1}{\varepsilon^2} g(\bar u^{n+\frac{1}{2}}) +\frac{1}{\varepsilon^2}f^{n+\frac{1}{2}},\label{eq:kgscheme2}\\
& \xi=\frac{\mathscr{F}(R^{n+1})}{\tilde E[\tilde u^{n+1},\tilde v^{n+1}]}, \label{eq:kgscheme4}\\
& E[\tilde u^{n+1},\tilde v^{n+1}]= \int_\Omega \Big( \frac{\alpha^2}{2} |\nabla \tilde u^{n+1}|^2+\frac{\varepsilon_1^2}{2}|\tilde u^{n+1}|^2+\frac{\varepsilon^2}{2}|\tilde v^{n+1}|^2+ G(\tilde u^{n+1}) \Big) d\Omega+C_0,  \label{eq:kgschemeE}  \\
& u^{n+1}=d_a^{n+1},\;\;\text{on} \;\Gamma, \label{eq:kgscheme5}
\end{align}
\end{subequations}
together with
\begin{equation}\label{eq:kgscheme3}
\begin{split}
  &D_{\mathscr{F}}(R)|^{n+\frac{1}{2}}\frac{R^{n+1}-R^n}{\Delta t}
  =\int_{\Omega}\Big(-\alpha^2 \nabla^2 u^{n+\frac{1}{2}} +\varepsilon_1^2 u^{n+\frac{1}{2}} +g(u^{n+\frac{1}{2}})\Big)\frac{u^{n+1}-u^n}{\Delta t}d\Omega \\
  &+\int_{\Omega} \varepsilon^2 v^{n+\frac{1}{2}}\frac{v^{n+1}-v^n}{\Delta t}
  d\Omega
 -\int_{\Omega} \Big(-\alpha^2 \nabla^2 u^{n+\frac{1}{2}} +\varepsilon_1^2 u^{n+\frac{1}{2}} +g(u^{n+\frac{1}{2}})\Big)v^{n+\frac{1}{2}} d\Omega\\
 &-\int_{\Omega} \varepsilon^2 v^{n+\frac{1}{2}}\Big\{ \Big( \frac{\alpha}{\varepsilon} \Big)^2 \nabla^2 u^{n+\frac{1}{2}}-\Big(  \frac{\varepsilon_1}{\varepsilon} \Big)^2 u^{n+\frac{1}{2}}-\xi \frac{1}{\varepsilon^2} g(\bar u^{n+\frac{1}{2}}) +\frac{1}{\varepsilon^2}f^{n+\frac{1}{2}}  \Big\}d\Omega \\
 &+\xi\Big(  \int_\Omega f^{n+\frac{1}{2}}\tilde v^{n+\frac{1}{2}}d\Omega
 +\alpha^2 \int_{\Gamma} \big(\bs n \cdot \nabla \tilde u^{n+\frac{1}{2}} \big) \tilde v^{n+\frac{1}{2}}d\Gamma \Big) \\
 &
 +(1-\xi)\Big|  \int_\Omega f^{n+\frac{1}{2}}\tilde v^{n+\frac{1}{2}}d\Omega
 +\alpha^2 \int_{\Gamma} \big(\bs n \cdot \nabla \tilde u^{n+\frac{1}{2}} \big) \tilde v^{n+\frac{1}{2}}d\Gamma \Big|.
\end{split}
\end{equation}
These equations are supplemented by the initial conditions
\begin{equation}\label{eq:kgscheme6}
u^0=u_{\rm in}(\bs x),\quad v^0=v_{\rm in}(\bs x),\quad R^0=\mathscr{G}( E^0),
\end{equation}
where $E^0$ is evaluated by
\begin{equation} \label{eq:kgE0}
E^0=\int_\Omega \Big( \frac{\alpha^2}{2} |\nabla u_{\rm in}|^2+\frac{\varepsilon_1^2}{2}|u_{\rm in}|^2+\frac{\varepsilon^2}{2}|v_{\rm in}|^2+ G(u_{\rm in}) \Big) d\Omega+C_0.
\end{equation}
In the above equations,
$D_{\mathscr{F}}(R)|^{n+\frac{1}{2}}$ is defined by \eqref{eq:directdirtheta}
with $\theta=1/2$, and
\begin{equation}\label{eq:CN}
  \left\{
  \begin{split}
    &
    u^{n+1/2} = \frac{1}{2}(u^{n+1} + u^n), \quad
    v^{n+1/2} = \frac{1}{2}(v^{n+1} + v^n),
    \\
    &
    \bar u^{n+\frac{1}{2}}=\frac{3}{2}u^n-\frac{1}{2}u^{n-1}, \quad
    \bar v^{n+\frac{1}{2}}=\frac{3}{2}v^n-\frac{1}{2}v^{n-1}.
  \end{split}
\right.
\end{equation}
$\tilde u^{n+1},$ $\tilde v^{n+1},$ $\tilde u^{n+\frac{1}{2}}$ and $\tilde v^{n+\frac{1}{2}}$ are second-order approximations of $u^{n+1},$ $v^{n+1},$ $u^{n+\frac{1}{2}}$ and $v^{n+\frac{1}{2}},$ respectively, defined later in \eqref{eq:kguvstar}-\eqref{eq:kguv32}.

 \begin{theorem}\label{thm:kgeq}
   In the absence of the external force $f=0,$
   and with homogeneous boundary condition ($d_a=0$)
   and suppose that the initial condition $v_{\rm in}$
   satisfies the compatibility condition $v_{in}|_{\Gamma}=0,$
   the scheme consisting of \eqref{eq;kgscheme}-\eqref{eq:kgscheme6} conserves the modified energy $\mathscr{F}(R)$ in the sense that:
 \begin{equation}\label{eq:kgdislaw}
{\mathscr{F}(R^{n+1}) -\mathscr{F}(R^{n})}=0.
 \end{equation}
 \end{theorem}
 \begin{proof}
   Multiplying $\big(-\alpha^2 \nabla^2 u^{n+\frac{1}{2}} +\varepsilon_1^2 u^{n+\frac{1}{2}} +g(u^{n+\frac{1}{2}})\big)$ to equation \eqref{eq:kgscheme1},
   $\varepsilon^2 v^{n+\frac{1}{2}}$ to equation \eqref{eq:kgscheme2}, taking the $L^2$ integrals, and summing up the resultant equations
   with equation \eqref{eq:kgscheme3}, we arrive at the relation,
 \begin{equation}\label{eq:kgsolver}
 \begin{split}
   \frac{\mathscr{F}(R^{n+1}) -\mathscr{F}(R^{n})}{\Delta{t}}&=\xi\Big(  \int_\Omega f^{n+\frac{1}{2}}\tilde v^{n+\frac{1}{2}}d\Omega
   +\alpha^2 \int_{\Gamma} \big(\bs n \cdot \nabla \tilde u^{n+\frac{1}{2}} \big) \tilde v^{n+\frac{1}{2}} d\Gamma\Big)\\
   &+(1-\xi)\Big|  \int_\Omega f^{n+\frac{1}{2}}\tilde v^{n+\frac{1}{2}}d\Omega
   +\alpha^2 \int_{\Gamma} \big(\bs n \cdot \nabla \tilde u^{n+\frac{1}{2}} \big) \tilde v^{n+\frac{1}{2}}d\Gamma \Big|,
 \end{split}
\end{equation}
 where we have used equations \eqref{eq:tempdist}-\eqref{eq:directdir}.
 If $d_a=0$, then $u^n|_{\Gamma}=0$ and $v^n|_{\Gamma}=0$ for all $n> 0$.
 Based on the definition of $\tilde v^{n+\frac{1}{2}}$
 in the equation \eqref{eq:kguv32} below,
 it is straightforward to verify that $\tilde v^{n+\frac{1}{2}}|_{\Gamma}=0$
 as long as $v^0|_{\Gamma}=0$.  Furthermore, if $f=0,$ the volume
 integrals in equation \eqref{eq:kgsolver} vanish. This leads to 
 equation \eqref{eq:kgdislaw}.
\end{proof}

\begin{rem}
  Since $\mathscr{F}(R)$ is an approximation of $E(t),$
  the discrete conservation for $\mathscr{F}(R)$ in equation \eqref{eq:kgdislaw}
  does not imply the conservation for $E(t)$ on the discrete level.
  However, it does lead to an unconditionally energy stable
  scheme for long time simulations.
\end{rem}

Despite the complication caused by the unknown scalar variable $\xi,$ the proposed scheme can be solved in a decoupled fashion.
Combining equations \eqref{eq:kgscheme1} and \eqref{eq:CN},
we get 
\begin{equation}\label{eq:kgvn1}
v^{n+1}=\frac{2}{\Delta{t}}u^{n+1}-\frac{2}{\Delta{t}}u^n-v^n,
\end{equation}
Inserting equation \eqref{eq:kgvn1} into \eqref{eq:kgscheme2} leads to
\begin{equation}\label{eq:kgugoven}
\begin{split}
\Big[ \Big(\frac{2\varepsilon }{\alpha \Delta{t}} \Big)^2+\Big( \frac{\varepsilon_1}{\alpha} \Big)^2 \Big]u^{n+1}-\nabla^2 u^{n+1}=&\Big( \frac{\varepsilon}{\alpha} \Big)^2   \Big\{  \Big[ \big(\frac{2}{\Delta{t}}\big)^2-\big(\frac{\varepsilon_1}{\varepsilon}\big)^2  \Big] u^n +\frac{4}{\Delta{t}}v^n+\frac{2}{\varepsilon^2}f^{n+\frac{1}{2}}            \Big \}\\
&-\xi \frac{2}{\alpha^2}g(\bar u^{n+\frac{1}{2}}) +\nabla^2 u^n.
\end{split}
\end{equation}
To solve this equations, we
introduce $u_1^{n+1}$ and $u_2^{n+1}$ as solutions of the following two equations:
\begin{align}
  &
  \Big[ \Big(\frac{2\varepsilon }{\alpha \Delta{t}} \Big)^2+\Big( \frac{\varepsilon_1}{\alpha} \Big)^2 \Big]u_1^{n+1}-\nabla^2 u_1^{n+1}=\Big( \frac{\varepsilon}{\alpha} \Big)^2   \Big\{  \Big[ \big(\frac{2}{\Delta{t}}\big)^2-\big(\frac{\varepsilon_1}{\varepsilon}\big)^2  \Big] u^n +\frac{4}{\Delta{t}}v^n+\frac{2}{\varepsilon^2}f^{n+\frac{1}{2}} \Big \}+\nabla^2 u^n, \label{eq:kgu1solu}\\
  &
u_1^{n+1}=d_a^{n+1}\;\;\text{on}\;\Gamma,\label{eq:kgu1solubd}
\end{align}
and
\begin{equation}\label{eq:kgu2solu}
 \Big[ \Big(\frac{2\varepsilon }{\alpha \Delta{t}} \Big)^2+\Big( \frac{\varepsilon_1}{\alpha} \Big)^2 \Big]u_2^{n+1}-\nabla^2 u_2^{n+1}=- \frac{2}{\alpha^2}g(\bar u^{n+\frac{1}{2}}),\quad u_2^{n+1}=0\;\;\text{on}\;\Gamma.
\end{equation}
Then the solution to equation \eqref{eq:kgugoven}, together with
the boundary condition \eqref{eq:kgscheme5}, is given by
\begin{equation}\label{eq:kgu1xiu2}
u^{n+1}=u_1^{n+1}+\xi u_2^{n+1}.
\end{equation}
where $\xi$ is to be determined.


We define
\begin{align}
  &\tilde{u}^{n+1}=u_1^{n+1}+u_2^{n+1},\quad
  \tilde{u}^{n+1/2}=\frac{1}{2}(\tilde{u}^{n+1}+ u^n), \label{eq:kguvstar}\\
  & \tilde{v}^{n+1}=\frac{2}{\Delta t}(\tilde u^{n+1}-u^n)-v^n,  \quad
  \tilde{v}^{n+1/2}=\frac{1}{2}(\tilde{v}^{n+1} + v^n).
  \label{eq:kguv32}
\end{align}
By combining equations \eqref{eq:kgscheme4} and \eqref{eq:kgsolver}, we can
determine $\xi$,
\begin{equation}
  \xi=\frac{\mathscr{F}(R^{n})
    +{\Delta}t |S_0|}{ E[\tilde{u}^{n+1},\tilde{v}^{n+1}]
    +\Delta{t}(|S_0|-S_0) },\;\;\ \text{with}\;S_0=\Big(  \int_\Omega f^{n+1/2}\tilde{v}^{n+1/2}d\Omega+\alpha^2 \int_{\Gamma} (\bs n\cdot\nabla \tilde u^{n+1/2}) \tilde{v}^{n+1/2}d\Gamma \Big).
\end{equation}
With $\xi$ known, 
$u^{n+1}$ and $v^{n+1}$ can be computed by 
equations \eqref{eq:kgu1xiu2} and \eqref{eq:kgvn1}, respectively.
$R^{n+1}$ can be computed by, 
\begin{equation}\label{eq:kgRdef}
R^{n+1}=\mathscr{G}(\xi E[\tilde u^{n+1},\tilde v^{n+1}]).
\end{equation}

The weak formulations for equations \eqref{eq:kgu1solu} and \eqref{eq:kgu2solu} are: Find $(u_1^{n+1},u_2^{n+1})\in H^1(\Omega)$ such that
\begin{equation}\label{eq:kgu1dist}
\begin{split}
& \big( \nabla u_{1}^{n+1}, \nabla \varphi  \big)_{\Omega}+ \Big[ \Big(\frac{2\varepsilon }{\alpha \Delta{t}} \Big)^2+\Big( \frac{\varepsilon_1}{\alpha} \Big)^2 \Big] \big( u_{1}^{n+1}, \varphi  \big)_{\Omega}=-(\nabla u^n, \nabla \varphi)_{\Omega} \\
&+ \Big( \frac{\varepsilon}{\alpha} \Big)^2   \Big( \Big[ \big(\frac{2}{\Delta{t}}\big)^2-\big(\frac{\varepsilon_1}{\varepsilon}\big)^2  \Big] u^n +\frac{4}{\Delta{t}}v^n+\frac{2}{\varepsilon^2}f^{n+\frac{1}{2}} , \varphi  \Big)_{\Omega} ,\quad \forall \varphi \in H^1_{0}(\Omega):=\Big\{ w\in H^1(\Omega) : \, w|_{\Gamma}=0 \Big\};
\end{split}
\end{equation}
\begin{equation}\label{eq:kgu2dist}
\begin{split}
 \big( \nabla u_{2}^{n+1}, \nabla \varphi  \big)_{\Omega}+ \Big[ \Big(\frac{2\varepsilon }{\alpha \Delta{t}} \Big)^2+\Big( \frac{\varepsilon_1}{\alpha} \Big)^2 \Big] \big( u_{2}^{n+1}, \varphi  \big)_{\Omega_h}=-\frac{2}{\alpha^2} \big( g(\bar u^{n+\frac{1}{2}}),\varphi \big)_{\Omega},\quad \forall \varphi \in H^1_{0}(\Omega).
\end{split}
\end{equation}
These can be implemented with $C^0$ spectral elements
in a straightforward fashion.

\subsection{Numerical Results}


We next provide numerical examples to demonstrate the accuracy and unconditional stability of the proposed scheme to the Klein-Gordon equation \eqref{eq:kgsys1}-\eqref{eq:kgbd1}. Specifically, we fix the parameters therein and the potential energy function as
\begin{equation}\label{eq:kgparam}
\varepsilon=\varepsilon_1=\alpha=1,\quad G(u)=1-\cos(u),\quad g(u)=G'(u)=\sin(u).
\end{equation}
This corresponds to the dimensionless relativistic  Sine-Gordon equation
(DRSG) (see e.g.~\cite{Bao2012}).

\subsubsection{Convergence Rates}

\begin{figure}[tbp]
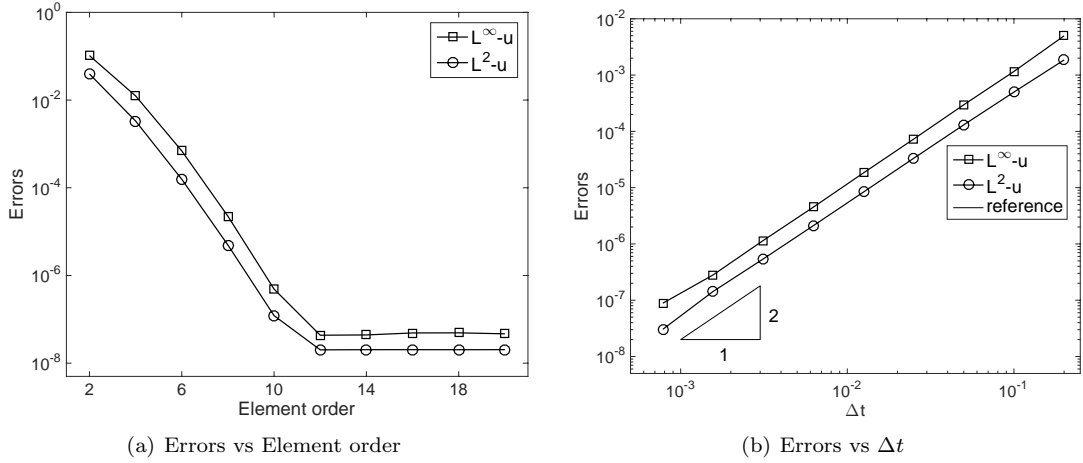
\centering
 \subfigure[Errors vs Element order]{ \includegraphics[scale=.39]{GNKGspatial.pdf}} \quad
 \subfigure[Errors vs ${\Delta}t$]{ \includegraphics[scale=.38]{GNKGtemporal.pdf}} \\
\caption{Spatial/temporal convergence tests for DRSG equation: $L^{2}$ and $L^{\infty}$ errors of $u$ versus (a) element order (fixed $\Delta t=0.001$ and $t_f=0.1$), and (b)  $\Delta t$ (fixed element order $18$ and $t_f=1$).}
\label{fig:DRSGstandtest}
\end{figure}

To study the convergence rates in space and time of the proposed method, we employ the following manufactured analytic solution 
\begin{equation}\label{eq:chemostd}
u=\cos(\pi x)\cos(\pi y)\sin(t),
\end{equation}
The external force $f(\bs x,t)$ in \eqref{eq:kgsys2} and the external boundary source term
$d_a(\bs x,t)$ are chosen such that the above expression \eqref{eq:chemostd} satisfies equations \eqref{eq:kgsys1}-\eqref{eq:kgbd1}. 

The computational domain $\Omega=[0,2]\times [-1, 1]$ is discretized using two equal-sized quadrilateral elements, with the element order and the time step size $\Delta{t}$ varied
systematically in the spatial and temporal tests. The  algorithm presented
in this section is employed to numerically integrate the DRSG equation from $t=0$ to $t=t_f.$ The mapping $\mathscr{F}(R)=R$ and $C_0=1$ are used
in these computations. The initial condition $u_{in}$ and $v_{in}$ are obtained by setting $t=0$ in the analytic expression \eqref{eq:chemostd} and using \eqref{eq:kgsys1}. We then record the numerical errors in different norms by comparing the numerical solution with the analytic solution at $t=t_f.$ 

To conduct the spatial convergence test, we vary systematically the element order from 2 to 20 and depict in Figure \ref{fig:DRSGstandtest}(a)  the $L^{\infty}$ and $L^2$ errors of $u$ as a function of the element order with a fixed $\Delta{t}=0.001$ and $t_f=0.1.$ It is observed that the numerical errors decay exponentially with increasing element order, and levels off beyond element order 12, caused by the saturation of temporal errors.  

To study the temporal convergence rate, we fix the element order at a large value 18 and $t_f=1.0$. The time step size $\Delta{t}$ is varied systematically from 0.2 to $7.8125\times 10^{-4}$ and the numerical errors in $L^{\infty}$ and $L^2$ norms are depicted in Figure \ref{fig:DRSGstandtest}(b). A second-order convergence rate in time is clearly observed. 

\subsubsection{Study of Method Properties}

\begin{figure}[tbp] \centering
  \subfigure[$t=0$]{ \includegraphics[scale=.39]{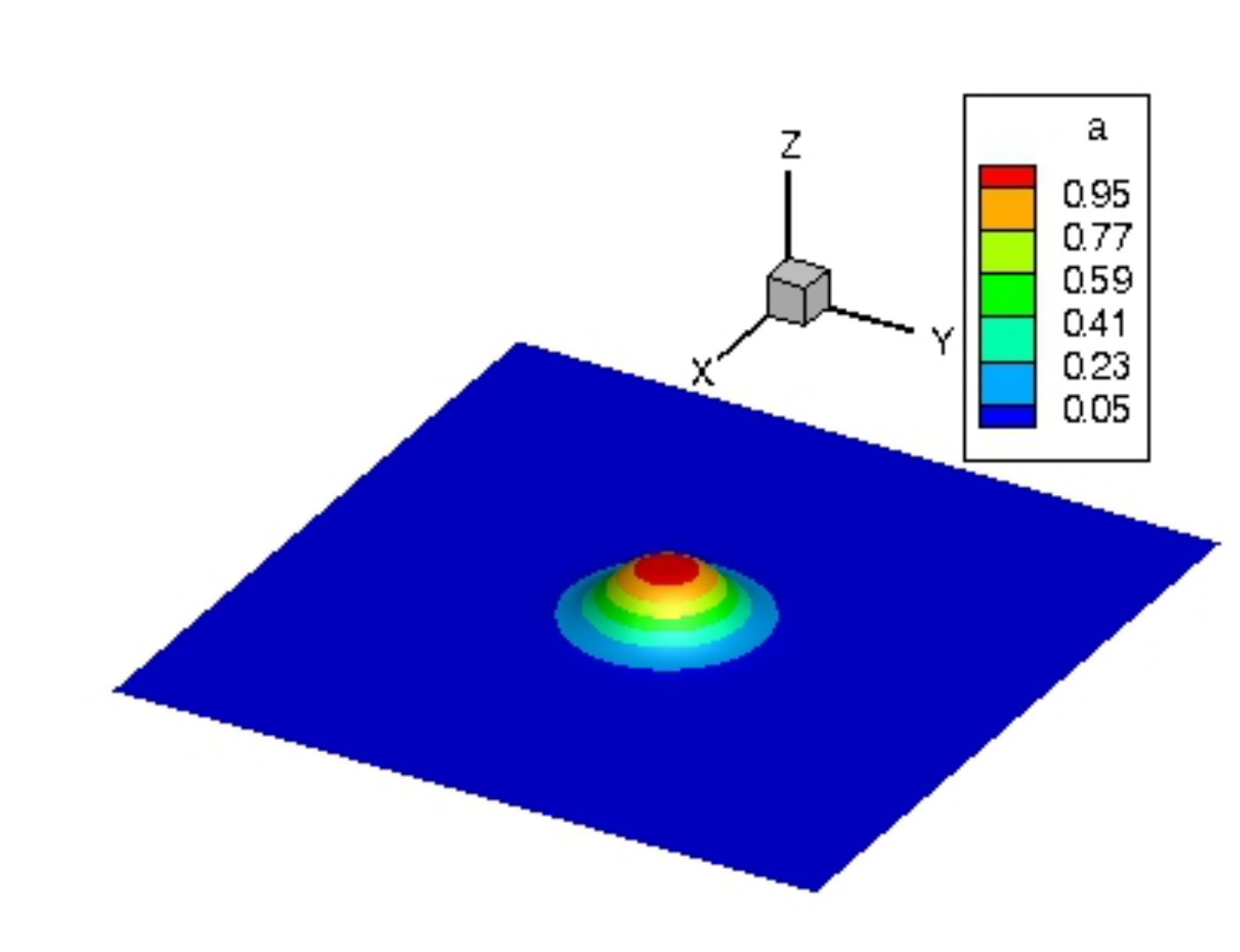}} \quad
  \subfigure[$t=3$]{ \includegraphics[scale=.39]{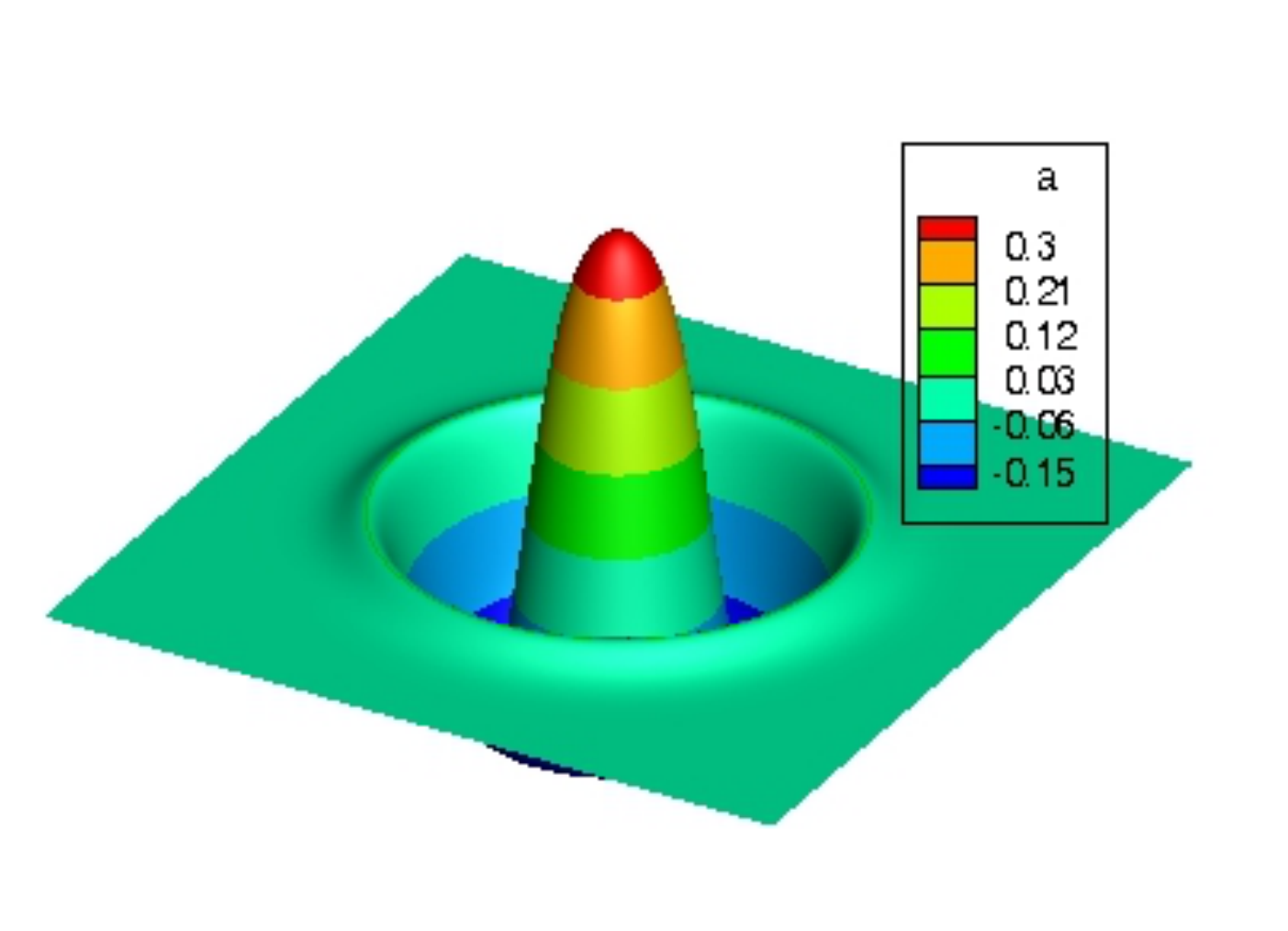}} \\
  \subfigure[$t=5$]{ \includegraphics[scale=.39]{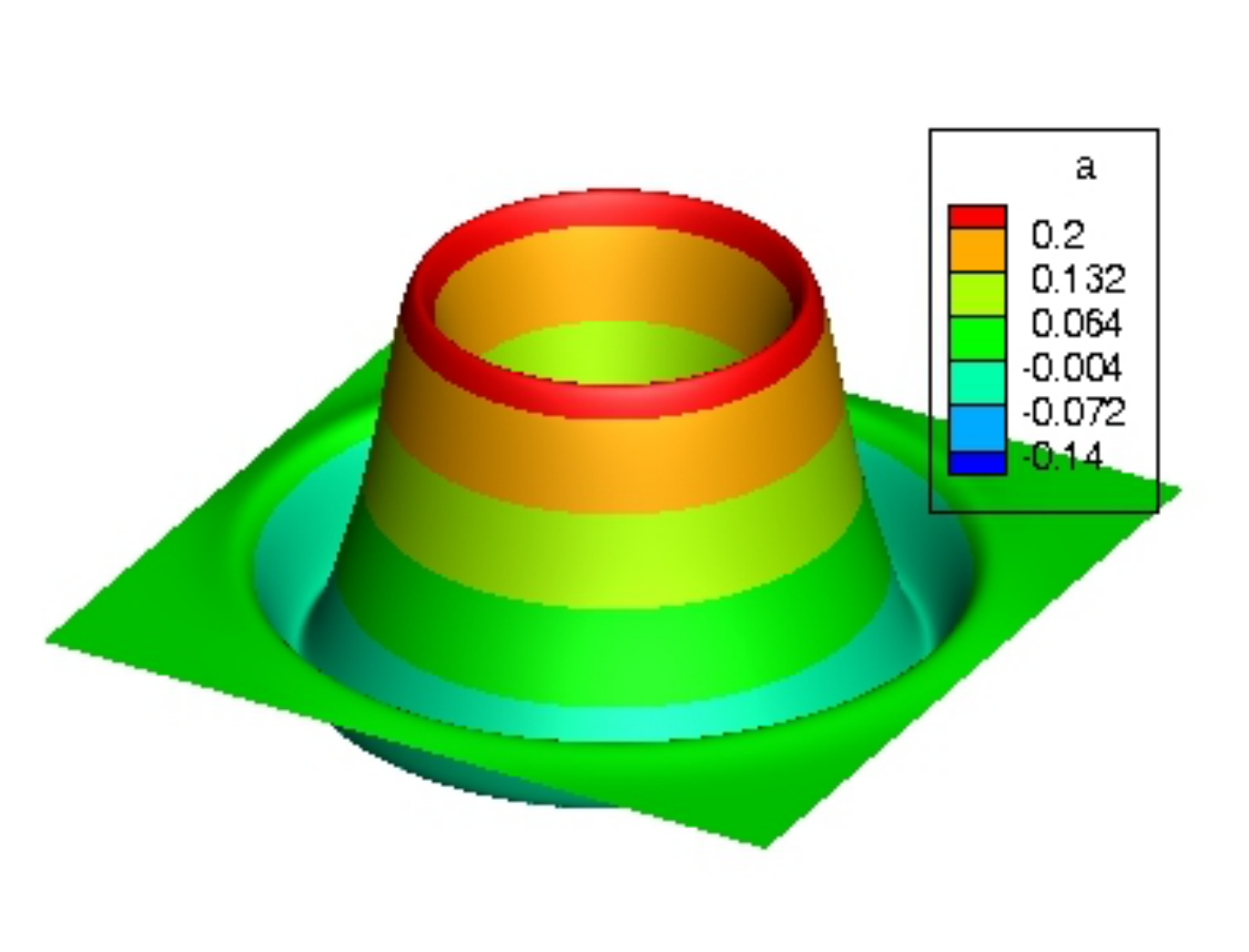}} \quad
  \subfigure[$t=10$]{ \includegraphics[scale=.39]{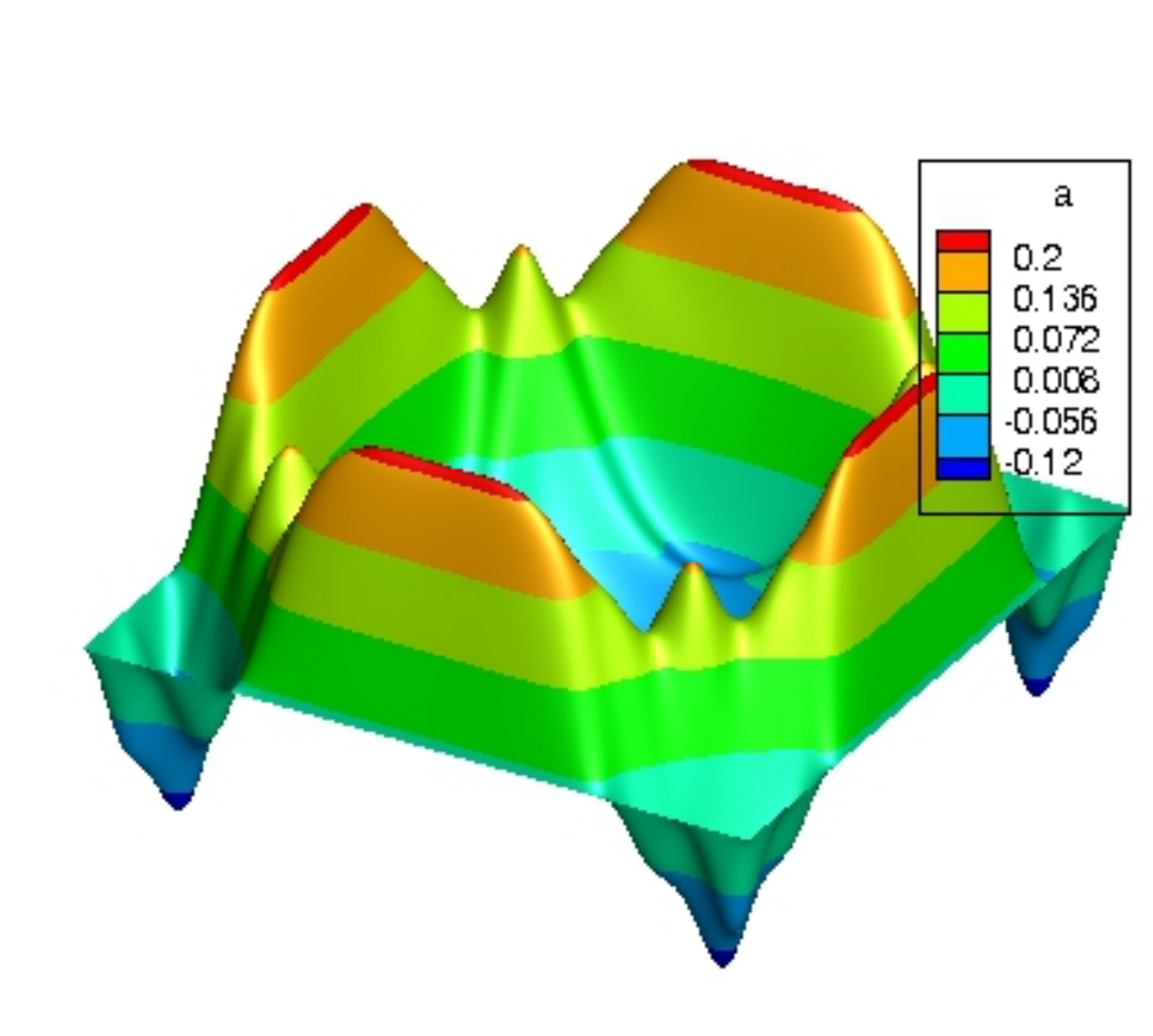}} \\
  \caption{ DRSG equation:
    Temporal sequence of snapshots for $u$ distribution.
    Simulation results are obtained
    with $\Delta t=10^{-4}$, and the mapping
    $\mathscr{F}(R)=\frac{e_0}{2}\ln\left( \frac{\kappa_0+R}{\kappa_0-R}\right)$
    (with $e_0=10$, $\kappa_0=100$).
  }
\label{fig:DRSGsnap}
\end{figure}

\begin{figure}[tbp] \centering
 \subfigure[]{ \includegraphics[scale=.39]{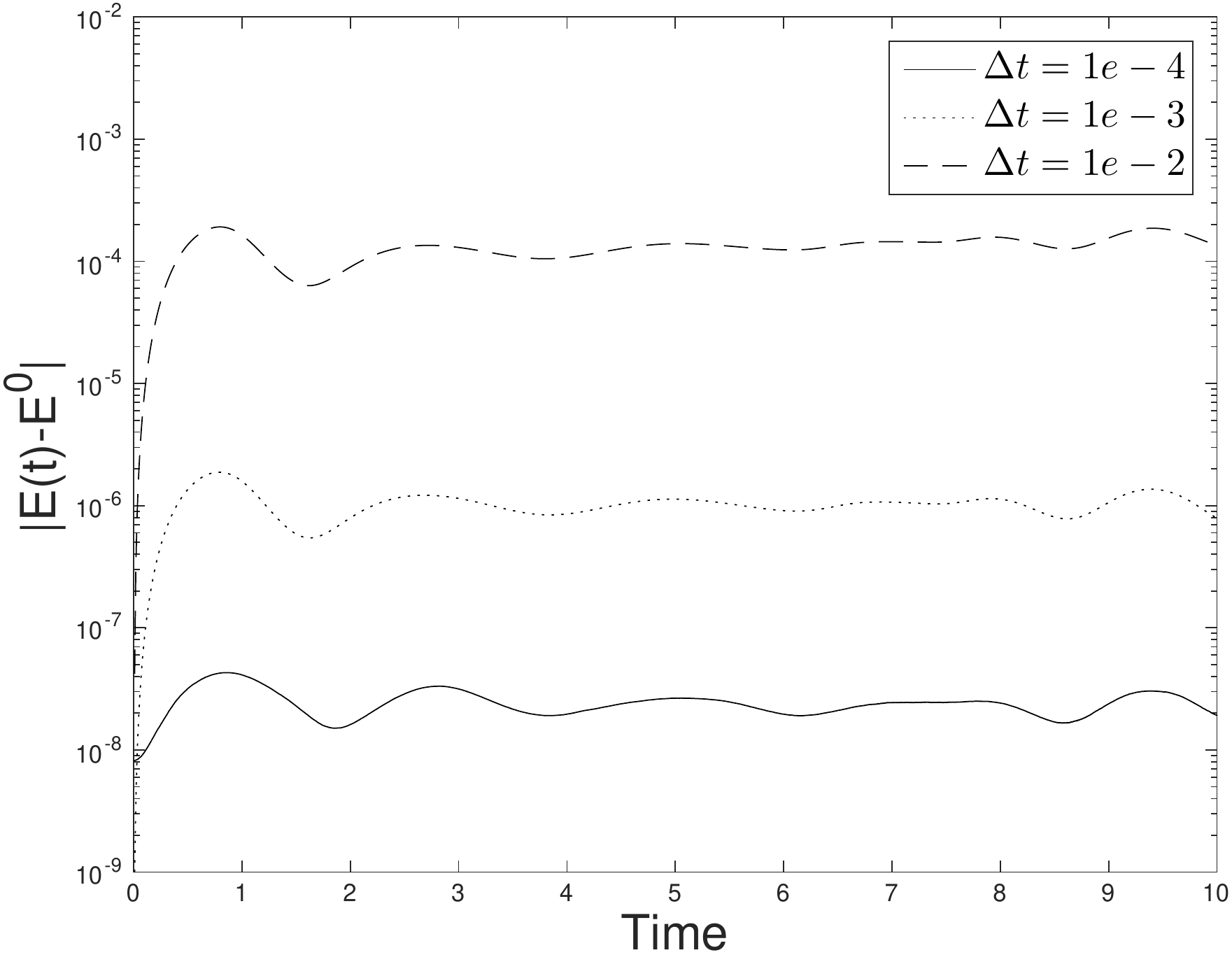}} 
  \subfigure[]{ \includegraphics[scale=.39]{GNKGxismall.pdf}} 
  \caption{DRSG equation: Time histories of (a) $|E(t)-E(0)|$ and (b) $\xi={\mathscr{F}(R)}/{E}$
    obtained with several time step sizes, $\Delta{t}=10^{-4}, 10^{-3}, 10^{-2}$.
    Numerical results correspond to $\mathscr{F}(R)=\frac{e_0}{2}\ln (\frac{\kappa_0+x}{\kappa_0-x})$ ($e_0=10,\, \kappa_0=100$).
  }
\label{fig:DRSGxismall}
\end{figure}

We next study the remarkable stability of the proposed method with
the DRSG equation.
Consider the DRSG equation on the domain $\Omega=[0,14]^2$,
with zero external force $f(\bs x,t)=0$ and zero boundary source term
$d_a(\bs x,t)=0$ in \eqref{eq:kgbd1}. The initial conditions are set to
\begin{equation}
u_{in}(\bs x)=\frac{2}{\exp\big((x-7)^2+(y-7)^2\big)+\exp\big(-(x-7)^2-(y-7)^2\big)},\quad v_{in}(\bs x)=0.
\end{equation}
With these initial and boundary conditions, the DRSG equation is energy conserving.

\begin{figure}[tbp] \centering
 \subfigure[]{ \includegraphics[scale=.39]{GNKGEnlarge.pdf}} 
  \subfigure[]{ \includegraphics[scale=.39]{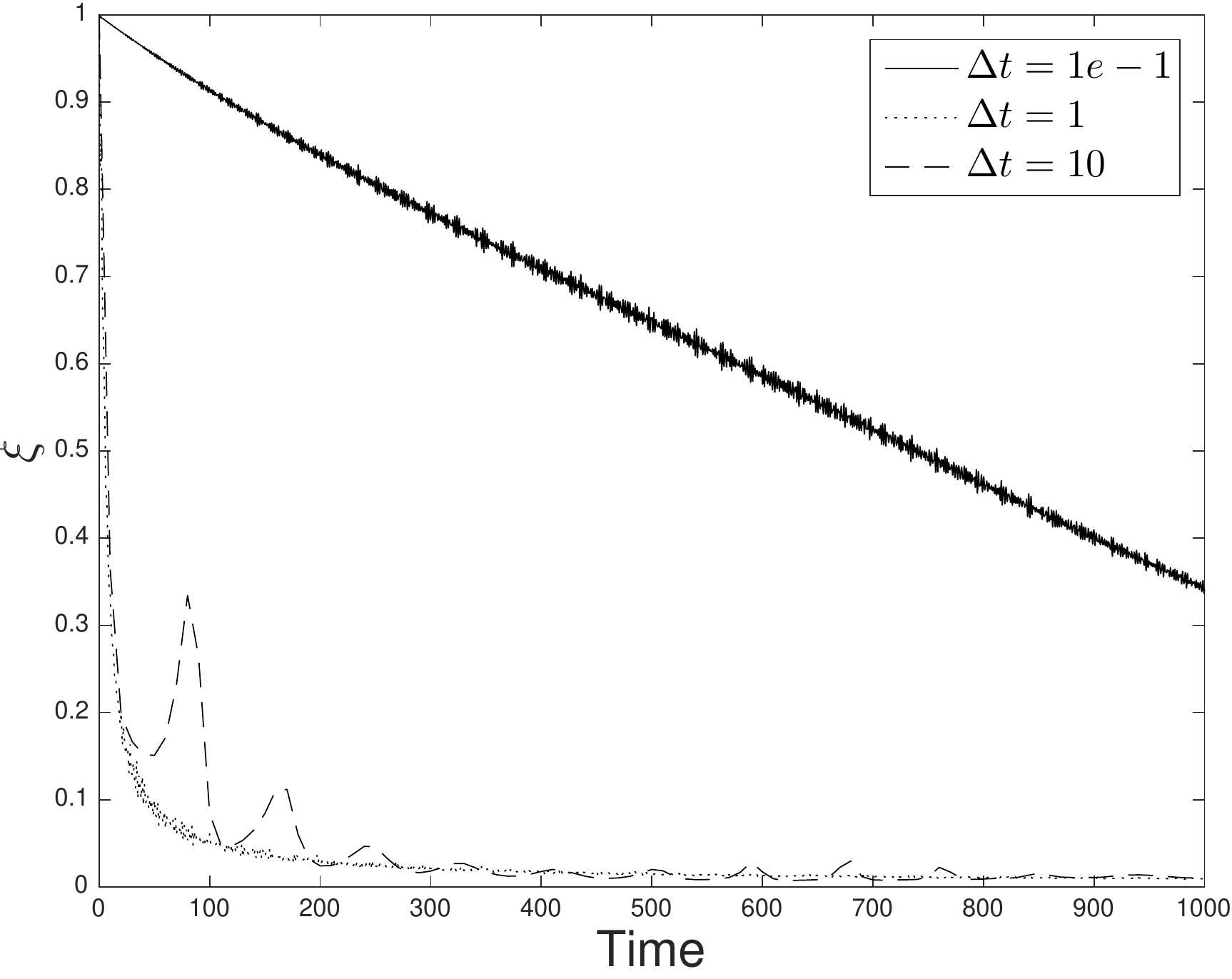}} 
  \caption{Time histories of (a) $E(t)$ and $\mathscr{F}(R)$ and (b) $\xi={\mathscr{F}(R)}/{E}$ versus large $\Delta{t}=0.1, 1, 10$ for DRSG equation.
    The numerical results are obtained with $\mathscr{F}(R)=\frac{e_0}{2}\ln (\frac{\kappa_0+x}{\kappa_0-x})$ ($e_0=10,\, \kappa_0=100$).
  }
\label{fig:DRSGxilarge}
\end{figure}

The domain $\Omega$ is discretized with 400 equal-sized quadrilateral elements with a fixed element order 10. We employ a mapping function
$\mathscr{F}(R)=\frac{e_0}{2}\ln (\frac{\kappa_0+R}{\kappa_0-R})$ ($e_0=10,\, \kappa_0=100$)
and the energy constant $C_0=1$ in the algorithm.
Figure \ref{fig:DRSGsnap} illustrates the evolution
of  $u$ by a sequence of snapshots of its contour levels.
One can observe a circular wave pattern starting from the center of the domain
and propagating outward toward the boundaries.
As the wave reaches the boundaries, the interaction with
the Dirichlet boundary ($u=0$) gives rise to an extremely complicated
wave pattern; see Figure \ref{fig:DRSGsnap}(d).

Figure \ref{fig:DRSGxismall}(a) shows the time histories of
the energy errors, $|E(t)-E(0)|$, obtained using several time step
sizes ($\Delta t=10^{-4}$, $10^{-3}$ and $10^{-2}$).
One can observe oscillations in the history curves about their respective mean
values that are consistent with
a second order accuracy in time.
It should again be noted that the current algorithm conserves
the modified energy $\mathscr{F}(R)$ discretely, not
the original energy $E(t)$. 
Figure \ref{fig:DRSGxismall}(b) shows
time histories of the ratio $\xi=\frac{\mathscr{F}(R)}{E}$
corresponding to these $\Delta t$ values.
The computed $\xi$ values are essentially $1$,
indicative of the accuracy of these simulations.

We then increase the time step size to $\Delta{t}=0.1, 1$ and  $10$,
and depict in Figure \ref{fig:DRSGxilarge}(a)
the time histories of $E(t)$ and $\mathscr{F}(R)$ for a
long time simulation to $t=1000$. 
Large discrepancies between the energy $E(t)$ and $\mathscr{F}(R)$ can be observed,
especially for $\Delta{t}=1$ and $10$, suggesting that $\mathscr{F}(R)$ no longer approximates
well the energy $E(t)$ with these time step sizes.
Note that the $\mathscr{F}(R)$ histories obtained by different large $\Delta t$ values
overlap with one another. This is consistent with Theorem \ref{thm:kgeq}
that the current scheme conserves the modified energy $\mathscr{F}(R)$.
It can be observed from Figure \ref{fig:DRSGxilarge}(b) that
the computed $\xi=\frac{\mathscr{F}(R)}{E}$ becomes significantly
smaller than 1, indicative of large errors in the simulations
with these large time step sizes.
However, the computations are evidently stable, 
even with these large $\Delta t$ values.

\section{Concluding Remarks}
\label{sec:summary}


In this paper
we have presented a framework (gPAV)  for developing unconditionally
energy-stable schemes for general dissipative systems.
The scheme is based on a generalized auxiliary variable (which is
a scalar number)
associated with the  energy functional of the system.
We find that the square root function, which is critical to previous
auxiliary-variable approaches,
is not essential to
devising energy-stable schemes.
In the current method,
the auxiliary
variable can be defined by a rather general class of functions,
not limited to the square-root function.
The gPAV method is applicable to general dissipative systems,
and a unified procedure for discretely treating the dissipative governing equations
and the generalized auxiliary variable 
has been presented.
The discrete energy stability of the proposed scheme
has been proven for general dissipative systems. 
%
The presented method has two attractive properties:
\begin{itemize}

\item
  The scheme requires only the solution of linear algebraic equations
  within a time step, and no nonlinear solver is needed.
  Furthermore,
  with appropriate choice of the $\bs F_L$ operator in the algorithm,
  the resultant linear algebraic systems upon discretization
  involve only constant and
  time-independent coefficient matrices, which only need to be
  computed once and can be pre-computed.
  In terms of computational cost,
  the scheme is computationally very competitive and attractive.

\item
  The generalized auxiliary variable can be computed directly
  by a well-defined explicit
  formula. The computed values for the auxiliary variable are
  guaranteed to be positive, regardless of the time step size
  or the external forces or source terms.

\end{itemize}

Three specific dissipative systems (a chemo-repulsion model,
Cahn-Hilliard equation with constant and variable mobility, and
the nonlinear Klein-Gordon equation) have been studied in relative detail
to demonstrate the gPAV framework developed herein.
Ample numerical experiments have been presented for each system
to demonstrate the performance of the method,
the effects of algorithmic parameters, and
the stability of the scheme with large time step sizes.

All physically meaningful systems in the real world are
energy dissipative (or conserving) due to the second law of
thermodynamics, and these systems are typically nonlinear.
The design of energy-stable and computationally-efficient
schemes for such systems is critical to their numerical
simulations, and this is in general a very challenging task.
The gPAV framework presented here
lays out a roadmap for devising discretely energy-stable
schemes for general dissipative systems.
The computational efficiency (e.g.~involving linear
equations with pre-computable coefficient matrices)
and the guaranteed positivity of the
computed auxiliary variable of the  method
are particularly attractive, in the sense that the gPAV method is not only
unconditionally energy-stable but also can be computationally
efficient and competitive.
We anticipate that the gPAV method will be useful and
instrumental in numerical simulations of a number of
computational science and engineering disciplines.


\section*{Acknowledgement}
This work was partially supported by
NSF (DMS-1522537). 

\section*{Appendix A. Approximation for the First Time Step}

We present a method on how to deal with the first time step
such that the approximation for the auxiliary variable
$R(t)$ at time step $\frac{1}{2}$ shall be positive. We consider below
only the  formulation based on $\frac{\mathscr{F}(R)}{E}$.
It is noted that for the alternative formulation
based on $\frac{R}{\mathscr{G}(E)}$ (see Section \ref{sec:alter}) one can modify
the following scheme in a straightforward fashion to
achieve the same property.
The notations here follow those employed in the main text.

Consider the system consisting of equations \eqref{eq:generaleq1},
\eqref{eq:generaleq2}, the boundary condition \eqref{eq:bc},
and the initial conditions \eqref{eq:ic} and \eqref{eq:ic_R}.
Define
\begin{equation}
  \left\{
  \begin{split}
    &
    \bs u^0 = \bs u_{in}(\bs x), \\
    &
    R^0 = \mathscr{G}(E^0), \quad
    \text{with} \ \ E^0 = \int_{\Omega}e(\bs u_{in})d\Omega + C_0.
  \end{split}
  \right.
\end{equation}
One notes that $E^0>0$ and $R^0>0$.

We compute the first time step in two substeps. In substep one
we compute an approximation of ($\bs u^1,R^1$), denoted by
($\bs u_a^1, R_a^1$),
and
in substep two we compute the final ($\bs u^1,R^1$).
More specifically, the scheme is as follows: \\
\noindent\underline{Substep One:}
\begin{subequations}
  \begin{align}
    &
    \frac{\bs u_a^1 - \bs u^0}{\Delta t} =
    \bs F_L(\bs u_a^1) +
    \xi_a\left[ \bs F(\bs u^0) - \bs F_L(\bs u^0)  \right]
    + \bs f^1, \label{eq:1st_s1_1} \\
    &
    \xi_a = \frac{\mathscr{F}(R_a^1)}{E[\tilde{\bs u}_a^1]}, \label{eq:1st_s1_2} \\
    &
    E[\tilde{\bs u}^1_a] = \int_{\Omega} e(\tilde{\bs u}^1_a) d\Omega + C_0, \label{eq:1st_s1_3} \\
    &
    \bs B(\bs u_a^1) = \bs f_b^1, \quad \text{on} \ \Gamma, \label{eq:1st_s1_4}
  \end{align}
  \begin{equation}\label{eq:1st_s1_5}
    \begin{split}
      \frac{\mathscr{F}(R_a^1)-\mathscr{F}(R^0)}{R_a^1 - R^0}
      &\frac{R_a^1 - R^0}{\Delta t} =
      \int_{\Omega} e'(\bs u_a^1)\cdot \frac{\bs u_a^1 - \bs u^0}{\Delta t} d\Omega \\
      &
      -  \int_{\Omega} e'(\bs u_a^1)\cdot\left(
      \bs F_L(\bs u_a^1) + \xi_a\left[\bs F(\bs u^0) - \bs F_L(\bs u^0)  \right]
      + \bs f^1
      \right) d\Omega \\
      &
      + \xi_a \left[
        -\int_{\Omega} V(\tilde{\bs u}_a^1)d\Omega
        + \int_{\Omega} V_s(\bs f^1, \tilde{\bs u}_a^1)d\Omega
        + \int_{\Gamma} B_s(\bs f_b^1, \tilde{\bs u}_a^1)d\Gamma
        \right] \\
      &
      + (1-\xi_a)\left|
      \int_{\Omega} V_s(\bs f^1, \tilde{\bs u}_a^1)d\Omega
      + \int_{\Gamma} B_s(\bs f_b^1, \tilde{\bs u}_a^1)d\Gamma
      \right|.
    \end{split}
  \end{equation}
\end{subequations}
\noindent\underline{Substep Two:}
\begin{subequations}
  \begin{align}
    &
    \frac{\bs u^1 - \bs u^0}{\Delta t} =
    \bs F_L(\bs u^1) +
    \xi\left[ \bs F(\bs u^0) - \bs F_L(\bs u^0)  \right]
    + \bs f^1, \label{eq:1st_s2_1} \\
    &
    \xi = \frac{\mathscr{F}(R^{3/2})}{E[\tilde{\bs u}^{3/2}]},\label{eq:1st_s2_2} \\
    &
    E[\tilde{\bs u}^{3/2}] = \int_{\Omega} e(\tilde{\bs u}^{3/2}) d\Omega + C_0,
    \label{eq:1st_s2_3} \\
    &
    \bs B(\bs u^1) = \bs f_b^1, \quad \text{on} \ \Gamma, \label{eq:1st_s2_4}
  \end{align}
  \begin{equation} \label{eq:1st_s2_5}
    \begin{split}
      \frac{\mathscr{F}(R^{3/2})-\mathscr{F}(R^{1/2})}{R^{3/2} - R^{1/2}}
      &\frac{R^{3/2} - R^{1/2}}{\Delta t} =
      \int_{\Omega} e'(\bs u^1)\cdot \frac{\bs u^1 - \bs u^0}{\Delta t} d\Omega \\
      &
      -  \int_{\Omega} e'(\bs u^1)\cdot\left(
      \bs F_L(\bs u^1) + \xi\left[\bs F(\bs u^0) - \bs F_L(\bs u^0)  \right]
      + \bs f^1
      \right) d\Omega \\
      &
      + \xi \left[
        -\int_{\Omega} V(\tilde{\bs u}^1)d\Omega
        + \int_{\Omega} V_s(\bs f^1, \tilde{\bs u}^1)d\Omega
        + \int_{\Gamma} B_s(\bs f_b^1, \tilde{\bs u}^1)d\Gamma
        \right] \\
      &
      + (1-\xi)\left|
      \int_{\Omega} V_s(\bs f^1, \tilde{\bs u}^1)d\Omega
      + \int_{\Gamma} B_s(\bs f_b^1, \tilde{\bs u}^1)d\Gamma
      \right|.
    \end{split}
  \end{equation}
\end{subequations}

Note that in the above equations the superscript of a variable
such as $(\cdot)^{1/2}$ and $(\cdot)^{3/2}$ denotes
the time step index.
In \eqref{eq:1st_s1_2} and \eqref{eq:1st_s1_5} $\tilde{\bs u}_a^1$ is an approximation of
$\bs u_a^1$ and will be specified later in \eqref{eq:1st_defutilde}.
In \eqref{eq:1st_s2_5} $\tilde{\bs u}^1$ is an approximation of $\bs u^1$
and will be specified later also in \eqref{eq:1st_defutilde}.
In \eqref{eq:1st_s2_2}, \eqref{eq:1st_s2_3} and \eqref{eq:1st_s2_5},
$\tilde{\bs u}^{3/2}$, $R^{1/2}$ and $R^{3/2}$ are defined by
\begin{equation}\label{eq:1st_def32}
  \left\{
  \begin{split}
    &
    \tilde{\bs u}^{3/2} = \frac{3}{2}\bs u_a^1 - \frac{1}{2}\bs u^0, \\
    &
    R^{3/2} = \frac{3}{2} R^1 - \frac{1}{2}R^0, \\
    &
    R^{1/2} = \frac{1}{2}\left(R_a^1 + R^0 \right).
  \end{split}
  \right.
\end{equation}
It can be noted that the above scheme represents
a first-order approximation of ($\bs u^1,R^1$) for
the first time step.

Combine equations \eqref{eq:1st_s1_1} and \eqref{eq:1st_s1_5}
and we have
\begin{equation}
  \mathscr{F}(R_a^1) - \mathscr{F}(R^0)
  = -\xi_a\Delta t\int_{\Omega} V(\tilde{\bs u}_a^1)d\Omega
  - \xi_a\Delta t(|S_a| - S_a) +|S_a|\Delta t
\end{equation}
where
$
S_a = \int_{\Omega} V_s(\bs f^1, \tilde{\bs u}_a^1)d\Omega
+ \int_{\Gamma} B_s(\bs f_b^1, \tilde{\bs u}_a^1)d\Gamma.
$
In light of \eqref{eq:1st_s1_2}, this leads to
\begin{equation}\label{eq:def_xia}
  \left\{
  \begin{split}
    &
    \xi_a = \frac{\mathscr{F}(R^0) + |S_a|\Delta t}
       {E[\tilde{\bs u}_a^1] + \Delta t\int_{\Omega} V(\tilde{\bs u}_a^1)d\Omega
         + (|S_a| - S_a)\Delta t }, \\
    &
    R_a^1 = \mathscr{G}(\xi_aE[\tilde{\bs u}_a^1]).   
  \end{split}
  \right.
\end{equation}
Since $R^0>0$, we conclude that $\xi_a>0$ and $R_a^1>0$ based on
these equations.
It follows that $R^{1/2} = \frac{1}{2}(R_a^1 + R^0)>0$
in light of equation \eqref{eq:1st_def32}.

Similarly, combining equations \eqref{eq:1st_s2_1} and \eqref{eq:1st_s2_5}
gives rise to
\begin{equation}
  \mathscr{F}(R^{3/2}) - \mathscr{F}(R^{1/2})
  = -\xi\Delta t\int_{\Omega} V(\tilde{\bs u}^1)d\Omega
  - \xi\Delta t(|S_0| - S_0) +|S_0|\Delta t
\end{equation}
where
$
S_0 = \int_{\Omega} V_s(\bs f^1, \tilde{\bs u}^1)d\Omega
+ \int_{\Gamma} B_s(\bs f_b^1, \tilde{\bs u}^1)d\Gamma.
$
In light of \eqref{eq:1st_s2_2} and \eqref{eq:1st_def32}, we have
\begin{equation}\label{eq:def_xi1st}
  \left\{
  \begin{split}
    &
    \xi = \frac{\mathscr{F}(R^{1/2}) + |S_0|\Delta t}
       {E[\tilde{\bs u}^{3/2}] + \Delta t\int_{\Omega} V(\tilde{\bs u}^1)d\Omega
         + (|S_0| - S_0)\Delta t }, \\
    &
    R^{3/2} = \mathscr{G}(\xi E[\tilde{\bs u}^{3/2}]), \\
    &
    R^1 = \frac{2}{3}R^{3/2} + \frac{1}{3}R^{0}.
  \end{split}
  \right.
\end{equation}
We therefore conclude that $\xi>0$, $R^{3/2}>0$
and $R^1>0$.

We still need to determine $\bs u_a^1$ and $\bs u^1$,
and specify $\tilde{\bs u}_a^1$ and $\tilde{\bs u}^1$.
Note that $\bs F_L(\bs u)$ and $\bs B(\bs u)$ are linear operators.
Equations \eqref{eq:1st_s1_1} and \eqref{eq:1st_s1_4},
and also equations \eqref{eq:1st_s2_1} and \eqref{eq:1st_s2_4},
can be solved as follows.
Define two variables $\bs u_1^1$ and $\bs u_2^1$ as solutions
to the following systems, respectively: \\
\noindent\underline{For $\bs u_1^1$:}
\begin{subequations}
  \begin{align}
    &
    \frac{1}{\Delta t}\bs u_1^1 - \bs F_L(\bs u_1^1) =
    \frac{\bs u^0}{\Delta t} + \bs f^1,
    \label{eq:1st_u1_1} \\
    &
    \bs B(\bs u_1^1) = \bs f_b^1, \quad \text{on} \ \ \Gamma.
    \label{eq:1st_u1_2}
  \end{align}
\end{subequations}
\noindent\underline{For $\bs u_2^1$:}
\begin{subequations}
  \begin{align}
    &
    \frac{1}{\Delta t}\bs u_2^1 - \bs F_L(\bs u_2^1) =
    \bs F(\bs u^0) - \bs F_L(\bs u^0),
    \label{eq:1st_u2_1} \\
    &
    \bs B(\bs u_2^1) = 0, \quad \text{on} \ \ \Gamma.
    \label{eq:1st_u2_2}
  \end{align}
\end{subequations}
Then it is straightforward to verify that,
for given $\xi_a$ and $\xi$, the following functions respectively
solve the equations \eqref{eq:1st_s1_1} and \eqref{eq:1st_s1_4},
and equations \eqref{eq:1st_s2_1} and \eqref{eq:1st_s2_4},
\begin{subequations}
  \begin{align}
    &
    \bs u_a^1 = \bs u_1^1 + \xi_a \bs u_2^1,
    \label{eq:1st_ua1} \\
    &
    \bs u^1 = \bs u_1^1 + \xi \bs u_2^1.
    \label{eq:1st_u1}
  \end{align}
\end{subequations}
We then specify $\tilde{\bs u}_a^1$ and $\tilde{\bs u}^1$ as follows,
\begin{align}
  &
  \tilde{\bs u}_a^1 = \tilde{\bs u}^1 = \bs u_1^1 + \bs u_2^1.
  \label{eq:1st_defutilde}
\end{align}

The solution for ($\bs u^1, R^1$) at the first time step consists of the
following procedure:
\begin{itemize}

\item
  Solve equations \eqref{eq:1st_u1_1}--\eqref{eq:1st_u1_2} for $\bs u_1^1$; \\
  Solve equations \eqref{eq:1st_u2_1}--\eqref{eq:1st_u2_2} for $\bs u_2^1$.

\item
  Compute $\tilde{\bs u}_a^1$ and $\tilde{\bs u}^1$ by
  equation \eqref{eq:1st_defutilde}; \\
  Compute $\xi_a$ and $R_a^1$ by equation \eqref{eq:def_xia}; \\
  Compute $\bs u_a^1$ by equation \eqref{eq:1st_ua1}.

\item
  Compute $\tilde{\bs u}^{3/2}$ and $R^{1/2}$ based on equation \eqref{eq:1st_def32}; \\
  Compute $\xi$ and $R^1$ based on equation \eqref{eq:def_xi1st}; \\
  Compute $\bs u^1$ by equation \eqref{eq:1st_u1}.
  
\end{itemize}
We can make the following conclusion based on the above discussions.
\begin{theorem}\label{thm:1st_step}

  The scheme represented by \eqref{eq:1st_s1_1}--\eqref{eq:1st_s2_5}
  for computing the first time step
  has the property that
  \begin{equation}
    R^1>0, \quad
    R^{1/2} > 0, \quad \text{and} \ \
    R^{3/2} > 0,
  \end{equation}
  where $R^{1/2}$ and $R^{3/2}$ are given by \eqref{eq:1st_def32},
  regardless of the time step size $\Delta t$
  and the external forces $\bs f$ and $\bs f_b$.
  
\end{theorem}

\bibliographystyle{plain}
\bibliography{adjstab,sem,mypub,interface,multiphase,engstab,contact_line}

\end{document}